\documentclass[journal]{vgtc}                     

\onlineid{1385}

\vgtccategory{Research}

\title{LLMs have Visualization Literacy: Now What? Experiments Exploring LLM Visualization Evaluation Capabilities}

\author{%
  \authororcid{Christian Seto}{0009-0009-0808-2470},
  \authororcid{Jacqueline Nguyen}{0009-0003-0479-1188},
  \authororcid{Jiayi Hong}{0000-0002-1332-5045}, and 
  \authororcid{Ross Maciejewski}{0000-0001-8803-6355}
}

\authorfooter{
  \item
  	Christian Seto, Jacqueline Nguyen, and Ross Maciejewski are with Arizona State University, USA.
  	E-mail: {\{cseto\,$|$\,jnguy201\,$|$\,rmacieje\}}@asu.edu.
  \item
  	Jiayi Hong is with Université Laval, CA.
  	E-mail: jiayi.hong@ift.ulaval.ca
}

\abstract{%
  As Large Language Models (LLMs) become more popular within the visualization community, researchers increasingly leverage them for diverse visualization tasks such as design guideline suggestions and visualization evaluation. However, in order for LLMs to act as trustworthy and fair evaluators, we argue that LLMs would need to possess visualization literacy, be capable of following user instructions and uphold graphical integrity. We test the latest versions of the most prominent LLMs, specifically Anthropic's Claude (Opus 4.5, \claudeFig), OpenAI's Generative Pretrained Transformers (GPT 5.2 Pro, \gptFig), and Google's Gemini (Gemini 3 Flash, \geminiFig) on these features and find that while these models now possess visualization literacy, they still struggle with other features necessary for instruction following and graphical integrity. Using a modified Visualization Literacy Assessment Test (VLAT), our findings show that these recent LLMs have achieved greater than human-levels of visualization literacy in contrast to prior research. In order to test the models' abilities to follow instructions, we used few-shot and chain-of-thought prompting as proxies for instruction following tasks on evaluating visualization literacy and find that these specialized prompting techniques are becoming obsolete with respect to improving visualization literacy. Additionally, we experiment with the inherent ability of LLMs to evaluate misleading visualizations to test the models' abilities for upholding graphical integrity and find that without specialized or leading prompting techniques, the models struggle with being able to accurately identify whether a visualization is misleading or not. Our results further break down the performance of each model on these tasks, but the culmination of our findings force us to reconsider the current effectiveness of LLMs as visualization evaluators. 
}


\graphicspath{{figs/}{figures/}{pictures/}{images/}{./}} 

\usepackage{tabu}                      
\usepackage{booktabs}                  
\usepackage{lipsum}                    
\usepackage{mwe}                       
\usepackage{ccicons}                   

\usepackage{mathptmx}                  

\usepackage{xcolor}
\usepackage{soul}
\usepackage{multirow}
\usepackage{amsmath}
\usepackage{makecell}
\usepackage[table]{xcolor}

\usepackage{tabularray}
\UseTblrLibrary{siunitx}

\usepackage{siunitx}
\usepackage{adjustbox}
\usepackage{rotating}
\newcommand{\tablefontsize}{\fontsize{7}{7}\selectfont}
\AtBeginEnvironment{tabu}{\tablefontsize}

\newcommand{\eg}{e.\,g.}
\newcommand{\ie}{i.\,e.}

\definecolor{claudeColor}{HTML}{DE7356}
\definecolor{geminiColor}{HTML}{4796E3}
\definecolor{gptColor}{HTML}{74AA9C}

\definecolor{better}{HTML}{43aa8b}
\definecolor{close}{HTML}{f9c74f}
\definecolor{worse}{HTML}{f94144}

\definecolor{manuScale}{HTML}{a0daa9}
\definecolor{manuData}{HTML}{ffd580}
\definecolor{manuAnnotate}{HTML}{779ecb}
\definecolor{manuVisEncode}{HTML}{995c00}

\newcommand{\claude}{\textcolor{white}{\fontsize{8}{0}\selectfont\sethlcolor{claudeColor}\hl{Claude-4.5}}}
\newcommand{\gemini}{\textcolor{white}{\fontsize{8}{0}\selectfont\sethlcolor{geminiColor}\hl{Gemini-3}}}
\newcommand{\gpt}{\textcolor{white}{\fontsize{8}{0}\selectfont\sethlcolor{gptColor}\hl{GPT-5.2}}}

\newcommand{\claudeFig}{\textcolor{white}{\fontsize{7}{0}\selectfont\sethlcolor{claudeColor}\hl{Claude-4.5}}}
\newcommand{\geminiFig}{\textcolor{white}{\fontsize{7}{0}\selectfont\sethlcolor{geminiColor}\hl{Gemini-3}}}
\newcommand{\gptFig}{\textcolor{white}{\fontsize{7}{0}\selectfont\sethlcolor{gptColor}\hl{GPT-5.2}}}

\newcommand{\claudeTab}{\textcolor{white}{\fontsize{7}{0}\selectfont\sethlcolor{claudeColor}\hl{Claude}}}
\newcommand{\geminiTab}{\textcolor{white}{\fontsize{7}{0}\selectfont\sethlcolor{geminiColor}\hl{Gemini}}}
\newcommand{\gptTab}{\textcolor{white}{\fontsize{7}{0}\selectfont\sethlcolor{gptColor}\hl{GPT}}}

\newcommand{\gptFour}{\inlinetext{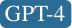}}
\newcommand{\geminiOne}{\inlinetext{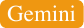}}

\newcommand{\better}[1]{\textbf{\textcolor{better}{#1}}}
\newcommand{\close}[1]{\textcolor{close}{#1}}
\newcommand{\worse}[1]{\textcolor{worse}{#1}}

\newcommand{\inlinetext}[1]{\raisebox{0pt}[0pt][0pt]{\raisebox{-0.6ex}{\includegraphics[height=2.4ex]{#1}}}}

\begin{document}


\firstsection{Introduction}

\maketitle

The use of Large Language Models (LLMs) within the visualization community has become more popular as models become more advanced and better at aiding with visualization tasks \cite{brossier:2026:sota}. As a result, we are seeing the emergence of LLMs as chatbot assistants in visual analytic systems (VA systems) \cite{Li:2025:RDS, Zhao:2025:lightva, guo:2025:medVRA, zhao:2025:ProactiveVA, goswami:2025:plotgen} where LLMs are serving as designers~\cite{zhu:2026:paperbanana}, interpreters~\cite{han:2023:chartllama}, and evaluators~\cite{xie:2025:visjudge,aubinLeQuere:2024:llms} of data visualizations. 
In this paper, we focus on LLMs as visualization evaluators. Visualization evaluation occurs at different stages in the design process\cite{lam:2012:empirical} and even varies across different visualization sub-communities\cite{isenberg:2013:systematic}. LLMs as visualization evaluators would be useful for visualization sanity checks, which would provide feedback on visualization readability~\cite{cabouat:2024:PPA}. Typically, evaluating visualizations requires human labor, which incurs high costs. If LLMs could support visualization experts in evaluating visual representations and visual systems to a similar level of proficiency as the general public, they could offer significant cost savings\cite{hamalainen:2023:evaluating}. However, what are the fundamental skills needed to evaluate visualizations? We argue that three skills are necessary for visualization evaluation: visualization literacy, instruction-following, and graphical integrity. While literacy\cite{lee:2017:vlat} serves as the fundamental baseline for chart interpretation, instruction-following ensures that models produce consistent, prompt-aligned outputs across diverse evaluation contexts. Finally, graphical integrity\cite{tufte:1998:visual} requires models to serve as rigorous auditors capable of identifying both intentional `visual lies' and inadvertent design confusions. As design confusions are often subjective and context-dependent, we focus exclusively on misleading chart elements\cite{Zhang:2023:GVI,Mahbub:2025:PCD,Chen:2025:UDV} to establish a baseline of objective graphical integrity. To validate these competencies in current models, we systematically address the following research questions:

\begin{itemize}
    \item \textbf{RQ1}--Visualization Literacy: To what extent do the latest LLMs have visualization literacy?
    \item \textbf{RQ2}--Instruction-following: To what extent does prompt engineering influence the visualization literacy of LLMs?
    \item \textbf{RQ3}--Graphical Integrity: To what extent can LLMs identify misleading chart elements?
\end{itemize}

For visualization literacy, standardized benchmarks, such as the Visualization Literacy Assessment Test (VLAT)~\cite{lee:2017:vlat} and its abridged counterpart, the Mini-VLAT~\cite{Pandey:2023:MSE}, were originally developed to quantify human proficiency in interpreting visual representations. While initial evaluations suggested that LLMs struggled significantly with these human-centric tasks~\cite{Hong:2025:LHV, bendeck:2025:empirical, bendeck:2025:literate, Khan:2025:evaluate, li:2024:vizLit}, the rapid iteration of models necessitates constant re-assessment. Thus, we address \textbf{RQ1} by conducting a comprehensive re-evaluation of the latest frontier models, \ie, \claude\footnote{https://www.anthropic.com/news/claude-opus-4-5}, \gpt\footnote{https://openai.com/index/introducing-gpt-5-2/}, and \gemini\footnote{https://docs.cloud.google.com/vertex-ai/generative-ai/docs/models/gemini/3-flash}, to determine if recent technical advancements have translated into measurable improvements in models' visualization literacy. Our findings indicate that the most recent generation of LLMs demonstrates an incredible level of visualization literacy, marking a significant shift from previous state-of-the-art performance. For \textbf{RQ2}, we adopt the base prompt from Hong et al.~\cite{Hong:2025:LHV} to evaluate instruction-following through the lens of prompt engineering. We employ two widely utilized strategies, few-shot learning~\cite{brown:2020:fewshot} and Chain-of-Thought (CoT) prompting~\cite{wei:2023:cot}, as proxies to measure the models' ability to follow complex task constraints during extended visualization literacy experiments. Our results showed that these specialized prompting techniques did not significantly improve the literacy of the models we used. Finally, to address \textbf{RQ3}, we evaluated the models' proficiency in identifying deceptive visual elements using the Misleading ChartQA benchmark~\cite{Chen:2025:UDV}. The models were tasked with a two-stage classification and extraction process: detecting the presence (and not hallucinating in the absence) of misleading elements and providing a comprehensive list of the specific misleading types identified within the visualization. While \gpt\ and \gemini\ performed on par with each other, we surprisingly found that \claude\ outperformed both of them. Further, we found that \gpt\ and \gemini\ performed similarly with each other on this task. However, the overall analysis indicated that all of the models performed poorly on identifying misleading components in visualizations. In summary, our research contributions are 

\begin{itemize}
    \item A comprehensive study characterizing the visualization literacy of state-of-the-art LLMs.
    \item An empirical investigation into the mixed effects of few-shot and Chain-of-Thought (CoT) prompting on model performance across diverse visualization tasks.
    \item An evaluation of LLMs' capacity for detecting and reasoning about misleading visual elements.
\end{itemize}

\section{Related Work}
\subsection {Visualization Literacy in LLMs}
Our work is based on the definition of visualization literacy as the ability to comprehend and interpret a given visualization~\cite{Pandey:2023:MSE}. There already exist frameworks and assessments that aim to benchmark visualization literacy in humans~\cite{lee:2017:vlat, borner:2019:literacy, Pandey:2023:MSE}. One such test includes the Visualization Literacy Assessment Test (VLAT)\cite{lee:2017:vlat}, and its abridged version, Mini-VLAT\cite{Pandey:2023:MSE}. They are used to gauge individuals' ability to interpret visual representations. These tests and frameworks focus on different types of data visualizations that involve tasks such as interpreting values and trends, making comparisons, and identifying different components of data represented visually. To facilitate a direct comparison between human and LLMs' visualization literacy, we adapt these established metrics to evaluate LLMs in comprehending and interpreting visualizations.

Several studies have already been done to assess the visualization literacy of LLMs. Hong et al.\cite{Hong:2025:LHV} built on these tests by outlining an experimental template for modifying VLAT's 53-item test to assess LLMs' visualization literacy. The purpose of proposing an experimental template instead of a new test or reusing VLAT was to prevent LLMs from ``cheating'' (\ie, using their previous knowledge to answer questions) as well as to adapt to their continual training (\ie, LLMs could be trained on or memorize the new test, making it obsolete). Their results showed that LLMs had poor visualization literacy. Similarly, Khan et al. \cite{Khan:2025:evaluate} demonstrated that LLMs have difficulty generating certain types of visualizations and answering questions about them. Additionally, Shao et al.\cite{Shao:2025:LMA} found that LLMs are visually literate enough to provide feedback that aligns with human evaluators, but still note that the LLMs do not process the visualization details that a human evaluator can. However, there are other works that indicate that LLMs are moving toward visualization literacy \cite{Vazquez:2024:ALR} and that visualizations are useful in helping LLMs understand data \cite{Li:2025:DVH, li:2025:visual}. Especially as newer models are being released to the public, there is a notable increase in LLM performance in visualization, which is explored in our studies. 

\subsection {Analysis on Current State of Prompt Engineering}
While the visualization literacy of LLMs has been well explored, little emphasis has been given to the impact of prompting strategies.
In order to gauge the baseline of prompting strategies within the visualization community, we surveyed papers across the field to understand which prompting strategies are commonly used and the prevalence of the prompts circulating in the field. To this end, we conducted a review of full conference papers from IEEE VIS, ACL, and EMNLP in 2025 that use LLMs to either perform visualization tasks and/or answer questions based on visualizations. We purposely excluded papers that solely use LLMs as chatbot assistants in visualization systems, as they typically incorporate the models as modular components without employing specialized prompt engineering strategies, nor are the models typically tested using singular prompt engineering styles. Across all three conferences, we identified 23 papers that fit our criteria of researching LLMs in visualization within the scope of literacy. Of those papers, we compiled the following summary statistics: 
\begin{itemize}
    \item Of the 23 papers, 12/23 papers do not include their prompts. By conference, IEEE VIS had 9/10 papers that did not have prompts, 1/10 EMNLP papers did not have prompts and 2/3 ACL papers did not include their prompts.    
    \item The most common type of prompts (5/11) listed in our surveyed set of papers were \textit{prompt templates} to reflect the different prompting styles in the experiments. The remaining papers used \textit{zero-shot prompts} (4/11) and \textit{role-based zero-shot prompts} (2/11).
    \item Of the papers that did not include their full prompts, 10/12 mentioned the prompt in the paper, but did not provide explicit templates or the actual prompts used. The remaining 2/12 papers included only illustrative examples, which lacked the necessary details to reconstruct the explicit prompt or template.
\end{itemize}

\noindent Through our survey, we noted that across the three conferences, IEEE VIS reported the least number of prompts in their papers surrounding LLMs and visualizations, while EMNLP was more consistent in including prompts in their papers. In total, there was almost an even split between the papers that included their prompts versus papers that did not include their prompt. Most papers listed their \textit{prompt templates} to indicate what prompts they used, and the \textit{prompt templates} were typically of the form of a role-based zero-shot prompt. Of the papers that did not list the prompts, we note that there was discussion about the prompts the papers used and the examples provided. However, given the black-box nature of LLMs and the necessity of reproducibility in a research setting, having the explicit prompts is still beneficial.

\subsection{Prompt Engineering}
Just as human empirical studies showed that having precise instructions can ensure participants accurately read and interpret novel visualizations when doing evaluations~\cite{lam:2012:empirical,isenberg:2013:systematic}, LLMs require clear and effective prompts to function as reliable visualization evaluators. This analogy highlights the necessity of prompt engineering in future automated assessments. For this reason, we focused on testing different prompting techniques to quantify their effects on LLMs' visualization literacy.

Prompt engineering is the act of developing the most appropriate instructions to an LLM to elicit a desired result\cite{Liu:2023:PPP}. Some of the most popular prompting techniques are few-shot \cite{brown:2020:fewshot}, zero-shot \cite{Kojima:2023:LLM}, and chain-of-thought (CoT)~\cite{wei:2023:cot}, which is unsurprising given the typical use cases of LLMs ~\cite{Schulhoff:2025:PRS}. In order to aid LLMs with visualization tasks, many researchers turn to prompting. Based on the taxonomy proposed by Schulhoff et al.~\cite{Schulhoff:2025:PRS}, visualization-related prompting can be viewed as a specialized subcategory of image prompting, which includes strategies such as multimodal CoT, multimodal in-context learning, negative prompts, and prompt modifiers. Within this framework, multimodal CoT is particularly relevant, as it covers scenarios where users provide the model with an image, \eg, a chart, alongside specific textual instructions. Some researchers are proposing novel prompting techniques to help LLMs with understanding different aspects of a visualization, such as a set of marks prompting for visual grounding \cite{yang:2023:SoM}. Others are adapting current prompting techniques with attributes that are known to improve visualization literacy, such as the work by Das et al.~\cite{Das:2026:cot}. They combine CoT prompting with data extraction, verification, and analysis. To prevent the LLMs from hallucinating results, Suri et al. propose a chart attribution algorithm combined with a set of marks prompting to help identify components of a visualization \cite{Suri:2025:CFV}. Beyond leveraging prompting to optimize model performance, researchers also investigate diverse prompting strategies to understand their specific influence on visualization-related tasks. For example, Wang et al. examine and compare the impact of different prompting techniques on visualization generation~\cite{wang:2025:VGLLM}. They find that few-shot prompting does very well in the task, while CoT and zero-shot CoT prompting strategies do not perform as well as expected.


Although prompting has shown to improve LLMs in some tasks, this may be due to a potentially detrimental trait: their agreeableness. While agreeableness allows LLMs to be cooperative and gives impetus to prompting techniques, LLMs may fail tasks that prioritize veracity over satisfying, coherence but potentially user-biased responses. Within the visualization space, graphic integrity is one such principle that prioritizes truthfulness over aesthetics.

\subsection{Graphical Integrity}
Famously emphasized by Tufte~\cite{tufte:1998:visual}, graphical integrity is the principle that data visualizations should accurately reflect the information they are communicating without misleading or confusing their viewers. This principle emphasizes the importance of truthfulness in data visualization, ensuring every aspect of a visualization conveys information clearly and honestly. While graphical integrity is important for visualization creation, achieving it requires the ability to identify misleading or distorted visualizations. If LLMs possess this capability, they could be used to provide feedback to designers in achieving graphical integrity in their visualizations. For this reason, we focused on testing LLMs' ability to identify misleading chart elements.

Several studies have delved into a framework for misleading visualizations as well as measuring the effects of misleading visualizations on humans. Lan et al.\cite{lan:2024:junk} categorized misleading visualizations into three broad categories: misinformation, uninformativeness (lacking meaningful or relevant data), and unsociability (making viewers feel uncomfortable). Our work is focused on the misinformation aspects of misleading visualizations as it closely relates to the ability of LLMs to operate in a capacity of informing users on their visual designs and integrity. A study by Rho et al.\cite{rho:2024:various} shows the impact of misleading visualizations on humans which further justified our motivation for studying the extent that LLMs can protect against this type of misinformation.

However, LLMs currently do not show strong performance in terms of recognizing misleading charts. Zhang et al. suggested that LLMs are capable of falling for visual illusions in the same way that humans do \cite{Zhang:2023:GVI}, and Chen et al. found that LLMs struggle with misleading charts \cite{Chen:2025:UDV}. The study done by Mahbub et al. suggested that current language models struggle more with interpreting spatial scales and structures than with improper data encodings \cite{Mahbub:2025:PCD}. There is also a difference in what types of misleading factors impact LLMs. Tonglet et al. showed that LLMs are more susceptible to tactics such as overplotting or dual-axes and perform worse than humans when faced with other misleading tactics \cite{tonglet:2026:protecting}. While these studies indicated that LLMs still struggle with misleading visualizations, other work has shown that LLMs can more accurately detect these misleading elements given there is strong guidance, typically done through rigorous prompting \cite{Das:2026:misvisfix, Alexander:2024,Lo:2025:HGB}. 

While other work in misleading visualizations focus on the potential ability of LLMs to evaluate misleaders through prompting or other user interventions, our work focuses primarily on how a model responds to misleading and corrected charts to identify what (if any) hallucinations might commonly occur without any bias from users. If LLMs were to replace human evaluators of visualizations or work as built in evaluators within a visual analytic system, it is important the LLMs are capable of giving unbiased feedback without leading/specific prompting that would implicate where misleaders may be present. Realistically, the LLMs would be used by non-visualization experts who may not know what to prompt their LLMs with in order to coax the model into determining misleaders. For this reason, our work relies on using neutral prompting strategies and allowing the  LLM to rely on the built-in reasoning. 

\section{Pilot Study}
Before performing a systematic investigation into LLMs' visualization evaluation capabilities, we ran pilot studies on ChatGPT with our modified VLAT. Specifically, we manually copied and pasted questions and choices into the latest ChatGPT interface\footnote{https://chatgpt.com/} and assessed the veracity of its answers. The purpose of this pilot study was two-fold: first, to validate our experimental procedure and our modified VLAT; and second, to preliminarily assess whether the visualization literacy of current models has advanced beyond that of previous iterations. The study involved asking the LLM each of the 53 questions 10 times and creating a new chat session for each request for a total of 530 requests but did not change the order of the presented choices. We found marked improvements in ChatGPT's performance compared to Hong et al.'s~\cite{Hong:2025:LHV} previous results, in which ChatGPT scored perfectly on 29 out of the 53 questions (see \autoref{tab:results_pilot} for more details). These results validate our evaluation framework and provide a robust foundation for our subsequent hypotheses regarding updated visualization literacy and prompting strategies.

\section{Revisiting LLMs' Visualization Literacy and Exploring Prompt Engineering}\label{sec:vlRevisit_prompt}
To address \textbf{RQ1}: \textit{To what extent do the latest LLMs have visualization literacy?}, we investigated three sub-research questions:

\begin{itemize}
    \item \textbf{RQ1.1}: To what extent do the latest LLMs’ visualization literacy differ from humans' visualization literacy?
    \item \textbf{RQ1.2}: To what extent do the latest LLMs’ visualization literacy differ from older LLMs' visualization literacy?
    \item \textbf{RQ1.3}: To what extent do the latest LLMs’ visualization literacy differ between models on different visualizations and tasks?
\end{itemize}

\noindent We compared LLMs’ performance with human results reported in previous work\cite{lee:2017:vlat}, which provided a baseline for the evaluation. Furthermore, we revisited the effects of different conditions on LLMs’ visualization literacy testing. By examining the latest LLMs’ visualization literacy, we aimed to quantify the current progress of visualization literacy in LLMs and highlight potential areas of improvement. Furthermore, as previously mentioned, an LLM's capability to adhere to prompts is analogous to a human's ability for instruction-following, which is essential for any visualization evaluator. Consequently, we explored the effects of the diverse prompting techniques, i.e., few-shot\cite{brown:2020:fewshot} and CoT\cite{wei:2023:cot}, to address \textbf{RQ2}: \textit{Does prompt engineering improve the visualization literacy of LLMs?}. 
By exploring prompt engineering techniques, we aim to illuminate LLMs' instruction-following capabilities with regard to visualization evaluation.

\subsection{Experiments}
We followed Hong et al.'s template~\cite{Hong:2025:LHV} by developing our own modified VLAT and tested on a handful of the latest LLMs. Since our paper is focused on visualization evaluation, it is necessary for LLMs to be multimodal, \ie, they allow multiple media inputs (text, images, and video). This allows us to feed PNG images of data visualizations directly to the LLMs as opposed to writing descriptions of the data visualization\cite{kim:2023:how} or providing the text of an SVG\cite{chen:2023:beyond}. This requirement limits the number of LLMs that can be investigated. We focused on Anthropic's Claude Opus 4.5 (\texttt{claude-opus-4-5}, \claude), OpenAI's GPT 5.2 Pro (\texttt{gpt-5.2-pro}, \gpt), and Google's Gemini 3 Flash (\texttt{gemini-3-flash-preview}, \gemini) due to their widespread use, ease of access to their APIs for large-scale visualization analysis, and their multimodality. Although open-source LLMs are rapidly catching up with their closed-source counterparts\cite{kim:2025:benchmarking, oketch:2025:bridging, Dong:2025:PVL}, we ultimately decided to focus on closed-source models due to the outsized time and money spent on training and improving them. Further, we initially intended to test all of the ``pro'' models each company offered. For instance, Anthropic's and OpenAI's largest models are Opus and Pro, respectively. Unfortunately, Gemini's ``pro'' model, Gemini 3 Pro\footnote{https://docs.cloud.google.com/vertex-ai/generative-ai/docs/models/gemini/3-pro}, has a minute limit of 25 requests per minute and a daily limit of 250 requests per day (in Paid Tier 1\footnote{https://aistudio.google.com/app/rate-limit}), making it prohibitively slow to run our proposed experiments. As a result, we opted to use the Gemini 3 Flash model for all of our experiments.

\subsubsection{Experimental Design}\label{subsec:vis_design}
\begin{figure*}
    \centering
    \includegraphics[width=\linewidth]{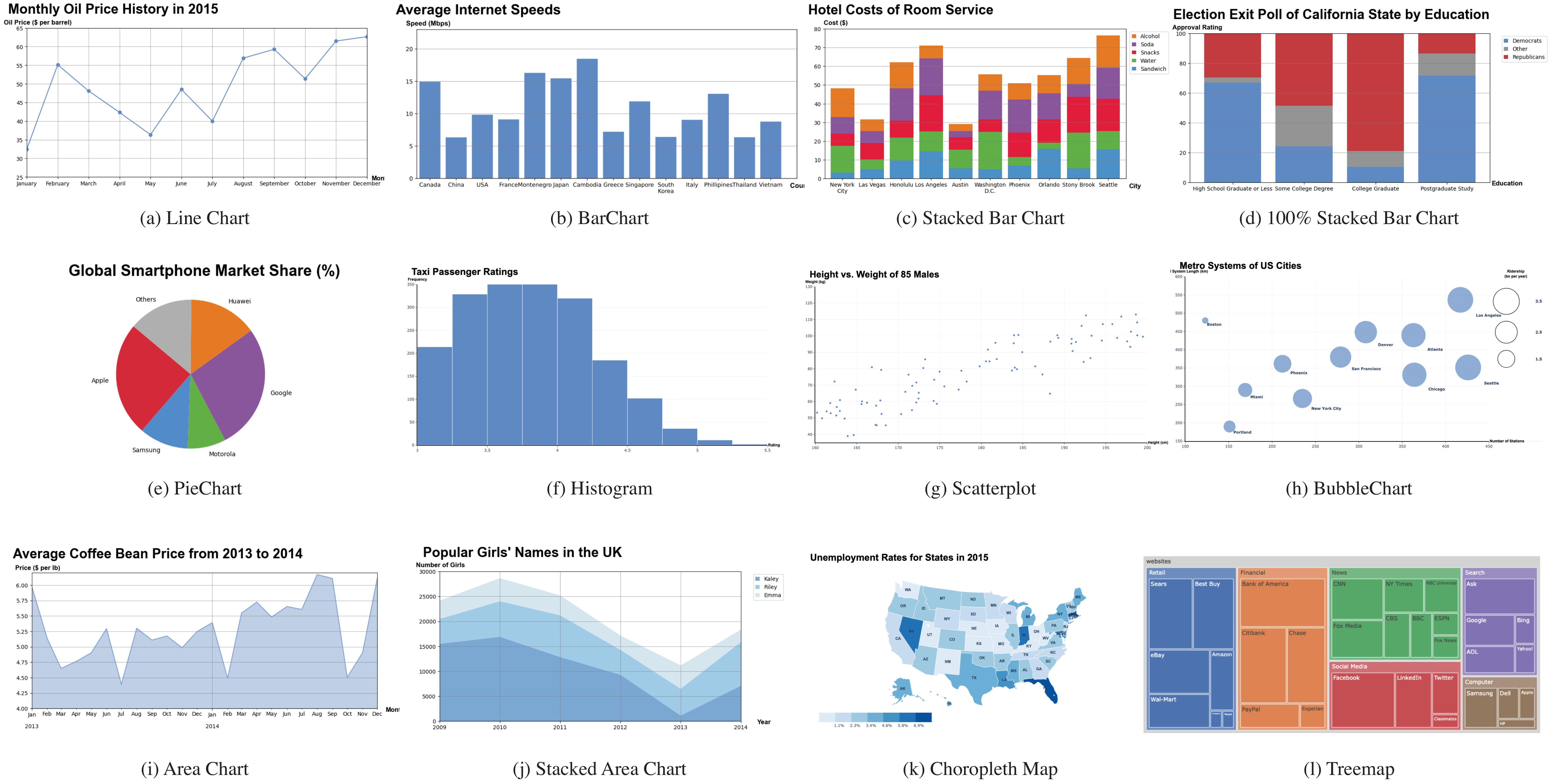}
    \caption{12 visualizations created based on VLAT examples and tested in our experiments.}
    \label{fig:visualizations}
    \vspace{-1em}
\end{figure*}

Inspired by previous work's experimental template~\cite{Hong:2025:LHV}, we created a modified version of the VLAT, including writing new questions and answers as well as creating new charts to test the LLMs. The purpose is to minimize the influence of the LLMs' prior parametric knowledge, thereby forcing the models to rely strictly on their in-context inference capabilities to derive answers. The visualizations we created and used in the experiments are shown in \autoref{fig:visualizations}. After setting up the visualizations with questions and answers, we used \claude, \gpt, and \gemini's APIs for our tests. Since all of the companies that own and maintain these LLMs claim that the data used for training is always being updated and refined, we did not publicly release our answers when conducting the studies. As such, \claude, \gpt, and \gemini\ are unlikely to have access to the answers to our modified VLAT, and we assumed that our newly updated charts were not included in the model's training data while we were conducting our tests.

\subsubsection{Experiment 1: Assess Latest LLMs' Visualization Literacy}
Using the questions, answers, and visualizations outlined in Section~\ref{subsec:vis_design}, we then evaluated the visualization literacy of \claude, \gpt, and \gemini. We used Hong et al.'s\cite{Hong:2025:LHV} base prompt for Experiment 1: \textit{``You are a helpful assistant for analyzing data visualizations. Please answer with the letter corresponding to the best option, or make a random guess if unsure. For example, if option (a) is correct, only reply with (a).''} A total of 53 questions were investigated across various types of visualizations and tasks. Each question featured 2 (True/False questions) to 4 answer options (multiple-choice questions). For each question, we conducted 120 tests with a counterbalanced order for all choices, meaning each combination of option orders was repeated between 120 / 4! = 5 and 120 / 2! = 60 times out of the total 120 tests. We shuffled all questions with a random seed to mitigate the effects of question order for the model. This results in a total of 53 $\times$ 120 = 6,360 trials for each LLM. Given that the answer options for later questions may influence LLMs' responses to earlier questions, we conducted each question in a separate session to ensure the independence of LLMs' answers.

\subsubsection{Experiment 2: Evaluate Effects of Prompting Techniques}
We used the same modified VLAT questions, answers, and visualizations outlined in Section~\ref{subsec:vis_design} but varied the prompts to evaluate two popular prompting techniques: few-shot~\cite{brown:2020:fewshot} and chain-of-thought (CoT)~\cite{wei:2023:cot}. The few-shot prompting technique involves providing an LLM with a small number of examples to guide its response to a specific task. This required providing examples of VLAT questions, answers, and charts for each corresponding request. We decided to use the original VLAT\cite{lee:2017:vlat} and Hong et al.'s\cite{Hong:2025:LHV} modified VLAT as our examples. The CoT prompting technique is similar to few shot but adds the solution instructions for each example. Again, we used the original VLAT and Hong et al.'s modified VLAT as our examples.

To validate our experimental design, we conducted preliminary trials across the two prompting strategies. Our results indicated that few-shot prompting involving two or more examples led to inconsistent outputs across the LLMs. This suggests that increasing the number of image-text pairs may introduce noise or exceed the models' effective context processing limits. Thus, we chose to provide one example for few-shot prompting. Based on these findings, we excluded the two-example prompt for our CoT trials, focusing on the one-example format to avoid the models' confusion. To maintain full experimental transparency, the data from our initial two-example few-shot trials are available in the supplemental materials.

To optimize the balance between computational cost and statistical reliability, we decided to only run 24 tests for each question with every order combination of choices. This meant that each combination of option orders was repeated between 24 / 4! = 1 and 24 / 2! = 12 times out of the total 24 tests for both the original and Hong et al.'s modified VLAT questions and answers. Again, we shuffled all questions with a random seed running a total of 2 $\times$ 53 $\times$ 24 = 2,544 trials for each LLM. As in Experiment 1, we conducted each question in a separate session to ensure the independence of LLMs' answers.

\subsection{Analysis Methods}
We used the general public's performance as a baseline to evaluate LLMs' visualization literacy. Further, we qualitatively compared human results to our selected LLMs' results from Experiment 1 as well as the results of the LLMs chosen in Hong et al.'s\cite{Hong:2025:LHV} experiments. To address \textbf{RQ1.2}, we developed the following hypothesis:

\begin{itemize}[leftmargin=*]
\label{hypothesis1.1}
    \item \textbf{H1}: \claude, \gpt, and \gemini\ will outperform both \gptFour\ and \geminiOne\ across all charts, tasks, and questions.
\end{itemize}

\noindent We formulated \textbf{H1} based on our pilot study on ChatGPT 5.1 which achieved a near perfect score on our modified VLAT. For \textbf{RQ1.3}, we formulated three additional hypotheses:

\begin{itemize}[leftmargin=*]
\label{hypothesis1.2}
    \item \textbf{H2}: \claude, \gpt, and \gemini\ will all be on par with each other in terms of performance.
    \item \textbf{H3}: \claude, \gpt, and \gemini\ will perform better on all visualizations except for \textit{scatterplot}.
    \item \textbf{H4}: \claude, \gpt, and \gemini\ will perform better on all tasks except \textit{Retrieve Value}.
\end{itemize}

\noindent When forming \textbf{H2}, we believed that all of the companies developing \claude, \gpt, and \gemini\ are spending the most time and money training these LLMs, and thus we believed they would perform similarly. Further, other studies have shown that these LLMs are performing similarly in reasoning tasks\cite{Khan:2025:evaluate, naveen:2025:comparative}. When forming \textbf{H3} and \textbf{H4}, we again relied on the pilot study to inform our predictions. Specifically, we noticed that the only visualization that performed worse than all of the others was the \textit{scatterplot} visualization and that the \textit{Retrieve Value} task had less than perfect scores 7 out of 13 times, more than any other task. To address \textbf{RQ2}, we developed the following hypothesis:

\begin{itemize}[leftmargin=*]
\label{hypothesis1.3}
    \item \textbf{H5}: CoT prompting will outperform few-shot prompting, and both techniques will outperform the base prompt for \claude, \gpt, and \gemini.
\end{itemize}

\noindent We developed this hypothesis based on the numerous studies showing the efficacy of these prompting techniques\cite{Das:2026:cot, wang:2025:VGLLM}. Further, we intuited that since CoT is providing more information to the LLM compared to few shot, we believed that it would outperform this prompting technique.

\begin{table*}[t]
    \centering
    \caption{Summary table of hypotheses, statistical methods, and results used to support/not support for \textbf{RQ1} and \textbf{RQ2}.}
    \begin{tabular}{l|c|l|l|l}
        \hline
        RQ              &Hypothesis &Methods             &Results &Result Section(s)\\
        \hline
        \textbf{RQ1.2}  &\textbf{H1}
                        &Percentage Correct Comparisons
                        &Supported 
                        &Sections~\ref{sec:RQ1_llm_human} and \ref{sec:oldVsNewLLMs}\\
        \textbf{RQ1.3}  &\textbf{H2}
                        &Percentage Correct Comparisons / Coefficient Analysis
                        &Partially supported 
                        &Section~\ref{sec:exp1_llmResults}\\
        \textbf{RQ1.3}  &\textbf{H3}
                        &Coefficient Analysis
                        &Partially supported 
                        &Section~\ref{sec:exp1_chartTaskResults}\\
        \textbf{RQ1.3}  &\textbf{H4}
                        &Coefficient Analysis
                        &Not supported
                        &Section~\ref{sec:exp1_chartTaskResults}\\
        \textbf{RQ2}    &\textbf{H5}
                        &Percentage Correct Comparisons / Coefficient Analysis
                        &Partially supported
                        &Section~\ref{sec:exp2_promptResults}\\
        \hline
    \end{tabular}
    \label{tab:hypotheses_exp1+2}
    \vspace{-0.8em}
\end{table*}

Based on these hypotheses, we conducted hypothesis tests to examine the impact of four dimensions (visualization type, task type, LLM type, and prompt technique) on the LLMs' performances in answering questions. To achieve this, we modeled the results using logistic regression over these dimensions since our dependent variable (whether a question was answered correctly or not) is binary. Since all variables are categorical or binary, exploring them and their interactions was sufficient. The model is of the following form:

\vspace{-1.2em}
\begin{equation}
\label{eq:promptModel}
    P(y=1|\textbf{x})=\text{logit}\left(\beta_0+\boldsymbol{\beta}_1^T\textbf{x}_1+\boldsymbol{\beta}_2^T\textbf{x}_2+\boldsymbol{\beta}_3^T\textbf{x}_3+\boldsymbol{\beta}_4^T\textbf{x}_4\right)
\end{equation}
\vspace{-1em}

\noindent where $y$ represents whether the question was answered correctly ($y=1$) or not ($y=0$), $\textbf{x}_k$ represents a vector of $k$-way interaction variables ($k \in \{1,2,3,4\}$ since we explored four dimensions), and $\boldsymbol{\beta}_k$ represents a vector of coefficients for each of these variables. For task types, we simplified the entries by removing text in the parentheses.  For instance, \textit{Retrieve Value (Absolute Value)} became \textit{Retrieve Value}. After this simplification, a total of 49 unique visualization and task-type interactions remained, leading to a total of 1,120 variables (see Appendix~\ref{app:varInterCalcPrompt}).

\subsection{Results}
We collected $53\text{ (questions)} \times 120\text{ (trials per question)} \times 3\text{ (LLMs)} \times 1\text{ (base prompt)} = 19,080$ trials for Experiments 1 and $53\text{ (questions)} \times 24\text{ (trials/question)} \times 2\text{ (VLAT example types)} \times 3\text{ (LLMs)} \times 2\text{ (prompts)}=15,264$ trials for Experiment 2 for a total of $19,080+15,264=34,344$ trials. Each trial was comprised of the index, visualization type, task type, preset question, correct answer, LLMs' response, and prompt type (base, few-shot, and CoT). Overall hypothesis results are summarized in \autoref{tab:hypotheses_exp1+2}.

\subsubsection{\textbf{RQ1.1} Visualization Literacy of LLMs vs. Humans}\label{sec:RQ1_llm_human}
We compared the performances of the latest LLMs with that of the general public and the LLM results found in Hong et al. After compiling the human results from the original VLAT paper, the LLM results found in Hong et al., and the latest LLMs' results from Experiments 1, we performed a qualitative comparison as presented in \autoref{tab:results_ex1+2}. We find that the latest LLMs outperformed humans in almost every question. Specifically, \claude\ performed better than humans (scored greater than 0.05 or perfect) in 40 out of the 53 questions, \gpt\ on 47 questions, and \gemini\ on 47 questions.

\begin{table*}[]
    \caption{The comparisons between the expected accuracy rates for randomly choosing an answer (e.g., if there are four choices, the expected accuracy rate for random choices is 0.25), the accuracy rates of the general public from VLAT~\cite{lee:2017:vlat}, the results of \gptFour and \geminiOne in Hong et al.'s experiment\cite{Hong:2025:LHV}, and the results of \claudeFig, \gptFig, and \geminiFig\ for the base prompt, few shot, and CoT prompting techniques in Experiment 1. We color-encoded LLMs' results: \better{green} for much better than the performance of people (more than 0.05 higher) or perfect (1.00), \close{yellow} for values that were close to people (higher or lower within 0.05), and \worse{red} for values much worse than those of people (more than 0.05 lower).}\vspace{-0.7em}
    \tabulinesep=1pt
    \begin{tabular}{>{\small}p{1.1cm}>{\small}p{2.23cm}>{\small}p{0.6cm}>{\small}p{0.6cm}>{\small}p{0.6cm}>{\small}p{0.7cm}|>{\small}p{0.6cm}>{\small}p{0.6cm}>{\small}p{0.8cm}|>{\small}p{0.6cm}>{\small}p{0.6cm}>{\small}p{0.8cm}|>{\small}p{0.6cm}>{\small}p{0.6cm}>{\small}p{0.8cm}}
        \toprule    
        \multirow{2}{*}{Visualization}  & \multirow{2}{*}{Task} & \multirow{2}{*}{Random}   & \multirow{2}{*}{VLAT} & \multirow{2}{*}{\gptFour}& \multirow{2}{*}{\geminiOne} & \multicolumn{3}{c|}{\underline{Base Prompt}}  & \multicolumn{3}{c|}{\underline{Few Shot}} & \multicolumn{3}{c}{\underline{Chain-of-Thought}}\\
                                        &                       &                           &                       &                       &                           & \claudeTab & \gptTab & \geminiTab                      & \claudeTab & \gptTab & \geminiTab                  & \claudeTab & \gptTab & \geminiTab\\[-0.2em]
        \midrule
        \multirow{5}{*}{Line Chart} & Retrieve Value & 0.25 & 0.95 & \worse{0.56} & \worse{0.25} & \better{1.00} & \better{1.00} & \better{1.00} & \worse{0.00} & \better{1.00} & \close{0.98} & \worse{0.02} & \better{1.00} & \close{0.96}\\[-0.2em]
                                    & Find Extremum & 0.25 & 0.97 & \worse{0.02} & \worse{0.22} & \better{1.00} & \better{1.00} & \better{1.00} & \worse{0.90} & \better{1.00} & \better{1.00} & \worse{0.67} & \better{1.00} & \better{1.00}\\[-0.2em]
                                    & Determine Range & 0.25 & 0.56 & \worse{0.23} & \worse{0.23} & \worse{0.22} & \better{1.00} & \better{1.00} & \worse{0.00} & \better{1.00} & \better{1.00} & \worse{0.00} & \better{1.00} & \better{1.00}\\[-0.2em]
                                    & Find Correlation/Trend & 0.33 & 0.98 & \worse{0.87} & \worse{0.42} & \better{1.00} & \better{1.00} & \better{1.00} & \better{1.00} & \better{1.00} & \better{1.00} & \better{1.00} & \better{1.00} & \better{1.00}\\[-0.2em]
                                    & Make Comparisons & 0.25 & 0.77 & \better{0.87} & \worse{0.28} & \better{1.00} & \better{1.00} & \better{1.00} & \better{0.88} & \better{1.00} & \better{1.00} & \better{1.00} & \better{1.00} & \better{1.00}\\[-0.2em]
        \hline
        \multirow{4}{*}{Bar Chart} & Retrieve Value & 0.25 & 0.88 & \worse{0.40} & \worse{0.56} & \better{1.00} & \better{1.00} & \better{0.97} & \worse{0.10} & \better{1.00} & \better{1.00} & \worse{0.08} & \better{1.00} & \better{0.94}\\[-0.2em]
                                   & Find Extremum & 0.25 & 0.98 & \worse{0.00} & \worse{0.01} & \better{1.00} & \better{1.00} & \close{0.97} & \worse{0.00} & \better{1.00} & \better{1.00} & \worse{0.08} & \better{1.00} & \worse{0.92}\\[-0.2em]
                                   & Determine Range & 0.25 & 0.54 & \worse{0.29} & \better{0.68} & \worse{0.05} & \better{1.00} & \better{1.00} & \worse{0.00} & \better{1.00} & \better{1.00} & \worse{0.00} & \better{1.00} & \better{0.96}\\[-0.2em]
                                   & Make Comparisons & 0.25 & 0.40 & \worse{0.20} & \worse{0.13} & \better{1.00} & \better{1.00} & \better{1.00} & \better{1.00} & \better{1.00} & \better{1.00} & \better{1.00} & \better{1.00} & \better{1.00}\\[-0.2em]
        \hline
        \multirow{5}{*}{\makecell[l]{Stacked \\ Bar Chart}} 
                                    & \makecell[l]{Retrieve Value \\ (Absolute Value)} & 0.25 & 0.38 & \worse{0.23} & \worse{0.25} & \worse{0.00} & \better{0.95} & \better{1.00} & \worse{0.00} & \better{0.94} & \better{1.00} & \worse{0.00} & \better{0.98} & \better{1.00}\\[-0.2em]
                                    & \makecell[l]{Retrieve Value \\ (Relative Value)} & 0.25 & 0.36 & \worse{0.04} & \worse{0.23} & \better{0.94} & \better{1.00} & \better{1.00} & \better{0.50} & \better{1.00} & \better{1.00} & \better{0.50} & \better{1.00} & \better{0.96}\\[-0.2em]
                                    & Find Extremum & 0.25 & 0.69 & \worse{0.17} & \worse{0.05} & \better{0.98} & \better{1.00} & \better{1.00} & \worse{0.33} & \better{1.00} & \better{1.00} & \worse{0.25} & \better{1.00} & \better{1.00}\\[-0.2em]
                                    & \makecell[l]{Make Comparisons \\ (Absolute Value)} & 0.50 & 0.59 & \better{0.70} & \better{0.78} & \better{0.83} & \better{1.00} & \better{1.00} & \better{0.79} & \better{1.00} & \better{1.00} & \better{0.98} & \better{1.00} & \better{0.98}\\[-0.2em]
                                    & \makecell[l]{Make Comparisons \\ (Relative Value)} & 0.50 & 0.47 & \better{0.94} & \close{0.43} & \worse{0.25} & \better{1.00} & \better{1.00} & \better{0.67} & \better{1.00} & \better{1.00} & \better{0.92} & \better{1.00} & \better{0.96}\\[-0.2em]
        \hline
        \multirow{3}{*}{\makecell[l]{100\% \\ Stacked \\ Bar Chart}} 
                                    & \makecell[l]{Retrieve Value \\ (Relative Value)} & 0.25 & 0.49 & \worse{0.00} & \better{0.79} & \better{1.00} & \better{1.00} & \better{1.00} & \worse{0.42} & \better{1.00} & \better{1.00} & \worse{0.19} & \better{1.00} & \better{1.00}\\[-0.2em]
                                    & \makecell[l]{Find Extremum \\ (Relative Value)} & 0.25 & 0.90 & \worse{0.00} & \worse{0.21} & \better{0.99} & \better{1.00} & \better{1.00} & \better{1.00} & \better{1.00} & \better{1.00} & \better{1.00} & \better{1.00} & \better{1.00}\\[-0.2em]
                                    & \makecell[l]{Make Comparisons \\ (Relative Value)} & 0.50 & 0.54 & \worse{0.15} & \better{0.98} & \better{1.00} & \better{1.00} & \better{1.00} & \better{1.00} & \better{1.00} & \better{1.00} & \better{1.00} & \better{1.00} & \better{1.00}\\[-0.2em]
        \hline
        \multirow{3}{*}{Pie Chart} 
                                & \makecell[l]{Retrieve Value \\ (Relative Value)} & 0.25 & 0.72 & \worse{0.00} & \worse{0.36} & \better{1.00} & \better{1.00} & \better{1.00} & \worse{0.00} & \better{1.00} & \better{1.00} & \worse{0.04} & \better{1.00} & \better{1.00}\\[-0.2em]
                                & \makecell[l]{Find Extremum \\ (Relative Value)} & 0.25 & 0.98 & \worse{0.00} & \worse{0.30} & \close{0.98} & \better{1.00} & \worse{0.88} & \worse{0.40} & \better{1.00} & \worse{0.58} & \worse{0.23} & \better{1.00} & \worse{0.56}\\[-0.2em]
                                & \makecell[l]{Make Comparisons \\ (Relative Value)} & 0.50 & 1.00 & \worse{0.00} & \worse{0.43} & \worse{0.33} & \better{1.00} & \better{1.00} & \better{1.00} & \better{1.00} & \better{1.00} & \better{1.00} & \better{1.00} & \better{1.00}\\[-0.2em]
        \hline
        \multirow{3}{*}{Histogram}
                                & \makecell[l]{Retrieve Value \\ (Derived Value)} & 0.25 & 0.84 & \worse{0.38} & \worse{0.18} & \better{1.00} & \better{1.00} & \better{1.00} & \better{0.96} & \better{1.00} & \better{1.00} & \worse{0.77} & \better{1.00} & \better{1.00}\\[-0.2em]
                                & \makecell[l]{Find Extremum \\ (Derived Value)} & 0.25 & 0.94 & \worse{0.06} & \worse{0.13} & \close{0.98} & \better{1.00} & \better{1.00} & \worse{0.79} & \better{1.00} & \better{1.00} & \worse{0.60} & \better{1.00} & \better{1.00}\\[-0.2em]
                                & \makecell[l]{Make Comparisons \\ (Derived Value)} & 0.50 & 0.86 & \close{0.89} & \worse{0.00} & \better{1.00} & \better{1.00} & \better{1.00} & \better{1.00} & \better{1.00} & \better{1.00} & \better{1.00} & \better{1.00} & \better{1.00}\\[-0.2em]
        \hline
        \multirow{7}{*}{Scatterplot}
                                & Retrieve Value & 0.25 & 0.85 & \worse{0.20} & \worse{0.49} & \worse{0.00} & \worse{0.33} & \worse{0.07} & \worse{0.02} & \worse{0.35} & \worse{0.15} & \worse{0.00} & \worse{0.21} & \worse{0.13}\\[-0.2em]
                                & Find Extremum & 0.25 & 0.76 & \better{0.85} & \better{1.00} & \better{1.00} & \better{1.00} & \better{0.98} & \worse{0.17} & \better{1.00} & \better{1.00} & \worse{0.25} & \better{1.00} & \better{0.98}\\[-0.2em]
                                & Determine Range & 0.25 & 0.53 & \better{0.86} & \better{0.76} & \better{1.00} & \better{1.00} & \better{1.00} & \worse{0.31} & \better{1.00} & \better{1.00} & \worse{0.23} & \better{1.00} & \better{0.98}\\[-0.2em]
                                & Find Anomalies & 0.25 & 0.42 & \worse{0.00} & \worse{0.00} & \better{0.83} & \better{1.00} & \better{0.98} & \worse{0.00} & \better{0.85} & \better{0.90} & \worse{0.02} & \better{0.73} & \better{0.52}\\[-0.2em]
                                & Find Clusters & 0.50 & 0.90 & \worse{0.06} & \worse{0.50} & \better{1.00} & \worse{0.80} & \better{1.00} & \close{0.94} & \worse{0.77} & \better{1.00} & \close{0.90} & \close{0.88} & \better{0.98}\\[-0.2em]
                                & Find Correlation/Trend & 0.50 & 0.52 & \better{1.00} & \worse{0.00} & \better{1.00} & \better{1.00} & \better{1.00} & \better{1.00} & \better{1.00} & \better{1.00} & \better{1.00} & \better{1.00} & \better{1.00}\\[-0.2em]
                                & Make Comparisons & 0.50 & 0.79 & \better{1.00} & \better{1.00} & \better{1.00} & \better{1.00} & \better{1.00} & \better{1.00} & \better{1.00} & \better{0.90} & \better{1.00} & \better{1.00} & \better{0.96}\\[-0.2em]
        \hline
        \multirow{4}{*}{Area Chart}
                                & Retrieve Value & 0.25 & 0.75 & \worse{0.28} & \worse{0.13} & \better{1.00} & \better{1.00} & \better{0.98} & \better{0.96} & \better{1.00} & \better{1.00} & \better{0.90} & \better{1.00} & \better{0.98}\\[-0.2em]
                                & Find Extremum & 0.25 & 0.44 & \better{0.57} & \better{0.73} & \better{1.00} & \better{1.00} & \better{1.00} & \better{0.73} & \better{1.00} & \better{1.00} & \better{0.67} & \better{1.00} & \better{1.00}\\[-0.2em]
                                & Determine Range & 0.25 & 0.38 & \worse{0.23} & \worse{0.13} & \better{0.90} & \better{1.00} & \better{0.85} & \worse{0.21} & \better{1.00} & \better{0.88} & \worse{0.04} & \better{1.00} & \better{0.85}\\[-0.2em]
                                & Find Correlation/Trend & 0.33 & 0.94 & \better{1.00} & \better{1.00} & \better{1.00} & \better{1.00} & \better{1.00} & \better{1.00} & \better{1.00} & \better{1.00} & \better{1.00} & \close{0.98} & \better{1.00}\\[-0.2em]
        \hline
        \multirow{6}{*}{\makecell[l]{Stacked \\ Area Chart}}
                                & \makecell[l]{Retrieve Value \\ (Absolute Value)} & 0.25 & 0.15 & \better{0.36} & \better{0.35} & \better{0.45} & \better{1.00} & \better{1.00} & \close{0.13} & \better{1.00} & \better{1.00} & \close{0.15} & \better{1.00} & \better{1.00}\\[-0.2em]
                                & \makecell[l]{Retrieve Value \\ (Relative Value)} & 0.25 & 0.25 & \worse{0.17} & \close{0.30} & \better{1.00} & \better{0.50} & \better{0.96} & \better{1.00} & \better{0.88} & \better{0.94} & \better{1.00} & \better{1.00} & \better{0.83}\\[-0.2em]
                                & Find Extremum & 0.25 & 0.97 & \worse{0.00} & \worse{0.01} & \worse{0.00} & \worse{0.00} & \worse{0.35} & \worse{0.00} & \worse{0.02} & \worse{0.58} & \worse{0.02} & \worse{0.00} & \worse{0.27}\\[-0.2em]
                                & Find Correlation/Trend & 0.33 & 0.96 & \worse{0.00} & \worse{0.72} & \worse{0.00} & \worse{0.85} & \worse{0.70} & \worse{0.06} & \close{0.96} & \close{0.98} & \worse{0.31} & \worse{0.29} & \worse{0.63}\\[-0.2em]
                                & \makecell[l]{Make Comparisons \\ (Absolute Value)} & 0.50 & 0.20 & \better{0.51} & \better{0.99} & \better{1.00} & \better{1.00} & \better{1.00} & \better{1.00} & \better{1.00} & \better{1.00} & \better{0.94} & \better{1.00} & \better{1.00}\\[-0.2em]
                                & \makecell[l]{Make Comparisons \\ (Relative Value)} & 0.50 & 0.24 & \better{1.00} & \better{1.00} & \better{0.86} & \better{0.58} & \better{1.00} & \better{1.00} & \better{0.60} & \better{1.00} & \better{0.98} & \better{0.79} & \better{1.00}\\[-0.2em]
        \hline                            
        \multirow{7}{*}{\makecell[l]{Bubble \\ Chart}}
                                & Retrieve Value & 0.25 & 0.41 & \worse{0.02} & \worse{0.15} & \better{1.00} & \better{1.00} & \better{0.81} & \better{0.90} & \better{1.00} & \better{1.00} & \better{0.52} & \better{1.00} & \better{0.96}\\[-0.2em]
                                & Find Extremum & 0.25 & 0.69 & \worse{0.00} & \worse{0.11} & \better{0.99} & \better{0.98} & \better{1.00} & \better{0.77} & \better{1.00} & \better{0.96} & \better{0.83} & \better{1.00} & \better{1.00}\\[-0.2em]
                                & Determine Range & 0.25& 0.29 & \close{0.25} & \close{0.26} & \worse{0.00} & \better{1.00} & \better{0.82} & \worse{0.00} & \better{1.00} & \better{1.00} & \worse{0.00} & \better{1.00} & \better{0.98}\\[-0.2em]
                                & Find Anomalies & 0.25 & 0.53 & \close{0.53} & \worse{0.06} & \better{1.00} & \better{1.00} & \better{0.99} & \worse{0.10} & \better{1.00} & \better{1.00} & \close{0.52} & \better{1.00} & \better{1.00}\\[-0.2em]
                                & Find Clusters & 0.50 & 0.59 & \worse{0.38} & \worse{0.48} & \better{0.94} & \worse{0.34} & \better{0.99} & \better{0.98} & \better{0.90} & \better{1.00} & \better{1.00} & \better{1.00} & \better{0.98}\\[-0.2em]
                                & Find Correlation/Trend & 0.50 & 0.26 & \better{0.93} & \better{1.00} & \better{1.00} & \better{1.00} & \better{1.00} & \better{1.00} & \better{1.00} & \better{1.00} & \better{1.00} & \better{1.00} & \better{1.00}\\[-0.2em]
                                & Make Comparisons & 0.50 & 0.80 & \worse{0.00} & \worse{0.09} & \better{1.00} & \better{1.00} & \better{0.99} & \better{1.00} & \better{1.00} & \better{1.00} & \better{1.00} & \better{1.00} & \better{1.00}\\[-0.2em]
        \hline
        \multirow{3}{*}{\makecell[l]{Chloropleth \\ Map}}
                                & \makecell[l]{Retrieve Value \\ (Approximate Value)} & 0.25 & 0.24 & \worse{0.00} & \worse{0.00} & \worse{0.18} & \better{1.00} & \better{0.4} & \close{0.25} & \better{1.00} & \better{0.63} & \worse{0.17} & \better{1.00} & \better{0.5}\\[-0.2em]
                                & \makecell[l]{Find Extremum \\ (Approximate Value)} & 0.25 & 0.97 & \close{0.93} & \worse{0.72} & \better{1.00} & \better{1.00} & \close{0.99} & \worse{0.44} & \better{1.00} & \better{1.00} & \worse{0.17} & \better{1.00} & \close{0.98}\\[-0.2em]
                                & \makecell[l]{Make Comparison \\ (Approximate Value)} & 0.50 & 0.92 & \worse{0.80} & \worse{0.00} & \close{0.96} & \worse{0.01} & \better{1.00} & \worse{0.13} & \worse{0.38} & \better{1.00} & \worse{0.75} & \worse{0.75} & \better{1.00}\\[-0.2em]
        \hline
        \multirow{3}{*}{Treemap}
                                & \makecell[l]{Find Extremum \\ (Relative Value)} & 0.25 & 0.68 & \worse{0.01} & \worse{0.00} & \better{0.99} & \better{0.96} & \better{0.97} & \worse{0.54} & \better{0.94} & \better{1.00} & \worse{0.40} & \better{1.00} & \better{0.98}\\[-0.2em]
                                & \makecell[l]{Make Comparison \\ (Relative Value)} & 0.50 & 0.42 & \worse{0.03} & \better{0.53} & \better{1.00} & \better{1.00} & \better{1.00} & \better{1.00} & \better{1.00} & \better{1.00} & \better{1.00} & \better{1.00} & \better{1.00}\\[-0.2em]
                                & \makecell[l]{Identify the \\ Hierarchical Structure} & 0.50 & 0.92 & \better{1.00} & \worse{0.00} & \better{1.00} & \better{1.00} & \better{1.00} & \better{1.00} & \better{1.00} & \better{1.00} & \better{1.00} & \better{1.00} & \better{1.00}\\[-0.2em]
        \hline
    \end{tabular}
    \label{tab:results_ex1+2}
\end{table*}

\begin{figure}[h]
    \centering
    \includegraphics[width=0.49\textwidth]{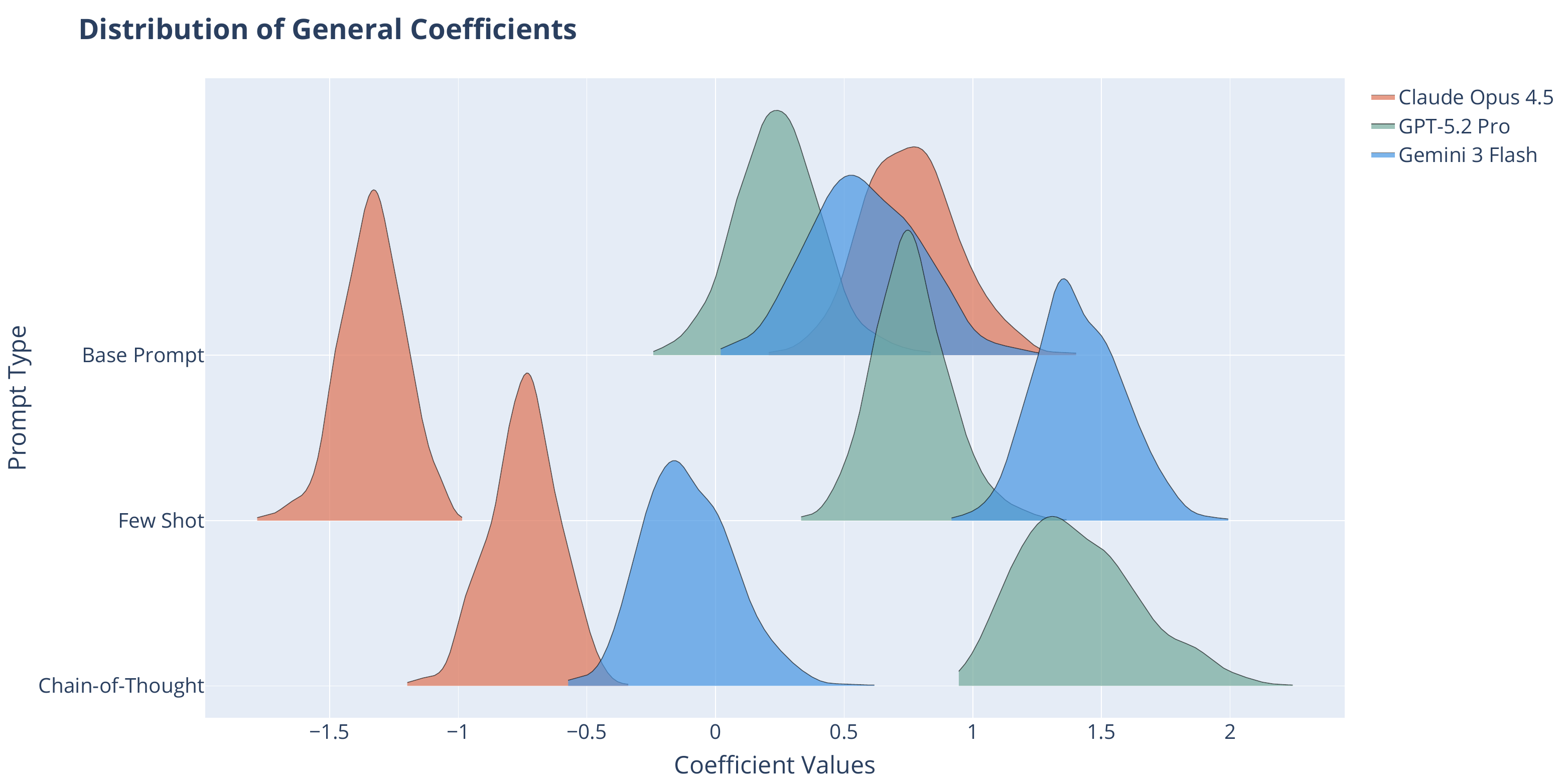}
    \caption{Ridge plot of the bootstrapped general coefficients from the logistic regression.}
    \label{fig:genCoef}
    \vspace{-1em}
\end{figure}

\subsubsection{\textbf{H1} Older vs. Newer LLMs}\label{sec:oldVsNewLLMs}
Based on our experiments for the different prompting styles, we found that \gpt\ and \gemini\ outperform their older versions as reported in the study done by Hong et al \cite{Hong:2025:LHV}. Comparing the results of the newer models with base prompting, \gpt\ performed better on every question type except for \textit{make comparisons (relative value) on stacked area chart}, \textit{find clusters on bubble chart}, and \textit{make comparison (approximate value) on chloropleth map} compared to the \gptFour. Meanwhile, \gemini\ performed better than \geminiOne\ on every question except \textit{retrieve value on scatterplot} and \textit{find correlation/text on stacked area chart}. Moreover, the differences in accuracy of the newer models compared to the older ones is large across the board, with \gpt\ and \gemini\ scoring perfectly on most VLAT question types. Both \gpt\ and \gemini\ performed better than humans on more questions than their older counter parts, only performing worse or on par with humans on 6 question types. Compared to \gptFour\ and \geminiOne, in which both performed worse than humans on a majority of the VLAT questions, this is a notable improvement. Given the exceptional performance of \gpt\ and \gemini, our experiments showed the latest models are visually literate as determined by the VLAT.

\subsubsection{\textbf{H2} Visualization Literacy of Latest LLMs}\label{sec:exp1_llmResults}
\noindent\textbf{Percent Correct Comparisons}
Our experiments showed the improved ability of the latest models to be visually literate per the VLAT. On the base prompt alone, \claude\ was 100\% accurate on 29/53 VLAT questions, \gpt\ on 41/53, and \gemini\ on 34/53. Further, we find that \claude\ was correct on 79.94\% of the questions, \gpt\ was 91.12\% correct and \gemini\ was 93.68\% correct. All three models performed better than humans on the VLAT across the board with only base prompting. See \autoref{tab:hypotheses_exp1+2} for more details.

\noindent\textbf{Coefficient Results}
Out of the 1,120 total combinations of visualizations, tasks, LLMs, and prompt types, 852 were found to be statistically significant (p-value$<0.05$, $H_0:\beta=0$ vs. $H_1:\beta\neq0$), meaning that a majority of these combinations affected the LLMs' performance. The coefficients for \claude, \gpt, and \gemini\ with the base prompt had statistically significant positive values ($\bar{\beta} \approx 0.7570$, $\bar{\beta} \approx 0.2448$, and $\bar{\beta} \approx 0.5798$ respectively, where $\bar{\beta}$ represents the mean bootstrapped coefficients) which indicate that all the LLMs were more likely to answer correctly. Although we did not perform difference tests on each of the LLMs, it seems clear from the first row of density plots in \autoref{fig:genCoef} that \claude, \gpt, and \gemini\ performed on par with each other, supporting our prediction in \textbf{H2}.

\subsubsection{\textbf{H3 \& H4} LLMs in Varied Visualizations and Tasks}\label{sec:exp1_chartTaskResults}
For the visualizations with the base prompt (no interactions), 8 out of the 12 coefficients were statistically significant with 4 positive and 4 negative values. The visualizations with positive coefficients were \textit{Line Chart} ($\bar{\beta} \approx 1.940$), \textit{Bar Chart} ($\bar{\beta} \approx 0.6612$), \textit{Scatterplot} ($\bar{\beta} \approx 0.4861$), and \textit{Area Chart} ($\bar{\beta} \approx 0.5470$), while the visualizations with negative coefficients were \textit{Stacked Bar Chart} ($\bar{\beta} \approx -0.0774$), \textit{Bubble Chart} ($\bar{\beta} \approx -0.6727$), \textit{Stacked Area Chart} ($\bar{\beta} \approx -1.818$), and \textit{Choropleth Map} ($\bar{\beta} \approx -0.7322$). These results did not align with our predictions in \textbf{H3}, especially noting that \textit{Scatterplot} has a positive coefficient.

For the tasks with the base prompt, 4 out of the 8 tasks were statistically significant with 3 positive coefficients and 1 negative. The tasks with positive coefficients were \textit{retrieve value} ($\bar{\beta} \approx 0.4494$), \textit{find anomalies} ($\bar{\beta} \approx 1.135$), and \textit{identify the hierarchical structure} ($\bar{\beta} \approx 0.2941$), while the task with a negative coefficient was \textit{determine range} ($\bar{\beta} \approx -0.2232$). Again, these results did not align with our predictions in \textbf{H4}, especially with \textit{retrieve value} being positive.

Finally, for visualization/task interactions with the base prompt, 32 out of the 49 interaction coefficients were statistically significant, of which 22 were positive and 10 were negative.

\subsubsection{\textbf{H5} Prompt Techniques}\label{sec:exp2_promptResults}
\noindent\textbf{Percent Correct Comparisons}
The results from our experiments suggested that prompting had a mixed impact on visualization literacy. For \claude, few-shot prompting resulted in worse performance in 29 questions, better performance in 8, and the same for 16 questions. CoT prompting had the exact same results as few-shot. Meanwhile, for \gpt, few-shot prompting resulted in worse performance in 4 questions, better performance in 8, and the same for 41. CoT prompting for \gpt\ also had the exact same performance as few-shot. Finally, for \gemini, few-shot prompting resulted in worse performance in 6 questions, better performance in 16, and the same for 31 questions. In contrast, CoT prompting for \gemini\ resulted in worse performance in 19 questions, better performance in 8, and the same for 26 questions. Overall, \claude\ achieved an abysmal 57.27\% with few-shot and an equally poor 56.76\% with CoT. \gpt\ achieved 93.55\% with few shot and 93.59\% with CoT. Finally, \gemini achieved an impressive 95.20\% with few-shot and less stellar 91.90\% with CoT. Recalling that \claude, \gpt, and \gemini\ scored 79.94\%, 91.11\%, and 93.68\% respectively with the base prompt, we find that there does not seem to be a discernible pattern with few shot or CoT on the visualization literacy of our tested LLMs. Based on \autoref{fig:genCoef}, we see that \gemini\ with few shot and \gpt\ with CoT performed on par with each other and were the best combinations of LLMs and prompting strategies while \claude\ with the base prompt and \gpt\ with the few shot prompt performed a close second. As a result, these results do not seem to support our predictions in \textbf{H5}.

\noindent\textbf{Coefficient Results}
Delving into the coefficients, all three prompt types had statistically significant positive coefficients where the coefficients for base prompt, few shot, and CoT were $\bar{\beta} \approx 1.582$, $\bar{\beta} \approx 0.8422$, and $\bar{\beta} \approx 0.5641$ respectively. Examining the individual LLMs, we see that the coefficients for \claude\ across the prompts were all statistically significant with few shot and CoT being negative ($\bar{\beta} \approx -1.197$ and $\bar{\beta} \approx -0.5705$ respectively). For \gpt, all the coefficients were statistically significant and positive with values of $\bar{\beta} \approx 0.7619$ and $\bar{\beta} \approx 1.414$ for few shot and CoT respectively. Finally, for \gemini, only few shot was statistically significant with a coefficient of $\bar{\beta} \approx 1.413$. \claude's performance is the most baffling, as prompting seems to actively hinder its visualization literacy. We discuss our theories as to why this is affecting \claude\ specifically in Section~\ref{sec:limit}.

\section{Identifying Misleading Chart Elements}
To address \textbf{RQ3}: \textit{To what extent can LLMs identify misleading chart elements?}, we utilized a recently developed dataset called the Misleading ChartQA benchmark\cite{Chen:2025:UDV}. One of the benefits of using this benchmark is that it was fully published on 2026-03-08, reducing the likelihood that any LLM was trained on it. The intent for the Misleading ChartQA benchmark is to gauge the chart question answering capabilities of LLMs when presented with a misleading visualization. However, we believe that chart question answering on misleading visualizations requires an even more essential skill: the ability to identify misleading chart elements. Repurposing their benchmark, we set out to test LLMs' capability to identify misleading chart elements. While exploring this benchmark, we were led to explore another research question:

\begin{itemize}
    \item \textbf{RQ3.1}: To what extent can the latest LLMs' ability to identify misleading chart elements inform us about the subjectivity of various misleaders?
    \item \textbf{RQ3.2}: To what extent can the latest LLMs identify misleading chart elements across different misleaders, categories, and visualizations?
\end{itemize}

\noindent By testing LLMs' ability to identify misleading chart elements with this benchmark, we hoped to specify our selected LLMs' identification strengths and weaknesses within Chen et al.'s framework.

\subsubsection{Experimental Design}
The Misleading ChartQA benchmark\footnote{https://github.com/CinderD/MisleadingChartQA/} contains a total of 3,055 figures, HTML code, data, and question/answer pairs\footnote{The README says there are 3,060 but we only found 3,055.}. Each misleader/visualization pair contained 50 examples except for \textit{Cherry Picking} misleader with \textit{scatterplot} figures, which had 55 examples. The dataset differs slightly from the description in figure 7 of their paper\cite{Chen:2025:UDV}. Specifically, the benchmark was missing the \textit{Dual Axes} misleader, \textit{Dual Encoding} with \textit{bar chart}, \textit{Inappropriate Aggregation} with \textit{100\% stacked bar chart}, and \textit{Inappropriate Scale Range} with \textit{100\% stacked bar chart}.  The benchmark had additional data for \textit{Dual Encoding} with \textit{area chart} and \textit{scatterplot}, \textit{Inappropriate Aggregation} with \textit{stacked bar chart}, and \textit{Inappropriate Scale Range} with \textit{stacked bar chart}.

After thoroughly examining this benchmark, we decided to exclude two misleaders: \textit{Concealed Uncertainty} and \textit{Missing Data}. For \textit{Concealed Uncertainty}, although there is extensive literature on visualizing uncertainty\cite{Ojha:2025:NUU, Reyes:2025:TAD, Wang:2025:AEE}, there appears to be no clear consensus on when visualizations should include uncertainty information. Furthermore, without the context of their question/answer pairs, most of the misleading figures in the benchmark would be misleading.  However, the \textit{Concealed Uncertainty} and \textit{Missing Data} figures do not appear to be misleading without their corresponding question/answer pairs. For these reasons, we decided to exclude the \textit{Concealed Uncertainty} and \textit{Missing Data} figures.

In addition to the misleading figures, we decided to include modified figures to assess how much LLMs ``hallucinate'' misleading chart elements when no misleaders were present. We selected the ``test set'' of figures that were initially published on 2025-08-30. By modifying the HTML code to remove the misleading elements and rendering the new figures to PNG images, we created additional ``corrected'' figures for the LLMs to evaluate.

\subsubsection{Experiment 3}
After finalizing the figures within the Misleading ChartQA benchmark, we tested \claude\ , \gpt\ , and \gemini\ by prompting them to identify all misleading chart elements. Specifically, we asked the LLMs to give all applicable misleading chart elements, listing all of the misleader names and definitions from figure 7 of Chen et al.'s paper\cite{Chen:2025:UDV} in our prompt. Further, our prompt included instructions to provide their answer in the JSON format with an example with fields titled ``misleadElements'' and ``explanation'' for the array of misleading elements and the LLMs' explanation respectively. Finally, we added instructions for the case when they encountered a visualization with no misleading elements, asking them to return a blank array in the ``misleadElements'' field when this occurs. See Appendix~\ref{app:misleadPrompt} for the explicit prompt.

\subsection{Analysis Methods}
Since there were no human trials performed using the Misleading ChartQA benchmark, results primarily focused on comparisons between the LLMs. To address \textbf{RQ3.1}, we explored the responses of LLMs' hallucinated misleaders. We define hallucinated misleaders as the misleaders that the LLMs identified that were not the correct response. We then created Pareto charts of these hallucinated misleaders based on different LLMs and between misleading and corrected chart. To address \textbf{RQ3.2}, we developed the following hypotheses:

\begin{itemize}[leftmargin=*]
\label{hypothesis1.1}
    \item \textbf{H6}: \gpt\ and \gemini\ will be on par for identifying misleading chart elements but both will outperform \claude.
    \item \textbf{H7}: \claude, \gpt, and \gemini\ will perform better on misleading chart elements in the \textit{Manipulated Data} category.
    \item \textbf{H8}: \claude, \gpt, and \gemini\ will perform better on the \textit{Overplotting}, \textit{Inappropriate Order}, \textit{Lack of Legend}, \textit{Lack of Scales} misleaders.
    \item \textbf{H9}: \claude, \gpt, and \gemini\ will perform better on \textit{area chart}, \textit{bar chart}, \textit{line chart}, and \textit{pie chart} visualizations.
\end{itemize}

\noindent We formed \textbf{H6} based on our results in Section~\ref{sec:exp1_llmResults}, arguing that visualization literacy should be correlated with other visualization reasoning tasks. Since \claude\ performed the worst overall compared to \gpt\ and \gemini\ , we believed that \claude\ would perform similarly in this task. For \textbf{H7}, we believed that the majority of figures within the \textit{Manipulated Data} category were easy to identify. Specifically, the figures with the \textit{Overplotting} misleader were overstimulating due to the huge number of data points, which we believed would make it easy to identify. \textit{Inappropriate Order} figures had randomly ordered x-axes or legend categories which we believe would also make identification straightforward. Since both misleaders constituted the majority of figures within the \textit{Manipulated Data} category ($350/505 \approx 69.3\%$), we believed that the LLMs would perform well in this category compared to the others. When forming our hypothesis for \textbf{H8}, we believed these misleaders would be easy to identify. As previously mentioned, we believed that \textit{Overplotting} and \textit{Inappropriate Order} would be easy to identify. We also believe that the two \textit{Lack of Labeling} misleaders (\textit{Lack of Legend} and \textit{Lack of Scales}) would also be easy to identify. Despite these two misleaders falling under the \textit{Manipulated Annotation} category, we did not include it in \textbf{H7}. The reason for this is because the figures with the \textit{Lack of Legend} and \textit{Lack of Scales} misleaders did not constitute a majority within the \textit{Manipulated Annotation} category ($250/650=\approx 38.5\%$). Finally, for \textbf{H9}, we believed that simple visualizations (i. e., visualizations that have minimal marks and channels) would be easier for LLMs to analyze, and hence, be easier for them to identify misleading chart elements. Although \textit{scatterplot} may be considered a simple visualization, we did not include it in \textbf{H9}.

\begin{table*}[t]
    \centering
    \caption{Summary table of hypotheses, statistical methods, and results used to support/not support for \textbf{RQ3}}
    \begin{tabular}{l|c|l|l|l}
        \hline
        RQ              &Hypothesis &Methods             &Results &Result Section(s)\\
        \hline
        \textbf{RQ3.2}  &\textbf{H6}
                        &Coefficient Analysis
                        &Partially supported
                        &Section~\ref{sec:misleadLLMDiffResults}\\
        \textbf{RQ3.2}  &\textbf{H7}
                        &Boxplot comparisons / Coefficient Analysis
                        &Partially supported 
                        &Section~\ref{sec:misleadCoefResults}\\
        \textbf{RQ3.2}  &\textbf{H8}
                        &Boxplot Comparisons / Coefficient Analysis
                        &Partially supported 
                        &Section~\ref{sec:misleadCoefResults}\\
        \textbf{RQ3.2}  &\textbf{H9}
                        &Boxplot comparisons / Coefficient Analysis
                        &Partially supported 
                        &Section~\ref{sec:misleadCoefResults}\\
        \hline
    \end{tabular}
    \label{tab:hypotheses_exp3}
    \vspace{-0.5em}
\end{table*}

Based on these hypotheses, we conducted hypothesis tests to examine the impact of four dimensions (misleader category, misleader, visualization type, and LLM type) on the LLMs' performances in identifying misleading chart elements. Before we began modelling our data, we needed to transform the dependent variable (the set of misleaders given by the LLMs' responses) into a quantity. We chose to convert these result into two different metrics: the Jaccard Index\cite{costa:2021:further} and the Dice-S{\o}rensen Coefficient\cite{Jackson:1989:SCM} (see Appendix~\ref{app:responseTransform} for more details). These metrics lie within $(0,1)$, which can be interpreted as percentages or proportions. Thus, we modelled the results using beta regression over these dimensions since our transformed dependent variable lies within $(0,1)$\cite{geissinger:2022:case} (see Appendix~\ref{app:betaReg} for more details). Since all independent variables are categorical or binary, exploring them and their interactions was sufficient resulting in a total of 477 variables and interactions (more details in Appendix~\ref{app:varInterCalcPrompt}).

\subsection{Results}
We ran a total of $\left(2,855\text{ (misleading charts)}+285\text{ (corrected charts)}\right) \times 3\text{ (LLMs)}=9,420$ trials for Experiment 3, with each trial comprising the index, figure type (misleading or corrected), misleader category, misleader name, visualization type, and LLM response. The overall hypothesis results are summarized in \autoref{tab:hypotheses_exp3}. Our reported results for our hypotheses are for the Jaccard Indices while some of the results for the Dice-S{\o}rensen Coefficients can be seen in Appendix~\ref{app:dsCoefResults}. The rest are in our supplemental material.

\begin{figure}[h]
    \centering
    \includegraphics[width=0.49\textwidth]{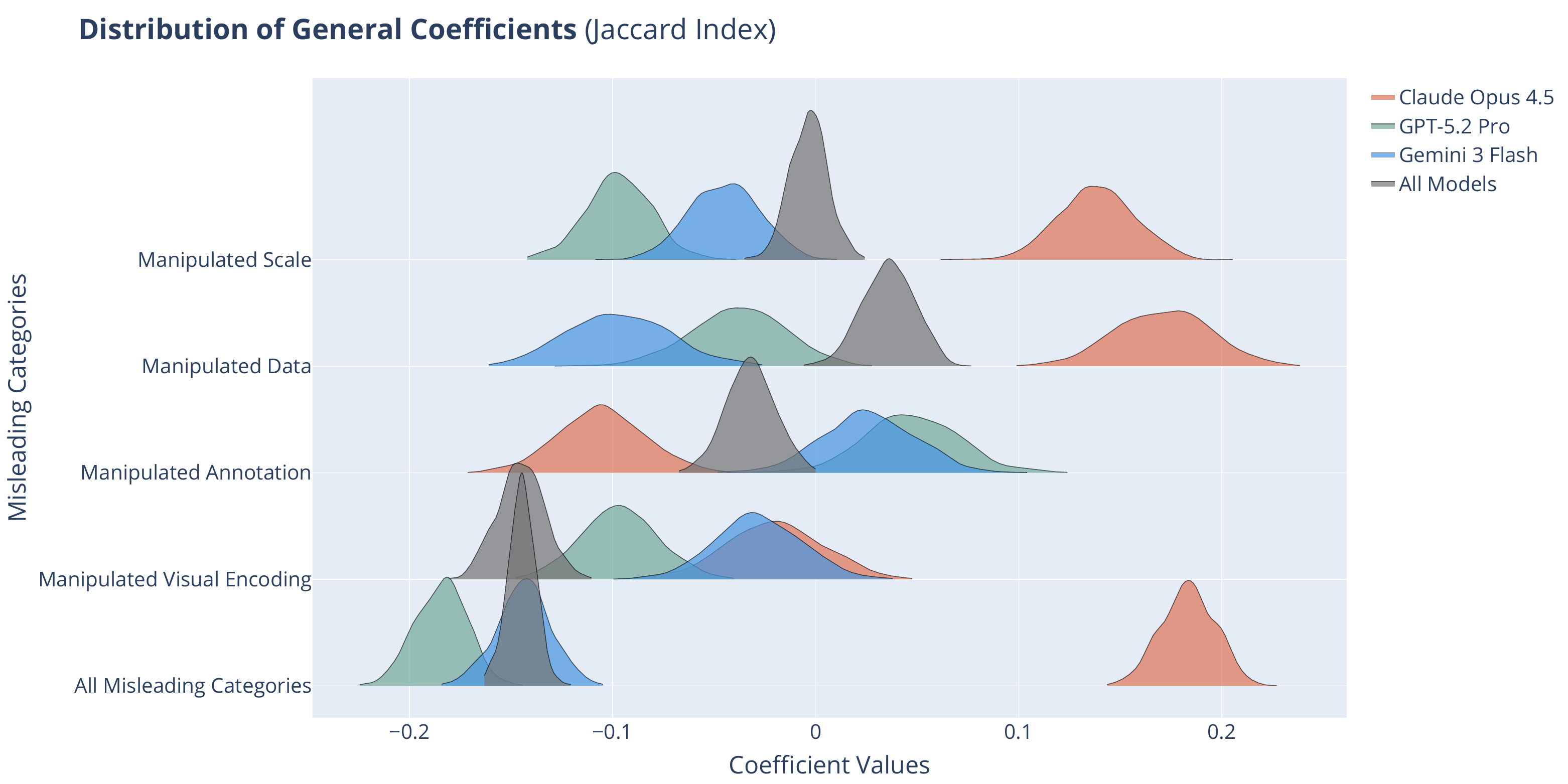}
    \caption{Ridge plot of the bootstrapped general coefficients from the beta regression of Jaccard Indices.}
    \label{fig:genCoefMislead}
    \vspace{-1em}
\end{figure}

\subsubsection{Hallucinated Pareto Charts}\label{sec:halParetoChart}
We collected the hallucinated responses across all misleader and visualization combinations and plotted Pareto charts by different LLMs. Focusing on \claude, we note that the most hallucinated misleader was \textit{Cherry Picking} followed by \textit{Deceptive Labeling} for both the misleading and corrected visualizations. For \gpt, \textit{Small Size} was the most hallucinated followed by \textit{Deceptive Labeling} again for both the misleading and corrected visualizations. For \gemini, \textit{Small Size} was the most hallucinated by a large margin followed by \textit{Data-visual Disproportion} for misleading visualizations and \textit{Inappropriate Scale Range} for corrected visualizations. Finally, for all LLMs, \textit{Small Size} was the most hallucinated followed by \textit{Deceptive Labeling} for both misleading and corrected visualizations. \claude\ had the least number of hallucinated misleaders for both misleading (2,291) and corrected (250) visualizations, while \gemini\ had the most for both misleading (5,191) and corrected (625). \gpt\ had 4,364 hallucinated misleaders for the misleading visualizations and 418 for corrected visualizations. To see all of the hallucinated misleaders broken down by LLM and misleading/corrected visualizations, see the Pareto charts in Figures~\ref{fig:paretosGen}, \ref{fig:paretosClaude}, \ref{fig:paretosGPT}, and \ref{fig:paretosGemini}.

\subsubsection{\textbf{H6} LLMs on Misleading Identification}\label{sec:misleadLLMDiffResults}
Comparing the coefficients of each LLM, \claude\ had a statistically significant positive coefficient ($\bar{\beta} \approx 0.2234$), which indicates the \claude\ was more likely to select the correct misleading chart element. Meanwhile, \gpt\ and \gemini\ had statistically significant negative coefficients ($\bar{\beta} \approx -0.2148$ and $\bar{\beta} \approx -0.1012$ respectively), indicating that they were more likely to select incorrect misleading chart elements. This seems evident in the bottom row of \autoref{fig:genCoefMislead}. See \autoref{tab:genMeanCoef_jIndex} in the Appendix for all coefficients.

\subsubsection{\textbf{H7 \& H8 \& H9} LLMs on Misleading Identification}\label{sec:misleadCoefResults}
\textbf{Boxplot comparison} Based on these boxplots, all LLMs seemed to perform poorly (\ie, median appears at or near zero) on the following misleader/visualization combinations: \textit{Cherry Picking} for \textit{line chart}; \textit{Deceptive Labeling} for \textit{pie chart}; \textit{Inappropriate Aggregation} for \textit{area chart}, \textit{bar chart}, and \textit{line chart}; \textit{Continuous Encoding of Categorical Data} for \textit{pie chart}, and \textit{Inappropriate Scale Functions} for \textit{pie chart}. None of the boxplots showed all the LLMs performing well at the same time (\ie, median appears at or near one). \claude\ performed poorly (\ie, median at zero) on 20 misleader/visualizations and performed well on 21 (\ie, median at one); \gpt\ performed poorly on 17 and performed well on 6; and \gemini\ performed poorly on 10 and performed well on 2. Thus, it appears that \claude\ outperformed both \gpt\ and \gemini. A full table of the Jaccard Indices boxplots can be found in \autoref{tab:results_jindex_ex3} of the Appendix.

\noindent\textbf{Coefficient Results}\label{sec:exp3_misleadResults}
Of the 477 variables, 461 were statistically significant, indicating that the majority of these combinations have an effect on LLMs' ability to identify misleading chart elements. A full reporting of all coefficients is provided in the Appendix \autoref{tab:jIndexInsig}. We report the detailed coefficients in our supplemental material. When focusing on misleading categories, all LLMs performed better on \textit{Manipulated Data} ($\bar{\beta} \approx 0.08251$) and \textit{Manipulated Scale} ($\bar{\beta} \approx 0.01662$), partially confirming our prediction in \textbf{H7}. The coefficients for all the misleaders were significant with positive values for \textit{Misuse of Cumulative Relationship}, \textit{Small Size}, \textit{Unconventional Scale Directions}, \textit{Deceptive Labeling}, \textit{Lack of Legend}, \textit{Lack of Scales}, \textit{Data-visual Disproportion}, \textit{Categorical encoding for continuous data}, \textit{Overplotting}, and \textit{Inappropriate Order}. Although \textbf{H8} correctly included four misleaders, our hypothesis did not include six of them. Finally, for the visualizations, all of their coefficients were significant with positive values for \textit{100\% stacked bar chart}, \textit{stacked bar chart}, \textit{stacked area chart}, \textit{scatterplot}, \textit{line chart}, and \textit{heatmap}. Our predictions in \textbf{H9} were abysmal: we only correctly predicted \textit{line chart}. See Tables~\ref{tab:genMeanCoef_jIndex}, \ref{tab:misleaderMeanCoef_jIndex}, \ref{tab:chartMeanCoef_jIndex}, and \ref{tab:misleaderChartMeanCoef_jIndex} in the Appendix for all coefficient values.

\section{Discussion}\label{sec:discuss}
Based on our experiments and analysis, we find that the state-of-the-art LLMs we tested possess visualization literacy. As previously reported, \gemini\ with few-shot and \gpt\ with CoT performed the best, and we recommend these LLM/prompt combinations for tasks that heavily rely on visualization literacy alone. Such tasks include cases where LLMs show superior performance where human perception is limited, such as comparing values within a stacked bar chart. As the VLAT scores illustrate, humans struggle to compare category values in stacked bar charts when those values are not on the base axis. In contrast, LLMs excel at this task and could be used to generate a table of the data or to query specific values. Our results suggest another promising direction for LLMs in visualization evaluation, which is to leverage their visualization literacy to detect data-visual discrepancies. By cross-referencing visualizations with raw datasets or external open-source data, they can be well-positioned to identify instances where a visualization contradicts its underlying numerical values or violates widely accepted public knowledge.

Our results demonstrate that \gpt\ shows the highest degree of robustness across all tested prompting strategies. Its performance remains largely consistent regardless of the specific prompt techniques, indicating a low sensitivity to prompt variation. This stability suggests that \gpt\ is a reliable choice for diverse users with varying levels of expertise in prompt engineering.
Although our mixed results with prompt strategies do not align with other research that shows CoT improved LLM performance \cite{Alexander:2024, Das:2026:misvisfix, Lo:2025:HGB}, we believe that this is because most research on CoT performs multiple requests from LLMs to coax them to the correct answer. In contrast, we only performed a neutral, single-turn, zero-shot approach with our prompting strategies, which may contribute to CoT's mixed results. Further research is necessary to illuminate these divergent outcomes.


Despite possessing visualization literacy, LLMs are inconsistent when identifying misleading chart elements. The purpose of using LLMs for visualization evaluation is for them to have expert-level judgment when criticizing visualizations. Thus, having perfect or near-perfect capabilities for identifying misleading chart elements is a crucial step towards LLMs becoming effective visualization evaluators. Despite the current models underperforming on this task, it is encouraging to see that at least one LLM was able to perform relatively well which shows that there is potential for LLMs to eventually reach the performance necessary for visualization evaluation. \claude, despite performing the worst on the modified VLAT, performed the best on identifying misleading chart elements. Furthermore, \claude\ hallucinated the least when identifying misleading chart elements for both misleading and corrected charts. On the topic of hallucinated misleaders, the hallucination Pareto charts could be interpreted as quantifying the subjectivity of the misleaders. This is supported by the fact that cumulatively for all LLMs, the most frequently hallucinated chart element was the \textit{Small Size} misleader. The classification of this element is fundamentally subjective, as it depends on relative perceptions, \ie, what constitutes `too small` varies significantly between automated evaluators and human viewers, display context, and visual acuity.

Finally, there are large discrepancies when we compare LLMs' costs. \gpt\ was the most expensive costing $\approx\$1,169.34$ across all experiments. This was followed by \claude\, which cost $\approx\$554.33$. Finally, \gemini\ cost $\$194.21$ across all these experiment. 
\gemini's current performance to cost ratio far exceeds the other two models and is the cheapest LLM to test. Ignoring cost, we believe \claude\ performed the best in terms of all visualization evaluation tasks tested.

\section{Limitations and Future Work}
\noindent\textbf{Limitations}\label{sec:limit}
In Experiments 1 and 2, \claude\ seemed to perform worse than expected, especially compared to \gpt\ and \gemini. Upon further investigation, we noticed that \claude\ seemed always to provide large explanations behind its responses rather than just the lettered answer. This may be due to the nature of the Opus line of models. Anthropic states that Opus should be used for ``[c]omplex reasoning that requires multiple steps'', while its Sonnet models should be used for ``[a]nalysis that needs reasoning but isn't extremely complex''\footnote{https://claude.com/resources/tutorials/choosing-the-right-claude-model}. We developed an automated parsing script to programmatically verify answer accuracy to facilitate the large-scale evaluation of LLM responses. While this approach enabled efficient analysis, its reliance on deterministic value matching may lack the nuanced interpretation that a manual review process might provide. Since \claude\ would thoroughly check the validity of each choice in the question, which would include all the lettered choices, our code would flag its response as incorrect even if it arrived at the correct answer. Interestingly, if we had used the lower-tier version of \claude\ (\texttt{sonnet-4.5}), \claude\ may have achieved a higher score. While adopting a structured output format, such as a JSON format, would have streamlined the evaluation and analysis process and mitigated the parsing issues of \claude, we intentionally maintained the original prompt structure for Experiments 1 and 2. We followed the prompt verbatim from previous work to ensure direct comparability with the baseline results established. This prioritization of experimental consistency allowed us to conduct a longitudinal comparison between LLMs of different generations, and we later relaxed our prompt in Experiment 3 to explore complicated NLP-based explanation analysis.

\noindent\textbf{Future Work}\label{sec:future}
Our work can be expanded upon in a variety of ways. Firstly, we are interested in exploring the impacts of prompting on other visualization tasks beyond literacy, such as chart editing and generation. On the topic of misleading visualizations, we plan to expand Experiment 3 from visualizations having one misleader to having multiple misleaders. Another potential future research lies in establishing a human baseline for misleading chart element identification. By quantifying human capability, we can conduct a rigorous comparison between LLMs' performance and humans'. Furthermore, while the current studies decomposed LLMs' proficiency in visualization evaluation into three skills, future investigations should explore holistic evaluation paradigms. Rather than assess isolated capabilities, researchers could ask LLMs to evaluate a random visualization to see their effectiveness in performing end-to-end quality assessments.

\clearpage



\bibliographystyle{abbrv-doi-hyperref}

\bibliography{template}

\clearpage

\appendix 
\crefalias{section}{appendix} 

\section{Initial Research and Pilot Study Results}
\autoref{tab:results_pilot} outlines our results for our pilot study on ChatGPT-5.1. We develop hypotheses \textbf{H3} and \textbf{H4} based on these results.

\section{Results of LLMs' Visualization Literacy and Prompting Experiments}
\subsection{Example of Prompts}\label{app:promptExamples}
\autoref{tab:fewShotPrompt} shows an example of a few shot prompt, whiel \autoref{tab:cotPrompt} shows example of a CoT prompt.

\subsection{Variables and Interactions}\label{app:varInterCalcPrompt}
For the logistic regression model, recall that we denote $\textbf{x}_k$ where $k \in \{1,2,3,4\}$ as the variables corresponding to the $k$-way interaction variables (see \autoref{eq:promptModel}). More explicitly, let $V$ be the set of visualization types, $T$ be the set of task types, $L$ be the set of LLMs, and $P$ be the set of prompting techniques. Then, let $D$ be the family of these dimension sets such that $\{V,T,L,P\} \equiv D$. Let 
$D_k=\begin{pmatrix}
    D\\
    k\\
\end{pmatrix}$ denote the $k$-subset of $D$ where $|D_k|=m$, and $D_{k,j}$ denote the $j$th element in $D_k$ where $j \in \{1,\dots,m\}$ for some ordering. For example, $D_{2,1}$ could be $\{V,T\}$. Then, let $C(D_{k,j})=\prod\limits_{S \in D_{k,j}}S$ be the Cartesian product of $D_{k,j}$, and $C(D_{k,j})_l$ denote the $l$th tuple in $C(D_{k,j})$ where $l \in \{1,\dots,n_j\}$ and $n_j=|C(D_{k,j})|$. For instance, $C(D_{2,1})=\prod\limits_{S \in \{V,T\}}S=V \times T$, which describes all two-way interactions between visualization and task types. This includes visualization/task interactions that were not tested, \eg, Line Chart and Find Anomalies, though in reality, we did not include them in our model\footnote{The same thing can be accomplished by including these variables but setting their coefficients to zero}. Then $\textbf{x}_k$ can be expressed as the following:

\begin{equation}
    \textbf{x}_k=
    \begin{bmatrix}
        \prod\limits_{i \in C(D_{k,1})_1}x_i\\
        \vdots\\
        \prod\limits_{i \in C(D_{k,1})_{n_1}}x_i\\
        \vdots\\
        \prod\limits_{i \in C(D_{k,m})_1}x_i\\
        \vdots\\
        \prod\limits_{i \in C(D_{k,m})_{n_m}}x_i\\
    \end{bmatrix}
\end{equation}

\noindent where $\textbf{x}_k$ is a $(n_1+\dots+n_m) \times 1$ vector, and $x_i$ is a binary variable that determines whether variable $i$ is included where $i$ is an element of the $C(D_{k,j})_l$ tuple. With 12 visualization types, 8 task types, 3 LLMs, 3 prompting types, and 49 unique visualization and task type interactions, the number of variables with their interactions is calculated as follows:\\

\noindent One-way: $12+8+3+3=26$\\
Two-way: $49+(12 \times 3)+(12 \times 3)+(8 \times 3)+(8 \times 3)+(3 \times 3)=178$\\
Three-way: $(49 \times 3)+(49 \times 3)+(12 \times 3 \times 3)+(8 \times 3 \times 3)=474$\\
Four-way: $49 \times 3 \times 3=441$\\

\noindent Thus, a total of $26+178+474+441=1,119$ variables and interactions were used for the logistic regression model in experiments 1 and 2 with an additional variable for the intercept.

\subsection{Bootstrapping}
Due to the prompt imbalance between the base prompt and few shot and CoT, we decided to downsample data from the data in Experiment 1 before bootstrapping the data. Specifically, the base prompt dataset contained 120 trials per question, which is 2.5 times more trials per question compared to few shot and CoT ($24\text{ trials} \times 2\text{ example types}=48\text{ total trials}$). Thus, each bootstrapped model had $53\text{ (questions)} \times 48\text{ (trials/question)} \times 3\text{ (LLMs)} \times 3\text{ (prompt type)}=22,896$ data points. Each time we fit a bootstrapped model, we stored the 1,120 coefficients that resulted from that model. We then analyzed the coefficients from a distribution perspective to see if they were different from zero, which allowed us to answer \textbf{RQ1.3} and \textbf{RQ2} and test hypotheses \textbf{H2}, \textbf{H3}, \textbf{H4}, and \textbf{H5}. \autoref{tab:bootstrap} shows the structure of how we stored the coefficient data and how the probabilities were associated with each bootstrap model.

\begin{table}[!h]
    \centering
    \caption{The table demonstrating the structure of our bootstrap methodology. The main model is based on all of the data from experiments 1 and 2, while the bootstrap models are based on sampling the same data with replacement.}
    \begin{tabu}{l|llll|l}
        \multirow{2}*{\textbf{Models}}  &   \multicolumn{4}{c|}{\textbf{Coefficients}}                                                              &   \multirow{2}*{\textbf{Probabilities}}\\
        \cline{2-5}
                                        &   \textbf{Intercept}      &   \textbf{Coef. 1}        &  \textbf{\dots}   &  \textbf{Coef. 1119}          &\\
        \hline
        \textbf{Main}                   &   $\hat{\beta}_0$         &   $\hat{\beta}_1$         &   $\dots$         &   $\hat{\beta}_{1119}$        &   $\hat{P}(y=1|\textbf{x}),\dots$\\
        \hline
        \textbf{Bootstrap 1}            &   $\hat{\beta}_{0,1}$     &   $\hat{\beta}_{1,1}$     &   $\dots$         &   $\hat{\beta}_{1119,1}$      &   $\hat{P}_{1}(y=1|\textbf{x}),\dots$\\
        $\vdots$                        &   $\vdots$                &   $\vdots$                &   $\ddots$        &   $\vdots$                    &   $\vdots$\\
        \textbf{Bootstrap 1000}         &   $\hat{\beta}_{0,1000}$  &   $\hat{\beta}_{1,1000}$  &   $\dots$         &   $\hat{\beta}_{1119,1000}$   &   $\hat{P}_{1000}(y=1|\textbf{x}),\dots$\\
        \hline
    \end{tabu}
    \label{tab:bootstrap}
\end{table}

\subsection{Bootstrapping Results}
As previously mentioned in Section~\ref{sec:exp1_chartTaskResults}, all 1,120 coefficients were found to be statistically significant (p-value$<0.05$) except for 268, which are detailed in our supplemental material. The general coefficients (\ie, LLM and prompt type variables and interactions) are visualized in \autoref{fig:genCoef} as a ridge plot and were found to be statistically significant (see \autoref{tab:genCoefResults}). A table of the mean visualization coefficients can be found in \autoref{tab:vizCoefResults}, and a similar table for the mean task coefficients can be found in \autoref{tab:taskCoefResults}.  The visualization/task interaction coefficients for the base prompt are visualized in \autoref{fig:chartTaskInterCoef} as a ridge plot with all of the mean values found in \autoref{tab:vizTaskCoefResults}.

\section{Results of LLMs' Misleading Chart Identification}
\subsection{Misleading Chart Identification Prompt}\label{app:misleadPrompt}
\autoref{fig:misleadPrompt} shows that prompt that we used for Experiment 3.

\subsection{Variables and Interactions}\label{app:varInterCalcMislead}
The beta regression model has similar variable structure as the logistic model outlined in Section~\ref{app:varInterCalcPrompt}. Let $M$ be the set of misleader categories, $T$ be the set of misleaders, $V$ be the set of visualization types, and $L$ be the set of LLMs. Thus, $D$ would be the family of these dimension sets such that $\{M,T,V,L\} \equiv D$. With 4 misleader categories, 18 misleaders, 10 visualization types, 3 LLMs, 57 unique misleader and visualization interactions, and $6+7+7+9=29$ unique misleader category and visualization interactions, the number of variables with their interactions is calculated as follows:\\

\noindent One-way: $4+18+10+3=35$\\
Two-way: $29+(4 \times 3)+57+(18 \times 3)+(10 \times 3)=182$\\
Three-way: $(29 \times 3)+(57 \times 3)=258$\\

\noindent Thus, a total of $35+182+258=475$ variables and interactions were used for the beta regression model in Experiment 3 with two additional variable for the intercept and precision (see Section~\ref{app:betaReg} for more details). Note that since there is a one-to-many relationship between misleader categories and misleaders, the four-way interactions would be redundant with the misleader/chart/LLM interactions.

\subsection{Beta Regression Model in Experiment 3}\label{app:betaReg}
Beta regression assumes that that dependent variable $y$ is distributed as a beta regression. Thus, its probability distribution would be as follows:

\begin{equation}
\label{eq:betaDist}
    f(y;\mu,\phi)=\frac{\Gamma(\phi)}{\Gamma(\mu\phi)\Gamma((1-\mu)\phi)}y^{\mu\phi-1}(1-y)^{(1-\mu)\phi-1}
\end{equation}

\noindent where $\Gamma(\cdot)$ is the gamma function, $0<y<1$, $0<\mu<1$, and $\phi>0$. Here, $\mu$ refers to the mean and $\phi$ to the precision. Like other generalized linear models (GLMs), beta regression relies on a link function. The most common link function used in beta regression is the logit link function, which is the same function used in logistic regression. Thus, like logistic regression, the independent variables and mean value $\mu$ are related as follows:

\begin{equation}
\label{eq:misleadModel}
    \begin{split}
        g(\mu_i)=\beta_0+\boldsymbol{\beta}_1^T\textbf{x}_{i,1}+\boldsymbol{\beta}_2^T\textbf{x}_{i,2}+\boldsymbol{\beta}_3^T\textbf{x}_{i,3}\\
        \longrightarrow \mu_i=g^{-1}\left(\beta_0+\boldsymbol{\beta}_1^T\textbf{x}_{i,1}+\boldsymbol{\beta}_2^T\textbf{x}_{i,2}+\boldsymbol{\beta}_3^T\textbf{x}_{i,3}\right)
    \end{split}
\end{equation}

\noindent where $g(\cdot)$ is the logit link function, $\textbf{x}_k$ represents the vector of $k$-way interaction variables ($k \in \{1,2,3\}$), and $\boldsymbol{\beta}_k$ represents a vector of coefficients for each of these variables. To solve for the $\boldsymbol{\beta}_k$ for $k \in \{1,2,3\}$ and $\phi$, we take the gradient of the log-likelihood function of \autoref{eq:betaDist}, set this function to zero, and solve for these values. The log-likelihood function is as follows:

\begin{equation}
\label{eq:loglike}
    \ell(\mu,\phi)=\sum\limits_{i=1}^{n}\ell(\mu_i,\phi)
\end{equation}

\begin{multline}
\label{eq:loglike_i}
    \ell_i(\mu_i,\phi)=\log\Gamma(\phi)-\log\Gamma(\mu_i\phi)-\log\Gamma((1-\mu_i)\phi)\\
    +(\mu_i\phi-1)\log y_i+\left((1-\mu_i)\phi-1\right)\log(1-y_i)
\end{multline}

\noindent Unlike logistic regression, there is no closed-form solution and requires numerical optimization to solve\cite{cribariNeto:2010:beta}.

\subsection{Variable Transformations in Experiment 3}\label{app:responseTransform}
The \textit{Jaccard Index}~\cite{costa:2021:further} is defined as the following:

\begin{equation}
\label{eq:Jaccard}
J(A,R)=\frac{|A \cap R|}{|A \cup R|}=y
\end{equation}

\noindent where $A$ is the answer set (\ie, the misleader associated with the misleading chart), $R$ is the response set (\ie, the array of misleaders in the LLM's response), $A \cap R$ is the intersection of $A$ and $R$, and $A \cup R$ is the union of $A$ and $R$. The Jaccard Index is always between 0 (no overlap) and 1 (identical sets) and compares the overall similarity between the answer set and the response set. For answer sets that are empty (\ie, when the chart has no misleading chart elements), it boils down to a binary result (\ie, if $|R|=0$, then $J(A,R)=1$; if $|R|>0$, then $J(A,R)=0$).

The \textit{Dice-S{\o}rensen Coefficient}~\cite{Jackson:1989:SCM} is defined as follows:

\begin{equation}
\label{eq:Sorensen}
D(A,R)=\frac{2|A \cap R|}{|A| + |R|}=y
\end{equation}

\noindent where the variables are the same as the Jaccard Index. This metric results in a measure that expresses the overlap in relation to the total size of both sets with a value ranging from 0 (indicating no overlap) to 1 (indicating identical intervals). The Dice-S{\o}rensen Coefficient results in the same calculation as the Jaccard Index when answer set is the empty set.

\subsection{Bootstrapping}
Each bootstrapped model had $2844\text{ (misleading charts)} \times 3\text{ (LLMs)}=8532$ data points. Each time we fit a bootstrapped model, we stored the 477 coefficients that resulted from that model. We then analyzed the coefficients from a distribution perspective to see if they were different from zero, which allowed us to answer \textbf{RQ3.2} and test hypotheses \textbf{H6}, \textbf{H7}, \textbf{H8}, and \textbf{H9}. We store the results similar to what is shown in \autoref{tab:bootstrap}, but instead of 1,120 variables, it is 477.

\subsection{Bootstrapping Results: \textit{Jaccard Index}}
As previously mentioned in Section~\ref{sec:misleadCoefResults}, 461 coefficients were found to be statistically significant (p-value$<0.05$) except for 16, which we list in \autoref{tab:jIndexInsig}. The general coefficients (\ie, LLM and visualization presence variables and interactions) are visualized in \autoref{fig:genCoef} as a ridge plot and were found to be statistically significant (see \autoref{tab:genCoefResults}). A table of the mean visualization coefficients can be found in \autoref{tab:vizCoefResults}, and a similar table for the mean task coefficients can be found in \autoref{tab:taskCoefResults}.  The visualization/task interaction coefficients when the visualization is present are visualized in \autoref{fig:chartTaskInterCoef} as a ridge plot with all of the mean values found in \autoref{tab:vizTaskCoefResults}.

\begin{table*}[h]
    \centering    
    \caption{Table of coefficients for LLMs and misleader category interactions.  \textbf{Coef.} refers to the coefficient in the logistic model using results in experiments 1 and 2, while the \textbf{Mean Coef.} refers to the mean of the bootstrapped coefficients.  When \textbf{Normal?} is true, a one-sample t-test is used, while if it is false, the Wilcoxon signed-rank test is used.  Note that when the Wilcoxon signed-rank test is used, the bounds are estimated from the ECDF.  Thus, bounds calculated from the ECDF that contain zero indicate poor estimations.}
    \begin{tabu}{l|l|S[table-format=1.4e1]|S[table-format=1.4e1]|S[table-format=1.4e3]|l|S[table-format=1.4e1]|S[table-format=1.4e1]}
        \hline
        \textbf{LLM}                &\textbf{Category}          &\textbf{Coef.} &\textbf{Mean Coef} &\textbf{p-value}   &\textbf{Normal?}   &\textbf{95\% LB}   &\textbf{95\% UB}\\
        \hline
        \multirow{5}*{\gptFig}      &Manipulated Data           &-0.02807       &-0.02789           &3.945E-201         &True               &-0.02930           &-0.02647\\
                                    &Manipulated Annotation     &0.08590        &0.08647            &0                  &True               &0.08497            &0.08798\\
                                    &Manipulated Visual Encoding&-0.1495        &-0.1512            &0                  &True               &-0.1525            &-0.1499\\
                                    &Manipulated Scale          &-0.1217        &-0.1222            &0                  &True               &-0.1232            &-0.1212\\
                                    &All                        &-0.213         &-0.2148            &0                  &True               &-0.2156            &-0.2141\\
        \hline
        \multirow{5}*{\geminiFig}   &Manipulated Data           &-0.1150        &-0.1160            &0                  &True               &-0.1176            &-0.1144\\
                                    &Manipulated Annotation     &0.04961        &0.04892            &0                  &True               &0.04750            &0.05034\\
                                    &Manipulated Visual Encoding&-0.01232       &-0.01131           &2.833E-46          &True               &-0.01278           &-0.009835\\
                                    &Manipulated Scale          &-0.02204       &-0.02282           &4.118E-238         &True               &-0.02383           &-0.02181\\
                                    &All                        &-0.09972       &-0.1012            &0                  &True               &-0.1021            &-0.1004\\
        \hline
        \multirow{5}*{\claudeFig}   &Manipulated Data           &0.2252         &0.2264             &0                  &True               &0.2250             &0.2278\\
                                    &Manipulated Annotation     &-0.1436        &-0.1449            &0                  &True               &-0.1463            &-0.1435\\
                                    &Manipulated Visual Encoding&-0.02033       &-0.01980           &1.118E-103         &True               &-0.02139           &-0.01820\\
                                    &Manipulated Scale          &0.1603         &0.1617             &0                  &True               &0.1605             &0.1628\\
                                    &All                        &0.2215         &0.2234             &0                  &True               &0.2226             &0.2242\\
        \hline
        \multirow{6}*{All}          &Manipulated Data           &0.08213        &0.08251            &0                  &True               &0.08178            &0.08325\\
                                    &Manipulated Annotation     &-0.008123      &-0.009491          &5.767E-113         &True               &-0.01021           &-0.008769\\
                                    &Manipulated Visual Encoding&-0.1822        &-0.1823            &0                  &True               &-0.1831            &-0.1816\\
                                    &Manipulated Scale          &0.01646        &0.01662            &7e-323             &True               &0.01606            &0.01718\\
                                    &All (Intercept)            &-0.09173       &-0.09268           &0                  &False              &-0.1065            &-0.07962\\
                                    &All (Precision)            &0.4879         &0.5062             &0                  &True               &0.5045             &0.5079\\
        \hline
    \end{tabu}
    \label{tab:genMeanCoef_jIndex}
\end{table*}

\begin{table*}[!h]
    \centering
    \caption{Mean coefficients of Jaccard Indices for misleaders (no interactions).}
    \begin{tabu}{l|l|S[table-format=1.5e1]|S[table-format=1.5e1]|S[table-format=1.5e1]}
        \hline
        \multirow{2}*{\textbf{Misleader}}       &\multicolumn{4}{c}{\textbf{LLM}}\\
        \cline{2-5}
                                                &\textbf{GPT}   &\textbf{Gemini}    &\textbf{Claude}    &\textbf{All}\\
        \hline
        Missing Normalization                   &-0.1355        &-0.007078          &-0.09353           &-0.2361\\
        Cherry Picking                          &-0.4627        &-0.01154           &-0.03731           &-0.5115\\
        Overplotting                            &0.4584         &-0.1449            &0.2161             &0.5296\\
        Inappropriate Order                     &0.1118         &0.04752            &0.1411             &0.3005\\
        Deceptive Labeling                      &0.03193        &0.3871             &-0.2617            &0.1574\\
        Lack of Legend                          &-0.1035        &0.2308             &0.3914             &0.5187\\
        Lack of Scales                          &0.4352         &-0.3865            &0.4829             &0.5316\\
        Inappropriate Aggregation               &-0.2771        &-0.1825            &-0.7575            &-1.217\\
        Data-visual Disproportion               &-0.3700        &0.2856             &0.3054             &0.2210\\
        Dual Encoding                           &-0.1041        &0.3069             &-0.4630            &-0.2601\\
        Continuous encoding of categorical data &0.3351         &-0.4228            &-0.1227            &-0.2103\\
        Categorical encoding of continuous data &-0.01228       &-0.1810            &0.2604             &0.06708\\
        Inappropriate Scale Range               &0.4936         &-0.2536            &-0.4974            &-0.2574\\
        Inappropriate Scale Functions           &-0.1779        &-0.1197            &-0.6405            &-0.9381\\
        Unconventional Scale Directions         &0.006932       &-0.3356            &0.9487             &0.6200\\
        Misuse of Cumulative Relationship       &-0.5255        &0.2857             &0.5011             &0.2613\\
        Exceeding the Canvas                    &-0.1481        &0.3333             &-0.3099            &-0.1246\\
        Small Size                              &0.2287         &0.06704            &0.1596             &0.4554\\
        \hline
    \end{tabu}
    \label{tab:misleaderMeanCoef_jIndex}
\end{table*}

\begin{sidewaystable*}[h]
    \sisetup{exponent-mode = scientific,
                     exponent-product=\times,
                     tight-spacing,
                     round-mode = figures,
                     round-precision=4
             }
    \centering    
    \caption{Mean coefficients of Jaccard Indices for visualizations (no interactions with misleader).}
    \scalebox{0.9}{
        \begin{tblr}{
            rowsep=0pt,
            colspec={@{} l|l|
                        Q[l, si={table-format=1.4e1}]|
                        Q[l, si={table-format=1.4e1}]|
                        Q[l, si={table-format=1.4e1}]|
                        Q[l, si={table-format=1.4e1}]|
                        Q[l, si={table-format=1.4e1}]|
                        Q[l, si={table-format=1.4e1}]|
                        Q[l, si={table-format=1.4e1}]|
                        Q[l, si={table-format=1.4e1}]|
                        Q[l, si={table-format=1.4e1}]|
                        Q[l, si={table-format=1.4e1}]
                   @{}},
            row{1} = {guard},
            cell{2}{1}={r=5}{c},
            cell{7}{1}={r=5}{c},
            cell{12}{1}={r=5}{c},
            cell{17}{1}={r=5}{c}
        }
            \hline
            \textbf{LLM}&\textbf{Category}          &\rotatebox[origin=bl]{70}{\textbf{100\% Stacked Bar Chart}}
                                                    &\rotatebox[origin=bl]{70}{\textbf{Area Chart}}
                                                    &\rotatebox[origin=bl]{70}{\textbf{Bar Chart}}
                                                    &\rotatebox[origin=bl]{70}{\textbf{Choropleth Map}}
                                                    &\rotatebox[origin=bl]{70}{\textbf{Heatmap}}
                                                    &\rotatebox[origin=bl]{70}{\textbf{Line Chart}}
                                                    &\rotatebox[origin=bl]{70}{\textbf{Pie Chart}}
                                                    &\rotatebox[origin=bl]{70}{\textbf{Scatterplot}}
                                                    &\rotatebox[origin=bl]{70}{\textbf{Stacked Area Chart}}
                                                    &\rotatebox[origin=bl]{70}{\textbf{Stacked Bar Chart}}\\
            \hline
            \gptFig     &Manipulated Data           &0.338516258544173  &-0.244644131284192     &-0.0633536456112792    &0.0379884808928039 &                   &0.181182828675234  &                   &-0.27757651703282  &                       &\\
                        &Manipulated Annotation     &                   &-0.0811098973343258    &0.124907031892084      &                   &                   &-0.0215487176573859&-0.114516107807579 &0.0119844627266093 &0.0972837745546291     &0.0694735815830058\\
                        &Manipulated Visual Encoding&                   &0.302029027597978      &-0.152180548669264     &0.0546097778407565 &-0.0668927395784427&0.00453325565230293&-0.258275905066646 &-0.0350387406187197&                       &\\
                        &Manipulated Scale          &-0.207805509440421 &0.147369891649429      &-0.0238672019614518    &-0.376418674970687 &                   &-0.0511799853469057&0.105655786185585  &0.460937006030475  &-0.0627788620352562    &-0.114124329096617\\
                        &All                        &                   &0.123644890628888      &-0.114494364349911     &-0.283820416237127 &-0.0668927395784427&0.112987381323245  &-0.26713622668864  &0.160306211105545  &0.034504912519373      &-0.0446507475136117\\
            \hline
            \geminiFig  &Manipulated Data           &0.148897387851265  &-0.000982805407774839  &0.0587704843269468     &0.0237455263385156 &                   &-0.126891904520492 &                   &-0.219567449915784 &                       &\\
                        &Manipulated Annotation     &                   &-0.0186393884757987    &-0.0649909752387753    &                   &                   &-0.0626279318676704&-0.360887196893473 &0.67096101379487   &-0.00737661988837329   &-0.107518579442794\\
                        &Manipulated Visual Encoding&                   &0.116072777835055      &-0.0592938667136615    &-0.131161681174548 &-0.0498487655789145&-0.132969508429882 &0.346031326698404  &-0.100140018184062 &                       &\\
                        &Manipulated Scale          &-0.216500563755442 &-0.230664777284233     &0.127469864911946      &0.115065418137084  &                   &0.146802865389231  &-0.104028068391464 &-0.0448283854908478&0.085602294777828      &0.0982604008751536\\
                        &All                        &                   &-0.134214193332752     &0.0619555072864557     &0.00764926330105145&-0.0498487655789145&-0.175686479428814 &-0.118883938586533 &0.306425160204175  &0.0782256748894546     &-0.0092581785676404\\
            \hline
            \claudeFig  &Manipulated Data           &-0.0510654340606012&0.400489869889184      &0.35953616130751       &-0.275548866290405 &                   &-0.250488548444137 &                   &0.0435062806036219 &                       &\\
                        &Manipulated Annotation     &                   &0.0027464585064668     &-0.0928953700890044    &                   &                   &0.308065716279222  &-0.119067507609066 &-0.297357948662339 &-0.055639157108321     &0.109262359721845\\
                        &Manipulated Visual Encoding&                   &-0.446592138055985     &0.129623595531875      &-0.10346411884198  &0.363842358732177  &0.156346607465651  &0.0202410519810953 &-0.139792691821082 &                       &\\
                        &Manipulated Scale          &0.312049848601657  &-0.00753954259507145   &-0.383559558414005     &0.08121367978531   &                   &-0.0289258366923208&-0.072231044152735 &0.0571897124622973 &0.143180007883388      &0.0602743358995287\\
                        &All                        &                   &-0.0508953522554062    &0.0127048283363769     &-0.297799305347075 &0.363842358732177  &0.184997938608414  &-0.171057499780705 &-0.336454647417502 &0.0875408507750671     &0.169536695621374\\
            \hline
            All         &Manipulated Data           &0.436348212334836  &0.154862933197216      &0.354953000023177      &                   &                   &-0.196197624289396 &                   &-0.453637686344982 &                       &\\
                        &Manipulated Annotation     &                   &-0.0970028273036578    &-0.032979313435695     &                   &                   &0.223889066754166  &-0.594470812310118 &0.385587527859139  &0.0342679975579343     &0.0712173618620567\\
                        &Manipulated Visual Encoding&                   &-0.0284903326229517    &-0.0818508198510509    &                   &0.247100853574819  &0.0279103546880721 &0.107996473612853  &-0.274971450623865 &                       &\\
                        &Manipulated Scale          &-0.112256224594204 &-0.0908344282298761    &-0.279956895463511     &                   &                   &0.066697043350004  &-0.0706033263586131&0.473298333001925  &0.16600344062596       &0.0444104076780652\\
                        &All                        &0.324091987740631  &-0.0614646549592699    &-0.0398340287270791    &-0.57397045828315  &0.247100853574819  &0.122298840502846  &-0.557077665055879 &0.130276723892217  &0.200271438183894      &0.115627769540122\\
            \hline
        \end{tblr}
    }
    \label{tab:chartMeanCoef_jIndex}
\end{sidewaystable*}

\begin{table*}[!h]
    \sisetup{exponent-mode = scientific,
                     exponent-product=\times,
                     tight-spacing,
                     round-mode = figures,
                     round-precision=4
             }
    \centering
    \caption{Mean coefficients of Jaccard Indices for misleaders and visualizations.}
    \begin{tabu}{l|l|S[table-format=1.5e1]|S[table-format=1.5e1]|S[table-format=1.5e1]|S[table-format=1.5e1]}
        \hline
        \multirow{2}*{\textbf{Misleader}}   &\multirow{2}*{\textbf{Visualization}}  &\multicolumn{4}{c}{\textbf{LLM}}\\
        \cline{3-6}
                                                                &                           &\textbf{GPT}   &\textbf{Gemini}        &\textbf{Claude}        &\textbf{All}\\
        \hline
        Missing Normalization                                   &Choropleth Map             &-0.1355        &-0.00707763298378427   &-0.0935254542944423    &-0.236114082130287\\
        \hline
        \multirow{2}*{Cherry Picking}                           &Line Chart                 &-0.1789        &-0.0363998023365307    &-0.862844695468029     &-1.07812520035362\\
                                                                &Scatterplot                &-0.2838        &0.024856041603999      &0.825539150797322      &0.5666095409193\\
        \hline
        Overplotting                                            &Scatterplot                &0.4584         &-0.144924792145789     &0.216122624612517      &0.529646284576794\\
        \hline
        \multirow{6}*{Inappropriate Order}                      &100\% Stacked Bar Chart    &0.3385         &0.148897387851265      &-0.0510654340606012    &0.436348212334836\\
                                                                &Area Chart                 &-0.2446        &-0.00098280540777484   &0.400489869889184      &0.154862933197216\\
                                                                &Bar Chart                  &-0.06335       &0.0587704843269468     &0.35953616130751       &0.354953000023177\\
                                                                &Choropleth Map             &0.1735         &0.0308231593223        &-0.182023411995963     &0.0222992230712016\\
                                                                &Line Chart                 &0.3601         &-0.0904921021839612    &0.612356147023893      &0.881927576064224\\
                                                                &Scatterplot                &-0.4522        &-0.0994986993739947    &-0.998155494806218     &-1.54989351184108\\
        \hline
        \multirow{3}*{Deceptive Labeling}                       &Bar Chart                  &-0.05866       &0.578193699952431      &-0.629016109341098     &-0.10948011680491\\
                                                                &Line Chart                 &0.2051         &0.169832776286906      &0.486389312943998      &0.861326515862879\\
                                                                &Pie Chart                  &-0.1145        &-0.360887196893473     &-0.119067507609066     &-0.594470812310118\\
        \hline
        \multirow{2}*{Lack of Legend}                           &Stacked Area Chart         &0.09728        &-0.00737661988837329   &-0.0556391571083211    &0.0342679975579345\\
                                                                &Stacked Bar Chart          &-0.2008        &0.238165611402217      &0.447046199668875      &0.484413185320227\\
        \hline
        \multirow{3}*{Lack of Scales}                           &Area Chart                 &0.07278        &0.157807278409417      &-0.123226455665785     &0.107362655610589\\
                                                                &Bar Chart                  &0.3503         &-0.341024976922795     &0.356243036728227      &0.365506677371898\\
                                                                &Line Chart                 &0.01211        &-0.203254943976167     &0.249839901029062      &0.0586963938205072\\
        \hline
        \multirow{5}*{Inappropriate Aggregation}                &Area Chart                 &-0.1539        &-0.176446666885216     &0.125972914172251      &-0.204365482914247\\
                                                                &Bar Chart                  &-0.1667        &-0.302159698268411     &0.179877702523867      &-0.289005874002683\\
                                                                &Line Chart                 &-0.2388        &-0.029205764178409     &-0.428163497693839     &-0.69613384292922\\
                                                                &Stacked Bar Chart          &0.2703         &-0.345684190845011     &-0.337783839947031     &-0.41319582345817\\
                                                                &Scatterplot                &0.01198        &0.67096101379487       &-0.297357948662339     &0.38558752785914\\
        \hline
        \multirow{4}*{Data-visual Disproportion}                &Bar Chart                  &0.09747        &-0.505701286450084     &0.764002757346752      &0.355775874077685\\
                                                                &Line Chart                 &0.09091        &0.247987436503397      &-0.74255891711807      &-0.403661653774437\\
                                                                &Pie Chart                  &-0.4098        &0.17447351834296       &1.22033152402107       &0.985004886484561\\
                                                                &Scatterplot                &-0.1486        &0.368827965979865      &-0.936332409095091     &-0.716072575058218\\
        \hline
        \multirow{3}*{Dual Encoding}                            &Bar Chart                  &-0.2497        &0.446407419736423      &-0.634379161814878     &-0.437626693928735\\
                                                                &Pie Chart                  &0.03207        &-0.468967984163927     &-0.625116792484442     &-0.263622114657194\\
                                                                &Scatterplot                &0.1135         &0.329424347941692      &0.796539717274009      &0.441101124434354\\
        \hline
        \multirow{3}*{Continuous encoding of categorical data}  &Area Chart                 &0.3021         &0.116072777835055      &-0.446592138055985     &-0.0284903326229517\\
                                                                &Line Chart                 &-0.08638       &-0.380956944933279     &0.898905524583721      &0.431572008462508\\
                                                                &Pie Chart                  &0.1195         &-0.157866539586248     &-0.574973679555535     &-0.613386298214514\\
        \hline
        \multirow{2}*{Categorical encoding of continuous data}  &Choropleth Map             &0.05461        &-0.131161681174548     &-0.10346411884198      &-0.180016022175771\\
                                                                &Heatmap                    &-0.06689       &-0.0498487655789145    &0.363842358732177      &0.247100853574819\\
        \hline
        \multirow{5}*{Inappropriate Scale Range}                &Bar Chart                  &0.4758         &0.0610324252970283     &-0.433071596535062     &0.103770250060323\\
                                                                &Choropleth Map             &0.09442        &0.0527567917551818     &-0.506952847620541     &-0.359779416258176\\
                                                                &Line Chart                 &-0.2174        &-0.0490261232359376    &0.4282114847144        &0.161771934807872\\
                                                                &Stacked Area Chart         &0.1925         &-0.412333478736584     &0.335880232261097      &0.116042156820793\\
                                                                &Stacked Bar Chart          &-0.05174       &0.093960458817511      &-0.32145432523883      &-0.279232884820793\\
        \hline
        \multirow{3}*{Inappropriate Scale Functions}            &Bar Chart                  &-0.07491       &0.00979034193631417    &-0.00601832778674242   &-0.0711377743336779\\
                                                                &Line Chart                 &-0.2087        &-0.0254893389353571    &-0.56221363541295      &-0.79635858355577\\
                                                                &Pie Chart                  &0.1057         &-0.104028068391464     &-0.072231044152735     &-0.0706033263586131\\
        \hline
        \multirow{5}*{Unconventional Scale Directions}          &Area Chart                 &-0.2380        &-0.46920606223556      &0.562615316315849      &-0.144620888512809\\
                                                                &Bar Chart                  &0.1684         &-0.244975448102158     &0.0809144307347845     &0.00433038427661968\\
                                                                &Choropleth Map             &-1.064         &0.310208398358954      &0.908180163945672      &0.154603684946358\\
                                                                &Line Chart                 &0.6204         &0.305049990963788      &-0.478716784882195     &0.44675422351004\\
                                                                &Scatterplot                &0.5199         &-0.23665946561129      &-0.124308353753473     &0.158966558680714\\
        \hline
        \multirow{3}*{Misuse of Cumulative Relationship}        &100\% Stacked Bar Chart    &-0.2078        &-0.216500563755442     &0.312049848601658      &-0.112256224594205\\
                                                                &Stacked Area Chart         &-0.2553        &0.497935773514412      &-0.192700224377709     &0.0499612838051669\\
                                                                &Stacked Bar Chart          &-0.06239       &0.00429994205764251    &0.381728661138357      &0.323643292498858\\
        \hline
        \multirow{3}*{Exceeding the Canvas}                     &Area Chart                 &0.3854         &0.238541284951327      &-0.57015485891092      &0.0537864602829327\\
                                                                &Bar Chart                  &-0.5932        &0.301622545780761      &-0.0253840648269848    &-0.316919755466776\\
                                                                &Line Chart                 &0.05970        &-0.20684227112721      &0.285662286959329      &0.138518565261563\\
        \hline
        \multirow{3}*{Small Size}                               &Choropleth Map             &0.5929         &-0.247899771977053     &-0.320013636539822     &0.0250361542635239\\
                                                                &Line Chart                 &-0.3052        &0.123110607723947      &0.298130811929094      &0.1160109033263\\
                                                                &Scatterplot                &-0.05900       &0.191831080120442      &0.181498066215771      &0.31433177432121\\
        \hline
    \end{tabu}
    \label{tab:misleaderChartMeanCoef_jIndex}
\end{table*}

\subsection{Bootstrapping Results: Dice-S{\o}rensen Coefficients}\label{app:dsCoefResults}

\subsection{Dice-S{\o}rensen Coefficients}\label{app:dsCoefResults}
\subsubsection{\textbf{H6} LLMs on Misleading Identification}
\begin{figure}[h]
    \centering
    \includegraphics[width=0.49\textwidth]{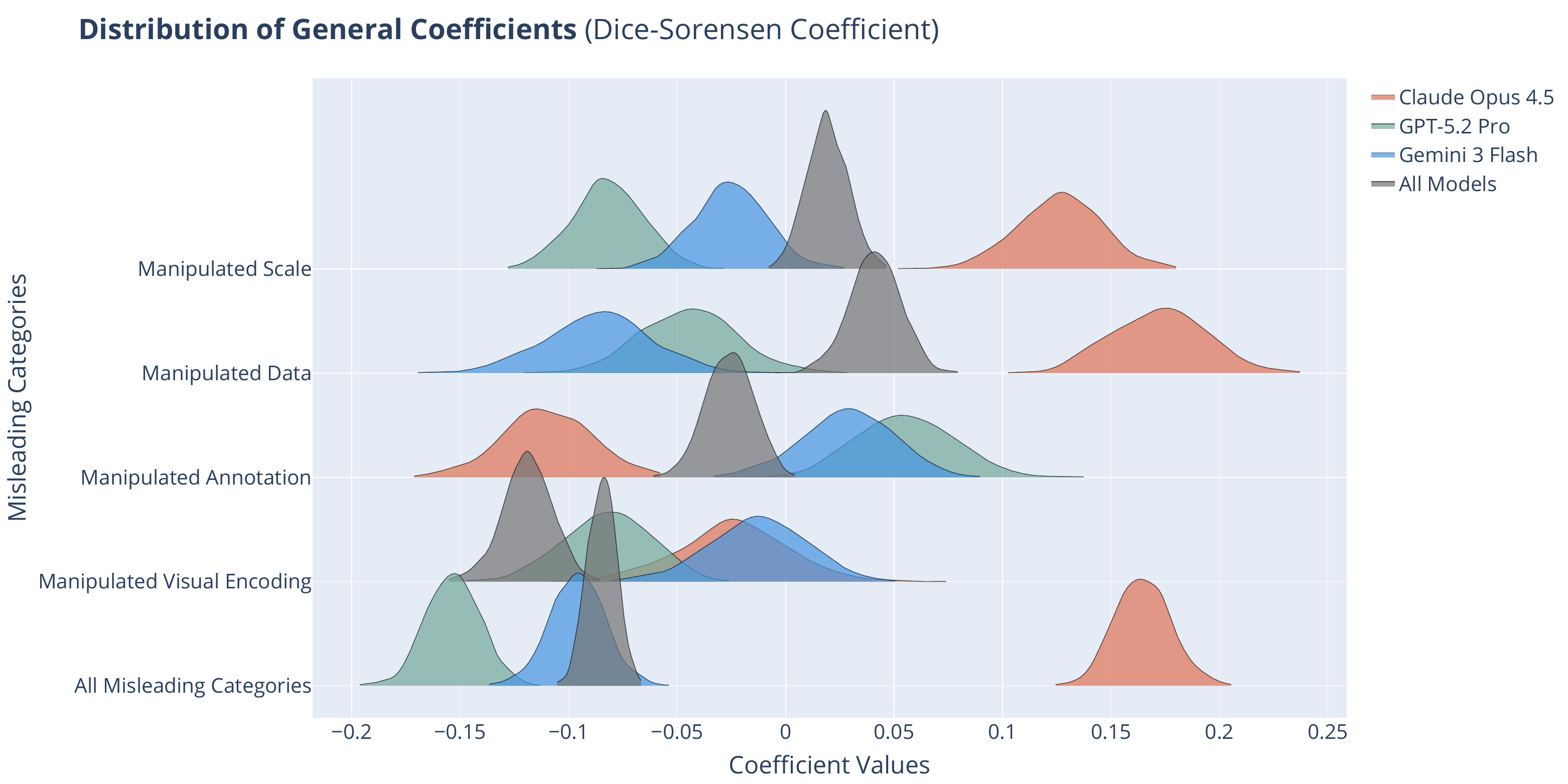}
    \caption{Ridge plot of the bootstrapped general coefficients from the beta regression of the Dice-S{\o}rensen Coefficients.}
    \label{fig:genCoefMislead_dsCoef}
\end{figure}

Comparing the coefficients of each LLM, \claude\ had a statistically significant positive coefficient ($\bar{\beta} \approx 0.1965$), which indicates the \claude\ was more likely to select the correct misleading chart element. Meanwhile, \gpt\ and \gemini\ had statistically significant negative coefficients ($\bar{\beta} \approx -0.1742$ and $\bar{\beta} \approx -0.03235$ respectively), indicating that they were more likely to select incorrect misleading chart elements. This seems evident in the bottom row of \autoref{fig:genCoefMislead_dsCoef}. See supplemental materials for more details of the Dice-S{\o}rensen Coefficients.





\subsubsection{\textbf{H7 \& H8 \& H9} LLMs on Misleading Identification}
\textbf{Boxplot comparison} Looking over the boxplots in \autoref{tab:results_dsCoef_ex3}, we observe large variations across all categories, misleaders, visualizations, and LLMs. Based on these boxplots, all LLMs seemed to perform poorly (\ie, median appears at or near zero) on the following misleader/visualization combinations: \textit{Cherry Picking} for \textit{line chart}; \textit{Deceptive Labeling} for \textit{pie chart}; \textit{Inappropriate Aggregation} for \textit{area chart}, \textit{bar chart}, and \textit{line chart}; \textit{Continuous Encoding of Categorical Data} for \textit{pie chart}, and \textit{Inappropriate Scale Functions} for \textit{pie chart}. None of the boxplots showed all the LLMs performing well at the same time (\ie, median appears at or near one). \claude\ performed poorly (\ie, median at zero) on 20 misleader/visualizations and performed well on 21 (\ie, median at one); \gpt\ performed poorly on 19 and performed well on 6; and \gemini\ performed poorly on 10 and performed well on 3. Thus, it appears that \claude\ outperformed both \gpt\ and \gemini.

\noindent\textbf{Coefficient Results}\label{sec:exp3_misleadResults}
Of the 477 variables, 460 were statistically significant, indicating that the majority of these combinations have an effect on LLMs' ability to identify misleading chart elements. The 17 statistically insignificant coefficients and results can be found in \autoref{tab:dsCoefInsig}. We report the detailed coefficients in our supplemental material. When focusing on misleading categories, all LLMs performed better on \textit{Manipulated Data} ($\bar{\beta} \approx 0.08998$) and \textit{Manipulated Scale} ($\bar{\beta} \approx 0.05182$), partially confirming our prediction in \textbf{H7}. The coefficients for all the misleaders were significant with positive values for \textit{Misuse of Cumulative Relationship}, \textit{Small Size}, \textit{Unconventional Scale Directions}, \textit{Deceptive Labeling}, \textit{Lack of Legend}, \textit{Lack of Scales}, \textit{Data-visual Disproportion}, \textit{Categorical encoding for continuous data}, \textit{Overplotting}, and \textit{Inappropriate Order}. Although \textbf{H8} correctly included four misleaders, our hypothesis did not include six of them. Finally, for the visualizations, all of their coefficients were significant with positive values for \textit{100\% stacked bar chart}, \textit{stacked bar chart}, \textit{stacked area chart}, \textit{scatterplot}, \textit{line chart}, and \textit{heatmap}. Our predictions in \textbf{H9} were abysmal: we only correctly predicted \textit{line chart}. See supplemental material for more details on the coefficients.

\clearpage
\begin{table*}[h]
    \caption{Boxplots of the Jaccard Indices\cite{costa:2021:further} calculated from \claudeFig, \gptFig, and \geminiFig's answers compared to the correct misleading chart elements from Experiment 3.}\vspace{-0.7em}
    \tabulinesep=1pt
    \begin{tblr}{
        rowsep=0pt,
        colspec={X[0.1,m]Q[h,3.1]|X[]X[]X[]X[]X[]X[]X[]X[]X[]X[]}
    }
        &   \textbf{Misleader Name} & \rotatebox[origin=bl]{25}{\textbf{100\% Stacked Bar Chart}} & \rotatebox[origin=bl]{25}{\textbf{Area Chart}} & \rotatebox[origin=bl]{25}{\textbf{Bar Chart}} & \rotatebox[origin=bl]{25}{\textbf{Choropleth Map}} & \rotatebox[origin=bl]{25}{\textbf{Heatmap}} & \rotatebox[origin=bl]{25}{\textbf{Line Chart}} & \rotatebox[origin=bl]{25}{\textbf{Pie Chart}} & \rotatebox[origin=bl]{25}{\textbf{Stacked Area Chart}} & \rotatebox[origin=bl]{25}{\textbf{Stacked Bar Chart}} & \rotatebox[origin=bl]{25}{\textbf{Scatterplot}} \\
        \hline
        \SetCell[r=4]{c,manuData} \rotatebox[origin=c]{90}{\textbf{Manipulated Data}}
                                    & \makecell[l]{\textbf{Missing}\\ \textbf{Normalization}}
                                        &
                                        &
                                        &
                                        & \includegraphics[scale=0.025]{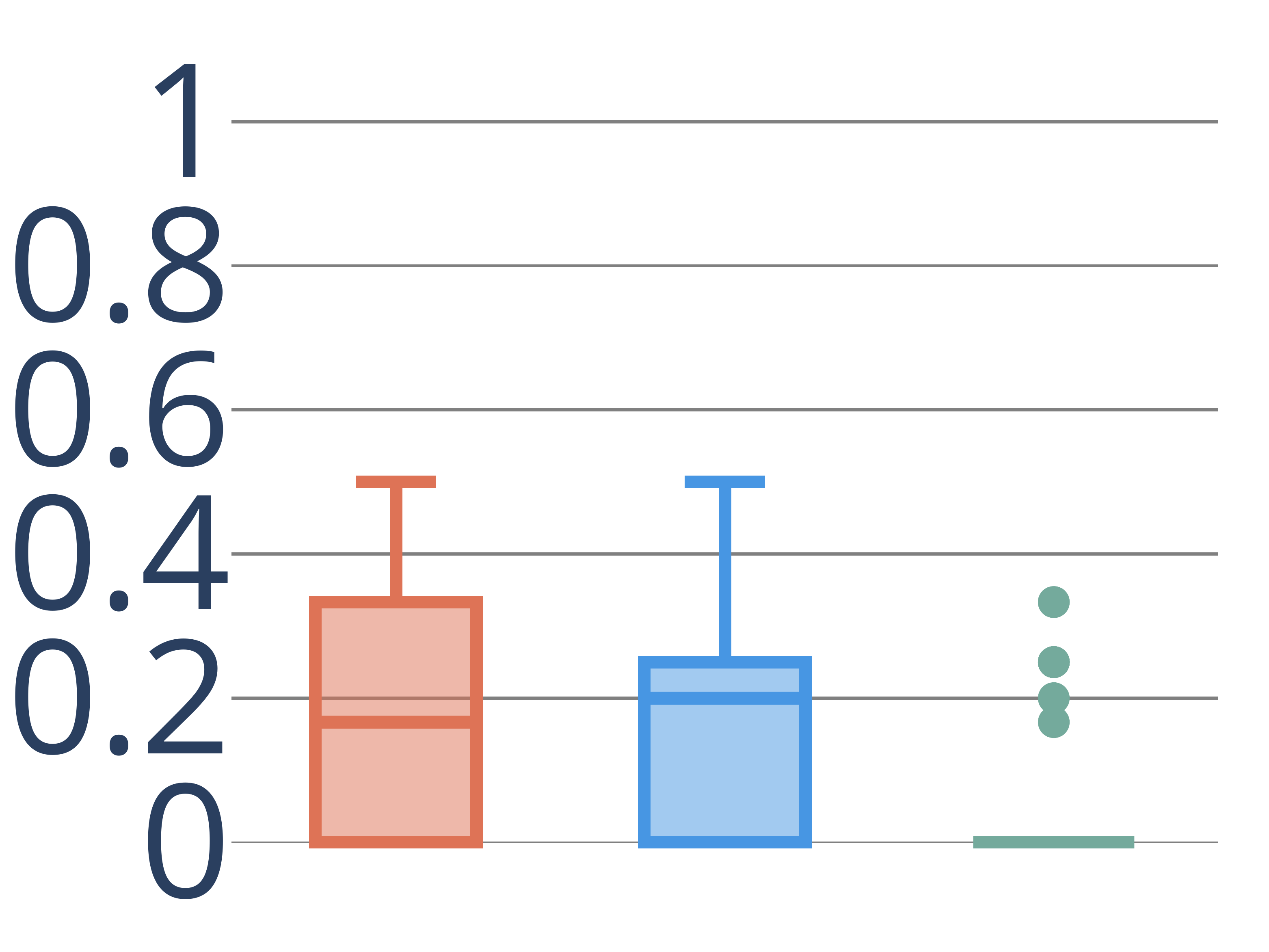}
                                        & 
                                        &
                                        &
                                        &
                                        &
                                        &
                                        &\\
                                    & \textbf{Cherry Picking} 
                                        &
                                        &
                                        &
                                        &
                                        &
                                        & \includegraphics[scale=0.025]{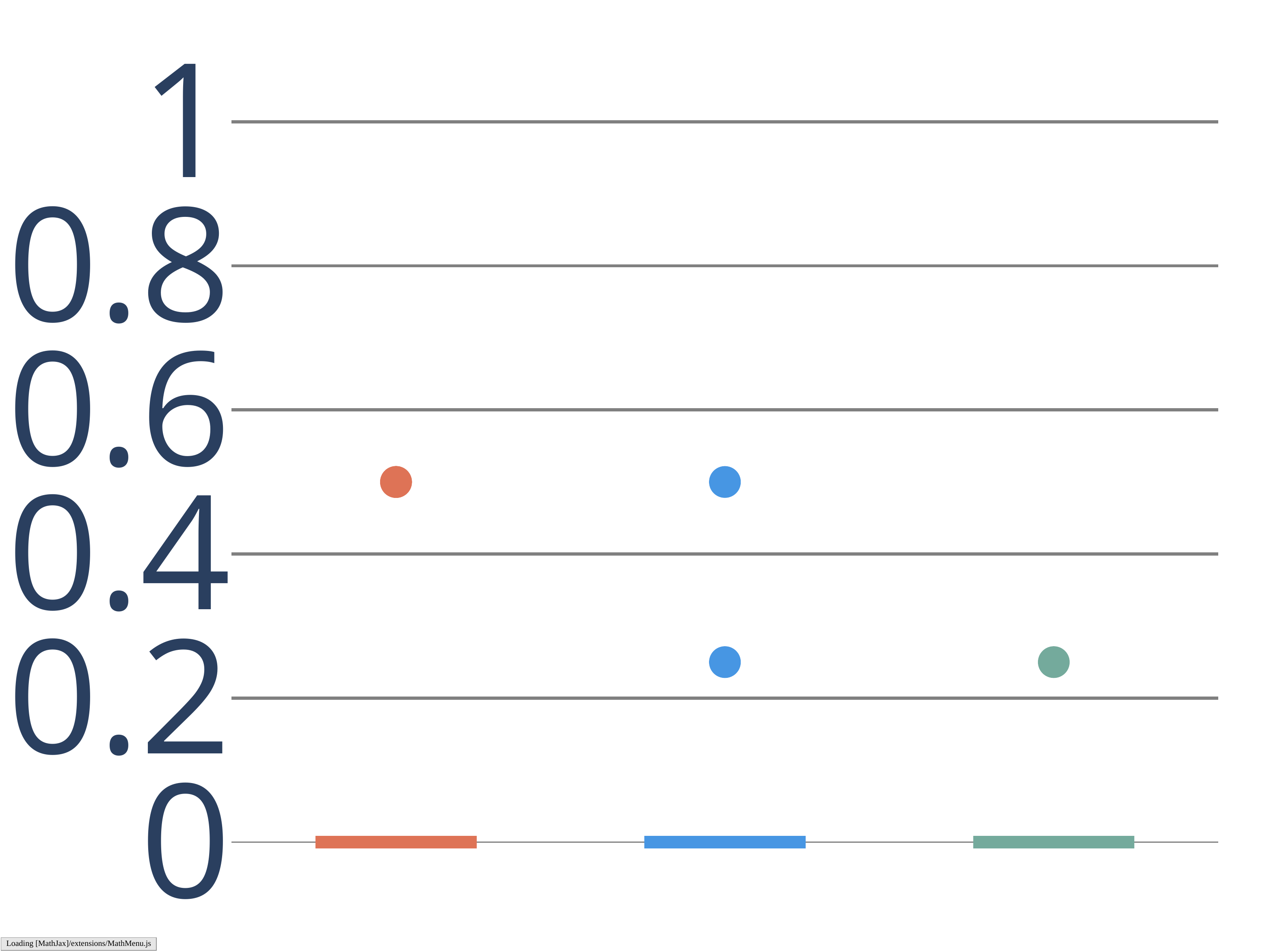}
                                        &
                                        &
                                        &
                                        & \includegraphics[scale=0.025]{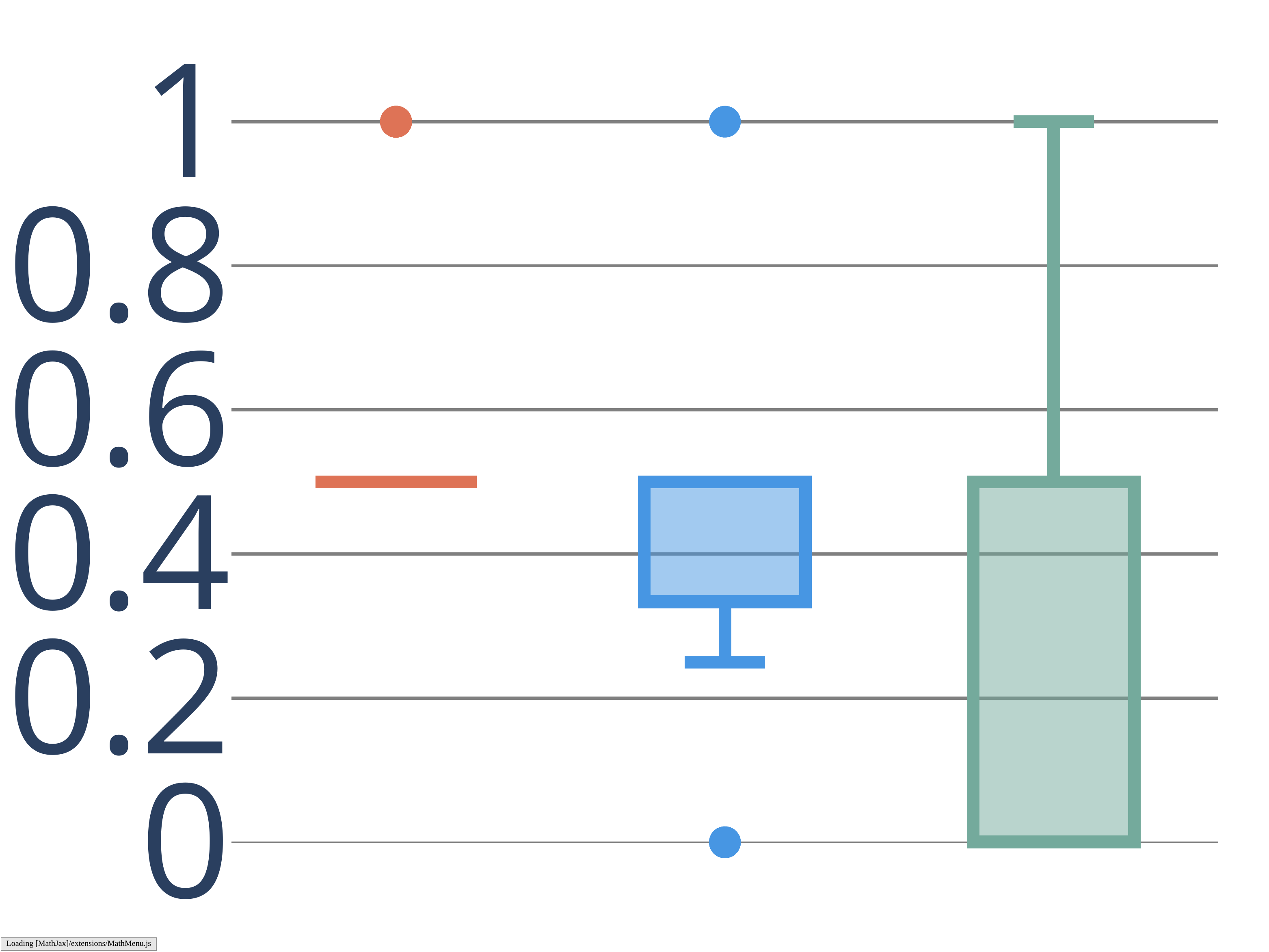}\\
                                    & \textbf{Overplotting}
                                        &
                                        &
                                        &
                                        &
                                        &
                                        &
                                        &
                                        &
                                        &
                                        & \includegraphics[scale=0.025]{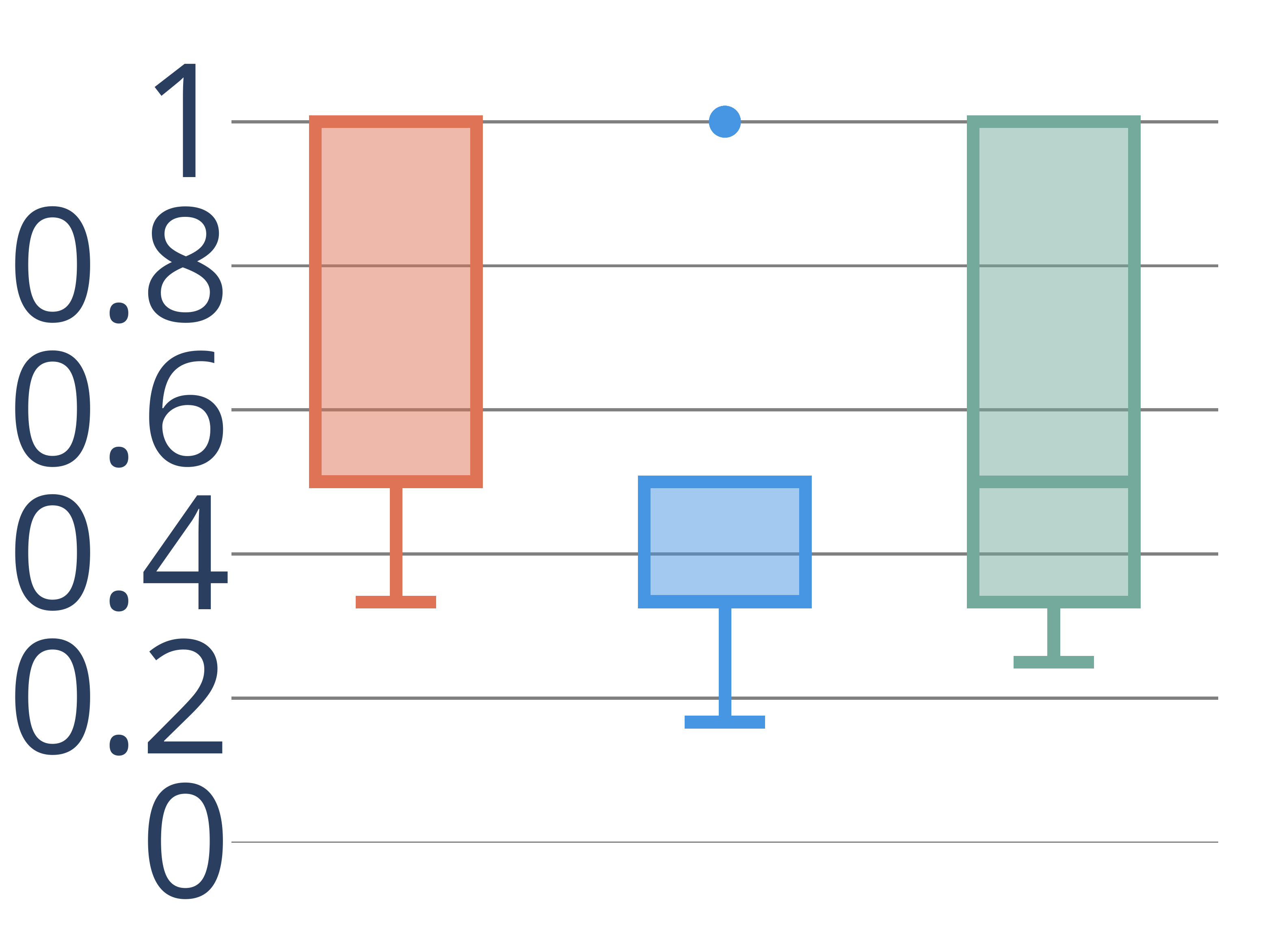}\\
                                    & \textbf{Inappropriate Order} 
                                        & \includegraphics[scale=0.025]{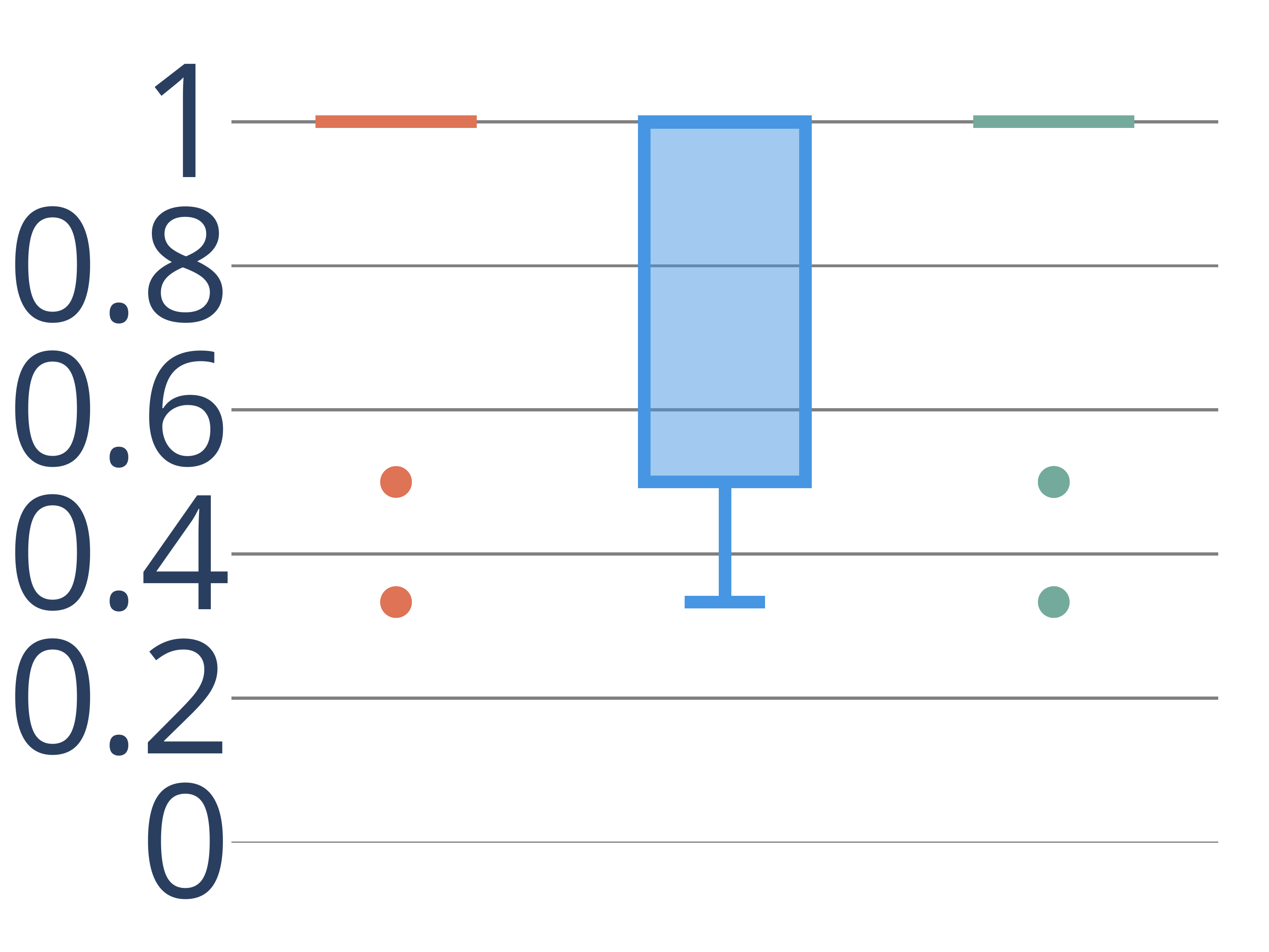} 
                                        & \includegraphics[scale=0.025]{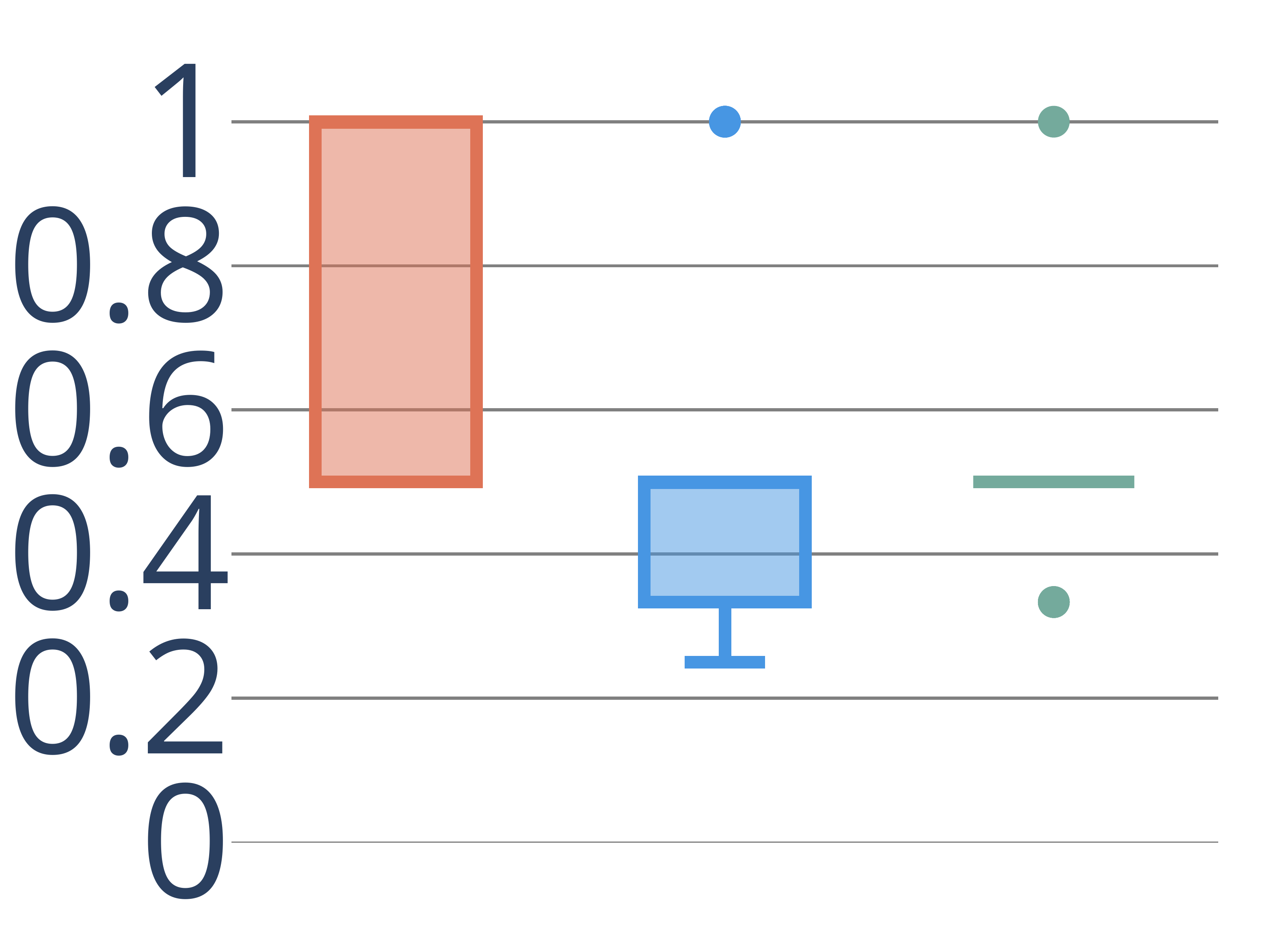} 
                                        & \includegraphics[scale=0.025]{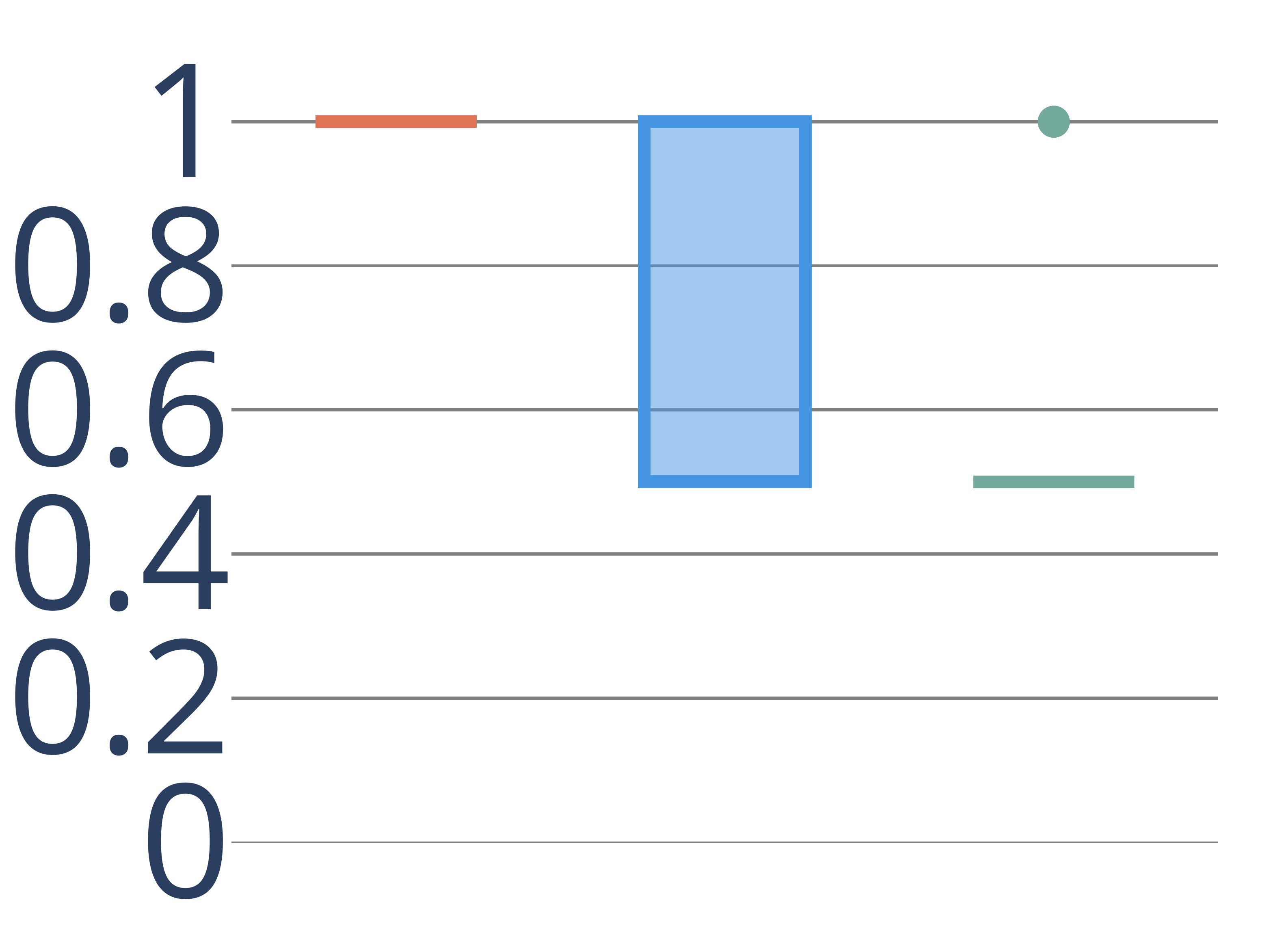} 
                                        & \includegraphics[scale=0.025]{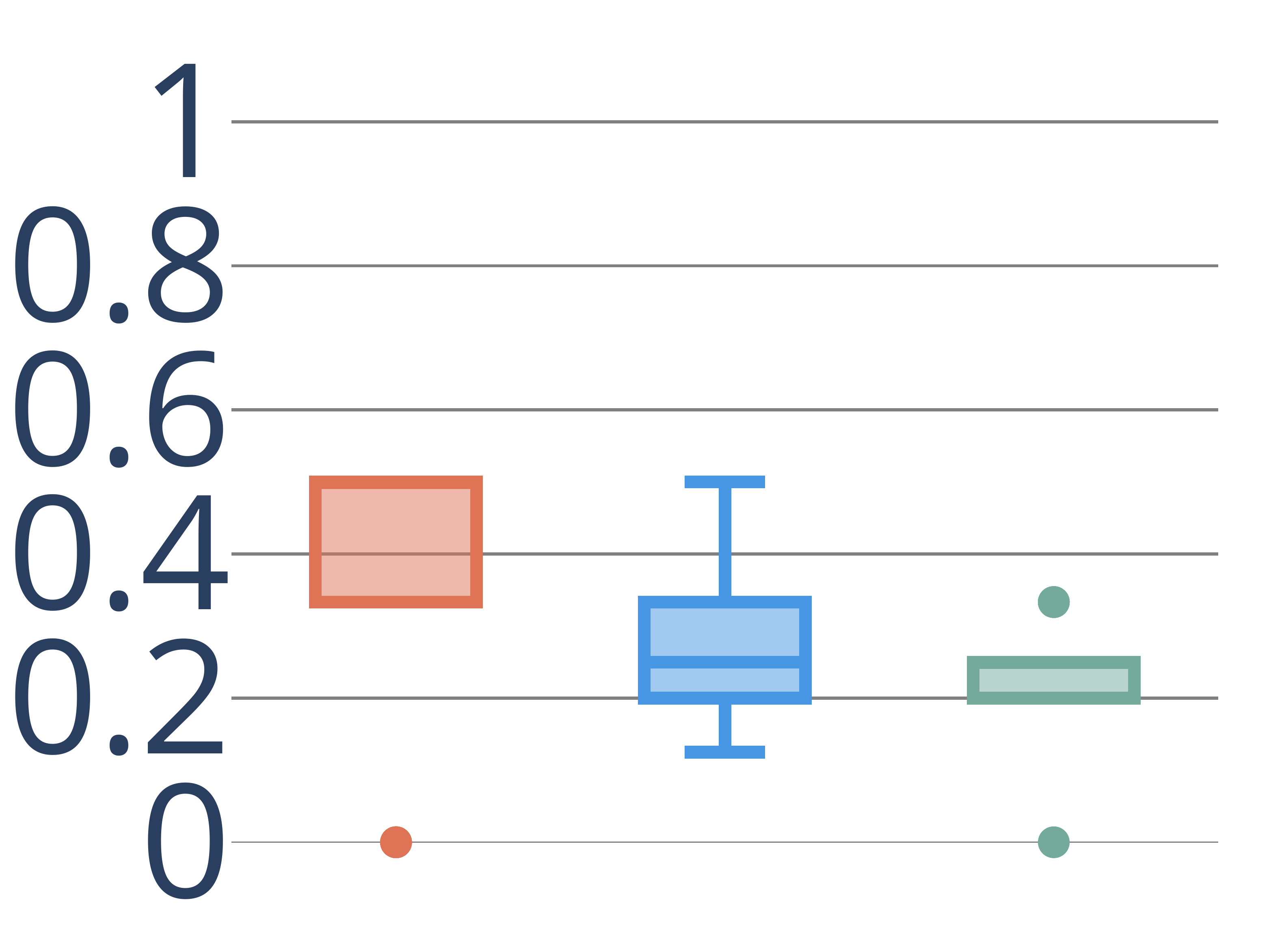} 
                                        & 
                                        & \includegraphics[scale=0.025]{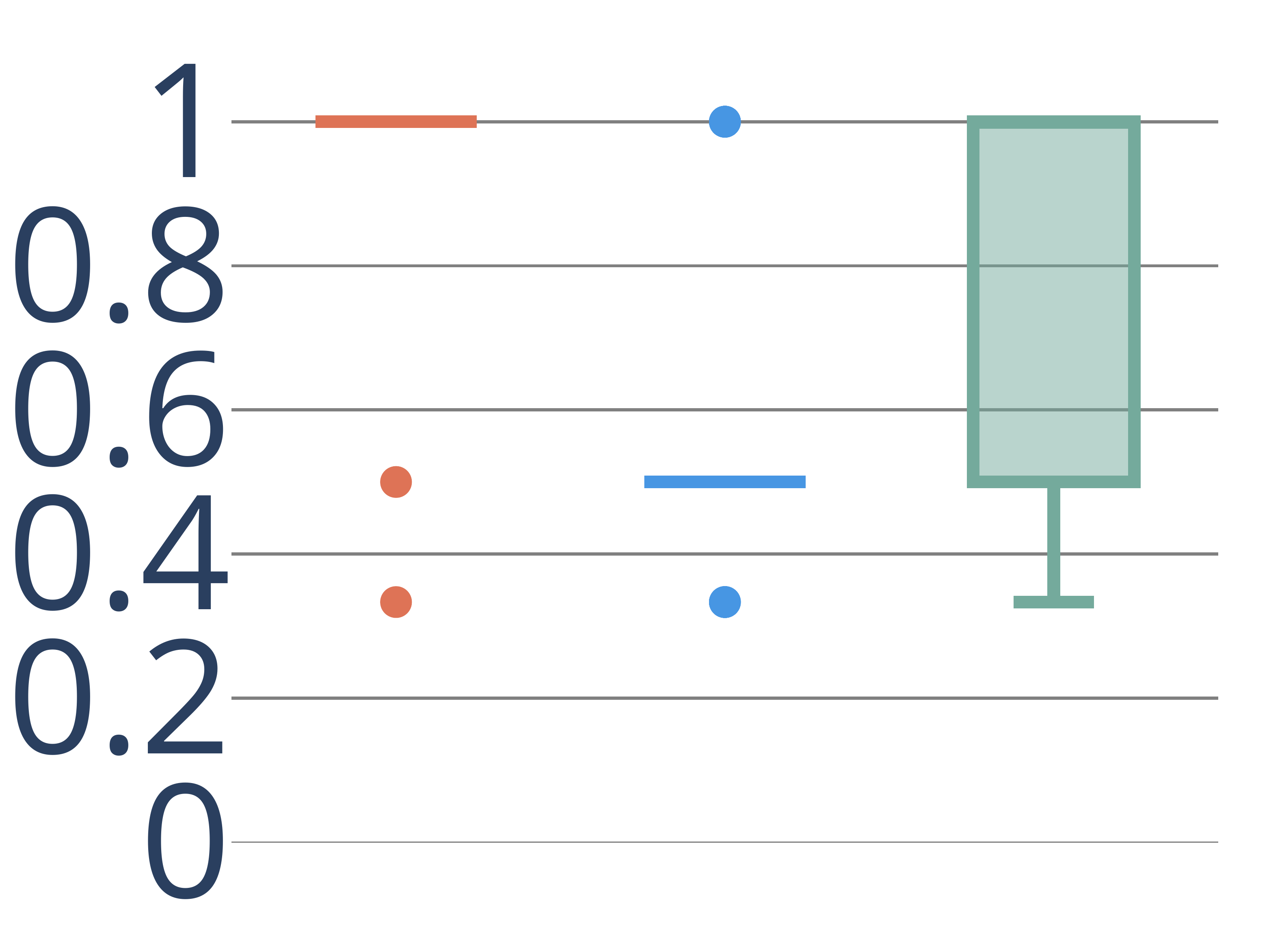} 
                                        & 
                                        & 
                                        & 
                                        & \includegraphics[scale=0.025]{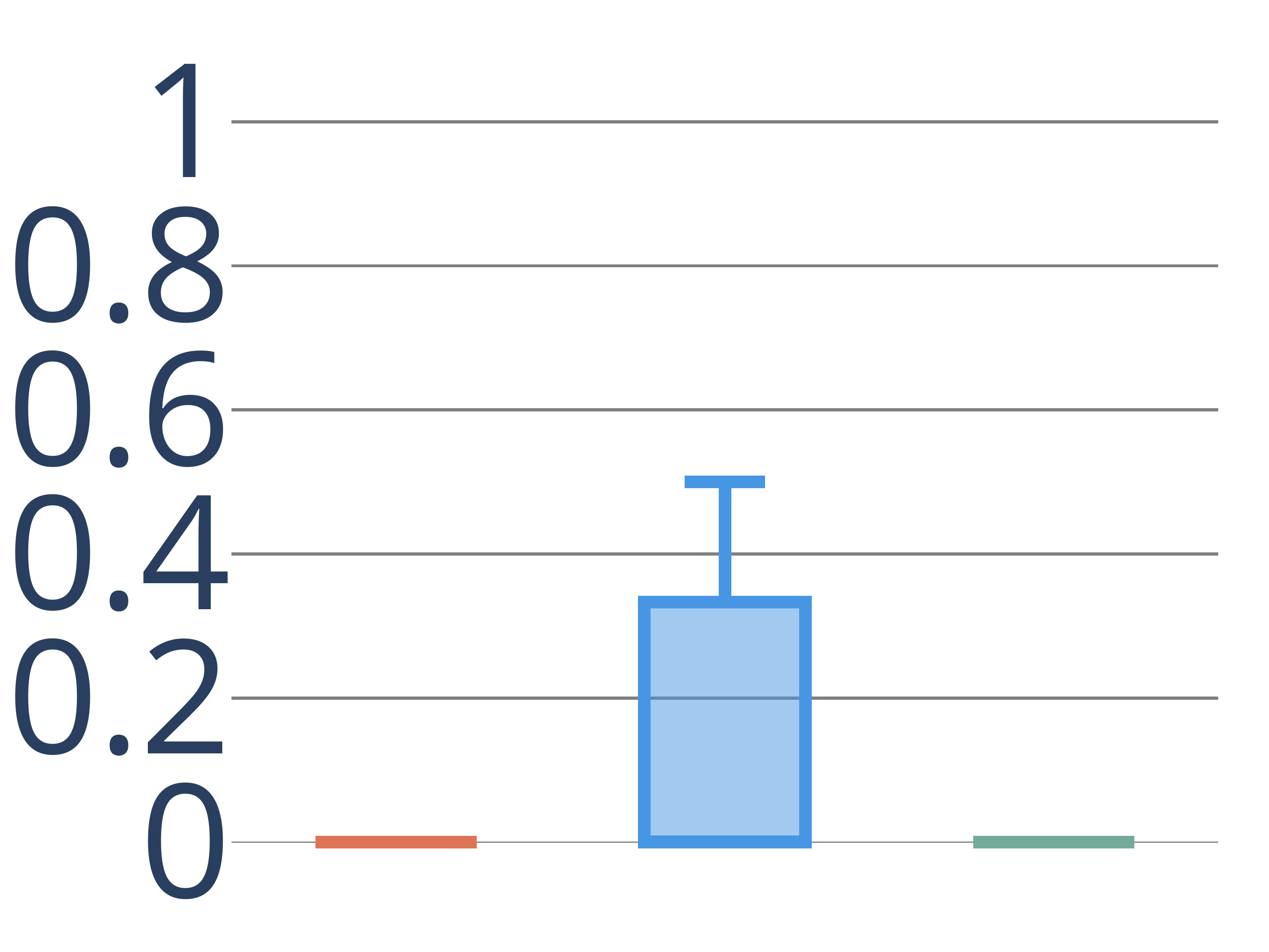}\\
        \hline
        \SetCell[r=4]{c,manuAnnotate} \rotatebox[origin=c]{90}{\textbf{Manipulated Annotation}}
                                    & \textbf{Deceptive Labeling} 
                                        & 
                                        & 
                                        & \includegraphics[scale=0.025]{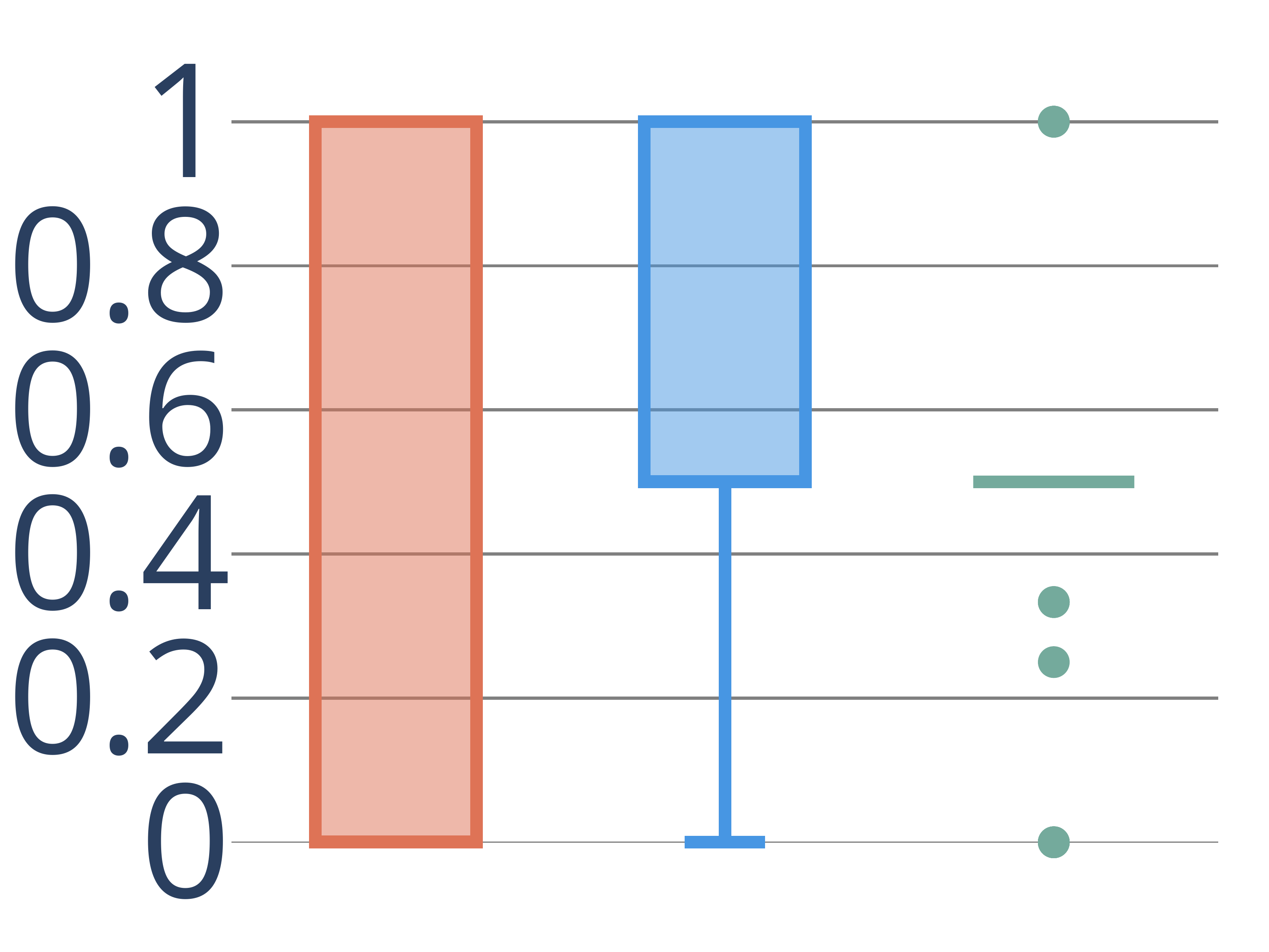} 
                                        & 
                                        & 
                                        & \includegraphics[scale=0.025]{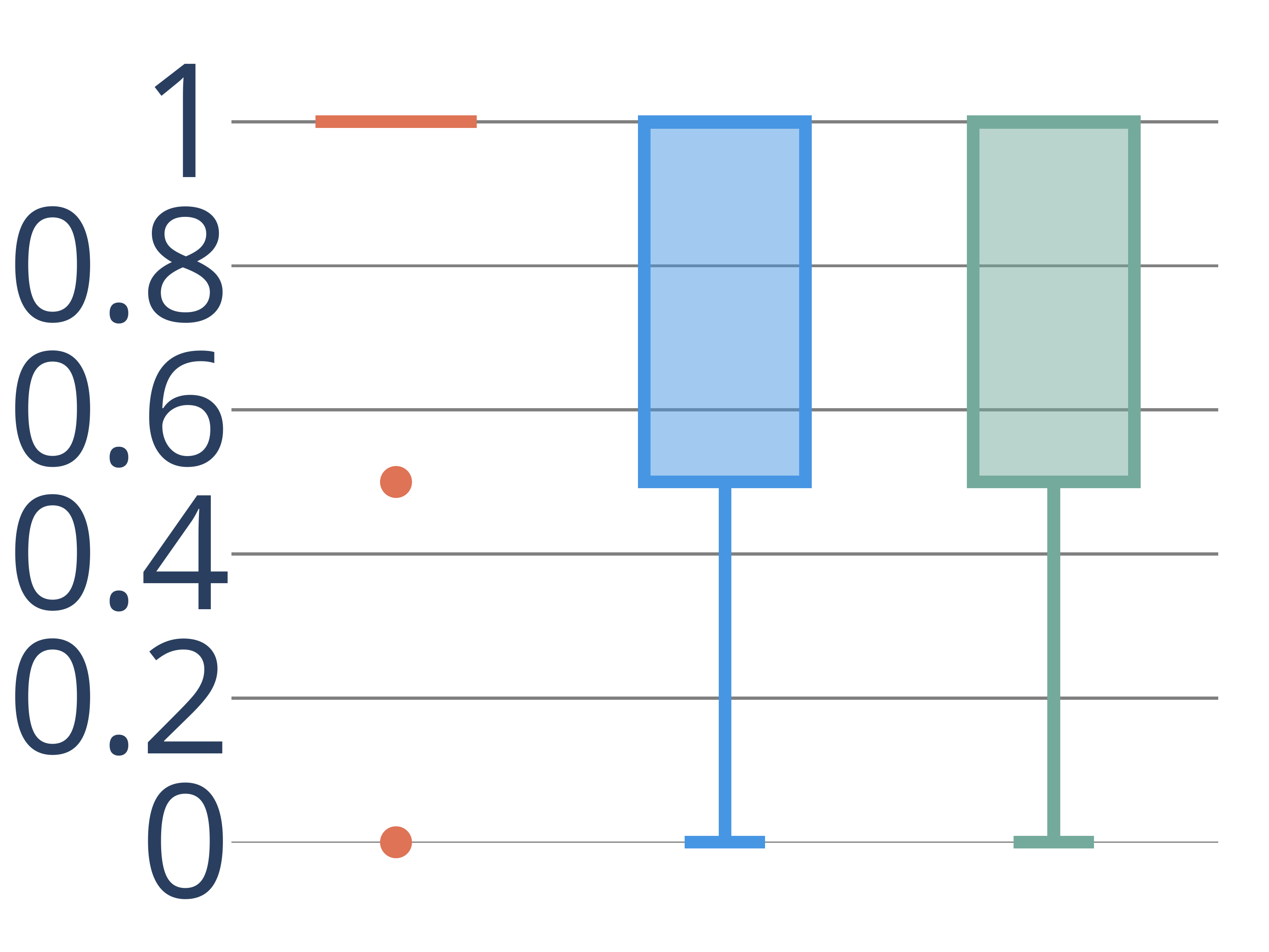} 
                                        & \includegraphics[scale=0.025]{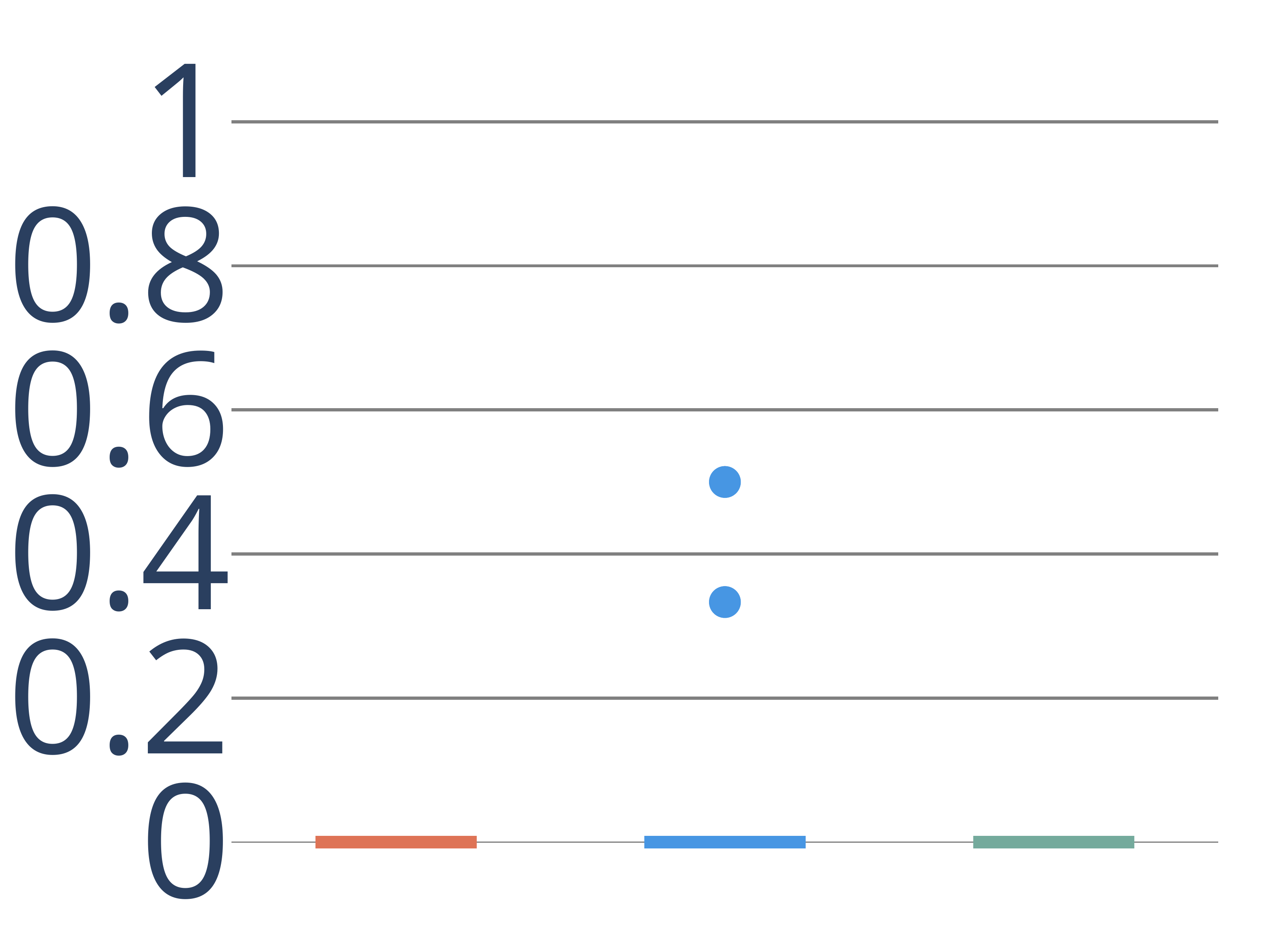} 
                                        & 
                                        & 
                                        &\\
                                    & \makecell[l]{\textbf{Lack of Labeling}\\ \textit{Lack of legend}} 
                                        & 
                                        & 
                                        & 
                                        & 
                                        & 
                                        & 
                                        & 
                                        & \includegraphics[scale=0.025]{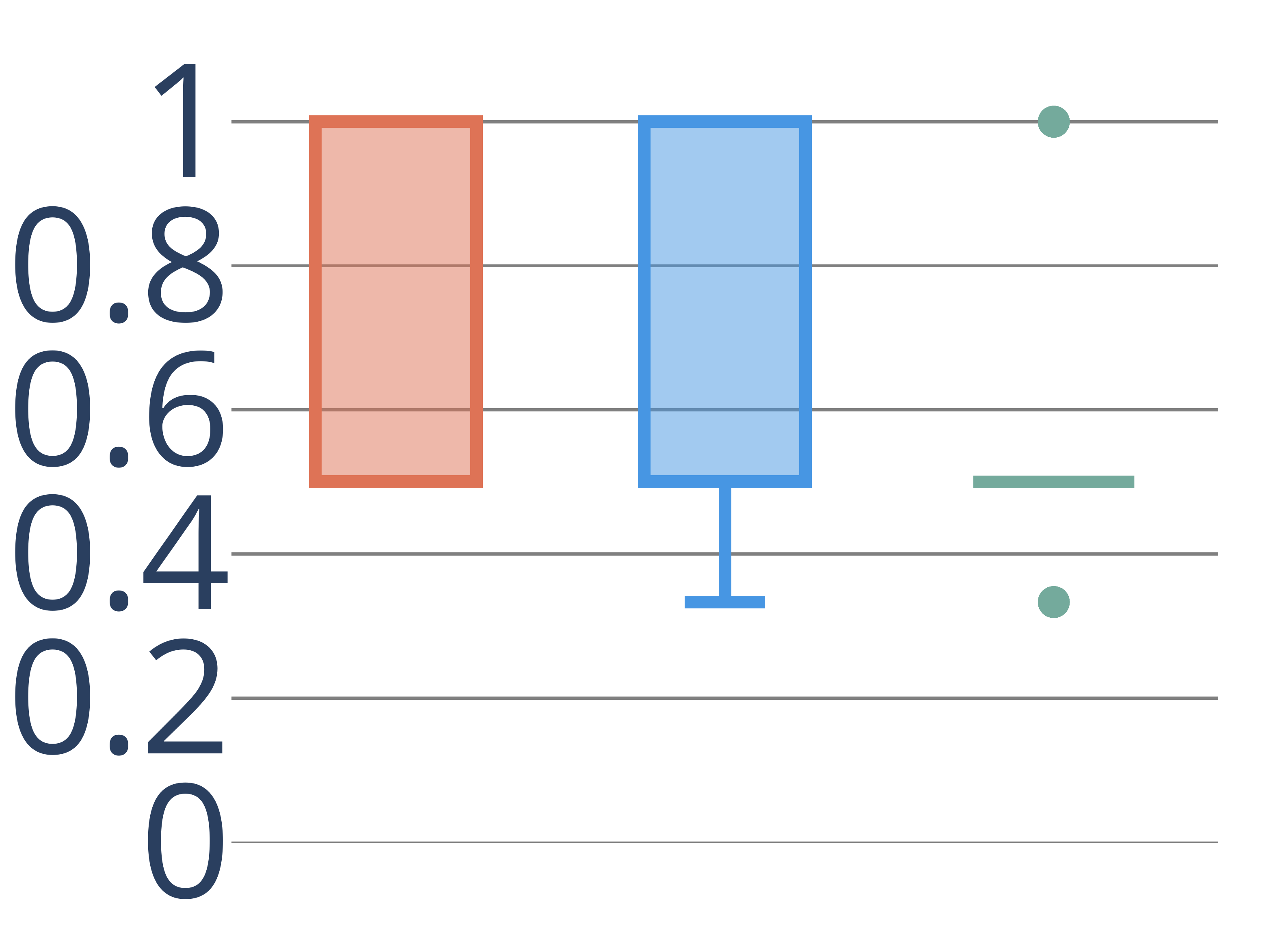}
                                        & \includegraphics[scale=0.025]{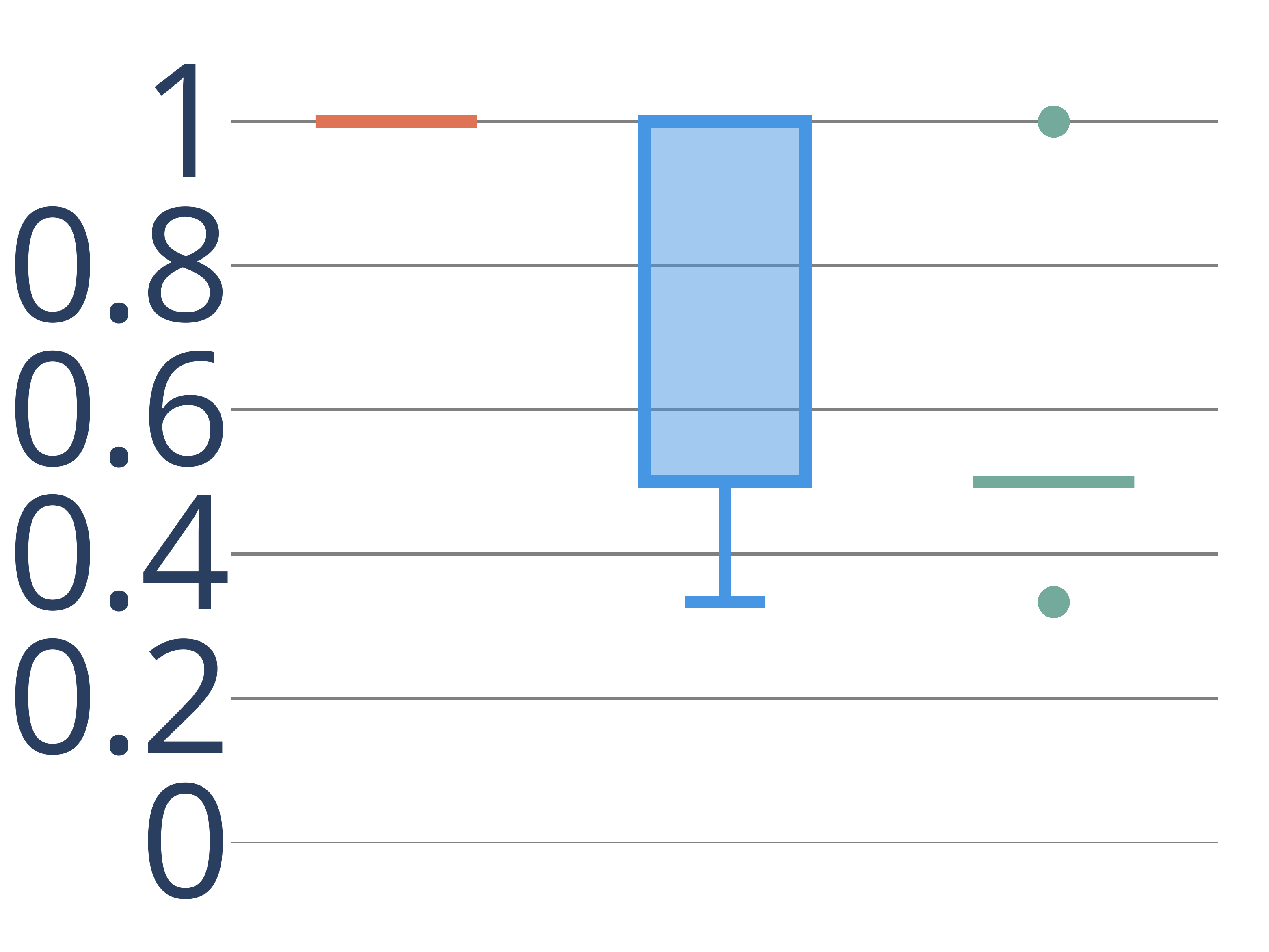} 
                                        &\\
                                    & \makecell[l]{\textbf{Lack of Labeling}\\ \textit{Lack of scales}} 
                                        & 
                                        & \includegraphics[scale=0.025]{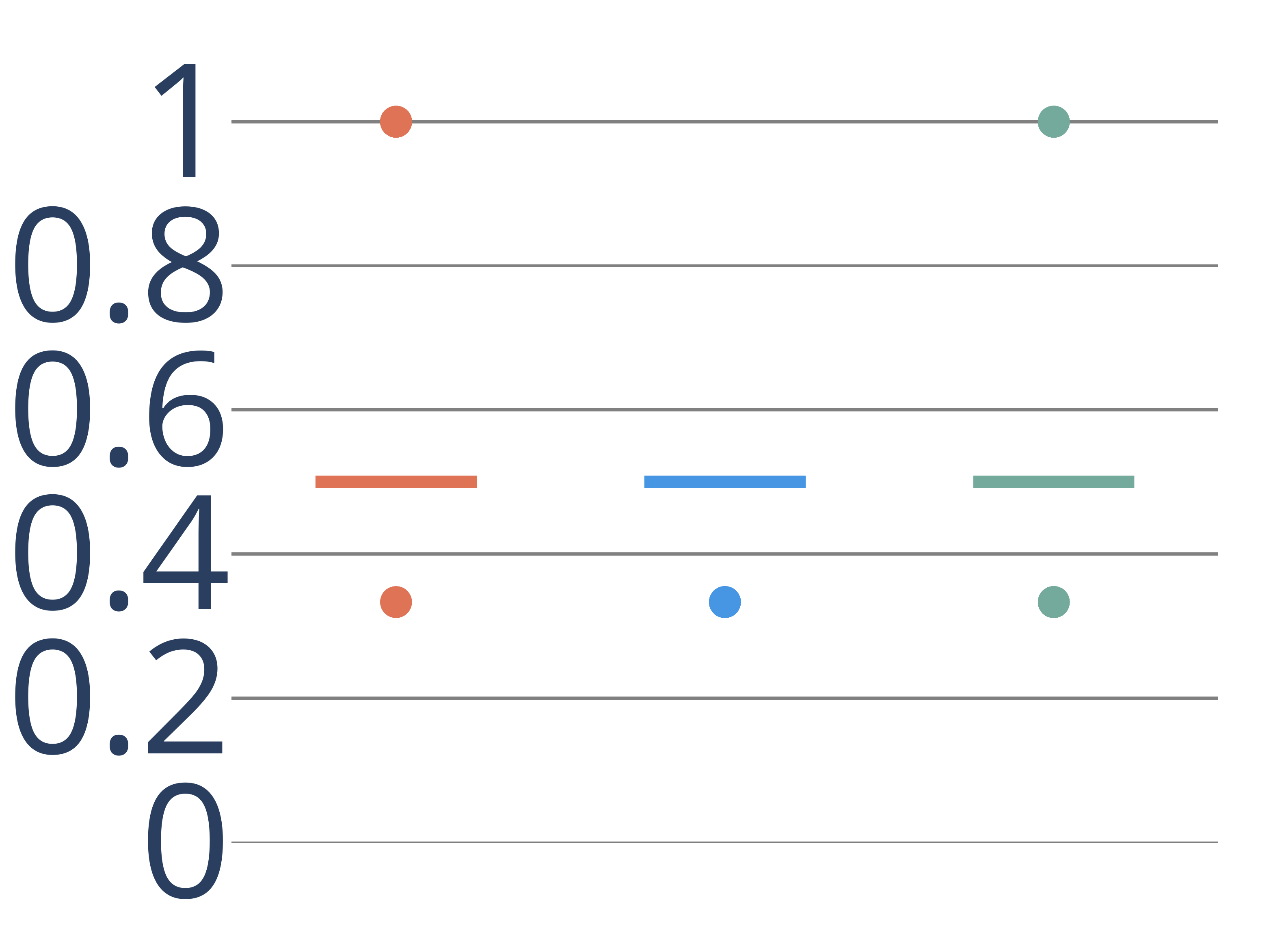} 
                                        & \includegraphics[scale=0.025]{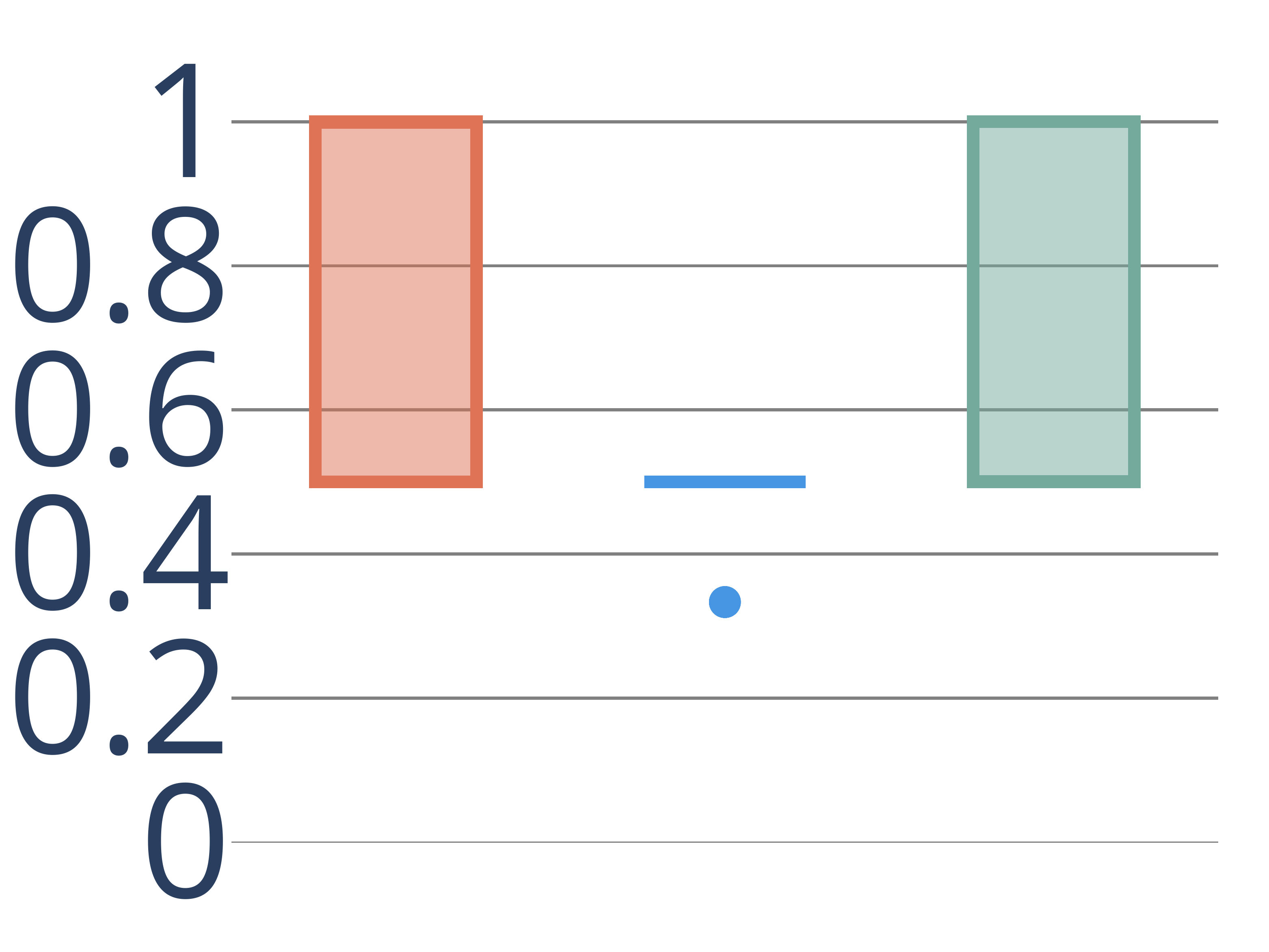} 
                                        & 
                                        & 
                                        & \includegraphics[scale=0.025]{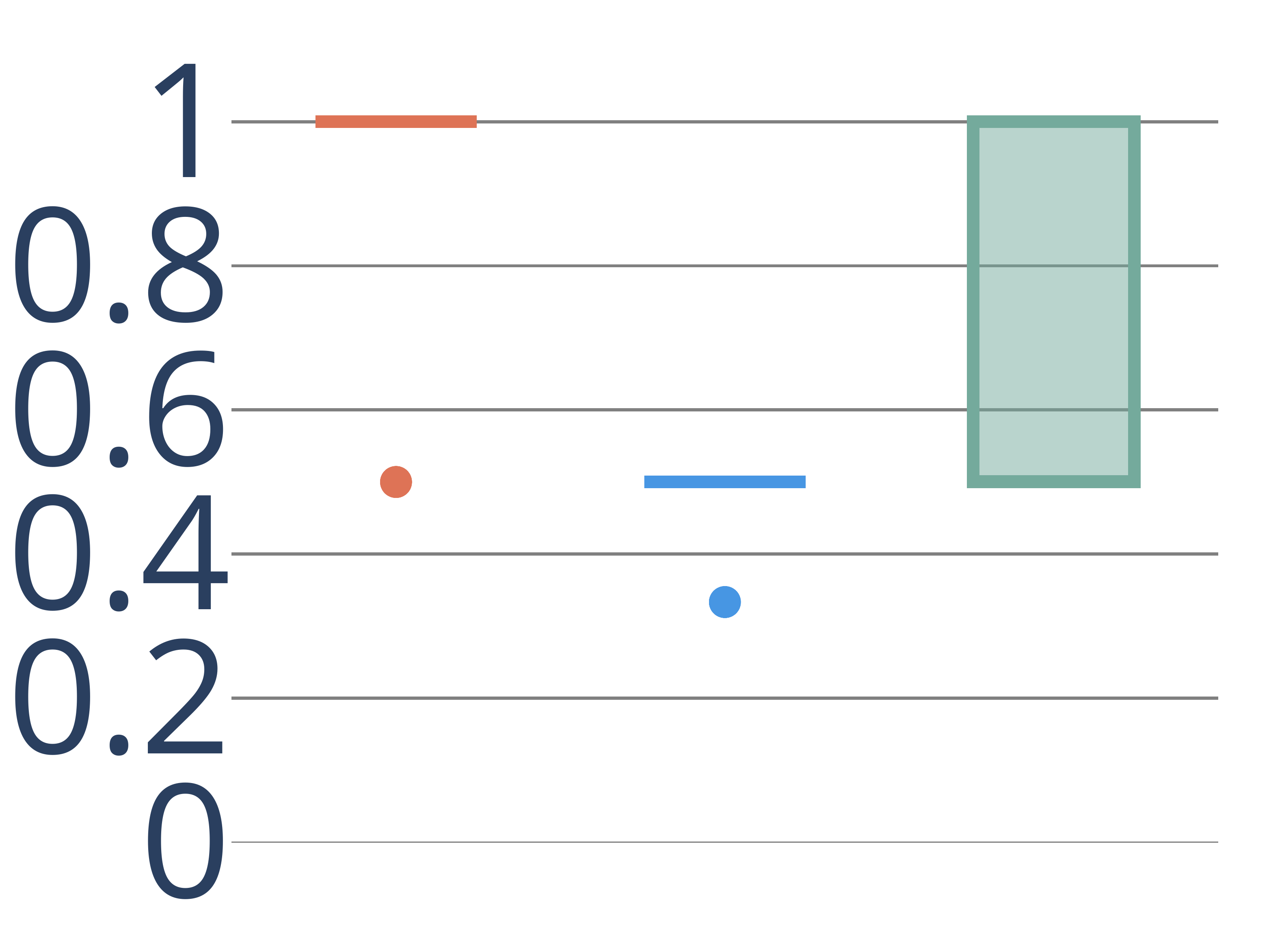} 
                                        & 
                                        & 
                                        & 
                                        &\\
                                    & \makecell[l]{\textbf{Inappropriate}\\ \textbf{Aggregation}} 
                                        & 
                                        & \includegraphics[scale=0.025]{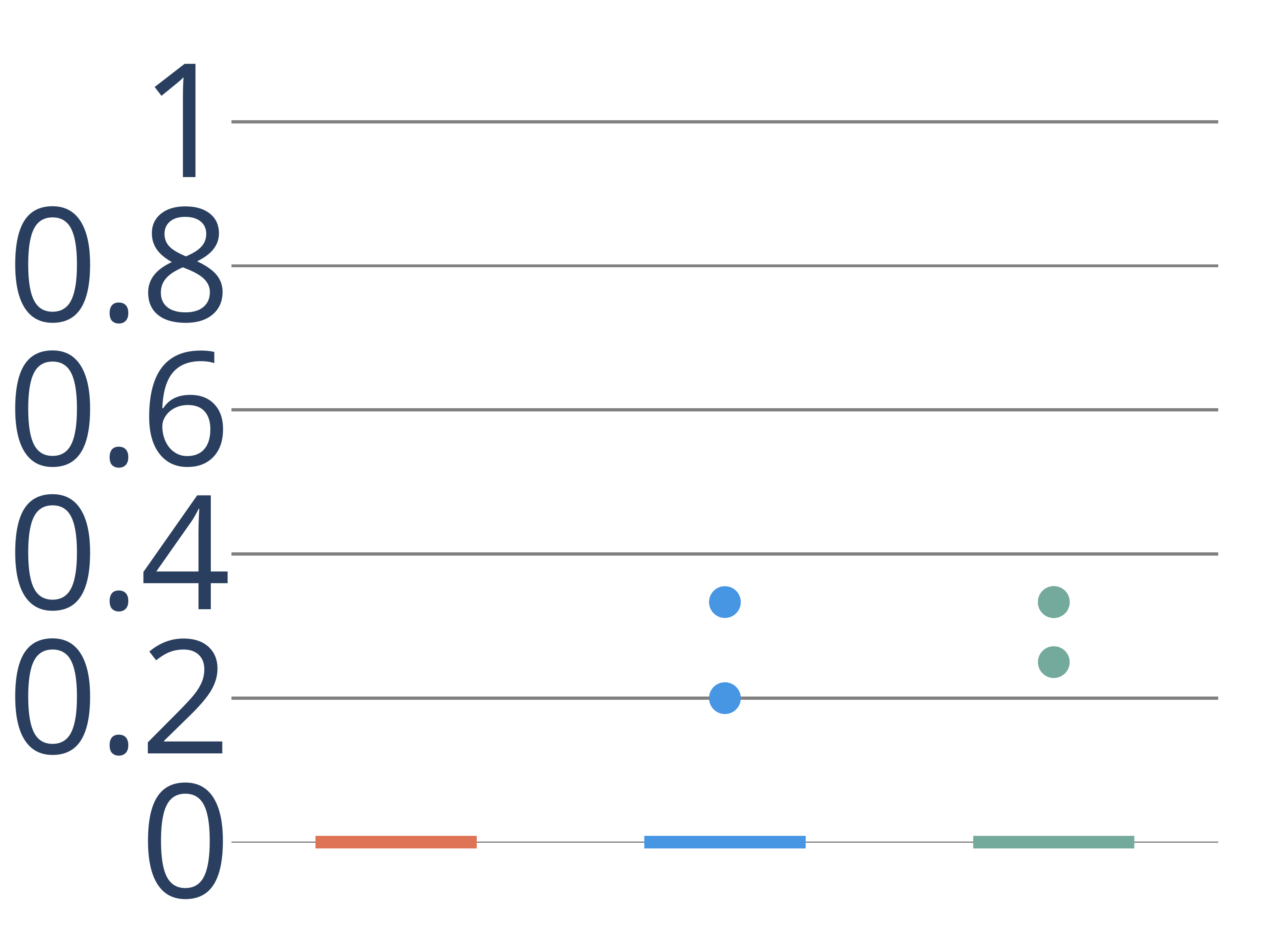} 
                                        & \includegraphics[scale=0.025]{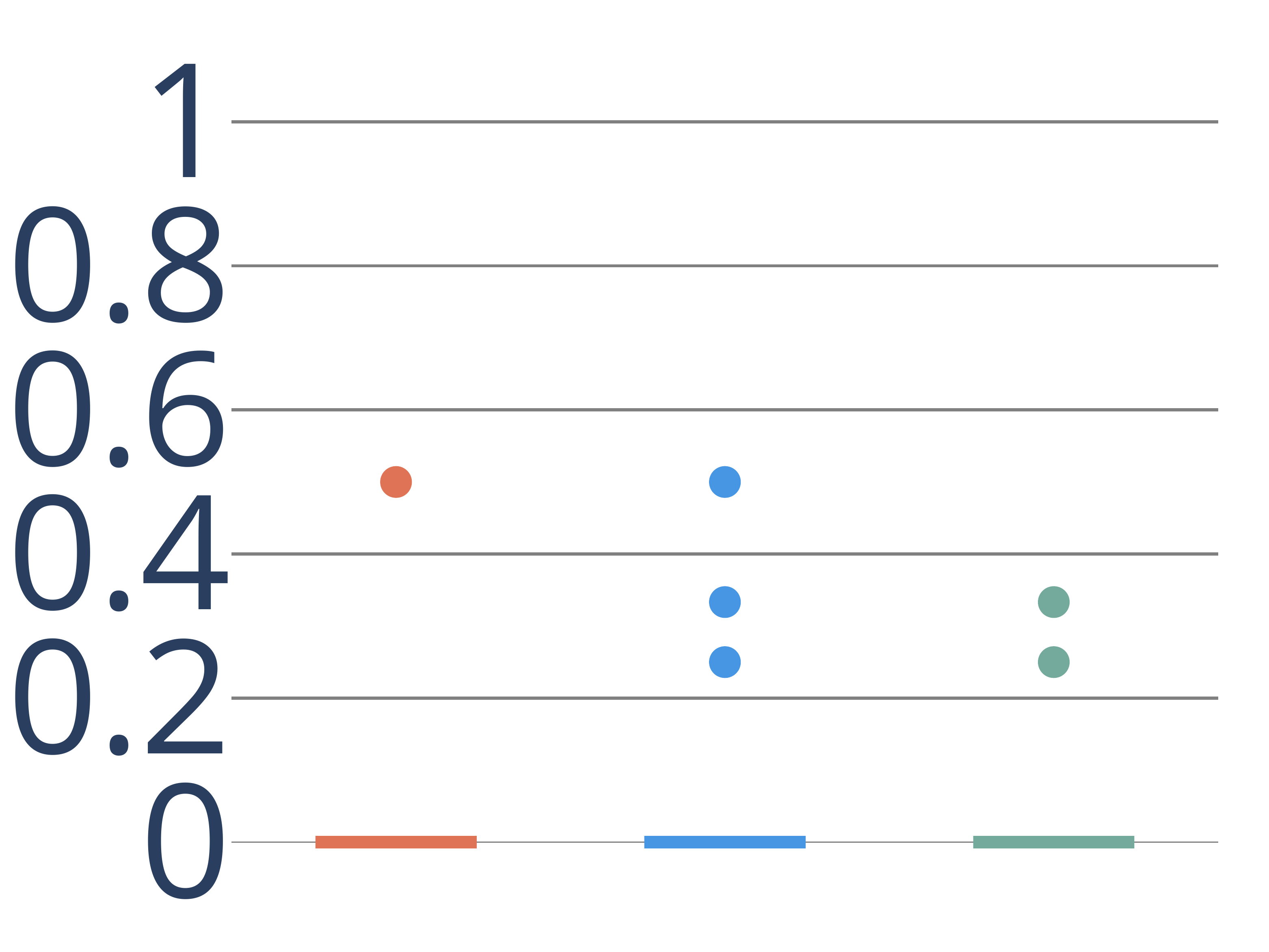} 
                                        & 
                                        & 
                                        & \includegraphics[scale=0.025]{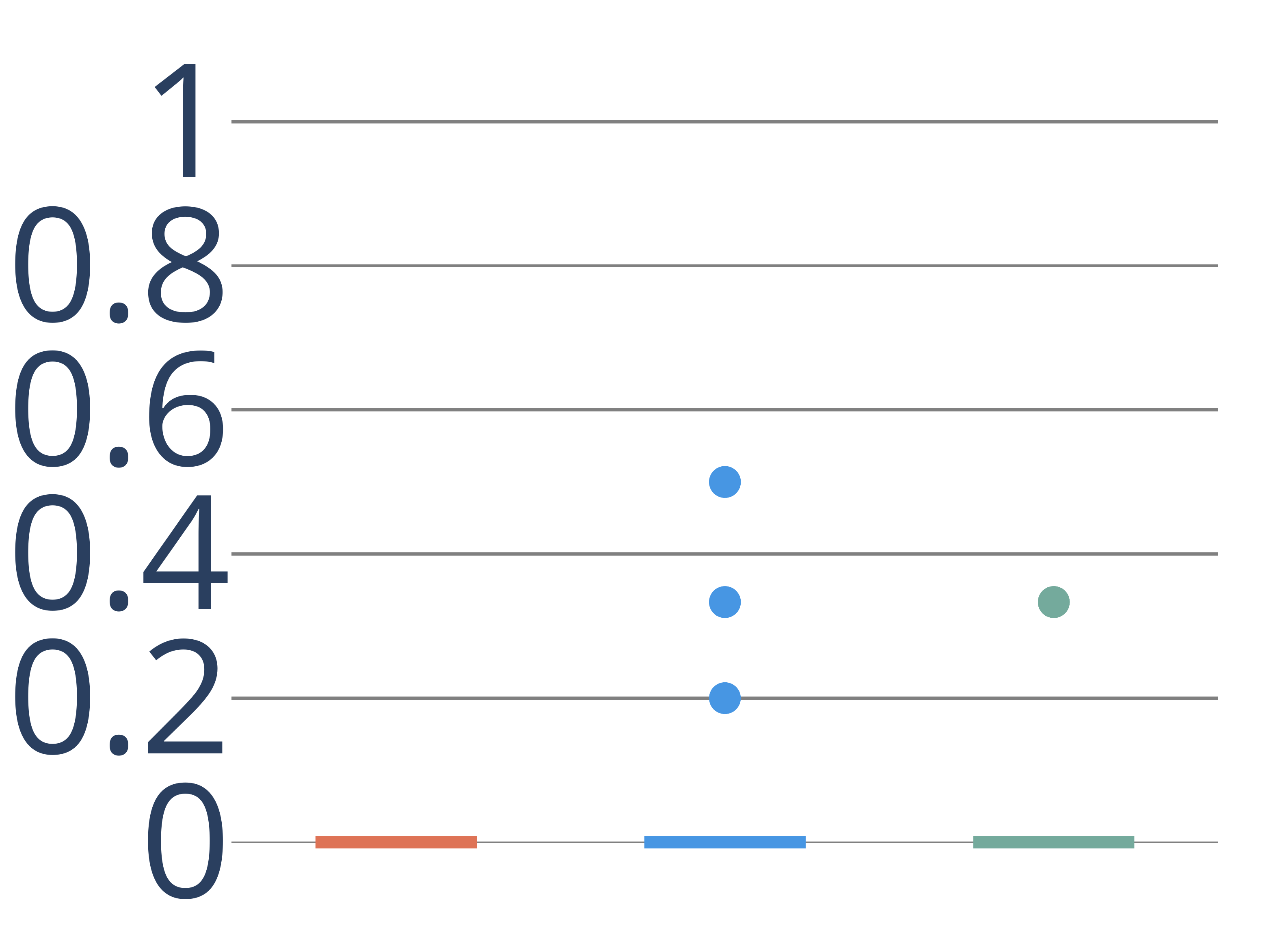} 
                                        & 
                                        & 
                                        & \includegraphics[scale=0.025]{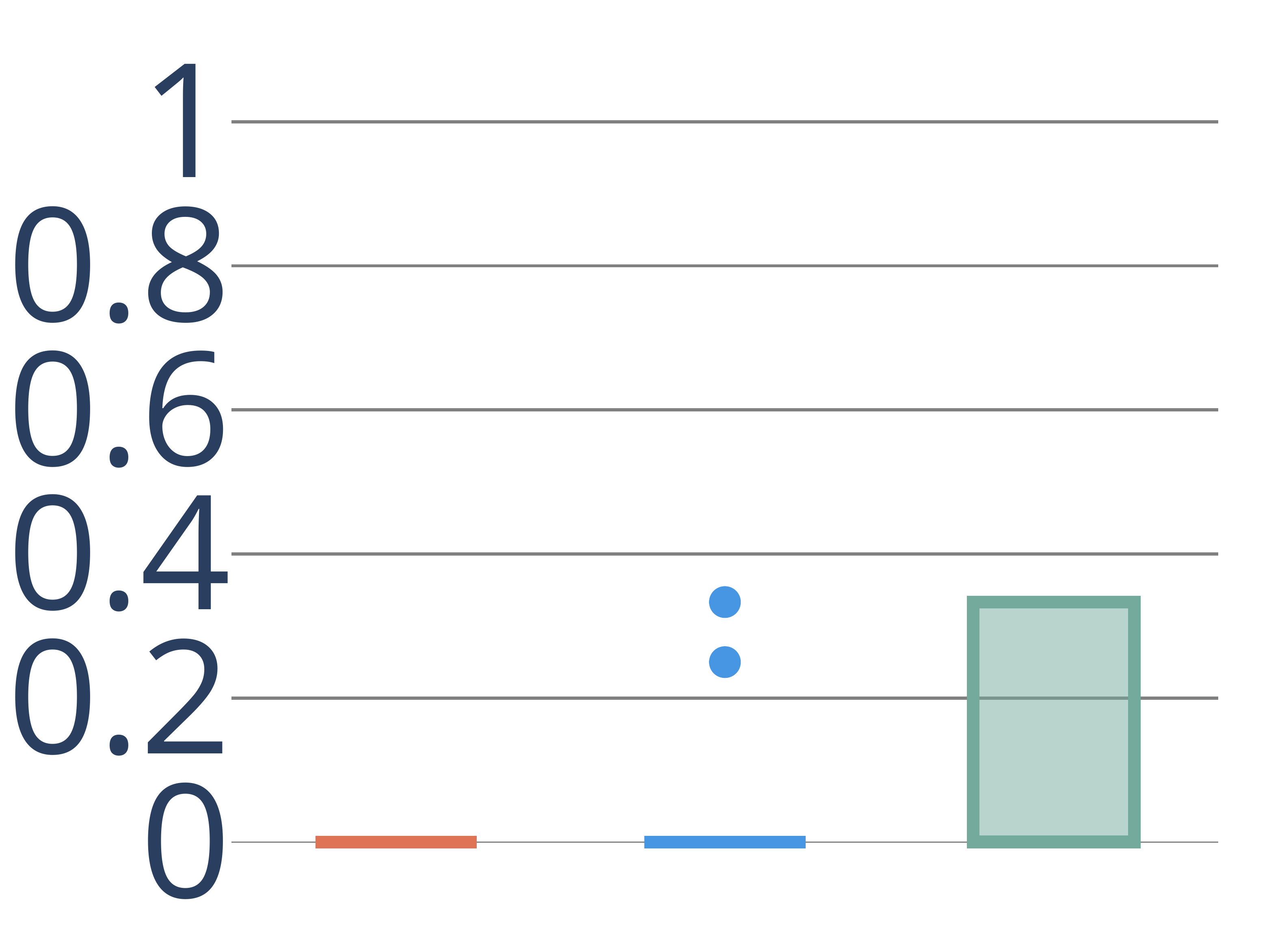} 
                                        & \includegraphics[scale=0.025]{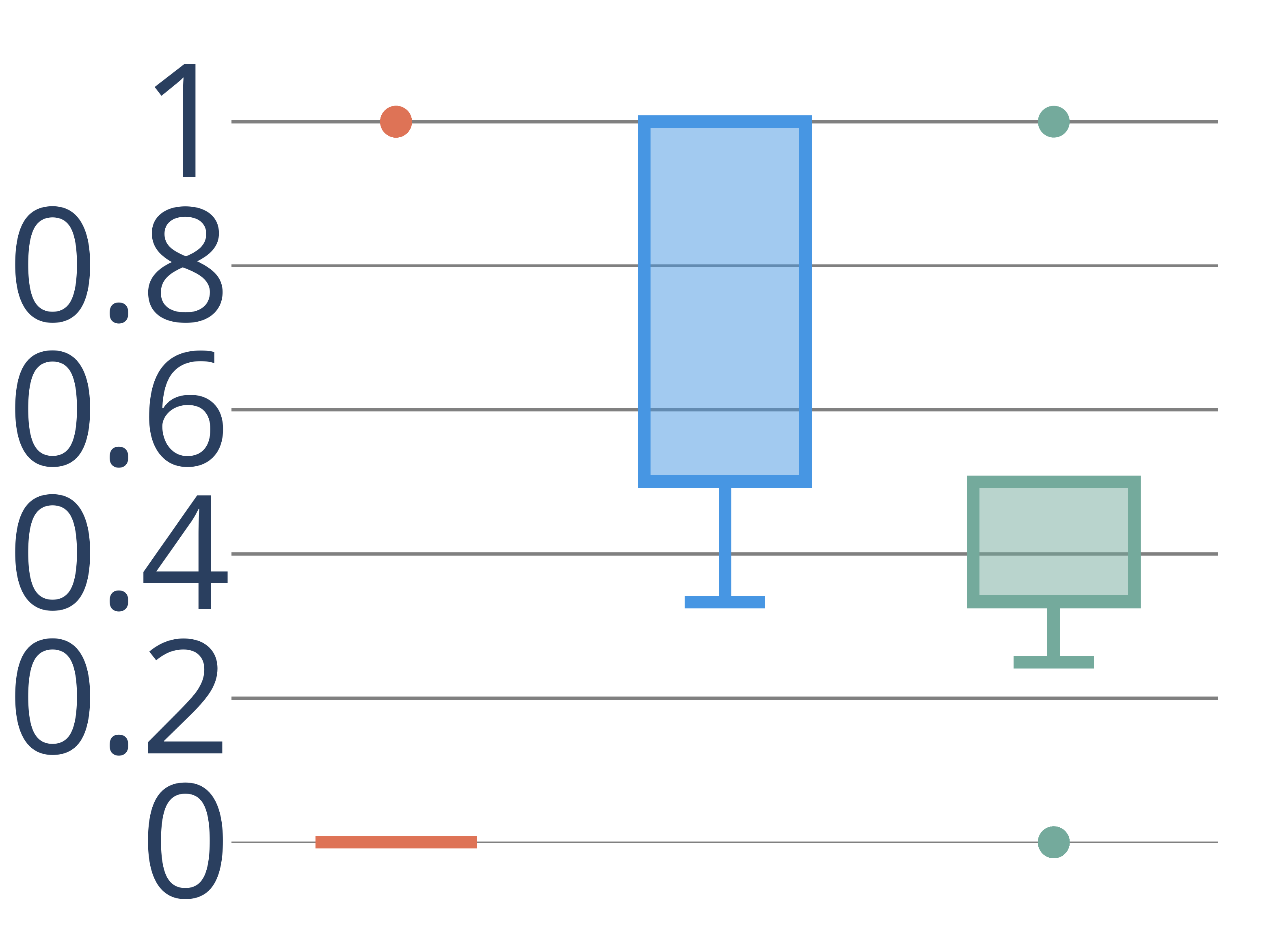}\\
        \hline
        \SetCell[r=4]{c,manuVisEncode} \rotatebox[origin=c]{90}{\textbf{Manipulated Visual Encoding}}
                                    & \makecell[l]{\textbf{Data-visual}\\ \textbf{Disproportion}} 
                                        & 
                                        & 
                                        & \includegraphics[scale=0.025]{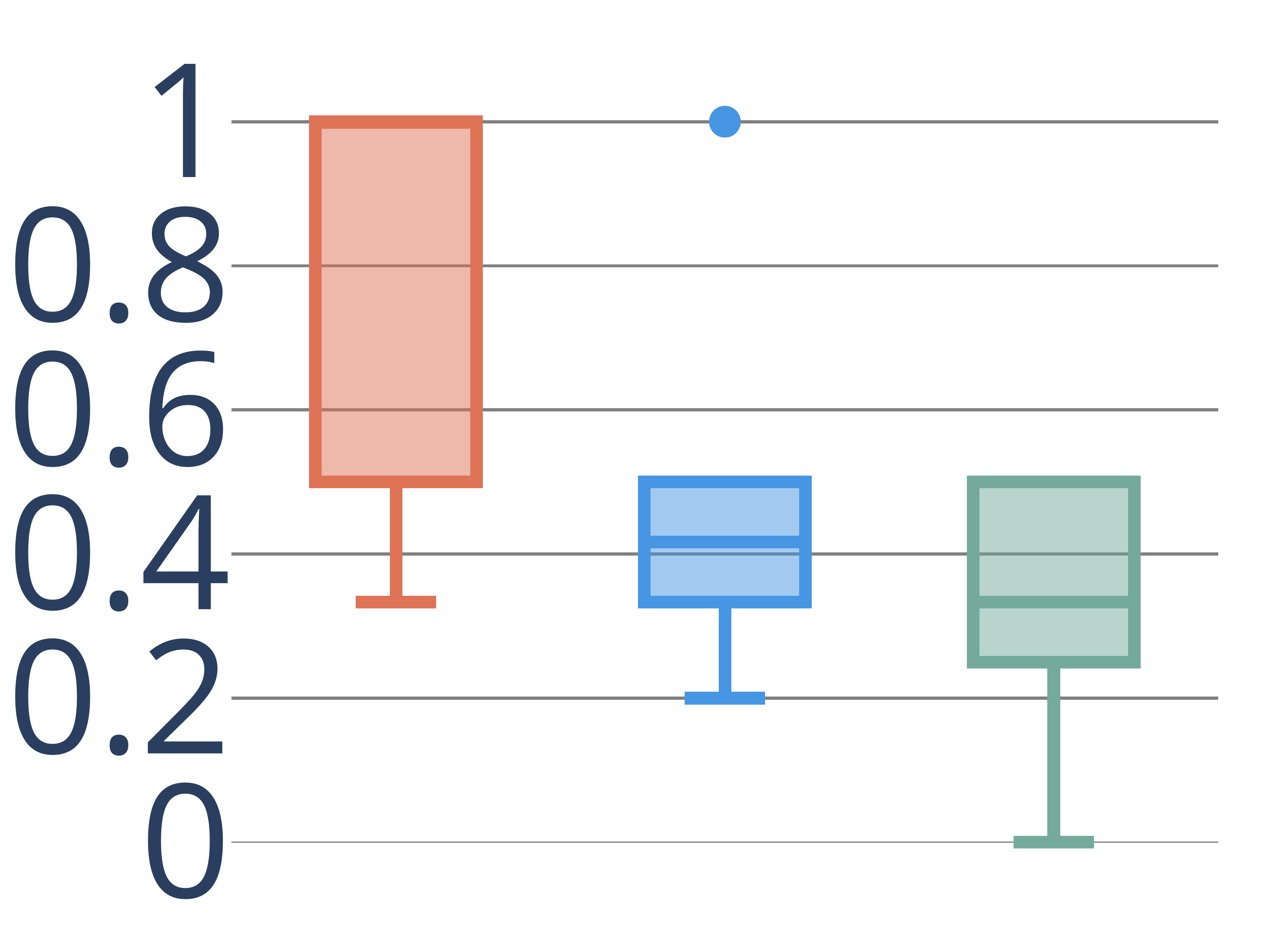} 
                                        & 
                                        & 
                                        & \includegraphics[scale=0.025]{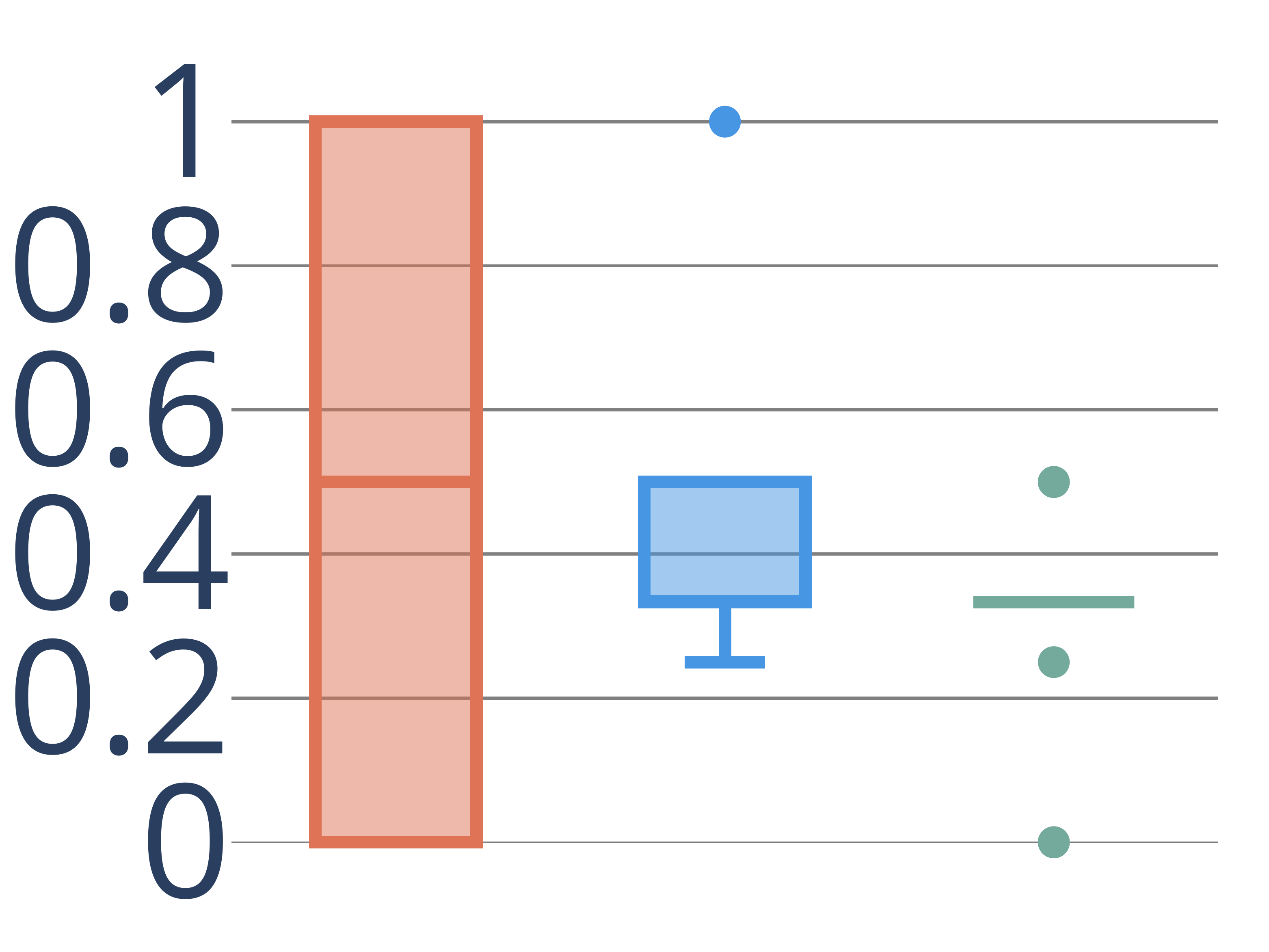} 
                                        & \includegraphics[scale=0.025]{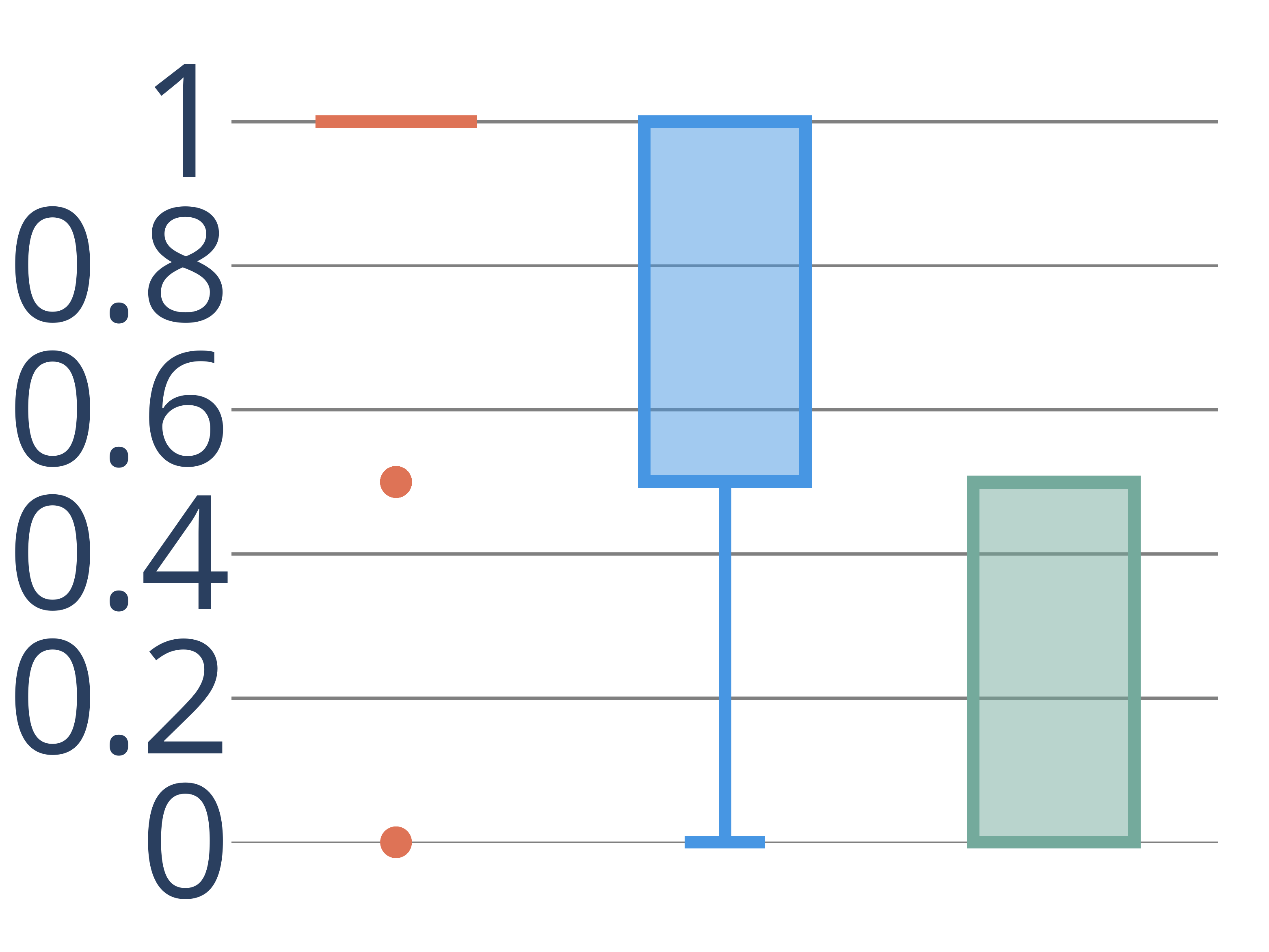} 
                                        & 
                                        & 
                                        & \includegraphics[scale=0.025]{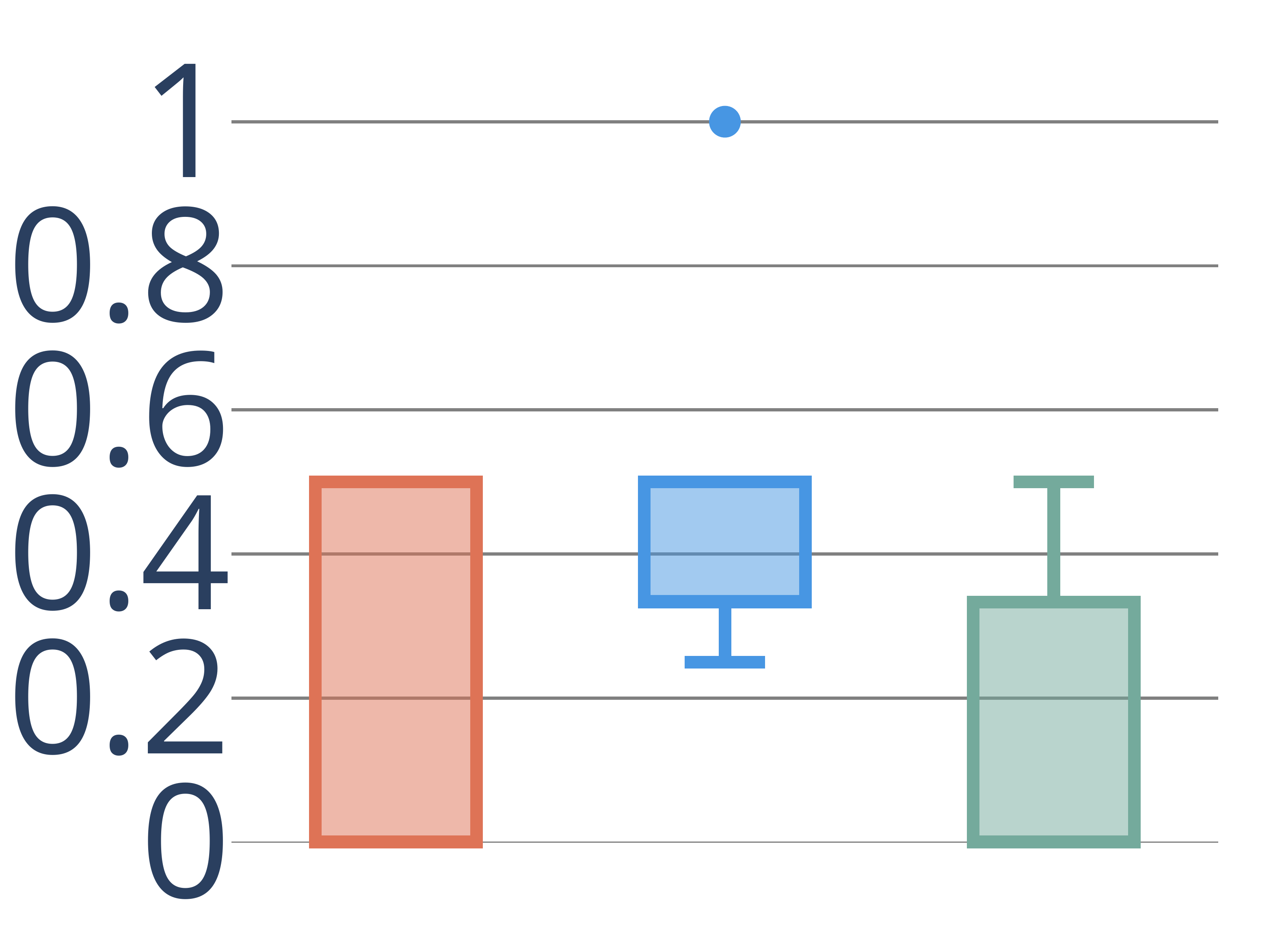}\\
                                    & \textbf{Dual Encoding} 
                                        & 
                                        & \includegraphics[scale=0.025]{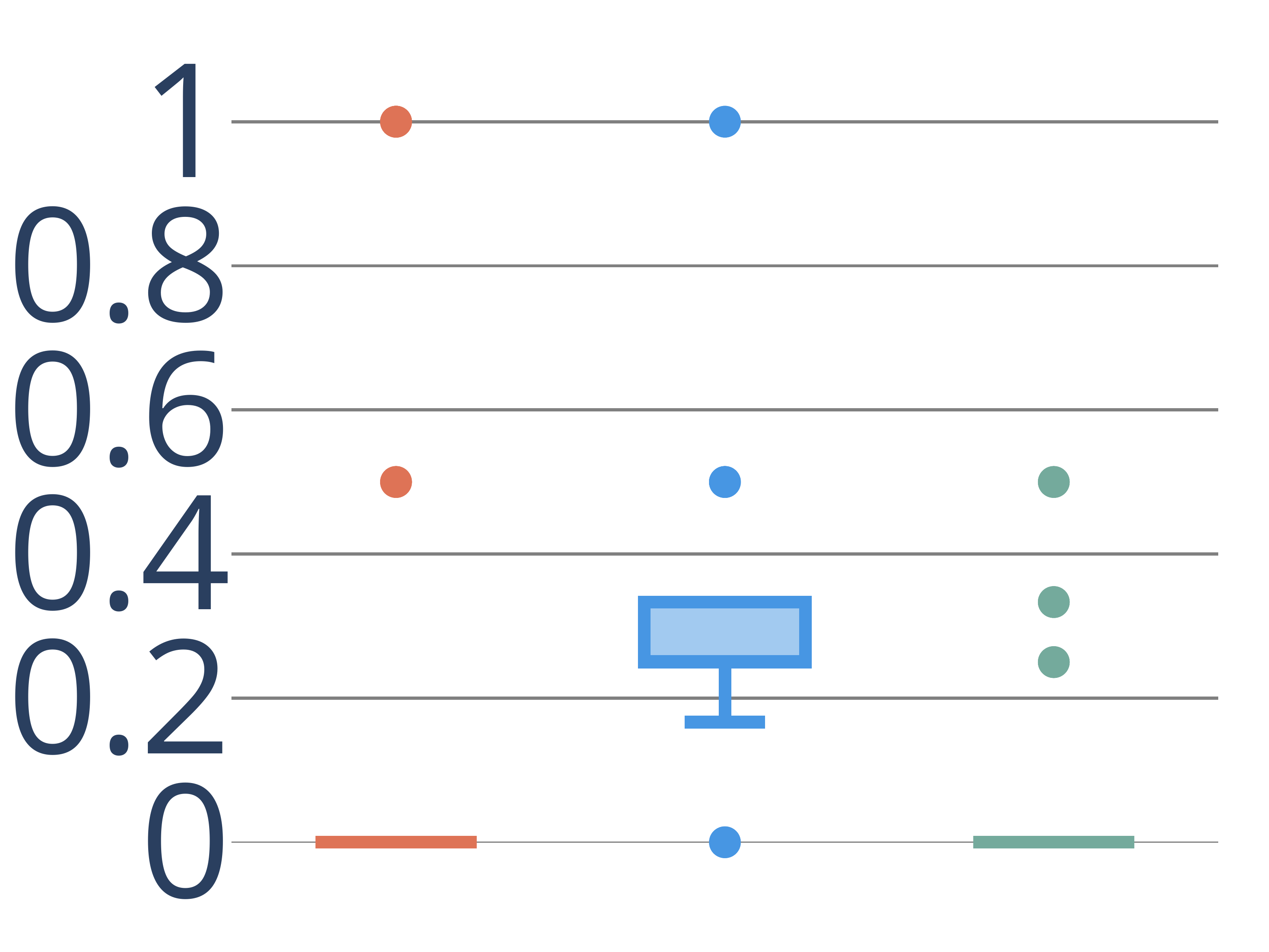} 
                                        & 
                                        & 
                                        & 
                                        & 
                                        & \includegraphics[scale=0.025]{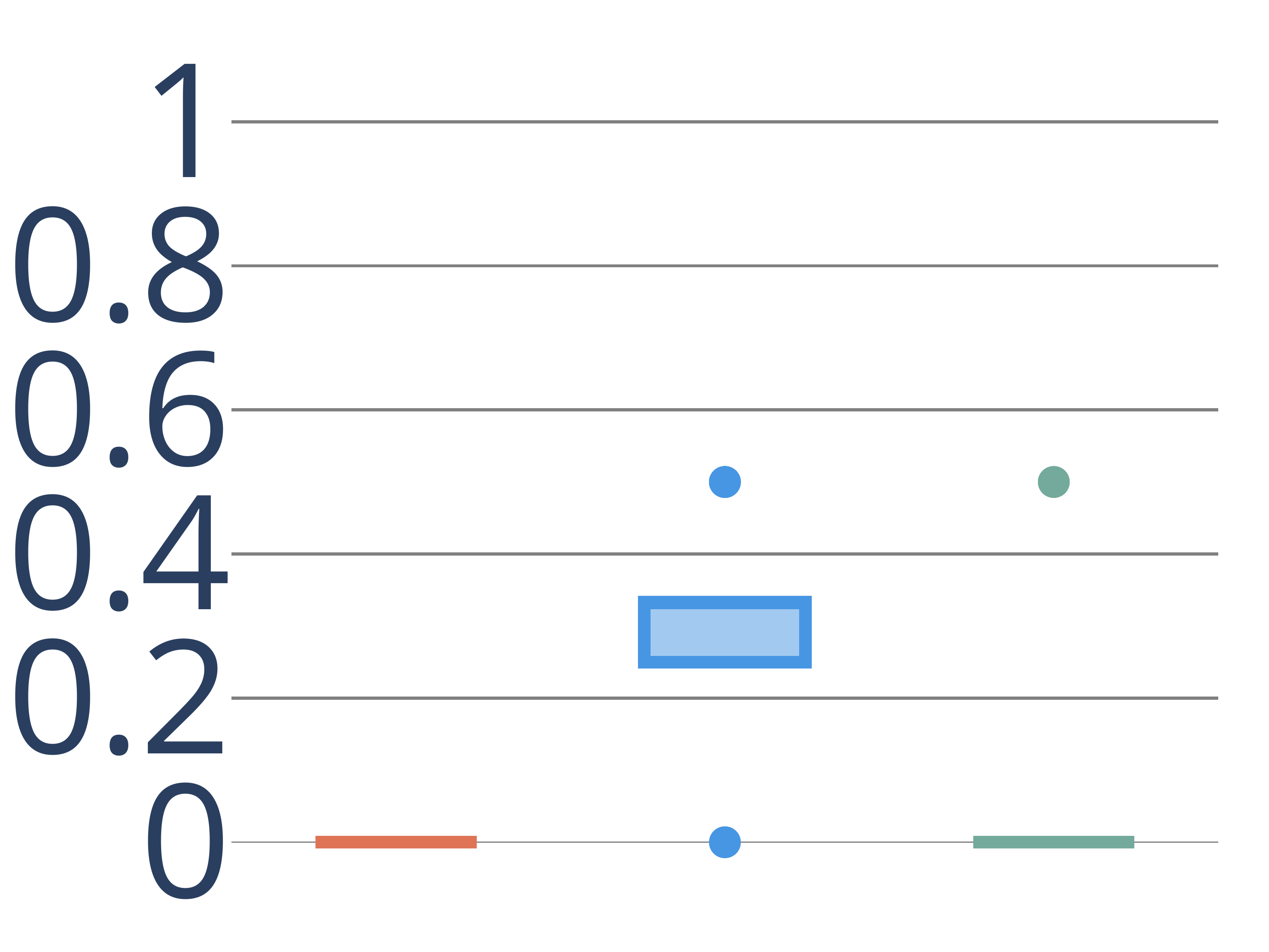} 
                                        & 
                                        & 
                                        & \includegraphics[scale=0.025]{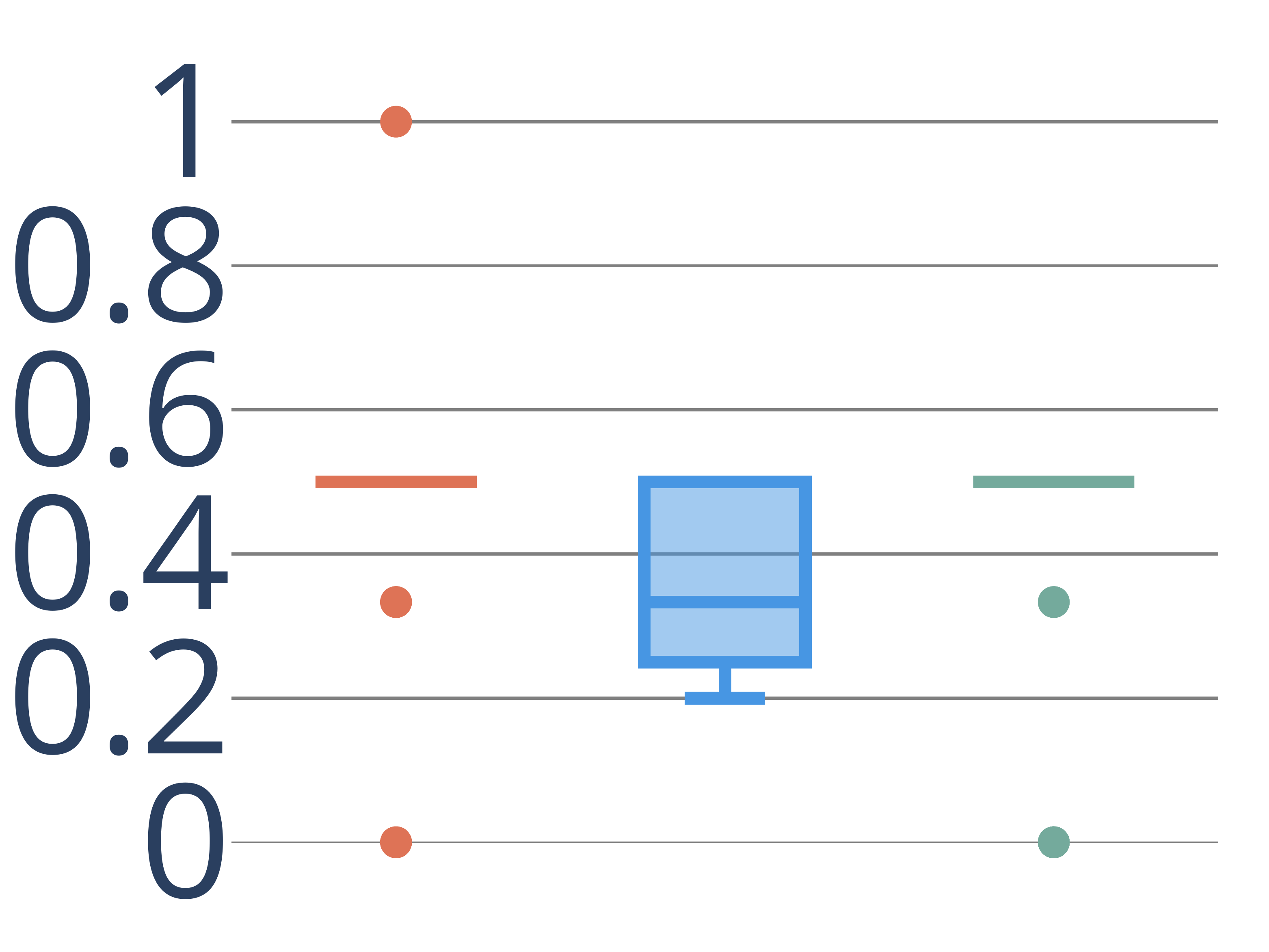}\\
                                    & \makecell[l]{\textbf{Mismatched Encoding:}\\ \textit{Continuous encoding of}\\ \textit{categorical data}} 
                                        & 
                                        & \includegraphics[scale=0.025]{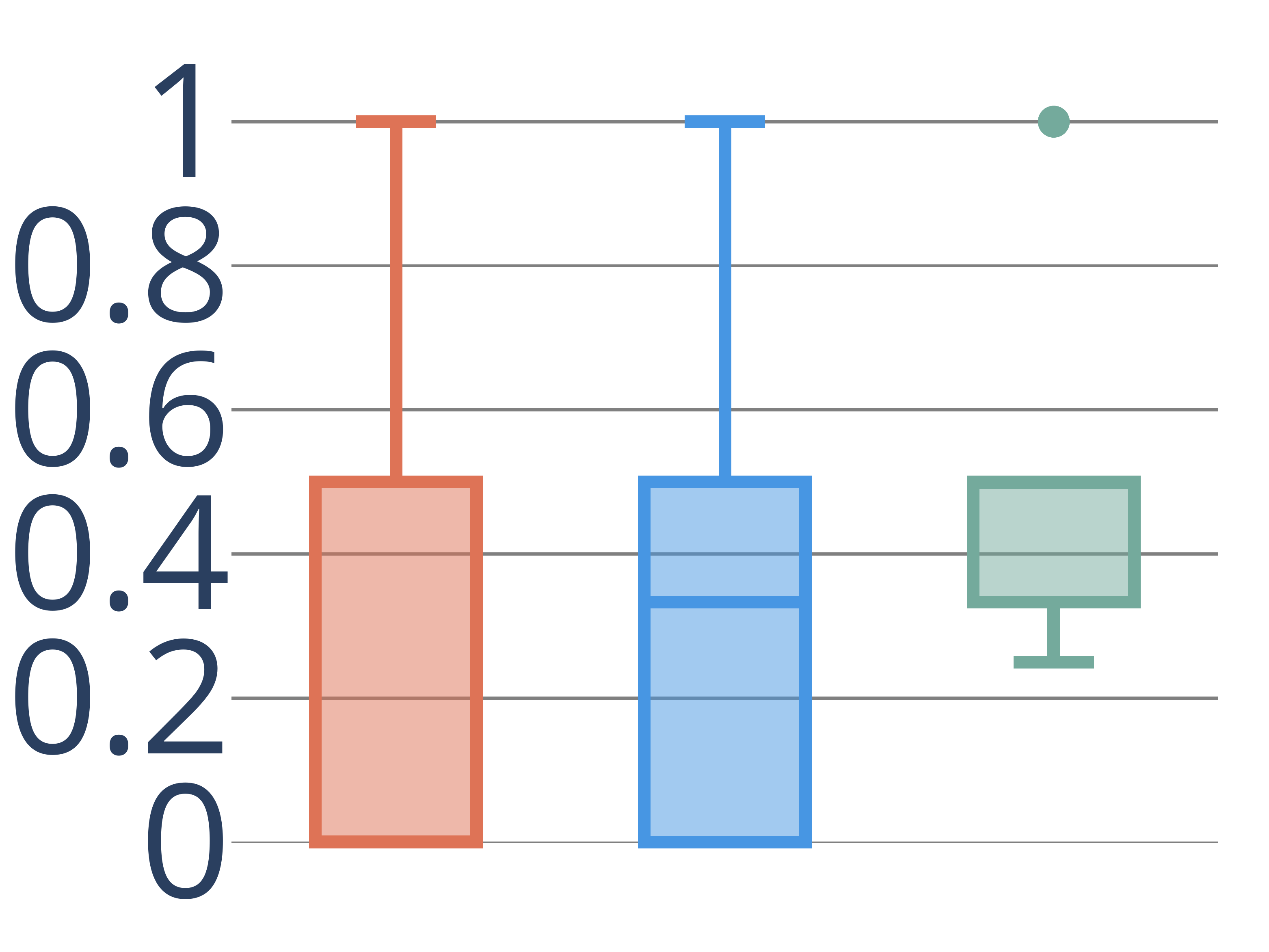} 
                                        & 
                                        & 
                                        & 
                                        & \includegraphics[scale=0.025]{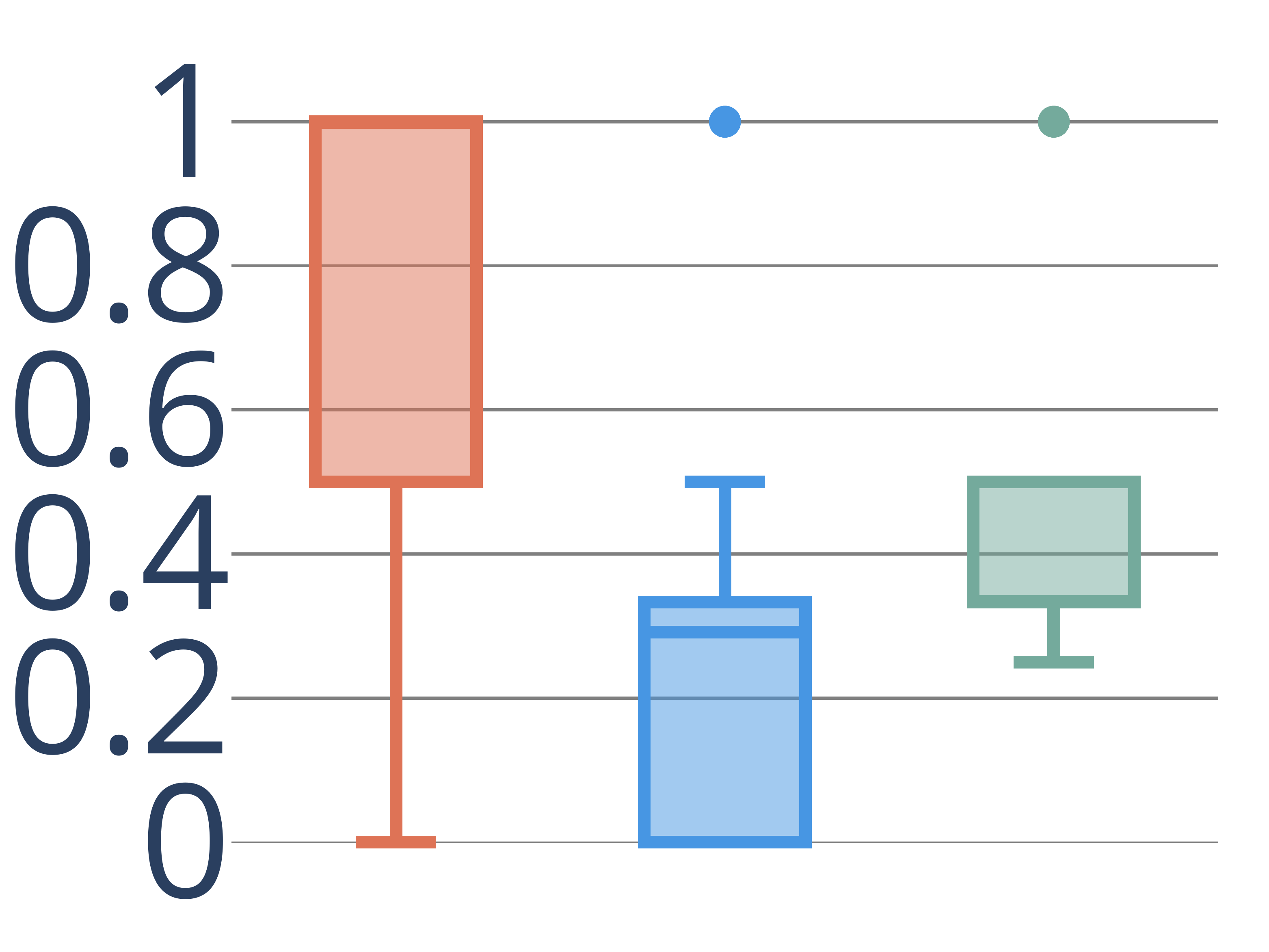} 
                                        & \includegraphics[scale=0.025]{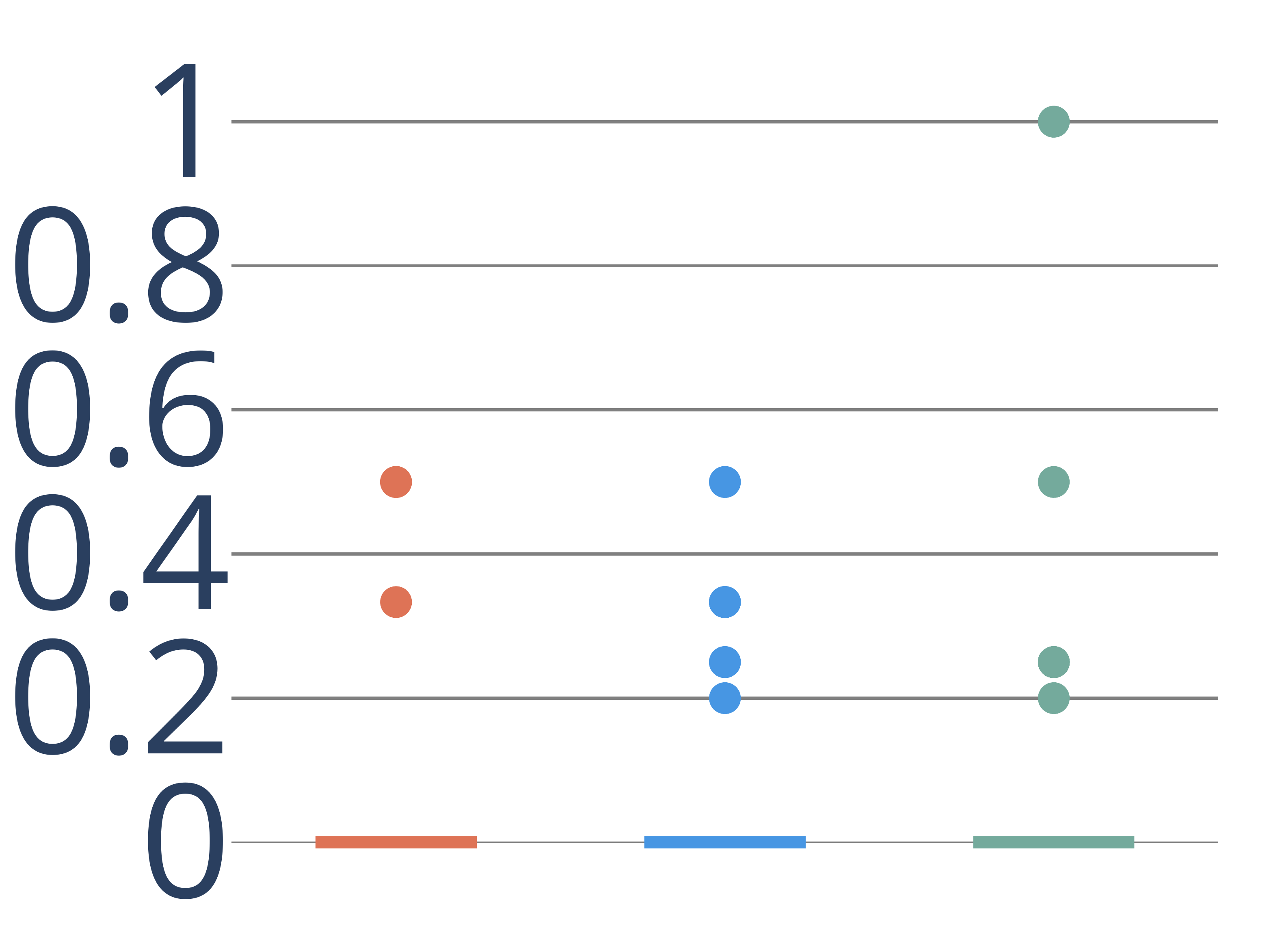} 
                                        & 
                                        & 
                                        &\\
                                    & \makecell[l]{\textbf{Mismatched Encoding:}\\ \textit{Categorical encoding of}\\ \textit{continuous data}} 
                                        & 
                                        & 
                                        & 
                                        & \includegraphics[scale=0.025]{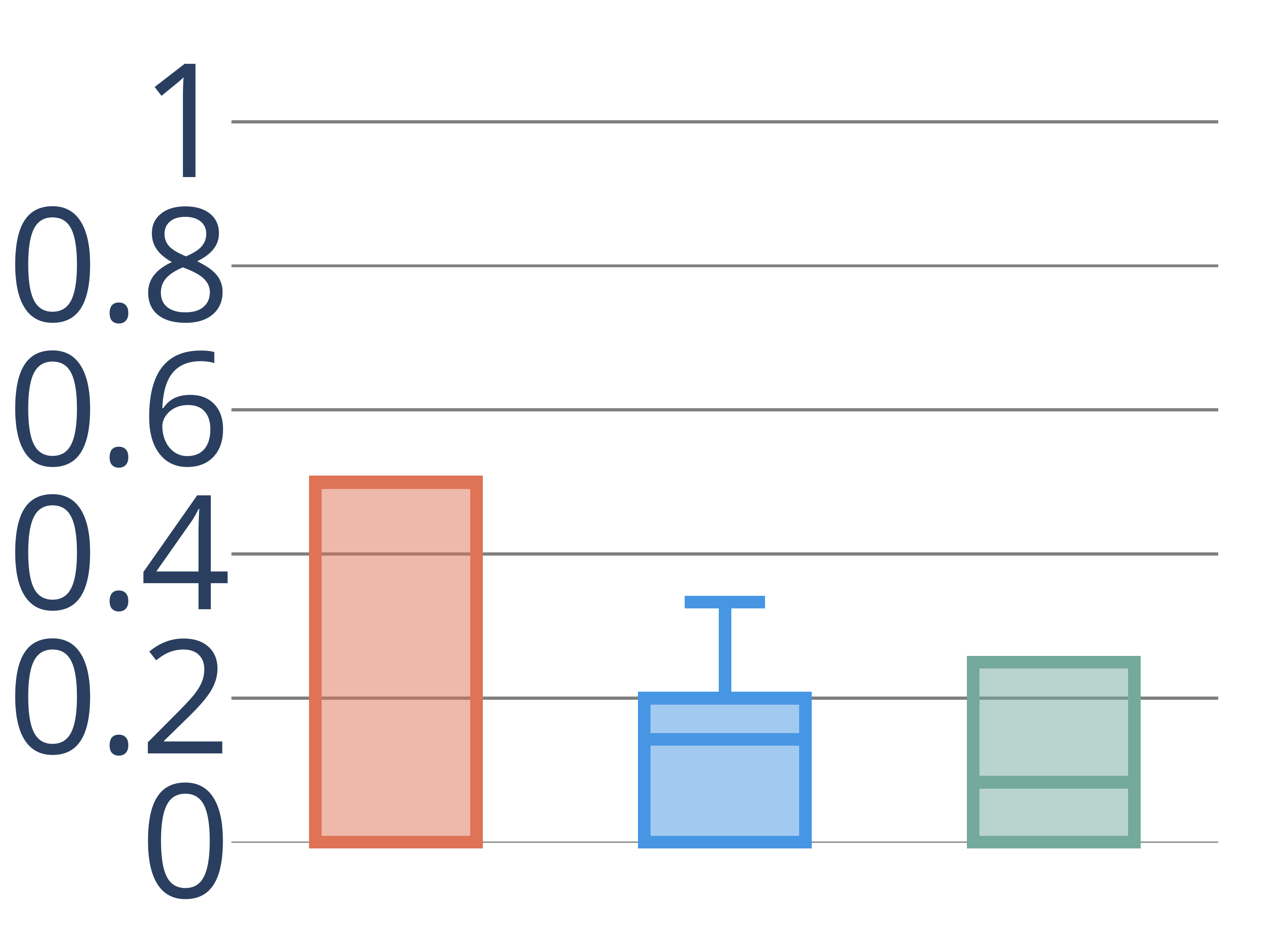} 
                                        & \includegraphics[scale=0.025]{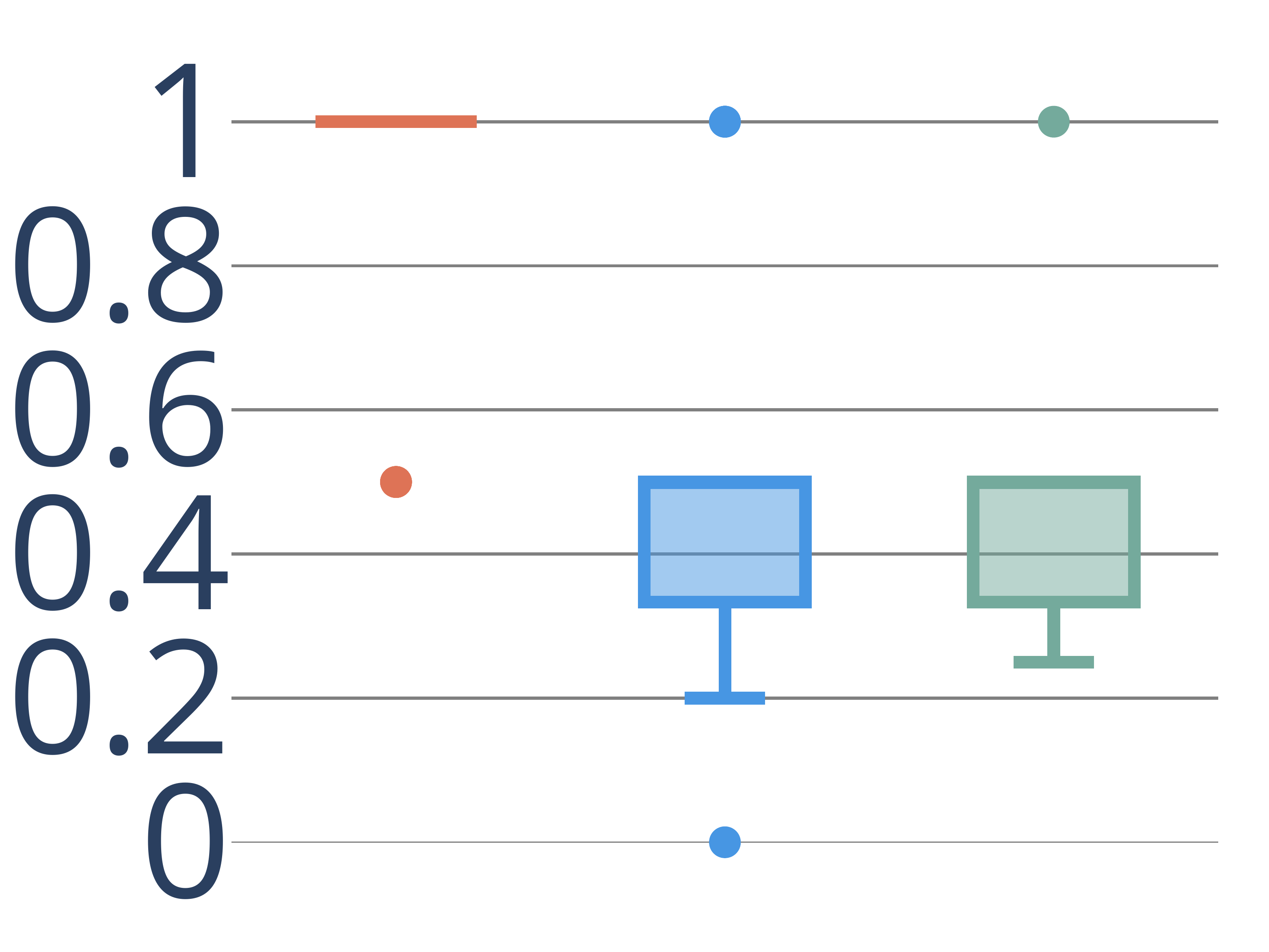} 
                                        & 
                                        & 
                                        & 
                                        & 
                                        &\\
        \hline
        \SetCell[r=6]{c,manuScale} \rotatebox[origin=c]{90}{\textbf{Manipulated Scale}}
                                    & \makecell[l]{\textbf{Inappropriate}\\ \textbf{Scale Range}} 
                                        & 
                                        & 
                                        & \includegraphics[scale=0.025]{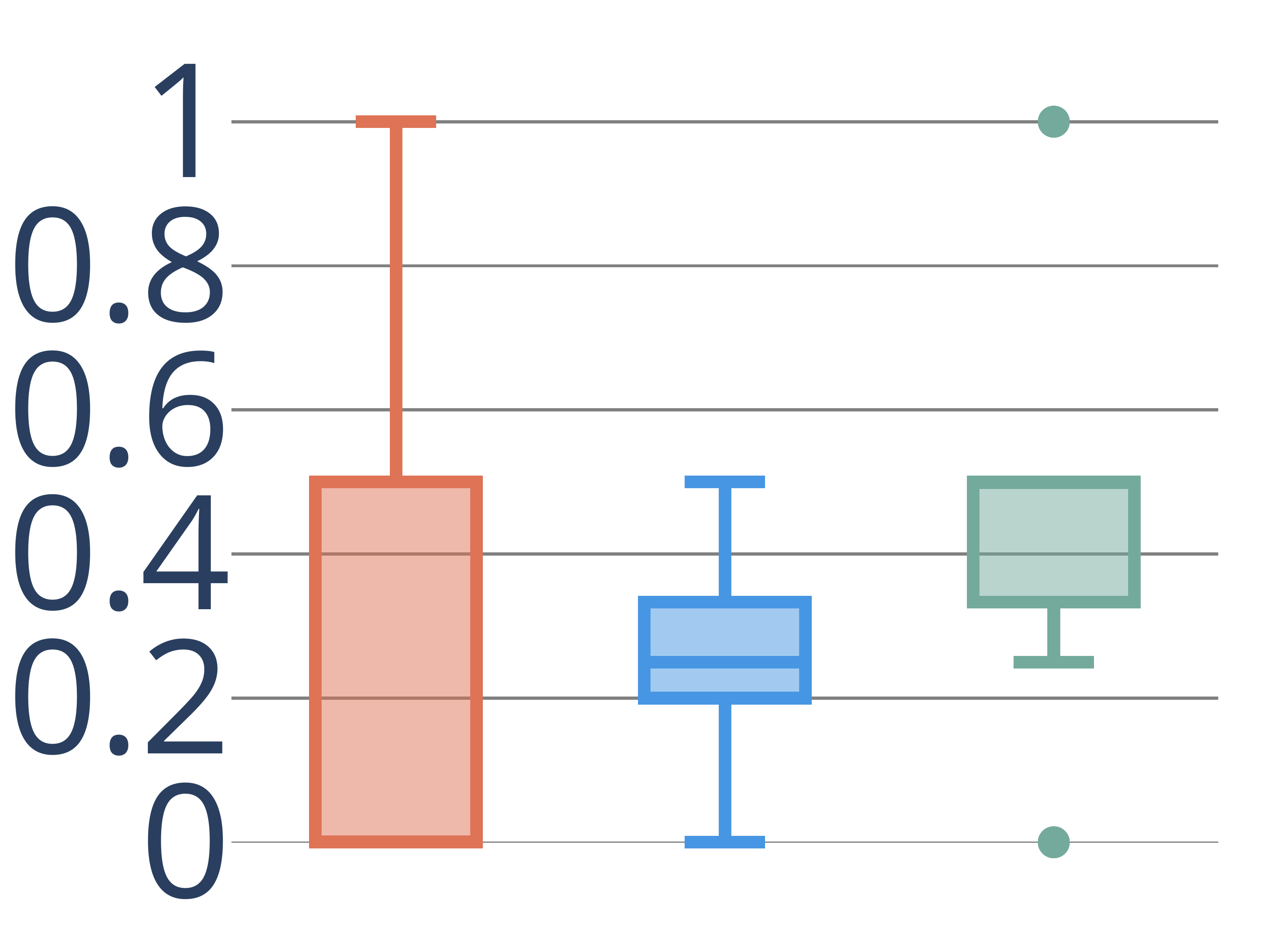} 
                                        & \includegraphics[scale=0.025]{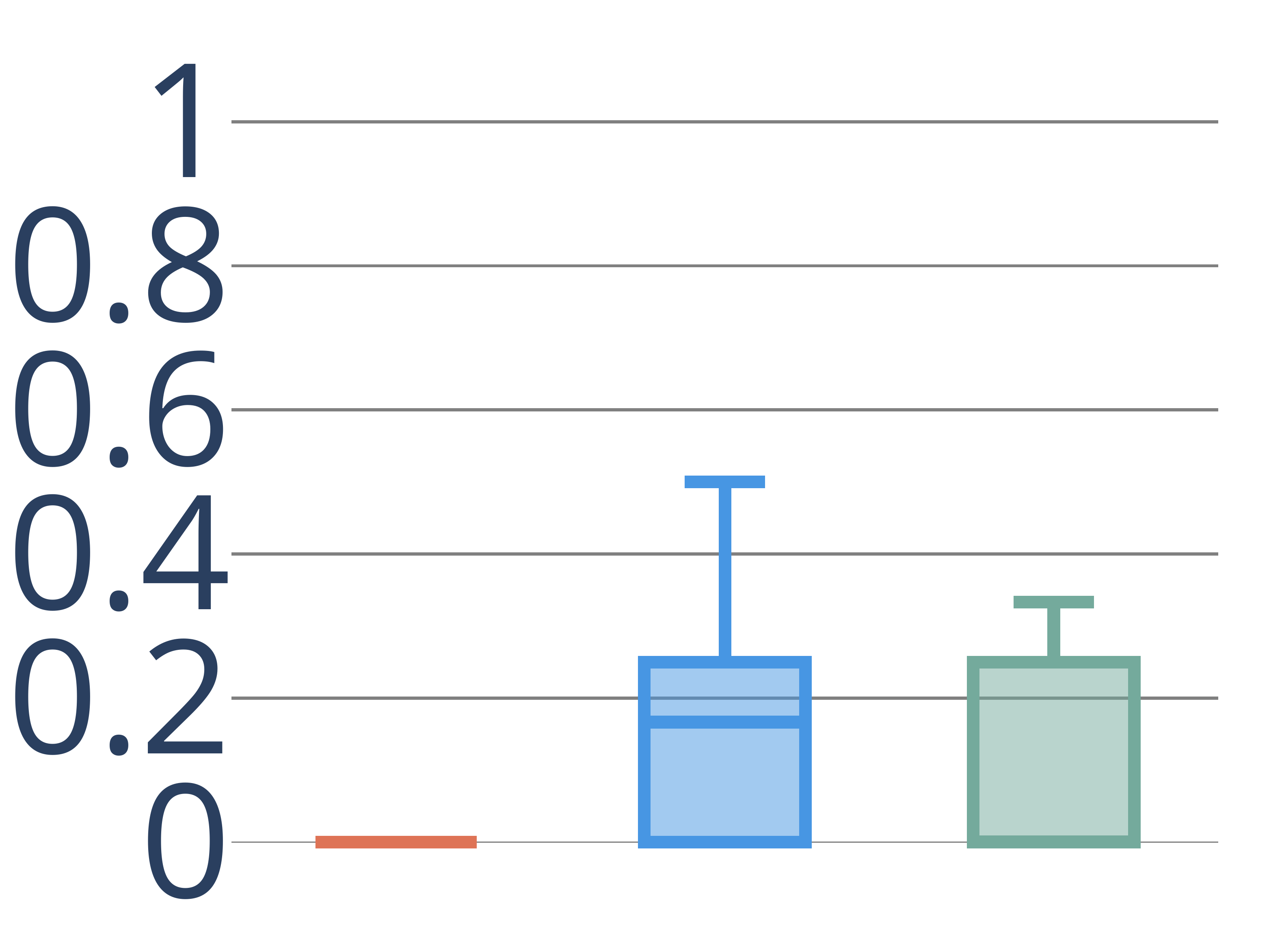} 
                                        & 
                                        & \includegraphics[scale=0.025]{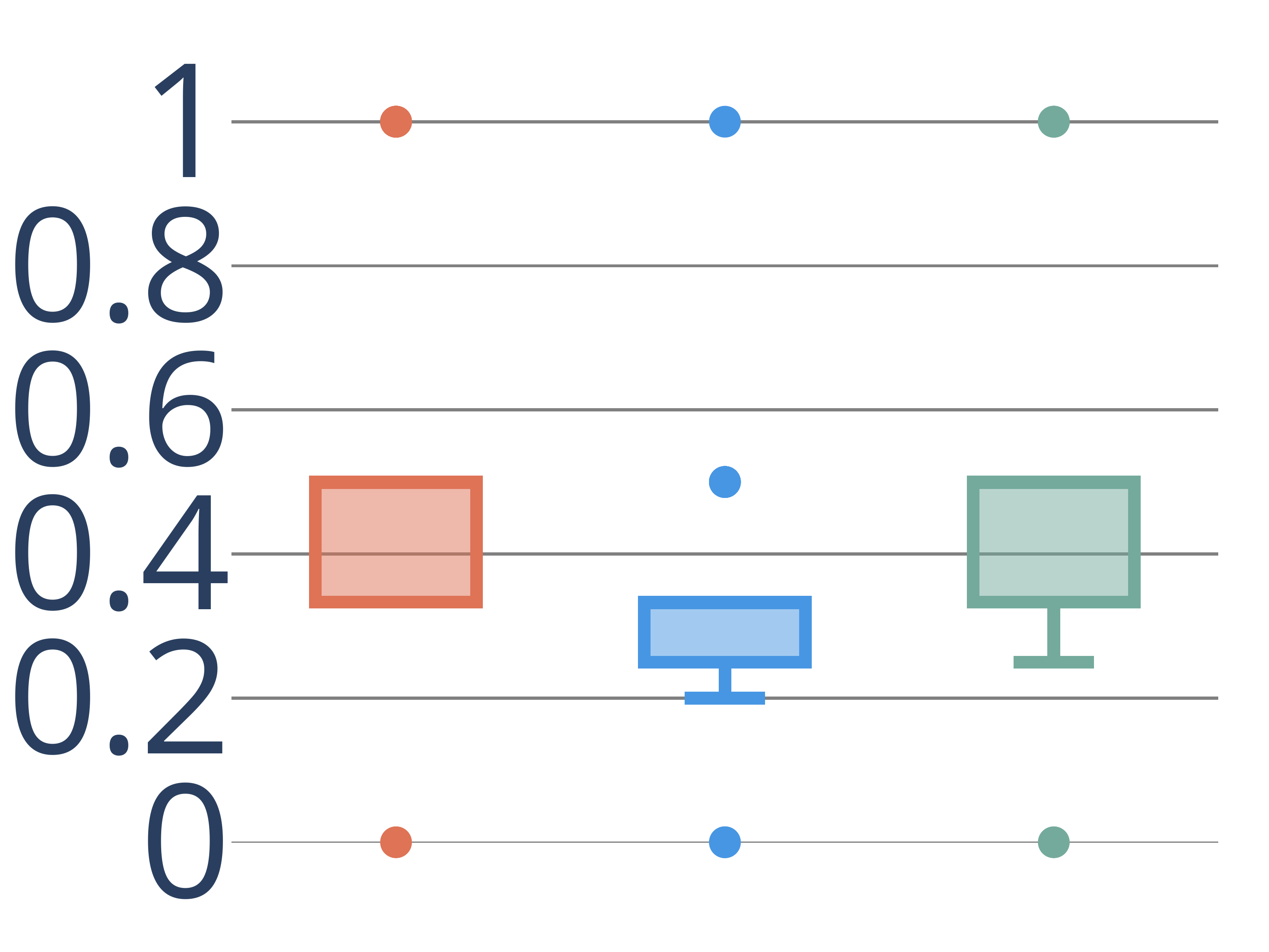} 
                                        & 
                                        & \includegraphics[scale=0.025]{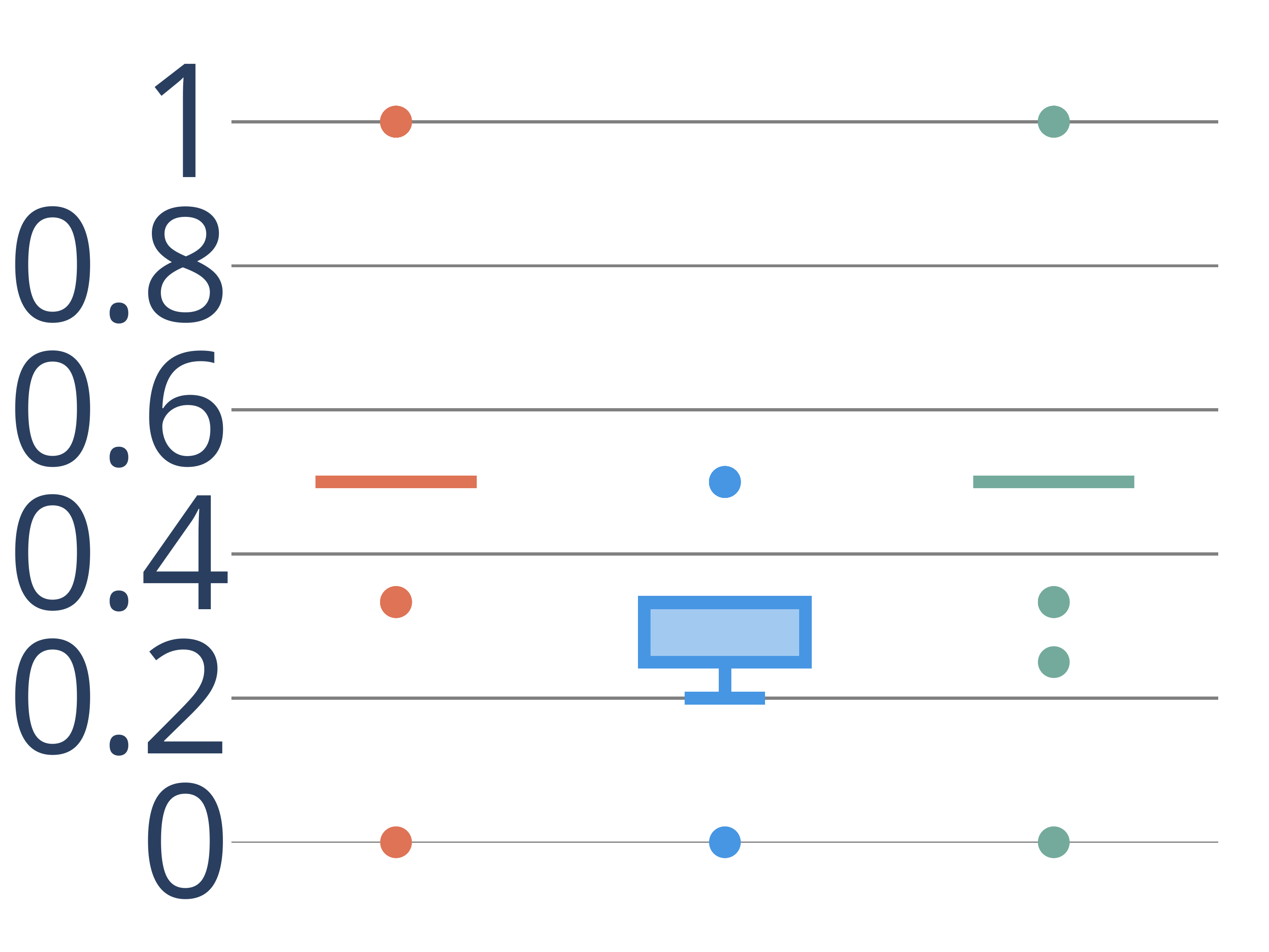} 
                                        & \includegraphics[scale=0.025]{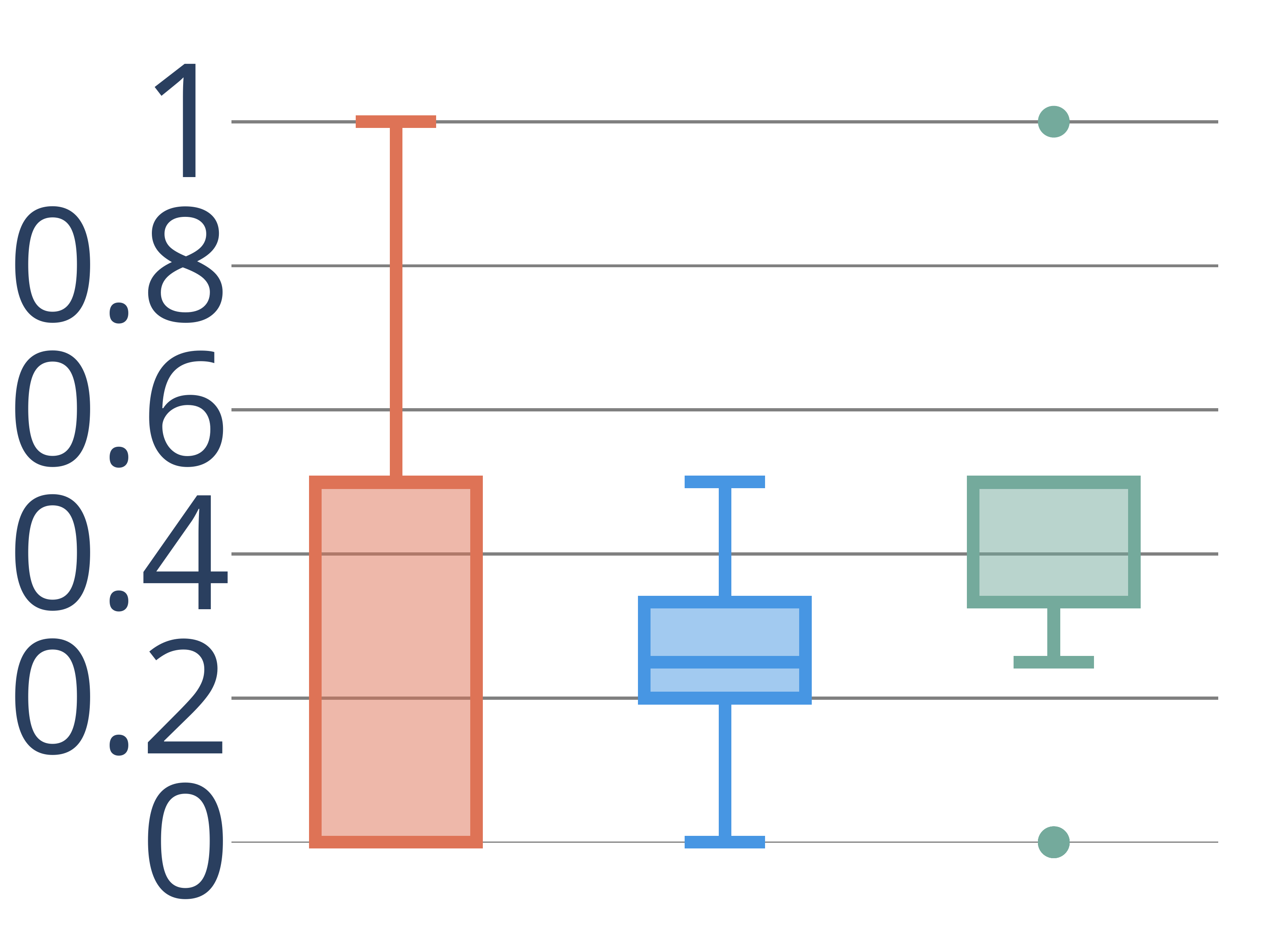} 
                                        &\\
                                    & \makecell[l]{\textbf{Inappropriate Scale}\\ \textbf{Functions}} 
                                        & 
                                        & 
                                        & \includegraphics[scale=0.025]{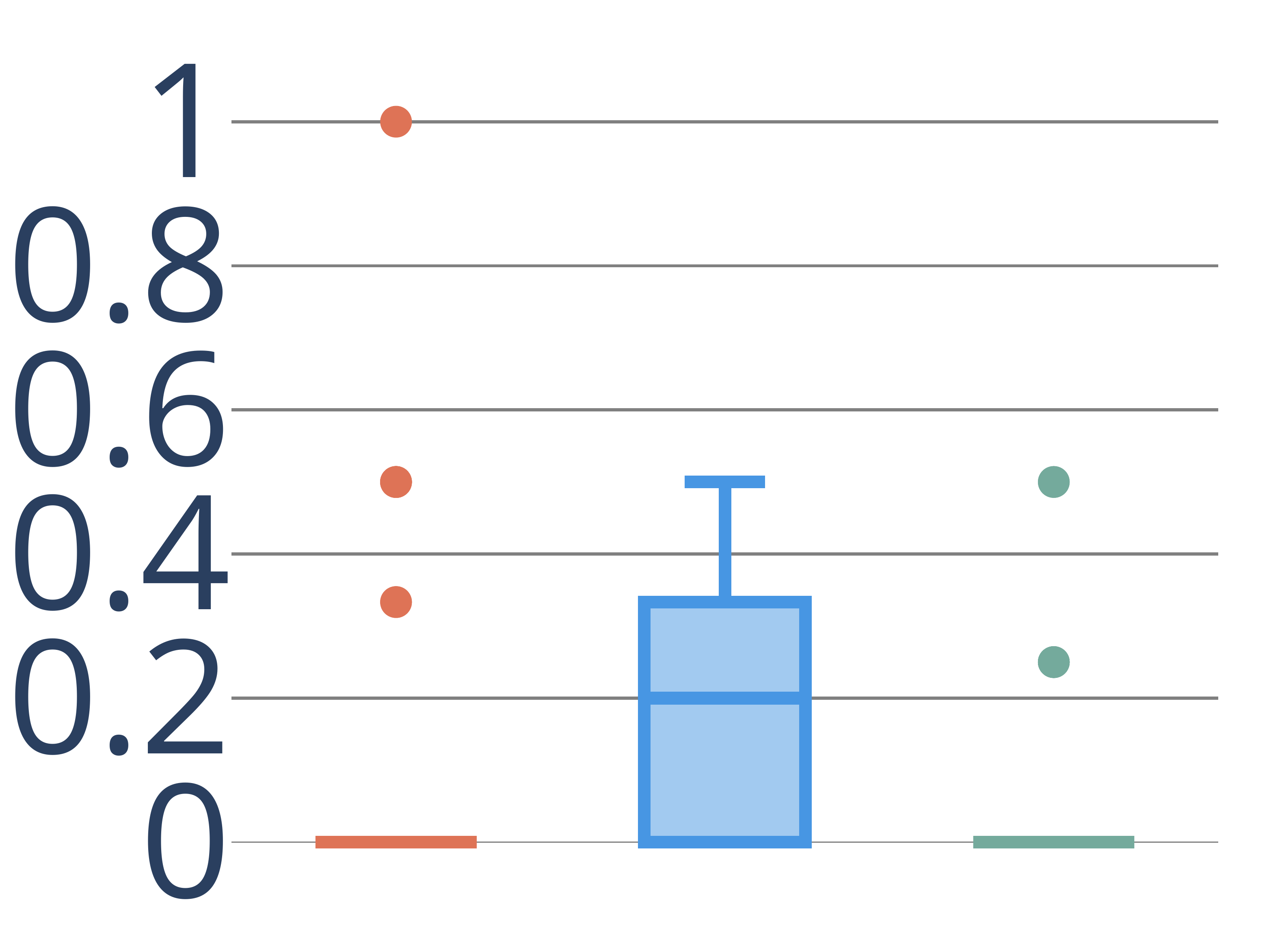} 
                                        & 
                                        & 
                                        & \includegraphics[scale=0.025]{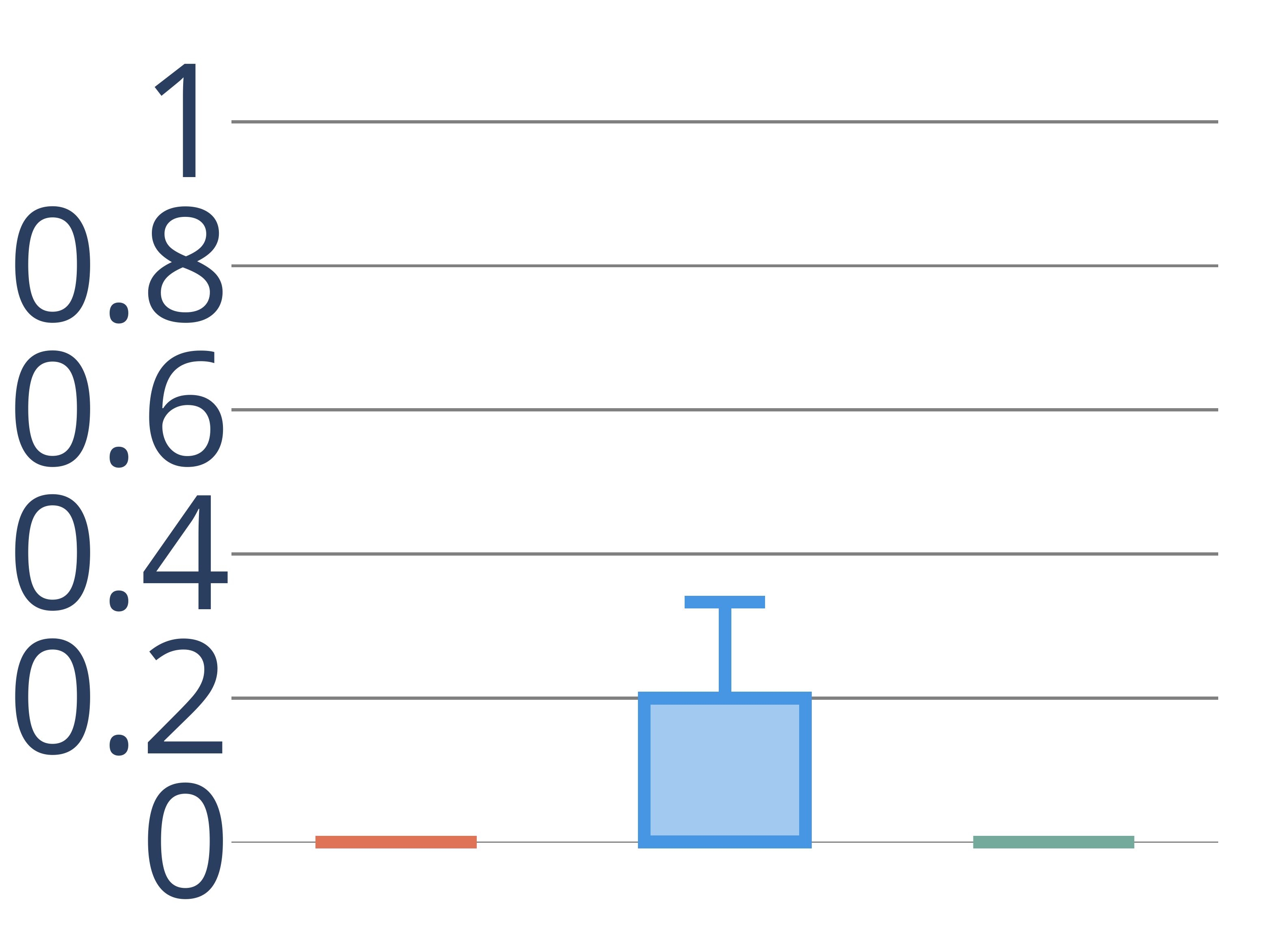} 
                                        & \includegraphics[scale=0.025]{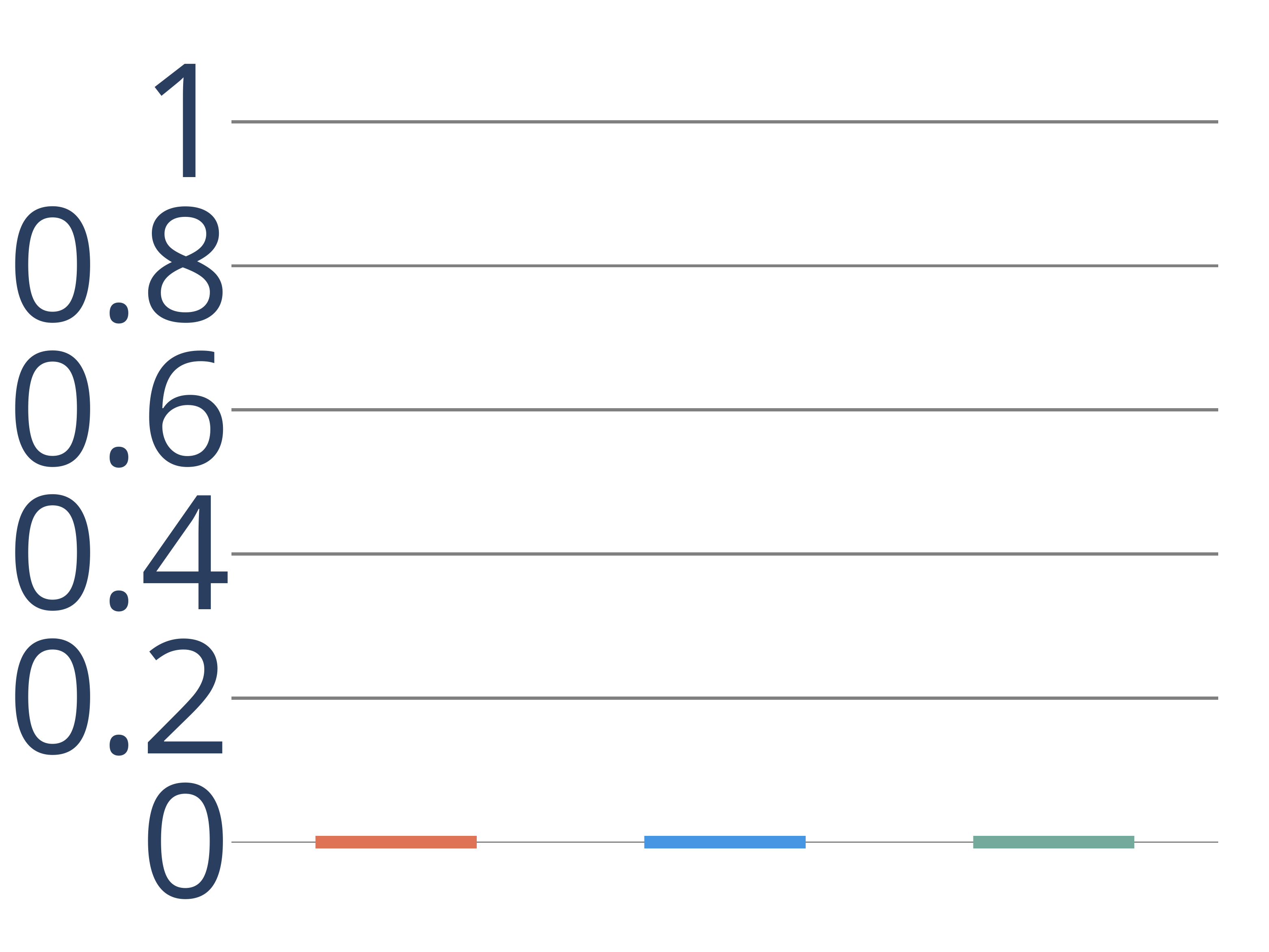} 
                                        & 
                                        & 
                                        &\\
                                    & \makecell[l]{\textbf{Unconventional Scale}\\ \textbf{Directions}} 
                                        & 
                                        & \includegraphics[scale=0.025]{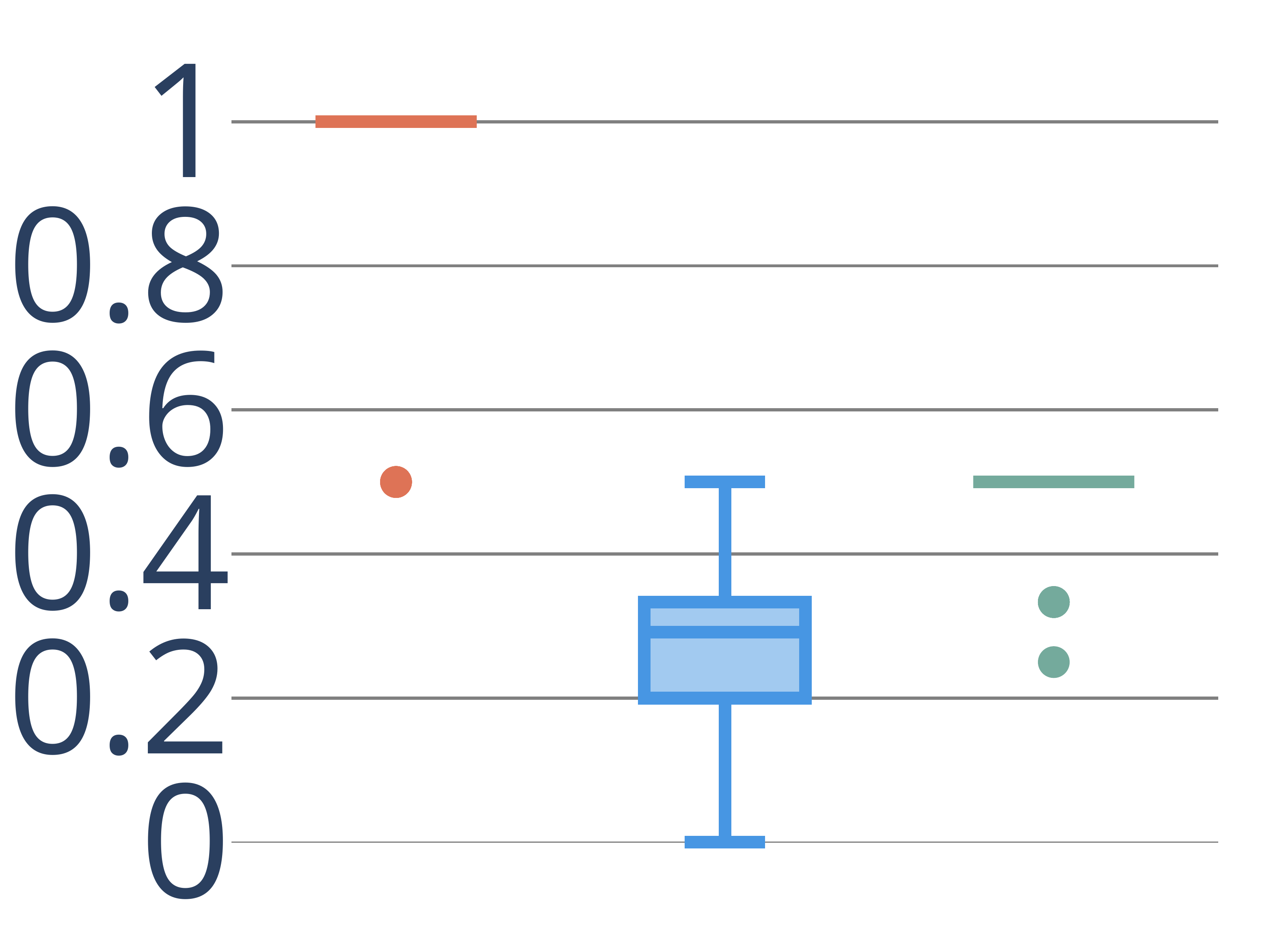} 
                                        & \includegraphics[scale=0.025]{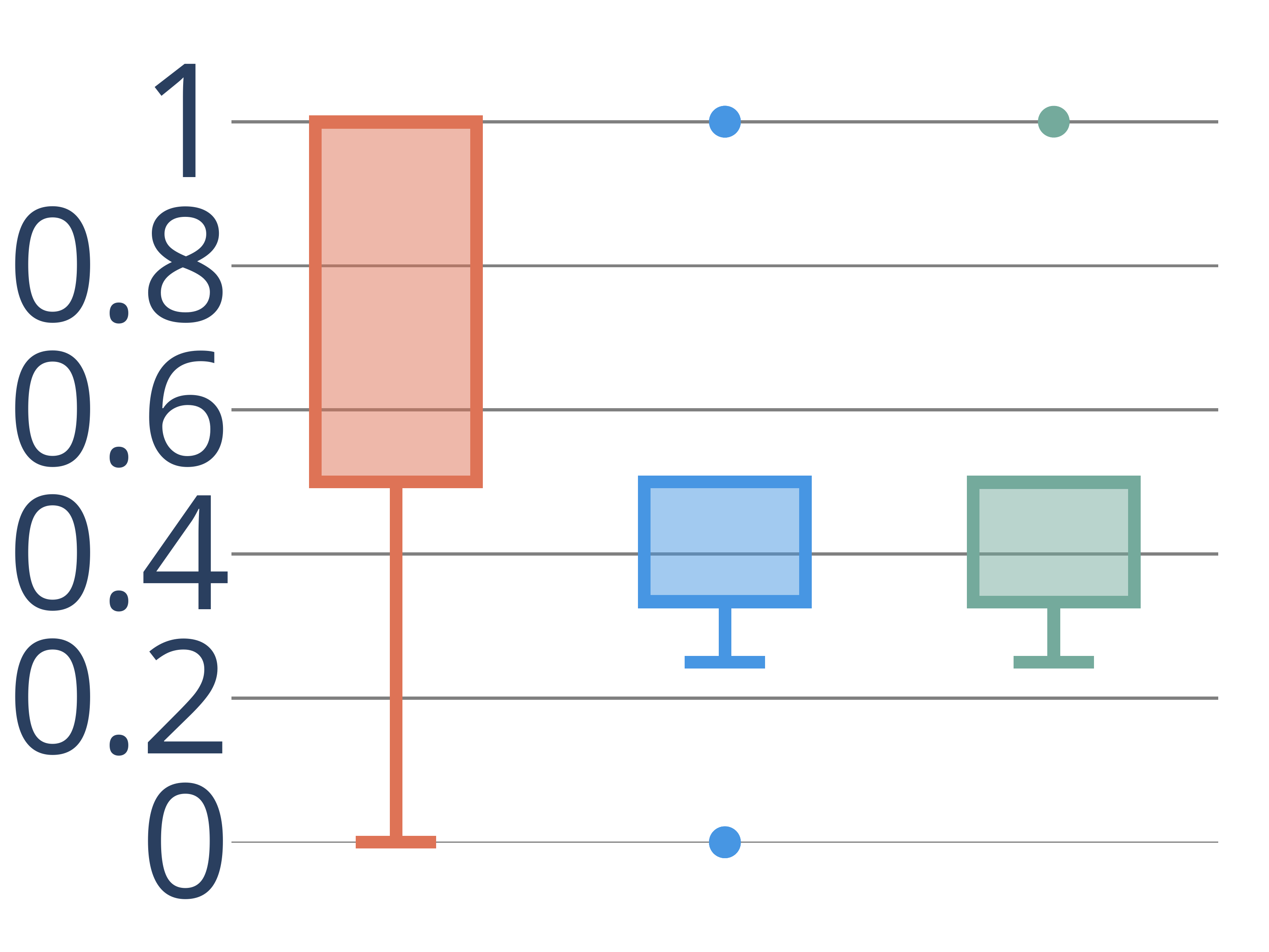} 
                                        & \includegraphics[scale=0.025]{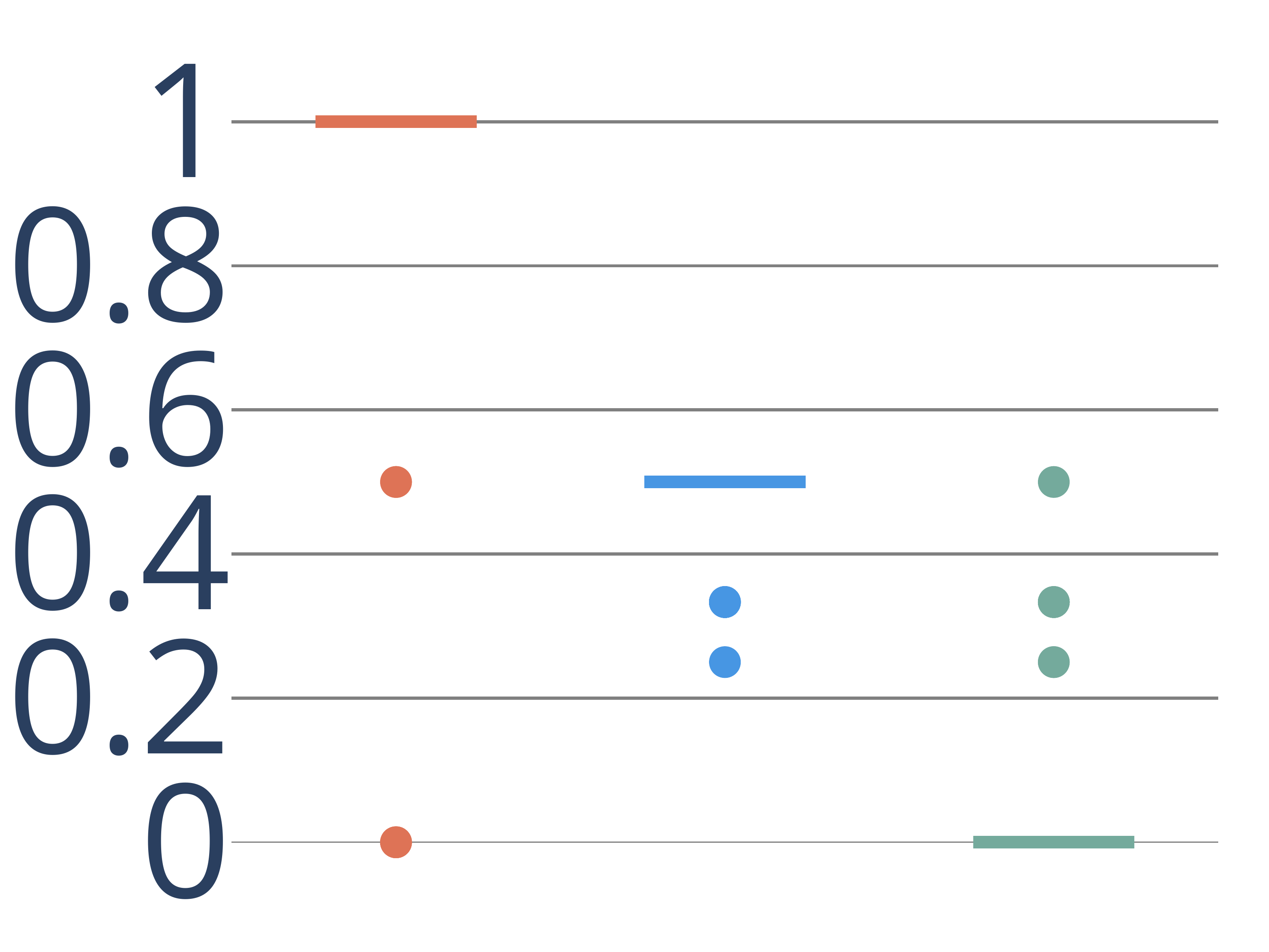} 
                                        & 
                                        & \includegraphics[scale=0.025]{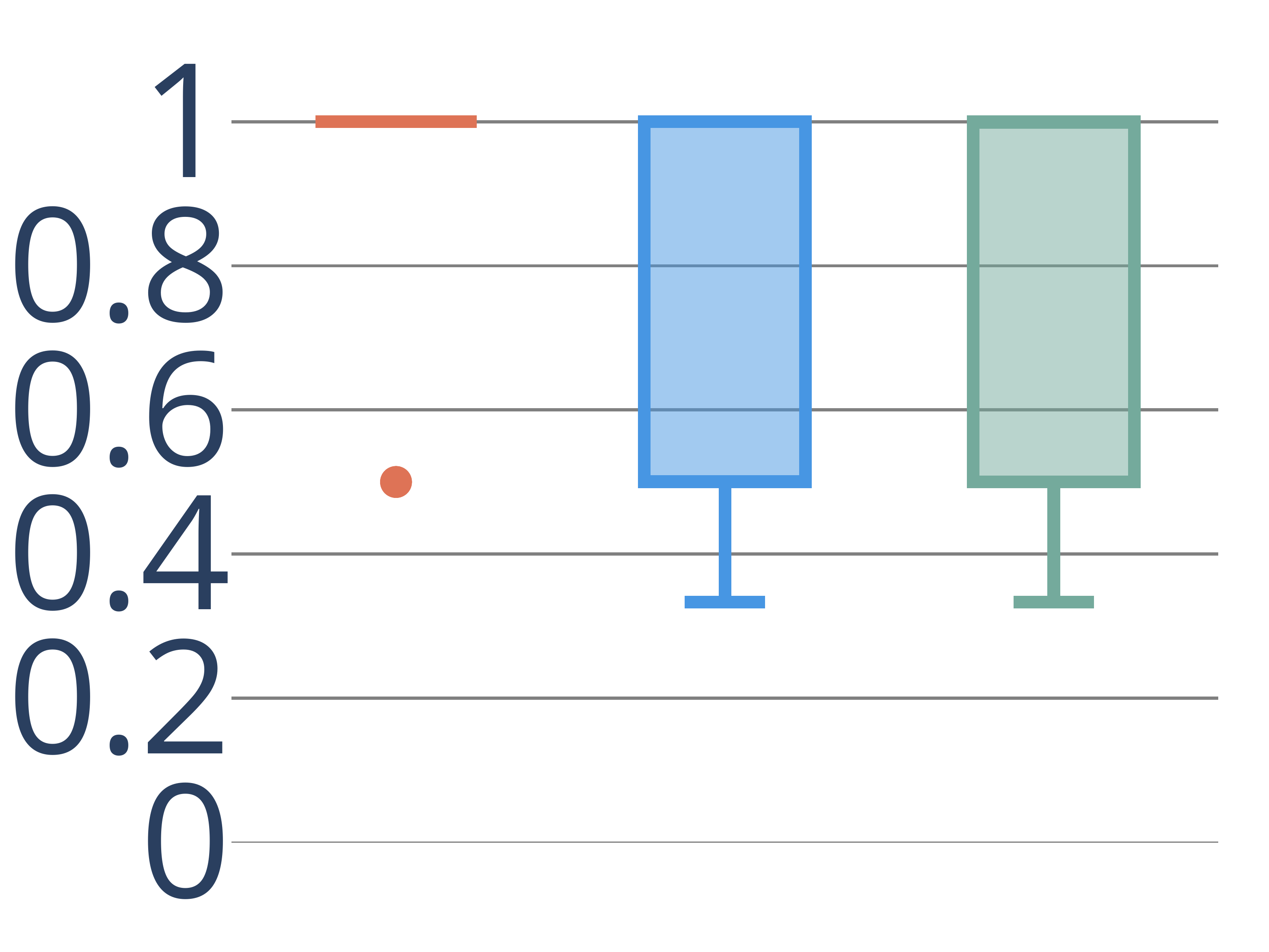} 
                                        & 
                                        & 
                                        & 
                                        & \includegraphics[scale=0.025]{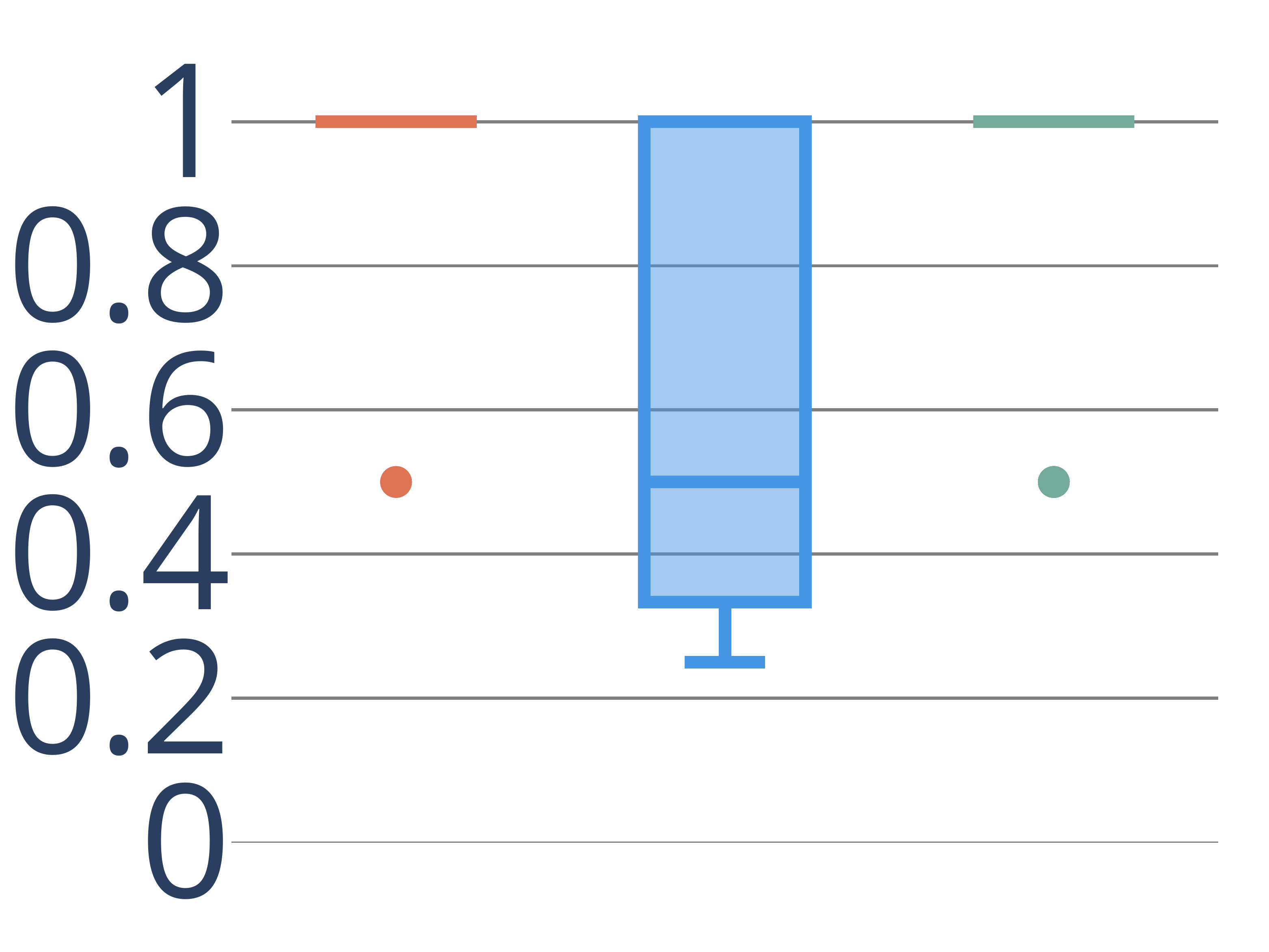}\\
                                    & \makecell[l]{\textbf{Misuse of Cumulative}\\ \textbf{Relationship}} 
                                        & \includegraphics[scale=0.025]{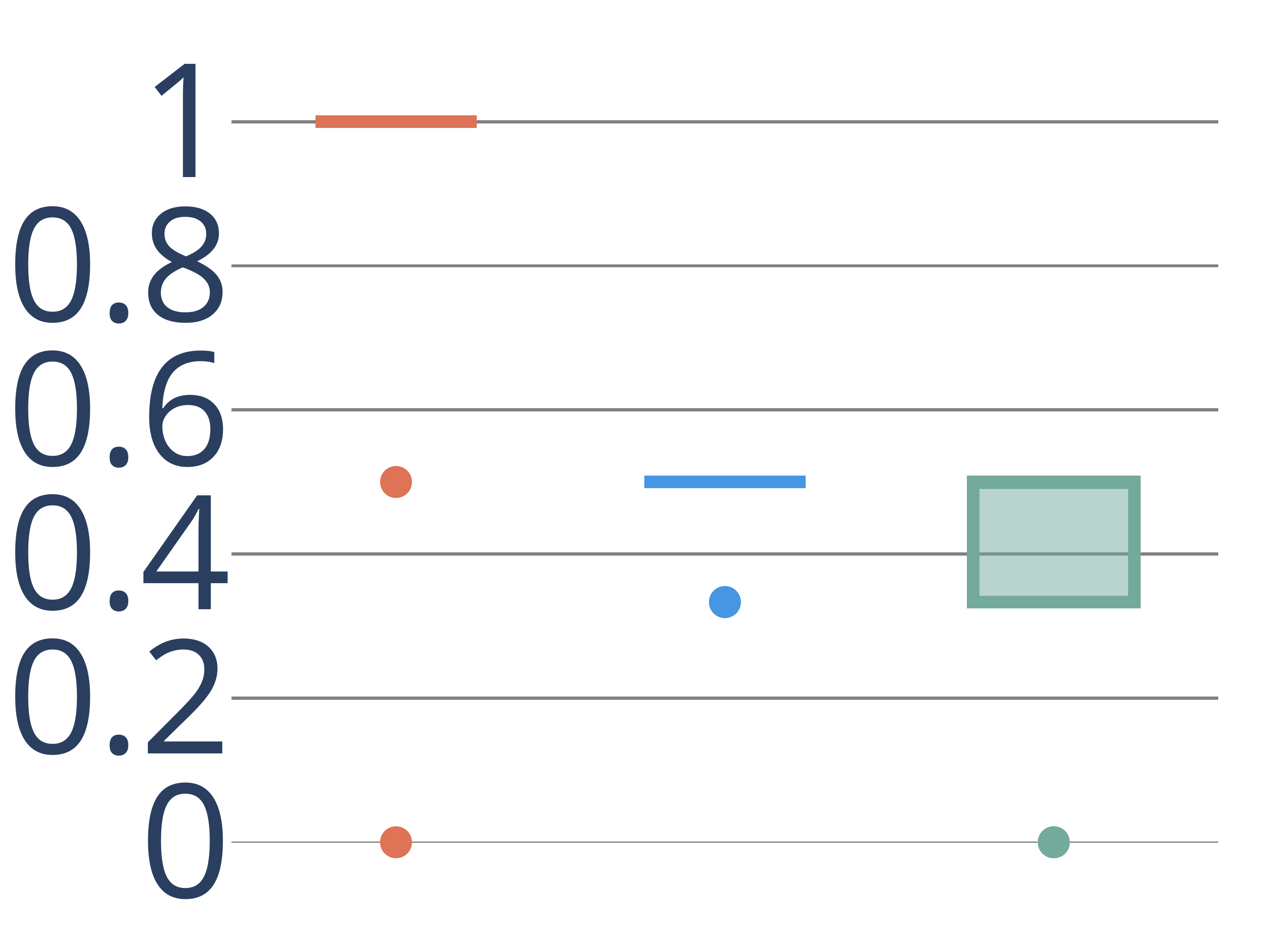} 
                                        & 
                                        & 
                                        & 
                                        & 
                                        & 
                                        & 
                                        & \includegraphics[scale=0.025]{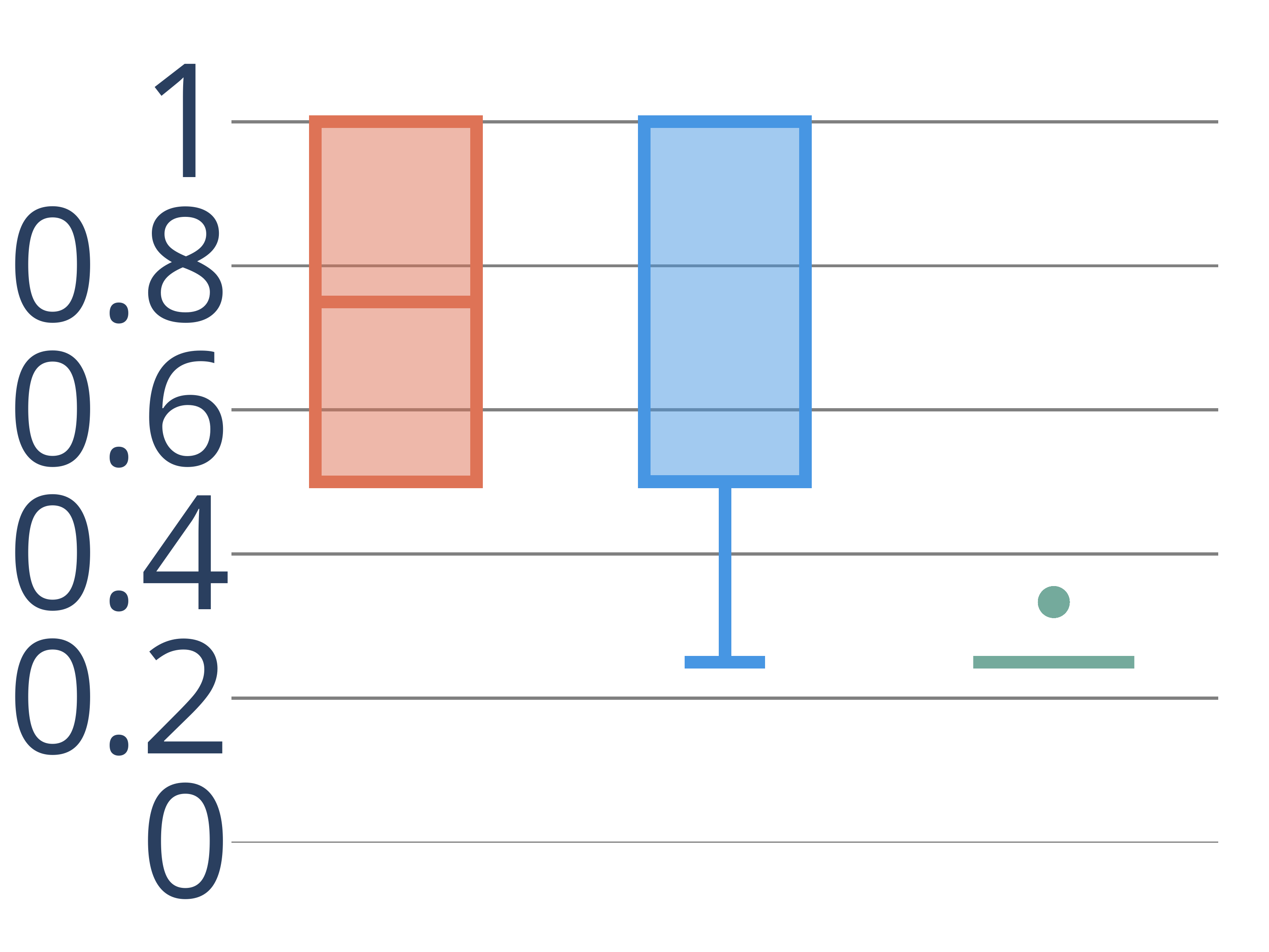} 
                                        & \includegraphics[scale=0.025]{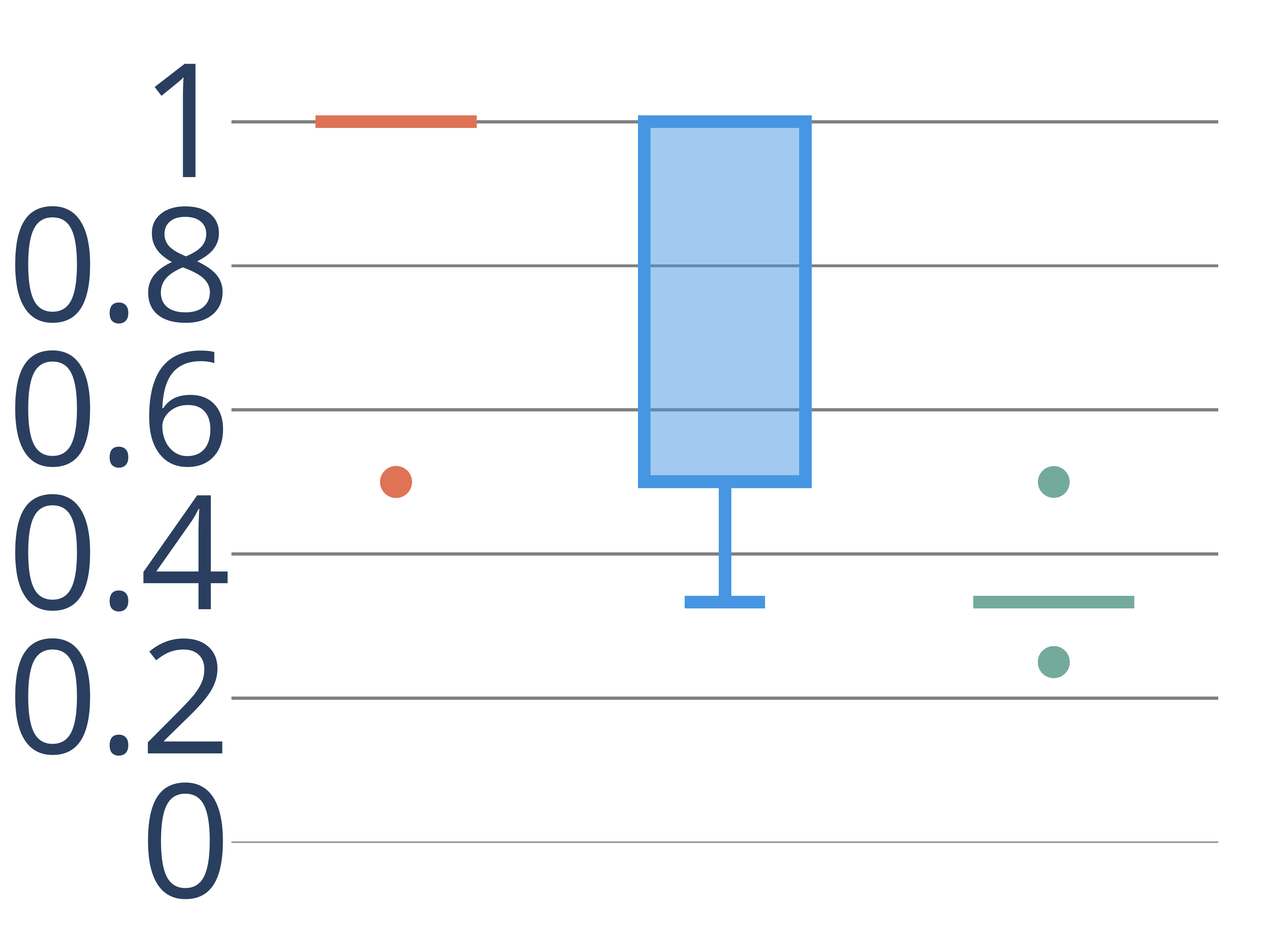} 
                                        &\\
                                    & \textbf{Exceeding the Canvas} 
                                        & 
                                        & \includegraphics[scale=0.025]{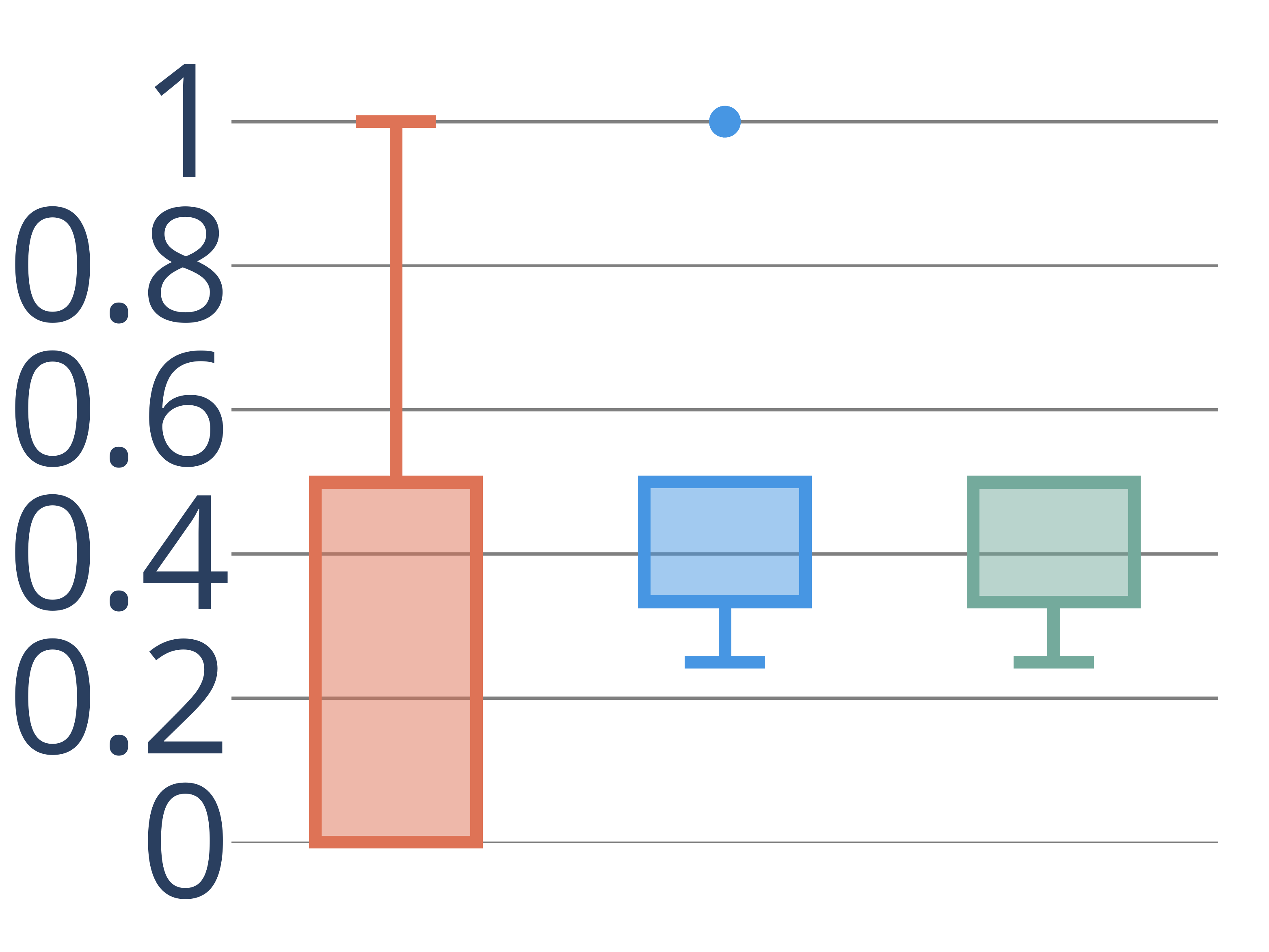} 
                                        & \includegraphics[scale=0.025]{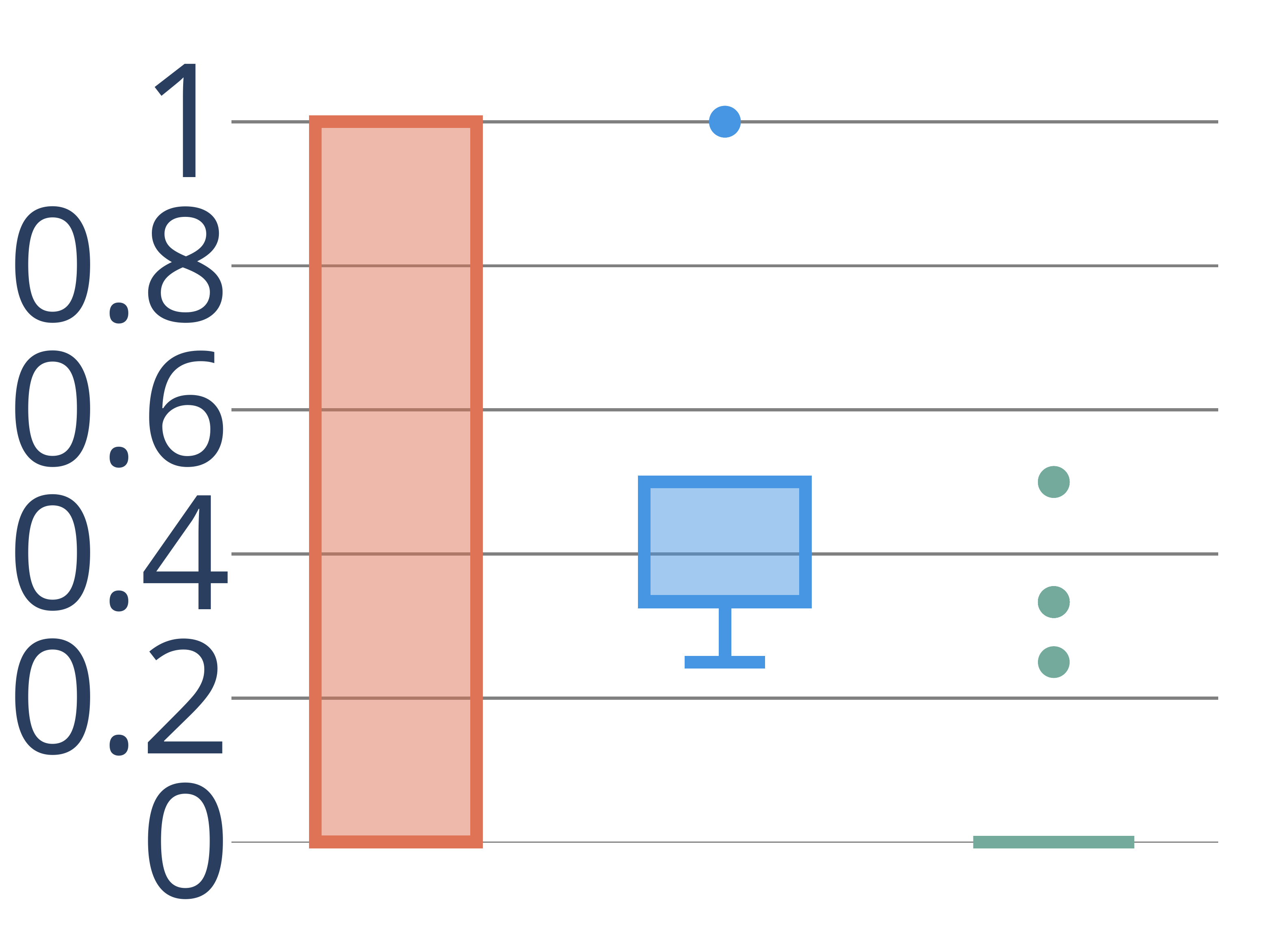} 
                                        & 
                                        & 
                                        & \includegraphics[scale=0.025]{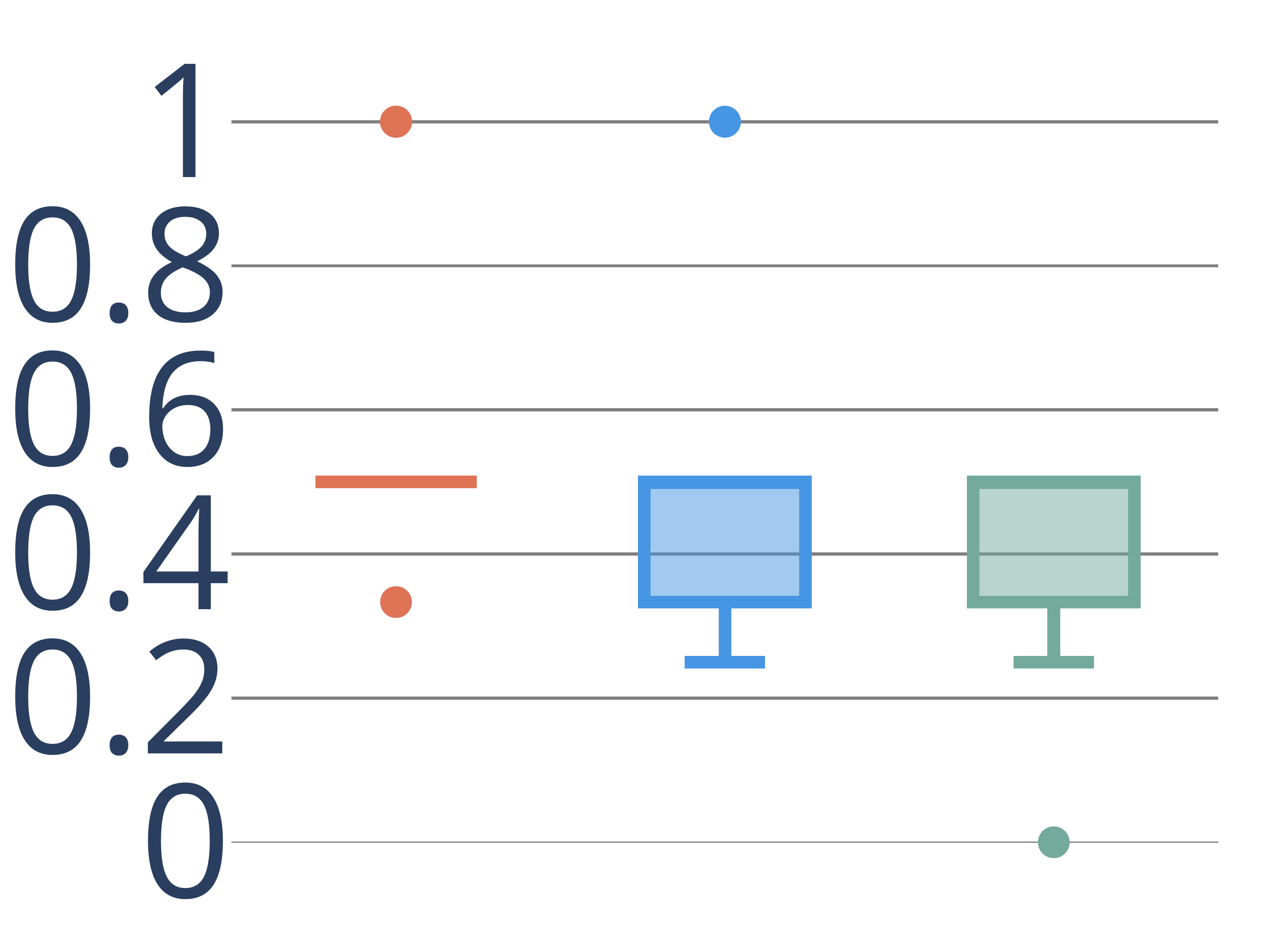} 
                                        & 
                                        & 
                                        & 
                                        & 
                                        & \\
                                    & \textbf{Small Size} 
                                        & 
                                        & 
                                        & 
                                        & \includegraphics[scale=0.025]{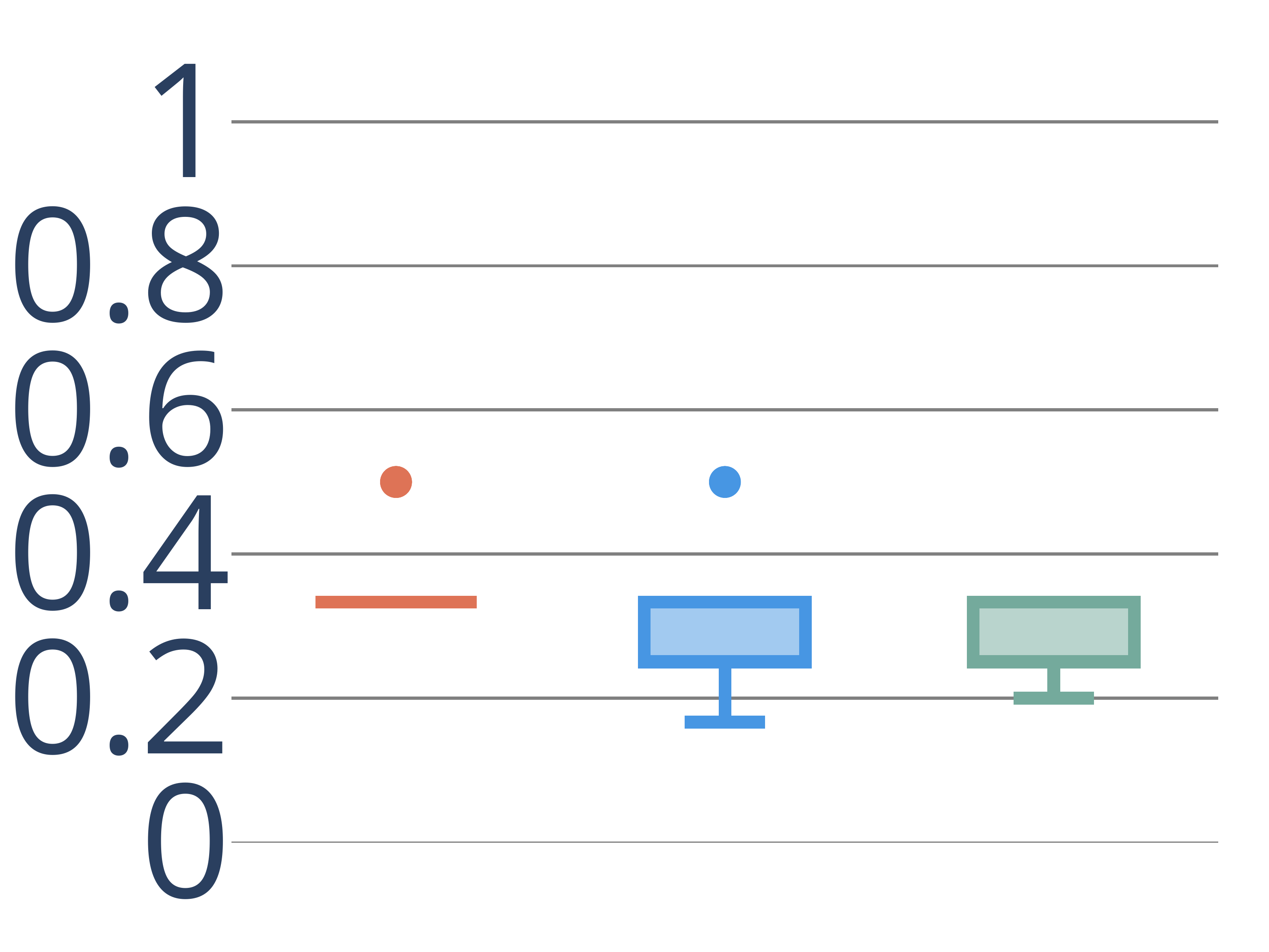} 
                                        & 
                                        & \includegraphics[scale=0.025]{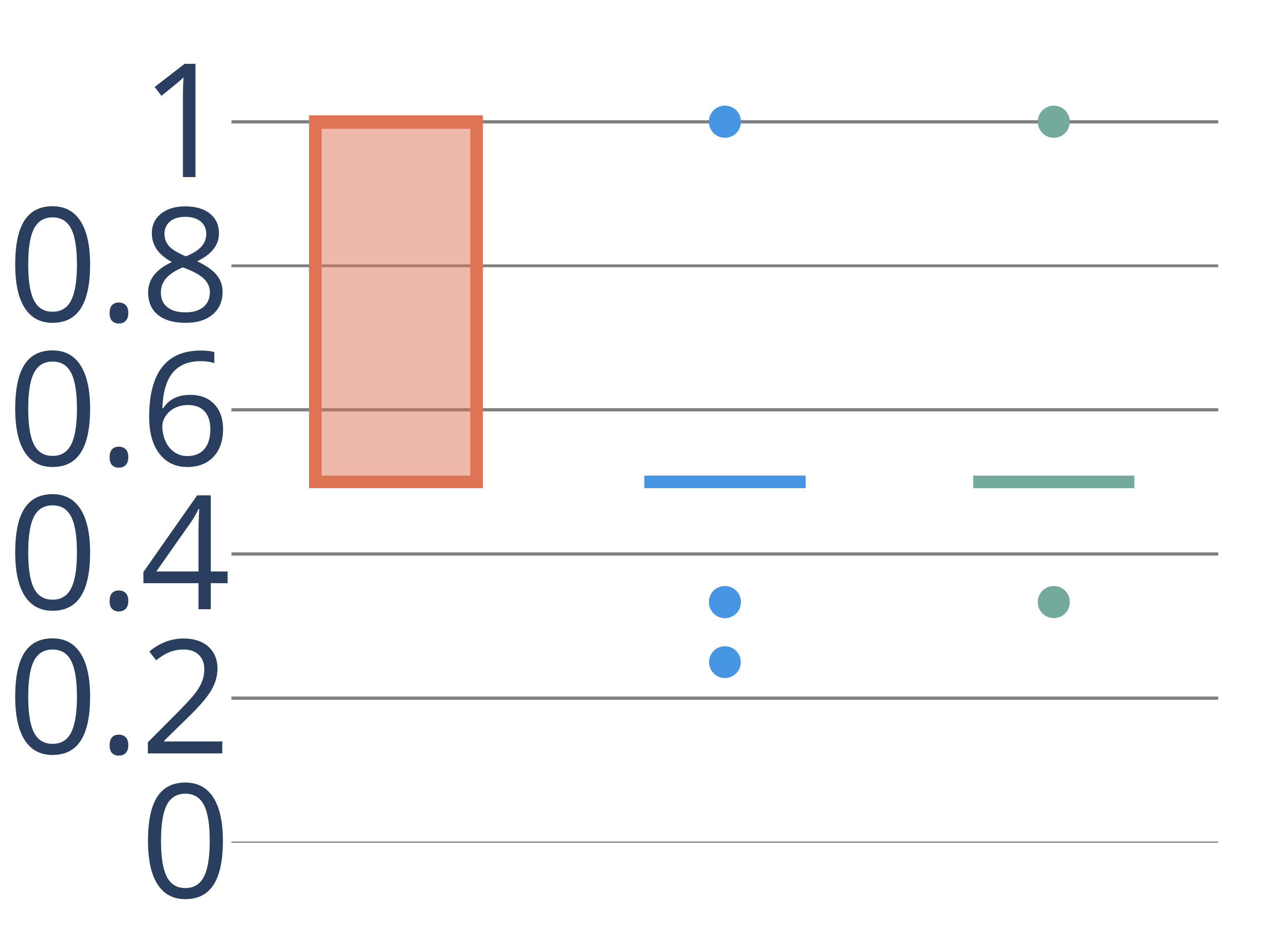} 
                                        & 
                                        & 
                                        & 
                                        & \includegraphics[scale=0.025]{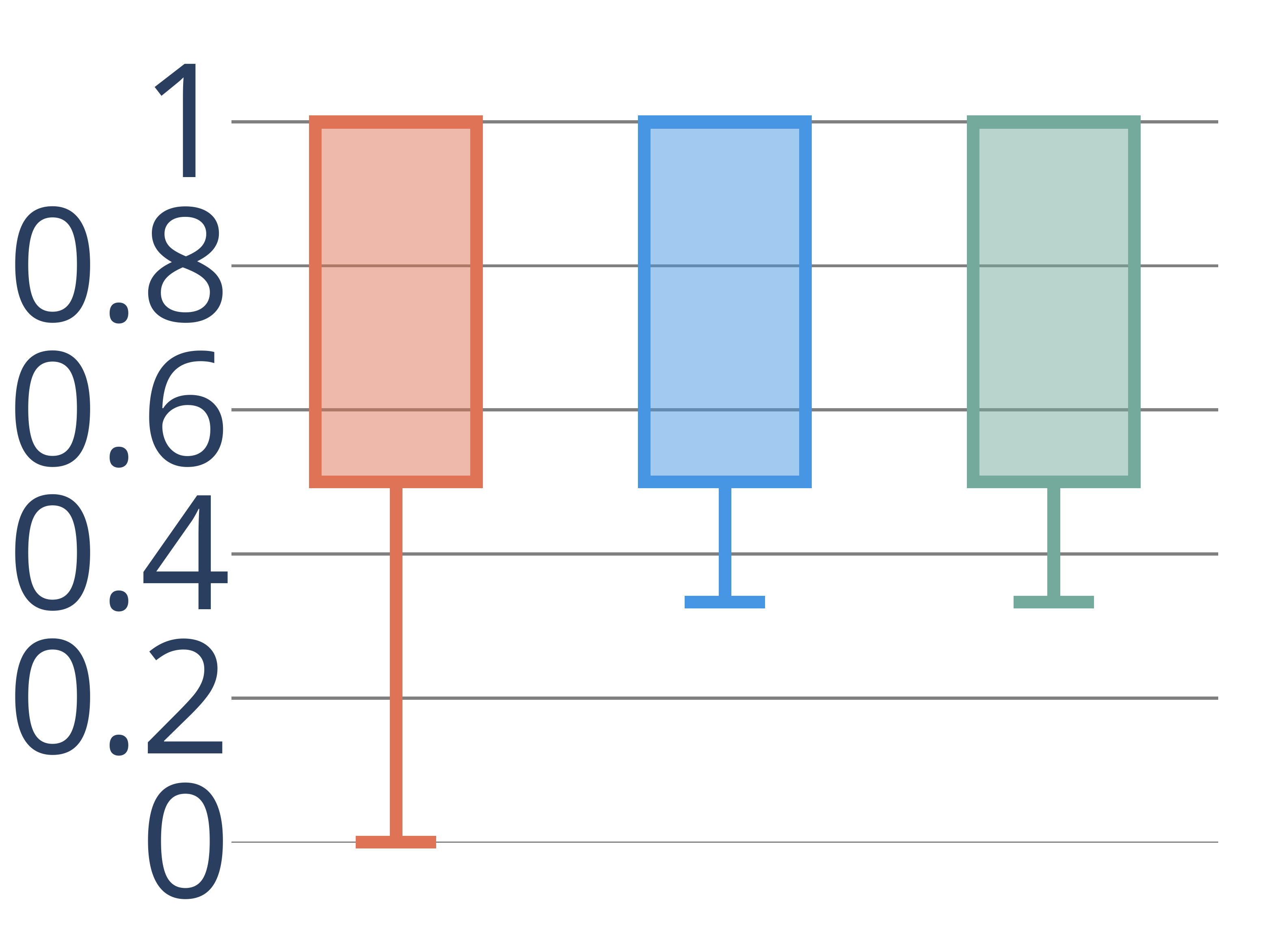}\\
        \hline
    \end{tblr}
    \label{tab:results_jindex_ex3}
\end{table*}

\begin{table*}[!h]
    \centering
    \caption{The comparisons between the expected accuracy rates for randomly choosing an answer (\eg, if there are four choices, the expected accuracy rate for random choices is 0.25), the accuracy rates of the general public from VLAT~\cite{lee:2017:vlat}, and accuracy rate of ChatGPT 5.1 in our pilot study. We color-encoded LLMs' results: \better{green} for much better than the performance of people (more than 0.05 higher) or perfect (1.00), \close{yellow} for values that were close to people (higher or lower within 0.05), and \worse{red} for values much worse than those of people (more than 0.05 lower).}\vspace{-0.7em}
    \tabulinesep=1pt
    \begin{tabu}{llccc}
        \toprule    
        Visualization  & Task & Random & VLAT & ChatGPT 5.1\\
        \midrule
        \multirow{5}{*}{Line Chart} & Retrieve Value & 0.25 & 0.95 & \better{1.00}\\
                                    & Find Extremum & 0.25 & 0.97 & \better{1.00}\\
                                    & Determine Range & 0.25 & 0.56 & \better{1.00}\\
                                    & Find Correlation/Trend & 0.33 & 0.98 & \better{1.00}\\
                                    & Make Comparisons & 0.25 & 0.77 & \better{1.00}\\
        \hline
        \multirow{4}{*}{Bar Chart} & Retrieve Value & 0.25 & 0.88 & \better{1.00}\\
                                   & Find Extremum & 0.25 & 0.98 & \better{1.00}\\
                                   & Determine Range & 0.25 & 0.54 & \better{1.00}\\
                                   & Make Comparisons & 0.25 & 0.40 & \better{1.00}\\
        \hline
        \multirow{5}{*}{\makecell[l]{Stacked \\ Bar Chart}}
                                    & Retrieve Value (Absolute Value) & 0.25 & 0.38 & \better{0.90}\\
                                    & Retrieve Value (Relative Value) & 0.25 & 0.36 & \better{1.00}\\
                                    & Find Extremum & 0.25 & 0.69 & \better{1.00}\\
                                    & Make Comparisons (Absolute Value) & 0.50 & 0.59 & \better{1.00}\\
                                    & Make Comparisons (Relative Value) & 0.50 & 0.47 & \better{1.00}\\
        \hline
        \multirow{3}{*}{\makecell[l]{100\% Stacked \\ Bar Chart}} 
                                    & Retrieve Value (Relative Value) & 0.25 & 0.49 & \better{1.00}\\
                                    & Find Extremum (Relative Value) & 0.25 & 0.90 & \better{1.00}\\
                                    & Make Comparisons (Relative Value) & 0.50 & 0.54 & \better{1.00}\\
        \hline
        \multirow{3}{*}{Pie Chart} 
                                & Retrieve Value (Relative Value) & 0.25 & 0.72 & \better{1.00}\\
                                & Find Extremum (Relative Value) & 0.25 & 0.98 & \better{1.00}\\
                                & Make Comparisons (Relative Value) & 0.50 & 1.00 & \better{1.00}\\
        \hline
        \multirow{3}{*}{Histogram}
                                & Retrieve Value (Derived Value) & 0.25 & 0.84 & \worse{0.60}\\
                                & Find Extremum (Derived Value) & 0.25 & 0.94 & \better{1.00}\\
                                & Make Comparisons (Derived Value) & 0.50 & 0.86 & \better{1.00}\\
        \hline
        \multirow{7}{*}{Scatterplot}
                                & Retrieve Value & 0.25 & 0.85 & \worse{0.10}\\
                                & Find Extremum & 0.25 & 0.76 & \worse{0.00}\\
                                & Determine Range & 0.25 & 0.53 & \better{1.00}\\
                                & Find Anomalies & 0.25 & 0.42 & \close{0.40}\\
                                & Find Clusters & 0.50 & 0.90 & \better{1.00}\\
                                & Find Correlation/Trend & 0.50 & 0.52 & \better{1.00}\\
                                & Make Comparisons & 0.50 & 0.79 & \better{1.00}\\
        \hline
        \multirow{4}{*}{Area Chart}
                                & Retrieve Value & 0.25 & 0.75 & \better{1.00}\\
                                & Find Extremum & 0.25 & 0.44 & \better{1.00}\\
                                & Determine Range & 0.25 & 0.38 & \better{1.00}\\
                                & Find Correlation/Trend & 0.33 & 0.94 & \close{0.90}\\
        \hline
        \multirow{6}{*}{\makecell[l]{Stacked \\ Area Chart}}
                                & Retrieve Value (Absolute Value) & 0.25 & 0.15 & \close{0.10}\\
                                & Retrieve Value (Relative Value) & 0.25 & 0.25 & \better{0.90}\\
                                & Find Extremum & 0.25 & 0.97 & \better{1.00}\\
                                & Find Correlation/Trend & 0.33 & 0.96 & \better{1.00}\\
                                & Make Comparisons (Absolute Value) & 0.50 & 0.20 & \better{0.90}\\
                                & Make Comparisons (Relative Value) & 0.50 & 0.24 & \worse{0.00}\\
        \hline                            
        \multirow{7}{*}{Bubble Chart}
                                & Retrieve Value & 0.25 & 0.41 & \better{0.80}\\
                                & Find Extremum & 0.25 & 0.69 & \better{1.00}\\
                                & Determine Range & 0.25& 0.29 & \better{0.90}\\
                                & Find Anomalies & 0.25 & 0.53 & \better{0.90}\\
                                & Find Clusters & 0.50 & 0.59 & \better{1.00}\\
                                & Find Correlation/Trend & 0.50 & 0.26 & \better{1.00}\\
                                & Make Comparisons & 0.50 & 0.80 & \better{1.00}\\
        \hline
        \multirow{3}{*}{Choropleth Map}
                                & Retrieve Value (Approximate Value) & 0.25 & 0.24 & \better{0.80}\\
                                & Find Extremum (Approximate Value) & 0.25 & 0.97 & \better{1.00}\\
                                & Make Comparison (Approximate Value) & 0.50 & 0.92 & \better{0.10}\\
        \hline
        \multirow{3}{*}{Treemap}
                                & Find Extremum (Relative Value) & 0.25 & 0.68 & \better{1.00}\\
                                & Make Comparison (Relative Value) & 0.50 & 0.42 & \better{1.00}\\
                                & Identify the Hierarchical Structure & 0.50 & 0.92 & \better{1.00}\\
        \hline
    \end{tabu}
    \label{tab:results_pilot}
\end{table*}

\begin{figure*}[h]
    \centering
    \begin{subfigure}{.494\textwidth}
        \begin{tabu}{l}
            \texttt{You are a helpful assistant for analyzing data visualizations.  }\\
            \texttt{As an example, the following chart and question have the following }\\
            \texttt{answer:}\\
            \\
            \texttt{About what is the global smartphone market share of }
            \texttt{Huawei?}\\
            \\
            \texttt{(a) 15\%}\\
            \texttt{(b) 20\%}\\
            \texttt{(c) 33\%}\\
            \texttt{(d) 50\%}\\
            \\
            \texttt{Answer: 20\%}\\
            \includegraphics[width=\linewidth]{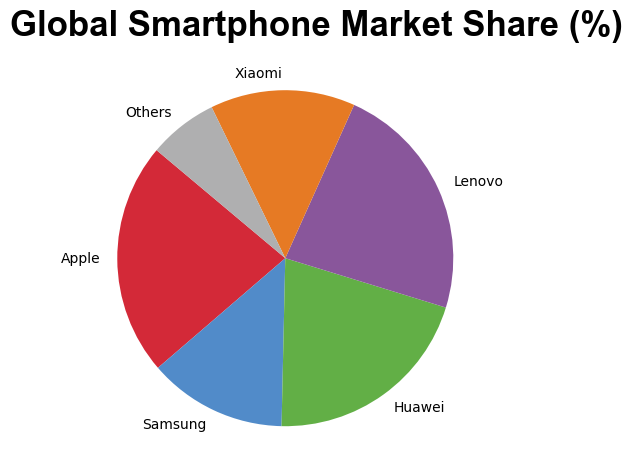}\\
            \texttt{Please answer with the letter corresponding to the best option, }\\
            \texttt{or make a random guess if unsure.  For example, if option (a) is}\\
            \texttt{ correct, only reply with (a).}\\
            \\
            \texttt{About what is the global smartphone market share of Apple?}\\
            \\
            \texttt{(a) 10\%}\\
            \texttt{(b) 25\%}\\
            \texttt{(c) 50\%}\\
            \texttt{(d) 60\%}\\
            \includegraphics[width=\linewidth]{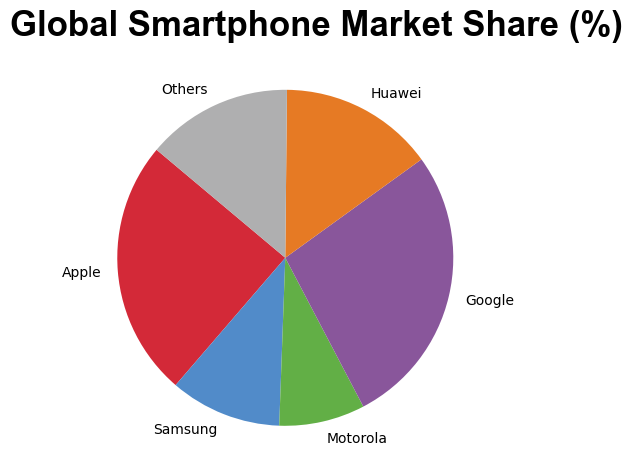}\\
        \end{tabu}
        \caption{Prompt used for Experiment 2 for few shot prompting.}
        \label{tab:fewShotPrompt}
    \end{subfigure}
    \begin{subfigure}{.494\textwidth}
        \begin{tabu}{l}
            \texttt{You are a helpful assistant for analyzing data visualizations.  }\\
            \texttt{As an example, the following chart and question have the following }\\
            \texttt{answer:}\\
            \\
            \texttt{About what is the global smartphone market share of }
            \texttt{Huawei?}\\
            \\
            \texttt{(a) 15\%}\\
            \texttt{(b) 20\%}\\
            \texttt{(c) 33\%}\\
            \texttt{(d) 50\%}\\
            \\
            \texttt{Solution: Look for Huawei slice is green. The percentage of the}\\
            \texttt{ chart the Huawei slice takes up is about 20\%.}\\
            \\
            \texttt{Answer: 20\%}\\
            \includegraphics[width=\linewidth]{figures/prompt_viz/pie-chart_modVLAT.png}\\
            \texttt{Please answer with the letter corresponding to the best option, }\\
            \texttt{or make a random guess if unsure.  For example, if option (a) is}\\
            \texttt{ correct, only reply with (a).}\\
            \\
            \texttt{About what is the global smartphone market share of Apple?}\\
            \\
            \texttt{(a) 10\%}\\
            \texttt{(b) 25\%}\\
            \texttt{(c) 50\%}\\
            \texttt{(d) 60\%}\\
            \includegraphics[width=\linewidth]{figures/prompt_viz/pie-chart_reVLAT.png}\\
        \end{tabu}
        \caption{Prompt used for Experiment 2 for chain-of-thought prompting.}
        \label{tab:cotPrompt}
    \end{subfigure}
    \caption{Example of prompts used in Experiment 2, specifically showing the (a) few shot and (b) chain-of-thought prompting techniques.}
    \label{fig:promptExamples}
\end{figure*}

\begin{table*}[]
    \caption{Comparison of papers that include prompts based on journal and prompt type. \textcolor{blue}{Blue papers} are from IEEE VIS2025, \textcolor{green}{green papers} are from EMNLP 2025, and \textcolor{red}{red papers} are from ACL 2025}\vspace{-0.7em}
    \tabulinesep=3pt
    \begin{tabu}{|m{4.5cm}|m{2.1cm}||m{4.5cm}|m{3.5cm}|}
        \toprule    
        \multicolumn{4}{|c|}{Table of Prompts} \\
        \midrule
        \multicolumn{2}{|c||}{Includes Prompt} & \multicolumn{2}{c|}{Does Not Include Prompt} \\
        \midrule
        Paper & Prompt Style & Paper & Prompt Explanation \\
        \midrule
        \color{blue} Can LLMs Bridge Domain and Visualization? A Case Study on High-Dimension Data Visualization in Single-Cell Transcriptomics \cite {Wang:2026:CLB}& Prompt Template & \color{blue} The Perils of Chart Deception: How Misleading Visualizations Affect Vision-Language Models \cite{Mahbub:2025:PCD}& mentions prompts, but not included \\[-0.2em]
        \hline
        \color {green} Text2Vis: A Challenging and Diverse Benchmark for Generating Multimodal Visualizations from Text \cite{Rahman:2025:TCD}& Prompt Template & \color{blue} NLI4VolVis: Natural Language Interaction for Volume Visualization via LLM Multi-Agents and Editable 3D Gaussian Splatting \cite{Ai:2025:NNL} & examples of prompts, but not explicit \\[-0.2em]
        \hline
        \color {green} MemeReaCon: Probing Contextual Meme Understanding in Large Vision-Language Models \cite{Zhao:2025:MPC}& Zero Shot & \color{blue} OW-CLIP: Data-Efficient Visual Supervision for Open-World Object Detection via Human-AI Collaboration \cite{Duan:2025:ODV} & mentions prompts, but not included \\[-0.2em]
        \hline
        \color {green} Fooling the LVLM Judges: Visual Biases in LVLM-Based Evaluation \cite{Hwang:2025:FLJ}& Prompt Template & \color{blue} CLAImate: AI-Enabled Climate Change Communication through Personalized and Localized Narrative Visualizations \cite{Rashik:2025:CAC}& mentions prompts, but not included \\[-0.2em]
        \hline
        \color {green} Med-VRAgent: A Framework for Medical Visual Reasoning-Enhanced Agents \cite{guo:2025:medVRA} & Zero Shot & \color{blue} LightVA: Lightweight Visual Analytics with LLM Agent-Based Task Planning and Execution \cite{Zhao:2025:lightva}& mentions prompts, but not included \\[-0.2em]
        \hline
        \color {green} VisualWebInstruct: Scaling up Multimodal Instruction Data through Web Search \cite{Jia:2025:VSM}& Prompt Template & \color{blue} Enhancing XAI Interpretation through a Reverse Mapping from Insights to Visualizations \cite{Nuthalapati:2025:EXI}& mentions prompts, but not included \\[-0.2em]
        \hline
        \color {green} Unmasking Deceptive Visuals: Benchmarking Multimodal Large Language Models on Misleading Chart Question Answering \cite{Chen:2025:UDV}& Zero Shot & \color{blue} ProactiveVA: Proactive Visual Analytics with LLM-based UI Agent \cite{zhao:2025:ProactiveVA}& mentions prompts, but not included \\[-0.2em]
        \hline
        \color {green} Identifying and Interactively Refining Ambiguous User Goals for Data Visualization Code Generation \cite{Inan:2025:IIR}& Role Based Zero Shot & \color{blue} BDIViz: An Interactive Visualization System for Biomedical Schema Matching with LLM-Powered Validation \cite{Wu:2025:BIV}& mentions prompts, but not included \\[-0.2em]
        \hline
        \color {green} Language-to-Space Programming for Training-Free 3D Visual Grounding \cite{Mi:2025:LPTa}& Role Based Zero Shot & \color {green} Seeing Culture: A Benchmark for Visual Reasoning and Grounding \cite{Satar:2025:SCB}& mentions prompts, but not included \\[-0.2em]
        \hline
        \color {green} HealthCards: Exploring Text-to-Image Generation as Visual Aids for Healthcare Knowledge Democratizing and Education \cite{Wu:2025:HET}& Zero Shot & \color{blue} ConceptViz: A Visual Analytics Approach for Exploring Concepts in Large Language Models \cite{Li:2025:CVA}& examples of prompts, but explicit \\[-0.2em]
        \hline
        \color{red} nvAgent: Automated Data Visualization from Natural Language via Collaborative Agent Workflow \cite{Ouyang:2025:NAD}& Prompt Template & \color{red} Make Imagination Clearer! Stable Diffusion-based Visual Imagination for Multimodal Machine Translation \cite{Chen:2025:MIC}& mentions prompts, but not included \\[-0.2em]
        \hline
        - & - & \color{red} ChartLens: Fine-grained Visual Attribution in Charts \cite{Suri:2025:CFV}& mentions prompts, but not included \\[-0.2em]
        \midrule
    \end{tabu}
    \label{tab:promptPapers}
\end{table*}

\begin{table*}[h]
    \centering    
    \caption{Table of coefficients for LLMs and prompt type interactions from Experiment 1 and 2.  \textbf{Coef.} refers to the coefficient in the logistic model using results in experiments 1 and 2, while the \textbf{Mean Coef.} refers to the mean of the bootstrapped coefficients.  When \textbf{Normal?} is true, a one-sample t-test is used, while if it is false, the Wilcoxon signed-rank test is used.  Note that when the Wilcoxon signed-rank test is used, the bounds are estimated from the ECDF.  Thus, bounds calculated from the ECDF that contain zero indicate poor estimations.}
    \begin{tabu}{l|l|S[table-format=1.4e1]|S[table-format=1.4e1]|S[table-format=1.4e3]|l|S[table-format=1.4e1]|S[table-format=1.4e1]}
        \hline
        \textbf{LLM}                &\textbf{Prompt}    &\textbf{Coef.} &\textbf{Mean Coef} &\textbf{p-value}   &\textbf{Normal?}   &\textbf{95\% LB}   &\textbf{95\% UB}\\
        \hline
        \multirow{4}*{\gptFig}      &Base               &0.5938         &0.2448             &1.95E-253          &True               &0.2345             &0.2551\\
                                    &Few Shot           &0.6250         &0.7619             &0                  &False              &0.4887             &1.095\\
                                    &Chain of Thought   &1.122          &1.414              &0                  &False              &1.034              &1.917\\
                                    &All                &2.341          &2.421              &0                  &False              &2.116              &2.834\\
        \hline
        \multirow{4}*{\geminiFig}   &Base               &0.4014         &0.5798             &0                  &True               &0.5667             &0.5930\\
                                    &Few Shot           &1.322          &1.413              &0                  &False              &-0.7088            &1.757\\
                                    &Chain of Thought   &-0.1191        &-0.1040            &0.538              &False              &-0.4063            &0.2577\\
                                    &All                &1.604          &1.889              &0                  &False              &1.623              &2.202\\
        \hline
        \multirow{4}*{\claudeFig}   &Base               &0.5607         &0.7570             &0                  &False              &0.4148             &1.144\\
                                    &Few Shot           &-1.197         &-1.333             &0                  &True               &-1.341             &-1.325\\
                                    &Chain of Thought   &-0.5705        &-0.7458            &0                  &True               &-0.7541            &-0.7376\\
                                    &All                &-1.206         &-1.322             &0                  &False              &-1.546             &-1.122\\
        \hline
        \multirow{4}*{All LLMs}     &Base               &1.556          &1.582              &0                  &True               &1.573              &1.590\\
                                    &Few Shot           &0.7502         &0.8422             &0                  &False              &0.6721             &1.038\\
                                    &Chain of Thought   &0.4329         &0.5641             &0                  &False              &0.3612             &0.8291\\
                                    &All (Intercept)    &2.739          &2.988              &0                  &False              &2.689              &3.339\\
        \hline
    \end{tabu}
    \label{tab:genCoefResults}
\end{table*}

\begin{sidewaystable*}[h]
    \centering
    \caption{Experiment 1 and 2 mean coefficients for visualizations (no interactions with tasks).}
    \begin{tabu}{l|S[table-format=1.5]|S[table-format=1.5]|S[table-format=1.5]|S[table-format=1.5]|S[table-format=1.5]|S[table-format=1.7]|S[table-format=1.5]|S[table-format=1.5]|S[table-format=1.5]|S[table-format=1.5]|S[table-format=1.5]|S[table-format=1.4]|S[table-format=1.5]|S[table-format=1.5]|S[table-format=1.5]|S[table-format=1.4]}
        \hline
        \multirow{2}{*}{\textbf{Visualization}} &\multicolumn{4}{c|}{\textbf{Base}}     &\multicolumn{4}{c|}{\textbf{Few Shot}}             &\multicolumn{4}{c}{\textbf{Chain of Thought}}  &\multicolumn{4}{c}{\textbf{All Prompts}}\\
        \cline{2-17}
                                &\textbf{GPT}   &\textbf{Gemini}    &\textbf{Claude}    &\textbf{All}   &\textbf{GPT}   &\textbf{Gemini}    &\textbf{Claude}    &\textbf{All}   &\textbf{GPT}   &\textbf{Gemini}    &\textbf{Claude}&\textbf{All}   &\textbf{GPT}   &\textbf{Gemini}    &\textbf{Claude}    &\textbf{All}\\
        \hline
        Line Chart              &0.3573         &0.4544             &1.128              &1.940          &0.4685         &0.3507             &-1.513             &-0.6933        &0.4153         &0.1592             &-0.6291        &-0.05445       &1.241          &0.9644             &-1.013             &1.192\\
        Bar Chart               &0.3926         &-0.7767            &1.045              &0.6611         &0.5255         &1.298              &-1.590             &0.2331         &0.5071         &-0.5937            &-0.2702        &-0.3567        &1.425          &-0.07239           &-0.8154            &0.5375\\
        Stacked Bar Chart       &0.3099         &1.020              &-1.407             &-0.07735       &-0.3574        &0.7529             &-0.2815            &0.1140         &-0.07815       &-0.9333            &0.6389         &-0.3726        &-0.1257        &0.8399             &-1.050             &-0.3360\\
        100\% Stacked Bar Chart &0.2043         &0.9617             &0.4008             &0.8376         &0.2124         &0.2637             &0.1862             &0.6622         &0.2121         &0.06982            &-0.007147      &0.6705         &0.6288         &0.9617             &0.5800             &2.170\\
        Pie Chart               &0.8625         &0.1832             &-1.179             &0.4990         &-0.1197        &-0.5064            &-0.0221            &0.4467         &0.3608         &0.4656             &-0.1092        &0.4912         &1.808          &-0.1694            &-1.297             &0.3421\\
        Histogram               &0.2069         &0.2176             &0.08933            &0.5139         &0.5172         &0.2139             &0.2919             &0.7070         &0.2062         &0.3754             &-0.3052        &0.2764         &0.6142         &0.8070             &0.07607            &1.497\\
        Scatterplot             &0.3406         &-0.1434            &0.2889             &0.4861         &0.1066         &-0.7101            &0.1607             &-0.4427        &-0.7969        &0.08431            &0.1183         &-0.5942        &-0.3497        &-0.7692            &0.5679             &-0.5509\\
        Area Chart              &0.2824         &-0.7652            &1.030              &0.5470         &0.3288         &-0.4822            &0.3892             &0.2358         &1.621          &0.4116             &-0.1297        &-0.1388        &0.1904         &-0.8358            &1.289              &0.6440\\
        Stacked Area Chart      &-0.8044        &0.9465             &-1.960             &-1.818         &-1.0005        &-0.1767            &0.8155             &-0.3617        &-0.6859        &-0.4498            &0.5722         &-0.5635        &-2.491         &0.3201             &-0.5726            &-2.743\\
        Bubble Chart            &-0.3305        &-0.7762            &0.4339             &-0.6727        &0.02361        &0.07711            &0.2330             &0.3337         &0.8899         &0.2480             &-0.3011        &0.8368         &0.5830         &-0.4511            &0.3658             &0.4978\\
        Choropleth Map          &-1.439         &0.2664             &0.4404             &-0.7322        &0.5817         &-0.1742            &-0.5726            &-0.1651        &0.3333         &-0.09894           &-0.3307        &-0.09633       &-0.5239        &-0.006739          &-0.4629            &-0.9935\\
        Treemap                 &-0.1379        &-0.2798            &0.4335             &0.01586        &-0.8276        &0.5064             &0.5697             &0.2486         &0.3851         &0.07383            &0.006975       &0.4658         &-0.5804        &0.3004             &1.010              &0.7303\\
        \hline
    \end{tabu}
    \label{tab:vizCoefResults}

    \vspace{4em}

    \caption{Mean coefficients for tasks (no interactions with visualizations).}
     \begin{tabu}{l|S[table-format=1.5]|S[table-format=1.5]|S[table-format=1.5]|S[table-format=1.5]|S[table-format=1.5]|S[table-format=1.7]|S[table-format=1.5]|S[table-format=1.5]|S[table-format=1.5]|S[table-format=1.5]|S[table-format=1.5]|S[table-format=1.4]|S[table-format=1.5]|S[table-format=1.5]|S[table-format=1.5]|S[table-format=1.4]}
        \hline
        \multirow{2}{*}{\textbf{Task}} &\multicolumn{4}{c|}{\textbf{Base}}           &\multicolumn{4}{c|}{\textbf{Few Shot}}        &\multicolumn{4}{c}{\textbf{Chain of Thought}}  &\multicolumn{4}{c}{\textbf{All Prompts}}\\
         \cline{2-17}
                                             &\textbf{GPT}  &\textbf{Gemini}    &\textbf{Claude} &\textbf{All}   &\textbf{GPT}  &\textbf{Gemini}    &\textbf{Claude}    &\textbf{All}   &\textbf{GPT} &\textbf{Gemini}    &\textbf{Claude}    &\textbf{All}   &\textbf{GPT}     &\textbf{Gemini}    &\textbf{Claude}    &\textbf{All}\\
         \hline
         Retrieve Value                      &0.02083       &-0.7851            &1.214           &0.4495         &-0.06776      &0.05959            &-0.5360            &-0.5442        &1.131          &0.03587            &-1.474             &-0.3079      &1.084            &-0.6896            &-0.7968            &-0.4026\\
         Find Extremum                       &-0.7492       &-0.5618            &1.809           &0.4980         &0.5191        &-0.02179           &-0.7939            &-0.2966        &0.6898         &0.3091             &0.8563             &-0.2279      &0.4597           &-0.2745            &-0.2118            &-0.02660\\
         Determine Range                     &1.004         &-0.2894            &-0.9381         &-0.2232        &0.7519        &0.2348             &-1.099             &-0.1121        &0.8234         &0.2650             &-1.591             &-0.6225      &2.580            &0.2105             &-3.628             &-0.8380\\
         Find Correlation/Trend              &1.450         &-0.2529            &-1.164          &0.03364        &0.7968        &0.3833             &0.3040             &1.484          &-1.547         &0.06811            &2.009              &0.5304       &0.7001           &0.1986             &1.150              &2.048\\
         Make Comparisons                    &-0.6677       &1.487              &0.3728          &-0.8632        &-0.2226       &-0.3361            &1.322              &0.7638         &0.1704         &-0.2720            &1.141              &1.040        &-0.7199          &0.8794             &1.601              &1.760\\
         Find Anomalies                      &0.5715        &-0.0409            &0.6040          &1.134          &-0.02185      &0.5544             &-1.532             &-0.9996        &-0.2518        &-0.0373            &-0.4318            &-0.7210      &0.2978           &0.4761             &-1.360             &-0.5861\\
         Find Clusters                       &-1.466        &0.9501             &-0.04517        &-0.5615        &-1.070        &0.4724             &0.6306             &0.1756         &0.3247         &-0.5545            &0.4541             &0.2244       &-2.212           &0.8681             &1.040              &-0.3040\\
        Identify the Hierarchical Structure &0.08118        &0.07227            &0.1407          &0.2941         &0.07611       &0.06646            &0.3709             &0.5134         &0.7874         &0.08164            &0.3734             &0.07400      &0.2313           &0.2204             &0.8850             &1.337\\
        \hline
    \end{tabu}
     \label{tab:taskCoefResults}
\end{sidewaystable*}

\begin{sidewaystable*}[!h]
    \centering
    \caption{Experiment 1 and 2 mean coefficients for visualization and task interactions.}
    \begin{tabu}{l|l|S[table-format=1.4]|S[table-format=1.4]|S[table-format=1.4]|S[table-format=1.4]|S[table-format=1.4]|S[table-format=1.4]|S[table-format=1.4]|S[table-format=1.4]|S[table-format=1.4]|S[table-format=1.4]|S[table-format=1.4]|S[table-format=1.4]|S[table-format=1.4]|S[table-format=1.4]|S[table-format=1.4]|S[table-format=1.4]}
        \hline
        \multirow{2}{*}{\textbf{Visualization}} &\multirow{2}{*}{\textbf{Task}} &\multicolumn{4}{c|}{\textbf{Base Prompt}}           &\multicolumn{4}{c|}{\textbf{Few Shot}}        &\multicolumn{4}{c|}{\textbf{Chain of Thought}}      &\multicolumn{4}{c}{\textbf{All}}\\
        \cline{3-18}
                                                &                               &\textbf{GPT-4}   &\textbf{Gemini}  &\textbf{Claude}      &\textbf{All}   &\textbf{GPT-4}   &\textbf{Gemini}    &\textbf{Claude}      &\textbf{All}    &\textbf{GPT-4}   &\textbf{Gemini}    &\textbf{Claude}      &\textbf{All}  &\textbf{GPT-4}   &\textbf{Gemini}    &\textbf{Claude}      &\textbf{All}\\
        \hline
        \multirow{5}{*}{Line Chart}             &Retrieve Value                 &0.07689          &0.1424           &1.215             &1.434              &0.1550            &-0.02812             &-1.5259              &-1.399            &0.1102  &-0.4136              &-0.5400            &-0.84332 &0.3421 &-0.2993 &-0.8507 &-0.8078\\
                                                &Find Extremum                  &0.05680          &0.05318          &0.1569            &0.2668             &0.07175           &0.07211             &1.094             &1.237             &0.07648     &0.1145              &-0.4230            &-0.2321 &0.2050 &0.2397 &0.8274 &1.272\\
                                                &Determine Range                &0.1050           &0.1665           &-0.5799            &-0.3083              &0.1191            &0.2090             &-0.5242             &-0.1961            &0.1220     &0.3525              &-0.4986            &-0.0240 &0.3460 &0.7281 &-1.603 &-0.5284\\
                                                &Find Correlation/Trend         &0.05506          &0.05233          &0.1444             &0.2518             &0.05354              &0.05091             &0.5017             &0.6062             &0.05990     &0.06178              &0.3916            &0.5133 &0.1685 &0.1650 &1.038 &1.371\\
                                                &Make Comparisons               &0.06359          &0.03999          &0.1916            &0.2952           &0.06917            &0.04682         &-1.058            &-0.9418           &0.04669    &0.04404              &0.4410            &0.5317 &0.1794 &0.1308 &-0.4252 &-0.1149\\
        \hline
        \multirow{4}{*}{Bar Chart}              &Retrieve Value                 &0.08835          &-0.7762          &1.129              &0.4411              &0.1139            &0.3781             &-0.3278              &0.1642            &0.1202  &-0.002481              &-0.4572            &-0.3395 &0.3224 &-0.4005 &0.3438 &-0.2657\\
                                                &Find Extremum                  &0.09081          &-0.6250          &0.7195             &0.1853             &0.1939              &0.5618            &-1.619             &-0.8635             &0.1659     &-0.5063              &0.1960            &-0.1444 &0.4506 &-0.5695 &-0.7037 &-0.8226\\
                                                &Determine Range                &0.1532           &0.5778           &-1.026             &-0.3099             &-0.3374            &0.3129             &-0.3530             &0.1131             &0.1690  &-0.1408              &-0.3387            &-0.3105 &0.4614 &0.7499 &-1.719 &-0.5073\\
                                                &Make Comparisons               &0.07428          &0.04664          &0.2238            &0.3447            &0.06448            &0.04519           &0.7095             &0.8192            &0.05199    &0.05598              &0.3297            &0.4377 &0.1907 &0.1478 &1.263 &1.601\\
        \hline
        \multirow{3}{*}{\makecell[l]{Stacked \\ Bar Chart}}
                                                &Retrieve Value                 &-0.5161          &0.7609           &-1.143            &-0.8981             &-0.9274           &0.4444             &0.8421             &0.3591             &-0.4013     &-0.07621              &0.8251            &0.3476 &-1.845 &1.129 &0.5243 &-0.1915\\
                                                &Find Extremum                  &0.1186           &0.09614          &0.5914             &0.8062             &0.1778            &0.1365            &-0.09718              &0.2171             &0.1382     &0.4089              &-0.5997            &-0.05260 &0.4347 &0.6415 &-0.1055 &0.9707\\
                                                &Make Comparisons               &0.7074           &0.1634           &-0.8561             &0.01464            &0.3921             &0.1720            &-1.026            &-0.4622             &0.1849    &-1.266              &0.4136            &-0.6676  &1.284 &-0.9306 &-1.469 &-1.115\\
        \hline
        \multirow{3}{*}{\makecell[l]{100\% \\ Stacked \\ Bar Chart}} 
                                                &Retrieve Value                 &0.07818          &0.1349           &0.5688            &0.7819            &0.08968           &0.1586            &-0.7263             &-0.4780              &0.09441    &0.3382              &-1.108            &-0.6754 &0.2623 &0.6318 &-1.265 &-0.3715\\
                                                &Find Extremum                  &0.05930          &0.05629          &-0.2678            &-0.1522            &0.06801           &0.06506            &0.7632             &0.8963            &0.07180     &0.08590              &0.9999            &1.157 &0.1991 &0.2072 &1.495 &1.902\\
                                                &Make Comparisons               &0.06678          &0.04132          &0.09990             &0.2080             &0.05471             &0.03997             &0.1493             &0.2439             &0.04587  &0.04143              &0.1010            &0.1883 &0.1674 &0.1227 &0.3501 &0.6402\\
        \hline
        \multirow{3}{*}{Pie Chart}              &Retrieve Value                 &0.1178           &0.4151           &2.429  &2.961            &0.1458             &0.8269             &-2.638             &-1.666             &0.1378             &0.9879             &-1.208             &-0.08284 &0.4013 &2.230 &-1.418 &1.213\\
                                                &Find Extremum                  &0.1433           &-0.6159          &0.5673             &0.09470              &0.2466            &-1.429              &0.8043              &-0.3778            &0.2379  &-0.9354              &-0.04154            &-0.7390 &0.6277 &-2.980 &1.330 &-1.022\\
                                                &Make Comparisons               &0.6014           &0.3839           &-4.161           &-3.176             &0.1067            &0.09543             &1.812             &2.014             &0.0710     &0.1012              &1.141            &1.313 &0.7791 &0.5806 &-1.209 &0.1510\\
        \hline
        \multirow{3}{*}{Histogram}              &Retrieve Value                 &0.06673          &0.1026           &0.2547             &0.4241              &0.06386             &0.08449             &0.3784             &0.5267             &0.07135  &0.1746              &-0.2236            &0.0223 &0.2019 &0.3617 &0.4094 &0.9731\\
                                                &Find Extremum                  &0.07195          &0.07380          &-0.2709              &-0.1251             &0.08402             &0.08930            &-0.2162            &-0.04284              &0.08899     &0.1587              &-0.1852            &0.06247 &0.2449 &0.3218 &-0.6723 &-0.1055\\
                                                &Make Comparisons               &0.06820          &0.04124          &0.1055             &0.2149            &0.05326             &0.04012            &0.1297              &0.2231             &0.04584     &0.04204              &0.1037            &0.1916 &0.1673 &0.1234 &0.3389 &0.6296\\
        \hline
        \multirow{7}{*}{Scatterplot}            &Retrieve Value                 &-0.1819          &-0.7558          &-2.053            &-2.991           &-0.3951             &-0.4292             &0.6366            &-0.1876             &-1.347    &0.8643              &-0.2833            &-0.7661 &-1.924 &-0.3207 &-1.700 &-3.945\\
                                                &Find Extremum                  &0.1586           &-0.2725          &0.4111              &0.2972              &0.4481             &-0.09457           &-1.011           &0.3931              &0.1416  &2.001              &-0.3409            &0.1938 &0.9998 &0.5586 &-0.9413 &0.6170\\
                                                &Determine Range                &0.1550          &0.2179           &1.470             &1.8432            &0.2077             &0.3156            &-0.2030            &0.3203             &0.2472     &-0.2490              &0.2146            &0.2129 &0.6099 &0.2846 &1.481 &2.376\\
                                                &Find Anomalies                 &0.4129           &0.1150          &-0.3970             &0.1308              &-0.2579             &0.2980            &-0.7693             &-0.7292             &-0.3865     &-0.3818              &-0.02468            &-0.7931 &-0.2315 &0.03119 &-1.191 &-1.391\\
                                                &Find Clusters                  &-0.5431           &0.3080            &0.4678            &0.2326            &-0.3584           &0.3403           &0.1899             &0.1718             &-0.03295    &0.1068              &-0.3378            &-0.2639 &-0.9345 &0.7551 &0.3199 &0.1406\\
                                                &Find Correlation/Trend         &0.09628            &0.1025           &0.1818             &0.3806             &0.09333              &0.08892             &0.4966              &0.6789             &0.1370     &0.1640              &0.5002            &0.8012 &0.3266 &0.3554 &1.179 &1.861\\
                                                &Make Comparisons               &0.2428            &0.1414            &0.2085              &0.5927              &0.3690             &-2.013            &0.8212              &-0.8229             &0.1925  &-0.5617              &0.3901            &0.02097 &0.8042 &-2.433 &1.420 &-0.2092\\
        \hline
        \multirow{4}{*}{Area Chart}             &Retrieve Value                 &0.06467          &-0.1528          &0.1863            &0.09818           &0.06527             &0.2910             &-0.1570            &0.5134            &0.07060    &-0.1582              &0.3446            &0.2570 &0.2005 &-0.02004 &0.6881 &0.8685\\
                                                &Find Extremum                  &0.06673           &0.1581          &0.1292             &0.3541             &0.09894            &0.3114            &-0.6232             &-0.2129            &0.09756     &0.4131              &-0.3524            &0.1582 &0.2632 &0.8826 &-0.8464 &0.2994\\
                                                &Determine Range                &0.09529          &-0.8458          &0.6145             &-0.1359            &0.1121             &-1.140            &0.6557             &-0.3724           &0.1432     &0.04735              &-0.3468            &0.1562 &0.3506 &-1.939 &0.9235 &-0.6645\\
                                                &Find Correlation/Trend         &0.05569           &0.07530            &0.0996              &0.2306              &0.05250             &0.05556             &0.1997              &0.3077             &-0.7321  &0.1095              &0.2248            &-0.3978 &-0.6239 &0.2403 &0.5242 &0.1405\\
        \hline
        \multirow{4}{*}{\makecell[l]{Stacked \\ Area Chart}}
                                                &Retrieve Value                 &-1.412            &-0.7969          &0.3967              &-0.006763              &0.03362            &-0.4872             &0.7143              &1.747             &1.759  &-0.8242              &0.4741            &1.397 &0.3688 &-0.3031 &1.585 &1.651\\
                                                &Find Extremum                  &-0.9980          &0.2514           &-1.071             &-1.817             &-0.3805              &-0.001069            &-0.3496             &-1.246            &-2.396     &0.1748              &0.6981            &-0.3728 &-2.624 &0.4252 &-0.7224 &-2.922\\
                                                &Find Correlation/Trend         &1.174           &-0.5729            &-1.774             &-1.173             &0.5458             &0.1390             &-1.164             &-0.4795              &-1.069     &-0.3288              &0.6594            &-0.7381 &0.6514 &-0.7628 &-2.279 &-2.391\\
                                                &Make Comparisons               &0.4311          &0.2597            &0.4882              &1.179              &-1.199             &0.1726             &1.615              &0.5884             &-0.1183  &0.5284              &-1.259            &-0.8493 &-0.8867 &0.9607 &0.8440 &0.9181\\
                                                
        \hline                            
        \multirow{7}{*}{\makecell[l]{Bubble \\ Chart}}
                                                &Retrieve Value                 &0.2013            &-1.393           &0.9349             &-0.2573             &0.07730              &0.3707             &0.2841             &0.07680            &-1.509     &-0.07388              &-0.3765            &-0.3736 &0.3554 &-1.097 &0.8424 &0.1012\\
                                                &Find Extremum                  &-0.5501           &0.6599           &0.02006             &0.1299              &0.1489            &-1.312             &0.4646            &-0.6987            &0.0787     &0.2758              &-0.03130            &0.3231  &-0.3225 &-0.3765 &0.4534 &-0.2456\\
                                                &Determine Range                &0.5097            &-0.4058             &-1.416             &-1.312             &0.1599             &0.5375             &-0.6743             &0.1419            &-0.4059    &0.2549              &-0.6217            &-0.2248 &0.8115 &0.3866 &-2.712 &-1.514\\
                                                &Find Anomalies                 &0.1585            &-0.1559            &1.001              &1.003            &0.2360             &0.2563            &-0.7628              &-0.2704             &0.1347  &0.3445              &-0.4072            &0.07207 &0.5293 &0.4449 &-0.1689 &0.8053\\
                                                &Find Clusters                  &-0.9233            &0.6420            &-0.5129             &-0.7942             &-0.7115           &0.1322             &0.4407              &-0.1386            &0.3576     &-0.6612              &0.7919            &0.4883 &-1.277 &0.1129 &0.7197 &-0.4445\\
                                                &Find Correlation/Trend         &0.06918           &0.08991              &0.1845             &0.3436              &0.05161             &0.04893              &0.2703              &0.3709              &0.05675  &0.06167              &0.2334            &0.3519 &0.1775 &0.2005 &0.6882 &1.066\\
                                                &Make Comparisons               &0.2042           &-0.2127            &0.2225             &0.2139            &0.06135              &0.04361             &0.2105             &0.3154             &0.04335     &0.04624              &0.1103            &0.1999 &0.3089 &-0.1229 &0.5432 &0.7292\\
        \hline
        \multirow{3}{*}{\makecell[l]{Choropleth \\ Map}}
                                                &Retrieve Value                 &1.437           &-0.2710             &-2.704             &-1.538             &0.5103              &-1.550            &1.670             &0.6300             &0.4509     &-0.7806              &1.079            &0.7491 &2.3979 &-2.602 &0.04443 &-0.1594\\
                                                &Find Extremum                  &0.3709           &0.0004937              &0.6571              &1.028              &0.3432             &0.3974             &-0.02418              &0.7164             &0.3262  &0.03170              &-0.6679            &-0.3100 &1.040 &0.4296 &-0.03503 &1.435\\
                                                &Make Comparison                &-3.247            &0.5368             &2.488             &-0.2221            &0.9498             &-0.2718             &-2.218              &-1.511             &-0.7429    &0.6500              &-0.4438            &-0.5354 &-3.962 &2.165 &-0.4723 &-2.269\\
        \hline
        \multirow{3}{*}{Treemap}                &Find Extremum                  &-0.3381            &-0.3978             &0.1660            &-0.5698             &-0.9816             &0.3973             &0.02131            &-0.5631             &0.2607     &-0.05419              &-0.4788            &-0.2722 &-1.059 &-0.05473 &-0.2914 &-1.405\\
                                                &Make Comparison                &0.1191           &0.04573              &0.1268            &0.2916              &0.07797            &0.04266              &0.1776             &0.2982              &0.0503     &0.04638              &0.1123            &0.2089 &0.2473 &0.1348 &0.4167 &0.7987\\
                                                &Identify the Hierarchical Structure
                                                                                &0.8118            &0.07227             &0.1406             &0.2941            &0.07610              &0.06647             &0.3709              &0.5134             &0.07400     &0.08164              &0.3734            &0.5291 &0.2313 &0.2204 &0.8850 &1.336\\
        \hline
    \end{tabu}
    \label{tab:vizTaskCoefResults}
\end{sidewaystable*}

\begin{figure*}[h]
    \centering
    \begin{subfigure}{.494\textwidth}
        \includegraphics[width=\textwidth]{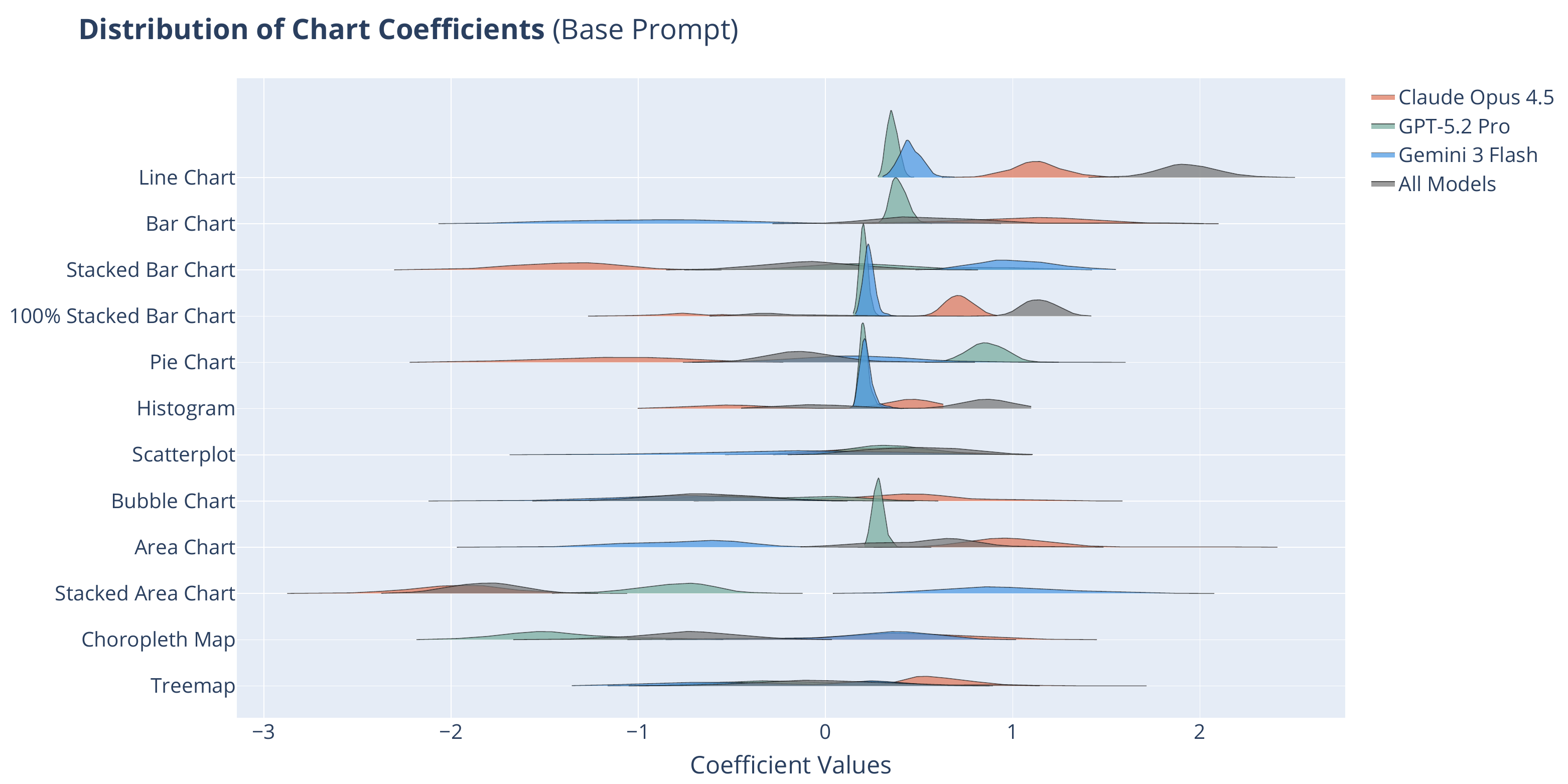}
        \caption{Visualization coefficients}
        \label{fig:chartCoef}
    \end{subfigure}
    \begin{subfigure}{.494\textwidth}
        \includegraphics[width=\textwidth]{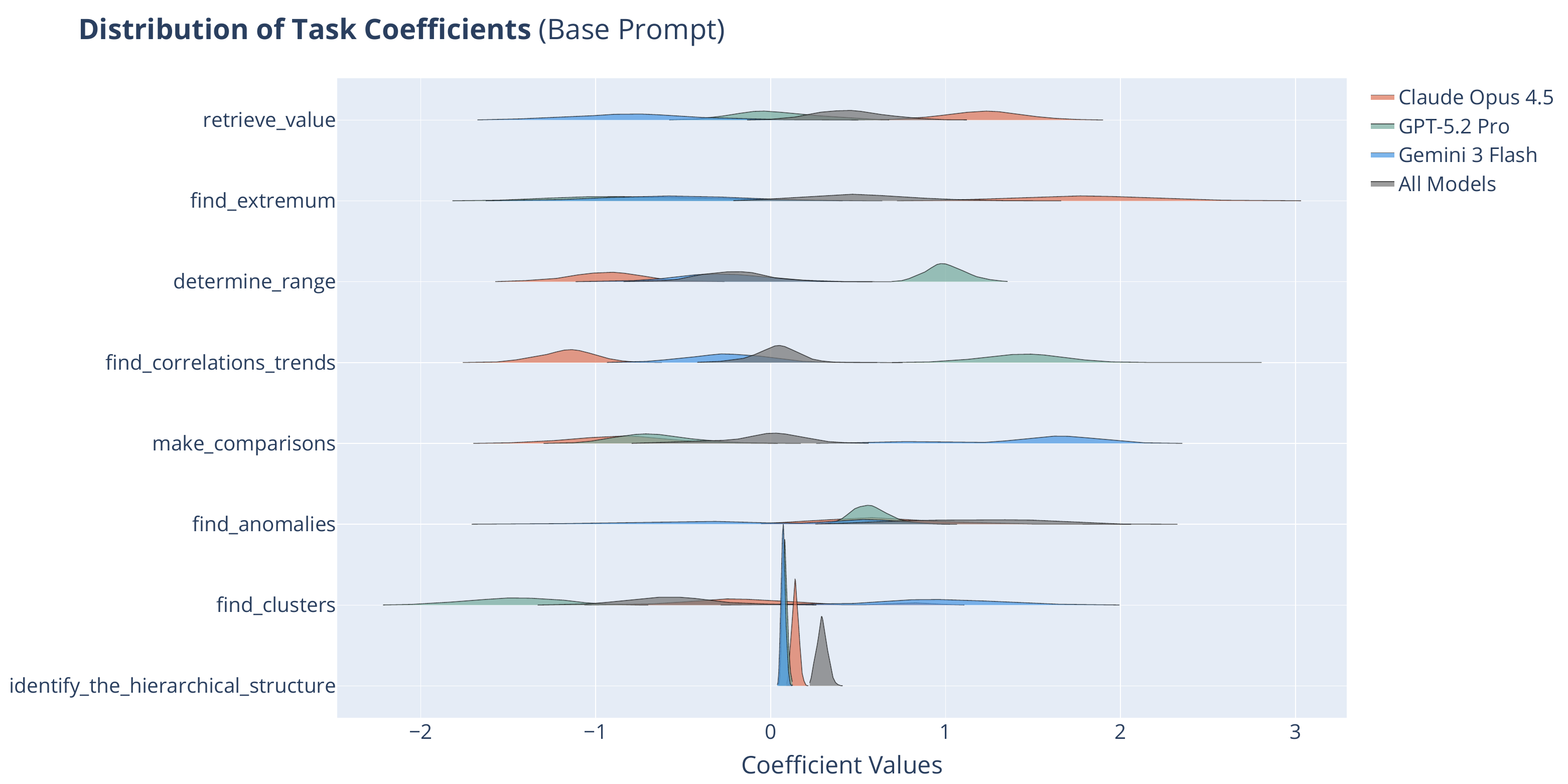}
        \caption{Task coefficients}
        \label{fig:taskCoef}
    \end{subfigure}
    \caption{Ridge plots of visualization and task bootstrapped coefficients with the base prompt from the logistic regression. (a) illustrates coefficients of \claudeFig, \gptFig, and \geminiFig\ for various visualization types. (b) shows the coefficients of the three LLMs for visualization tasks. A higher value indicates that the model performed better in this type of visualization (a) or visualization task (b).}
    \label{fig:chartTaskCoef}
\end{figure*}

\begin{figure}[h]
    \centering
    \includegraphics[width=\columnwidth]{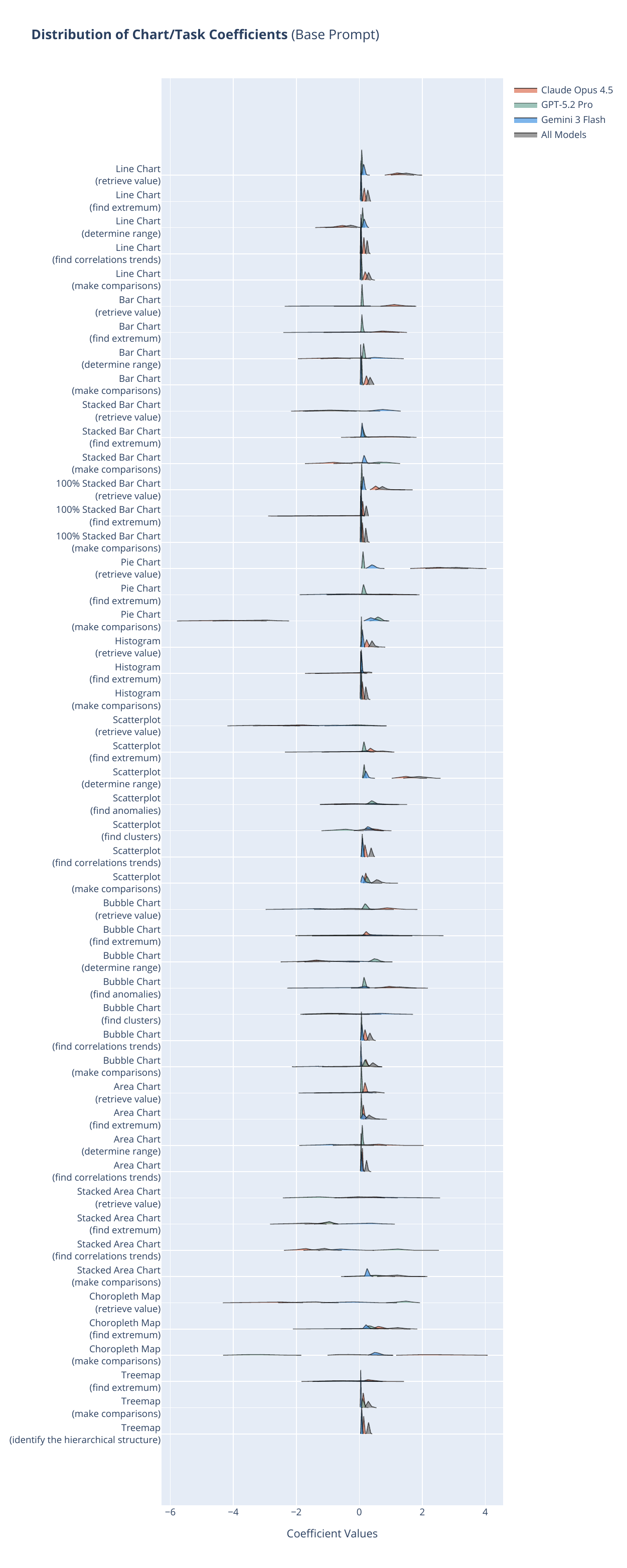}
    \caption{Ridge plots of the bootstrapped visualization/task interaction coefficients with the base prompt from the logistic regression for \claudeFig, \gptFig, and \geminiFig.}
    \label{fig:chartTaskInterCoef}
\end{figure}

\begin{table*}[!h]
    \centering    
    \caption{Table of insignificant coefficients for Experiment 3, Jaccard Indices.}
    \begin{tabu}{l|l|l|l}
        \hline
        \textbf{LLM}    &\textbf{Category}              &\textbf{Misleader}                         &\textbf{Visualization}\\
        \hline
        Claude          &Manipulated Scale              &Exceeding the Canvas 	                    &Bar Chart \\
        Claude          &Manipulated Scale              &Unconventional Scale Directions 	        &Bar Chart \\
        Claude          &Manipulated Annotation 	    &\textit{N/A}                               &Bar Chart \\  
        (All)           &Manipulated Annotation         &Deceptive Labeling 	                    &Bar Chart\\	
        Claude 	        &Manipulated Annotation         &\textit{N/A}    	                        &Stacked Area Chart  \\
        Claude 	        &Manipulated Annotation         &Lack of legend 	                        &Stacked Area Chart \\
        Gemini 	        &Manipulated Scale              &Misuse of Cumulative Relationship 	        &Stacked Area Chart \\
        Gemini 	        &Manipulated Visual Encoding 	&\textit{N/A}                               &Bar Chart \\
        Gemini 	        &Manipulated Visual Encoding    &Data Visual Disproportion 	                &Pie Chart \\
        Gemini 	        &Manipulated Visual Encoding 	&\textit{N/A}                               &Line Chart \\
        Gemini 	        &Manipulated Visual Encoding 	&\textit{N/A}                               &Area Chart \\
        Gemini 	        &Manipulated Visual Encoding    &Continous Encoding for Categorical Data 	&Area Chart \\
        Gemini 	        &Manipulated Data 	            &\textit{N/A}                               &Area Chart \\
        Gemini 	        &Manipulated Data               &Inappropriate Order 	                    &Area Chart \\
        GPT 	        &Manipulated Data               &Deceptive Labeling 	                    &Line Chart \\
        GPT 	        &Manipulated Data               &Dual Encoding 	                            &Scatter Plot \\
        \hline
    \end{tabu}
    \label{tab:jIndexInsig}
\end{table*}

\begin{table*}[!h]
    \centering    
    \caption{Table of insignificant coefficients for Experiment 3, Dice-S{\o}rensen Coefficients.}
    \begin{tabu}{l|l|l|l}
        \hline
        \textbf{LLM}    &\textbf{Category}              &\textbf{Misleader}                         &\textbf{Visualization}\\
        \hline
        Claude          &Manipulated Scale              &Exceeding the Canvas 	                    &Bar Chart\\
        Claude          &Manipulated Scale              &Unconventional Scale Directions 	        &Bar Chart\\
        Claude          &Manipulated Annotation 	    &\textit{N/A}                               &Bar Chart\\  
        (All)           &Manipulated Annotation         &Deceptive Labeling 	                    &Bar Chart\\	
        Claude 	        &Manipulated Annotation         &\textit{N/A}    	                        &Stacked Area Chart\\
        Claude 	        &Manipulated Annotation         &Lack of legend 	                        &Stacked Area Chart\\
        Claude 	        &Manipulated Visual Encoding 	&\textit{N/A}    	                        &Choropleth Map\\
        Claude 	        &Manipulated Visual Encoding 	&Categorical Encoding for Continuous Data   &Choropleth Map\\
        Claude 	        &Manipulated Visual Encoding    &Missing Normalization                      &(All)\\
        Claude 	        &Manipulated Visual Encoding    &Missing Normalization                      &Choropleth Map\\
        Gemini 	        &Manipulated Scale              &Inappropriate Scale Functions              &Bar Chart\\
        Gemini 	        &Manipulated Visual Encoding    &Data Visual Disproportion 	                &Pie Chart\\
        Gemini 	        &Manipulated Visual Encoding 	&\textit{N/A}                               &Line Chart\\
        Gemini 	        &Manipulated Visual Encoding 	&\textit{N/A}                               &Area Chart\\
        Gemini 	        &Manipulated Visual Encoding    &Continous Encoding for Categorical Data 	&Area Chart\\
        Gemini 	        &Manipulated Data               &Inappropriate Order 	                    &Line Chart\\
        GPT 	        &Manipulated Data               &Deceptive Labeling 	                    &Bar Chart\\
        \hline
    \end{tabu}
    \label{tab:dsCoefInsig}
\end{table*}

\begin{figure*}[h]
    \centering
    \caption{Prompt used for Experiment 3 on misleading chart element identification.}\vspace{-0.7em}
    \tabulinesep=1pt
    \begin{tabu}{l}
        \texttt{Given the following chart, please list all applicable misleading chart elements.  These elements include the following:}\\
        \\
        \texttt{Name: Misuse of Cumulative Relationship}\\
        \texttt{Definition: Incorrectly combining or accumulating data elements that do not logically sum or relate, distorting the true }
        \texttt{relationships.}\\
        \texttt{--------------------------------------------------------------------------------------------------------------------------------------------------------------------------------------------------------------------------------------------------------------------------------------------------------------------------------------------------------------------------------------------------------------------}\\
        \texttt{Name: Small Size}\\
        \texttt{Definition: Using excessively small text or graphical elements that hinder readability and make it difficult to interpret data.}\\
        \texttt{---------------------------------------------------------------------------------------------------------------------------------------------------------------------------------------------------------------------------------------------------------------------------------------------------------------------------------------------------------------------------------------------}\\
        \texttt{Name: Exceeding the Canvas}\\
        \texttt{Definition: Allowing data points, labels, or visual elements to extend beyond the display area, causing loss of critical information.}\\
        \texttt{---------------------------------------------------------------------------------------------------------------------------------------------------------------------------------------------------------------------------------------------------------------------------------------------------------------------------------------------------------------------------------------------------------------}\\
        \texttt{Name: Unconventional Scale Directions}\\
        \texttt{Definition: Using non-standard axis or legend orientations, such as inverting scales, which can confuse viewers and misrepresent }\\
        \texttt{relationships.}\\
        \texttt{------------------------------------------------------------------------------------------------------------------------------------------------------------------------------------------------------------------------------------------------------------------------------------------------------------------------------------------------------------------------------------------------------}\\
        \texttt{---------------------------------------}\\
        \texttt{Name: Inappropriate Scale Range}\\
        \texttt{Definition: Altering the scale of axes or legends by stretching, truncating, or using inconsistent binning, which distorts data }\\
        \texttt{representation.}\\
        \texttt{---------------------------------------------------------------------------------------------------------------------------------------------------------------------------------------------------------------------------------------------------------------------------------------------------------------------------------------------------------------------------------------------------}\\
        \texttt{------------------------------------------}\\
        \texttt{Name: Inappropriate Scale Functions}\\
        \texttt{Definition: Applying arbitrary or misleading non-linear transformations to the scale of an axis, affecting how viewers perceive the }\\
        \texttt{relationships within the data.}\\
        \texttt{---------------------------------------------------------------------------------------------------------------------------------------------------------------------------------------------------------------------------------------------------------------------------------------------------------------------------------------------------------------------------------------------------------------}\\
        \texttt{---------------------------------------------------------------------------------------}\\
        \texttt{Name: Deceptive Labeling}\\
        \texttt{Definition: Using annotations or labels that contradict the data or make the visualization difficult to interpret.}\\
        \texttt{------------------------------------------------------------------------------------------------------------------------------------------------------------------------------------------------------------------------------------------------------------------------------------------------------------------------------------------------------}\\
        \texttt{Name: Lack Of Legend}\\
        \texttt{Definition: Omitting a legend that explains colors, symbols, or other encodings, leaving viewers uncertain about the meaning of the }\\
        \texttt{visualization.}\\
        \texttt{---------------------------------------------------------------------------------------------------------------------------------------------------------------------------------------------------------------------------------------------------------------------------------------------------------------------------------------------------------------------------------------------------------------}\\
        \texttt{---------------------------------------}\\
        \texttt{Name: Lack Of Scales}\\
        \texttt{Definition: Failing to provide axis scales or units of measurement, which can oversimplify or obscure the interpretation of the data.}\\
        \texttt{---------------------------------------------------------------------------------------------------------------------------------------------------------------------------------------------------------------------------------------------------------------------------------------------------------------------------------------------------------------------------------------------------------------}\\
        \texttt{Name: Inappropriate Aggregation}\\
        \texttt{Definition: Combining or summarizing data in a way that distorts the true distribution or relationships, leading to inaccurate conclusions.}\\
        \texttt{---------------------------------------------------------------------------------------------------------------------------------------------------------------------------------------------------------------------------------------------------------------------------------------------------------------------------------------------------------------------------------------------------------------------------------}\\
        \texttt{Name: Dual Encoding}\\
        \texttt{Definition: Using multiple visual channels (e.g., both width and height) to encode the same variable, which exaggerates the data's visual }\\
        \texttt{impact.}\\
        \texttt{---------------------------------------------------------------------------------------------------------------------------------------------------------------------------------------------------------------------------------------------------------------------------------------------------------------------------------------------------------------------------------------------------------------------------------}\\
        \texttt{------------------}\\
        \texttt{Name: Data Visual Disproportion}\\
        \texttt{Definition: Creating a visual representation where the graphical elements (e.g., bar heights) do not accurately correspond to the actual }\\
        \texttt{data values, leading to misinterpretation.}\\
        \texttt{------------------------------------------------------------------------------------------------------------------------------------------------------------------------------------------------------------------------------------------------------------------------------------------------------------------------------------------------------------------------------------------------------------------------------}\\
        \texttt{---------------------------------------------------------------------------------------------------------------------------}\\
        \texttt{Name: Continous Encoding For Categorical Data}\\
        \texttt{Definition: Applying continuous encoding methods (e.g., color gradients or line connections) to categorical data, which can mislead }\\
        \texttt{viewers into perceiving relationships that do not exist.}\\
        \texttt{---------------------------------------------------------------------------------------------------------------------------------------------------------------------------------------------------------------------------------------------------------------------------------------------------------------------------------------------------------------------------------------------------------------}\\
        \texttt{---------------------------------------------------------------------------------------------------------------------------------------------------------------------}\\
        \texttt{Name: Categorical Encoding For Continuous Data}\\
        \texttt{Definition: Representing continuous data using discrete categories, potentially distorting trends and relationships.}\\
        \texttt{------------------------------------------------------------------------------------------------------------------------------------------------------------------------------------------------------------------------------------------------------------------------------------------------------------------------------------------------------------}\\
        \texttt{Name: Cherry Picking}\\
        \texttt{Definition: Selecting only a subset of data to display, potentially misleading viewers by implying conclusions about the entire dataset.}\\
        \texttt{------------------------------------------------------------------------------------------------------------------------------------------------------------------------------------------------------------------------------------------------------------------------------------------------------------------------------------------------------------------------------------------------------------------------}\\
        \texttt{Name: Overplotting}\\
        \texttt{Definition: Overcrowding a visualization with excessive data points or elements, making it difficult to discern meaningful patterns.}\\
        \texttt{------------------------------------------------------------------------------------------------------------------------------------------------------------------------------------------------------------------------------------------------------------------------------------------------------------------------------------------------------------------------------------------------------------}\\
        \texttt{Name: Inappropriate Order}\\
        \texttt{Definition: Manipulating the order of data by manipulating axis labels or legend items in a way that misleads viewers or creates a false }\\
        \texttt{impression of trends.}\\
        \texttt{------------------------------------------------------------------------------------------------------------------------------------------------------------------------------------------------------------------------------------------------------------------------------------------------------------------------------------------------------------------------------------------------------------------------------}\\
        \texttt{------------------------------------------------------------}\\
        \texttt{Name: Missing Normalization}\\
        \texttt{Definition: Displaying unnormalized absolute values when relative or normalized comparisons would be more appropriate for interpretation.}\\
        \texttt{---------------------------------------------------------------------------------------------------------------------------------------------------------------------------------------------------------------------------------------------------------------------------------------------------------------------------------------------------------------------------------------------------------------------------}\\
        \\
        \texttt{The following is an example of a sample output answer in the JSON format:}\\
        \\
        \texttt{\{}\\
                \quad\quad\quad\quad\texttt{"misleadElements":["Small Size","Cherry Picking"],}\\
                \quad\quad\quad\quad\texttt{"explanation":"<insert explanation here>"}\\
        \texttt{\}}\\
        \\
        \texttt{where <insert explanation here> should be replaced by the actual explanation of why each element was chosen, and the array ["Small Size",}\\
        \texttt{"Cherry Picking"] denotes that the chart contained the "Small Size" and "Cherry Picking" misleading elements.  If there are no misleading}\\
        \texttt{ chart elements, please return an empty array for "misleadElements".}\\
    \end{tabu}
    \label{fig:misleadPrompt}
\end{figure*}

\begin{table*}[]
    \caption{Boxplots of the Dice-S{\o}rensen Coefficients\cite{Jackson:1989:SCM} calculated from \claudeFig, \gptFig, and \geminiFig's answers compared to the correct misleading chart elements from Experiment 2.}\vspace{-0.7em}
    \tabulinesep=1pt
    \begin{tblr}{
        rowsep=0pt,
        colspec={X[0.1,m]Q[h,3.1]|X[]X[]X[]X[]X[]X[]X[]X[]X[]X[]}
    }
        &   \textbf{Misleader Name} & \rotatebox[origin=bl]{25}{\textbf{100\% Stacked Bar Chart}} & \rotatebox[origin=bl]{25}{\textbf{Area Chart}} & \rotatebox[origin=bl]{25}{\textbf{Bar Chart}} & \rotatebox[origin=bl]{25}{\textbf{Choropleth Map}} & \rotatebox[origin=bl]{25}{\textbf{Heatmap}} & \rotatebox[origin=bl]{25}{\textbf{Line Chart}} & \rotatebox[origin=bl]{25}{\textbf{Pie Chart}} & \rotatebox[origin=bl]{25}{\textbf{Stacked Area Chart}} & \rotatebox[origin=bl]{25}{\textbf{Stacked Bar Chart}} & \rotatebox[origin=bl]{25}{\textbf{Scatterplot}} \\
        \hline
        \SetCell[r=4]{c,manuData} \rotatebox[origin=c]{90}{\textbf{Manipulated Data}}
                                    & \makecell[l]{\textbf{Missing}\\ \textbf{Normalization}}
                                        &
                                        &
                                        &
                                        & \includegraphics[scale=0.025]{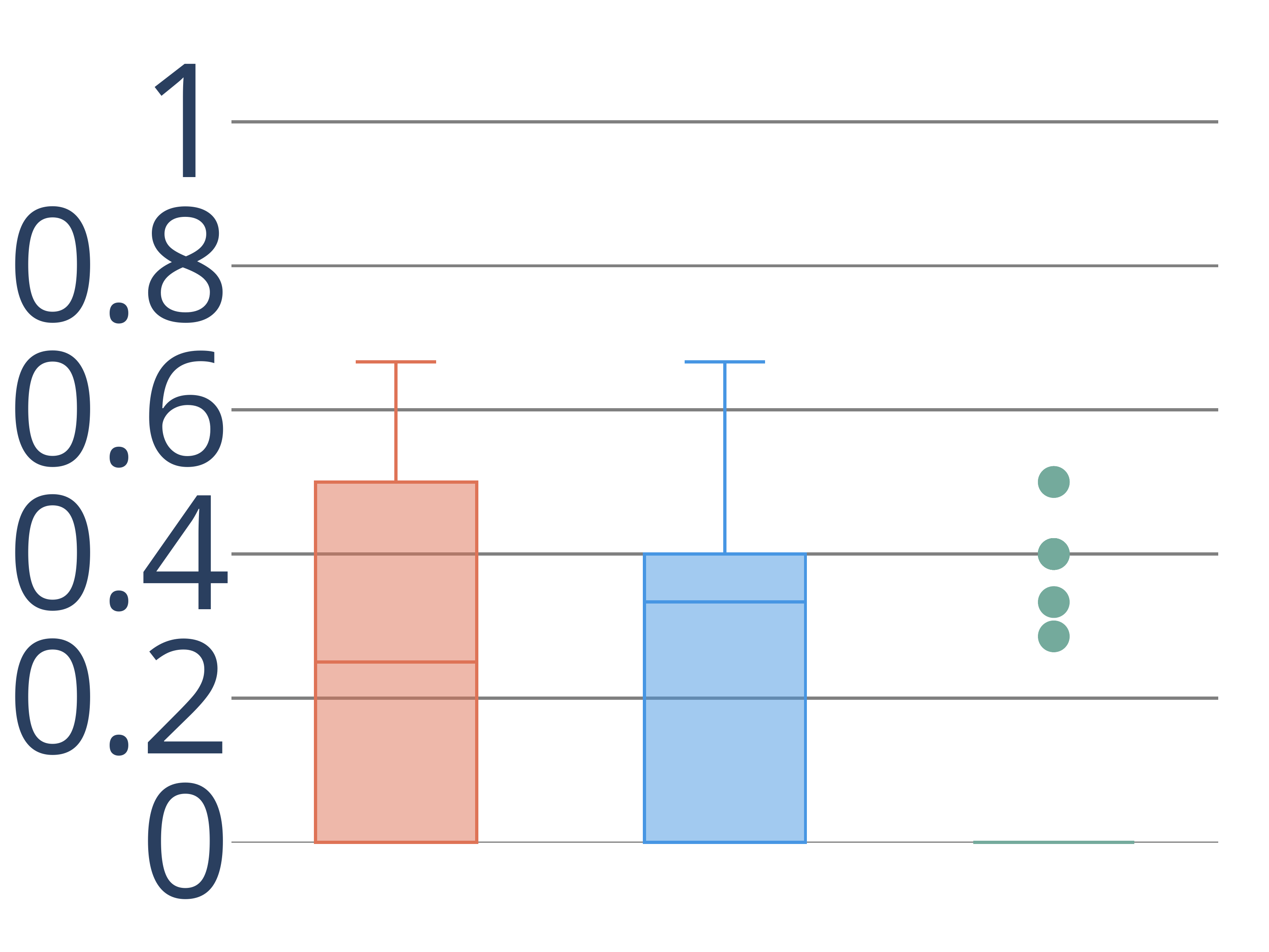}
                                        & 
                                        &
                                        &
                                        &
                                        &
                                        &
                                        &\\
                                    & \textbf{Cherry Picking} 
                                        &
                                        &
                                        &
                                        &
                                        &
                                        & \includegraphics[scale=0.025]{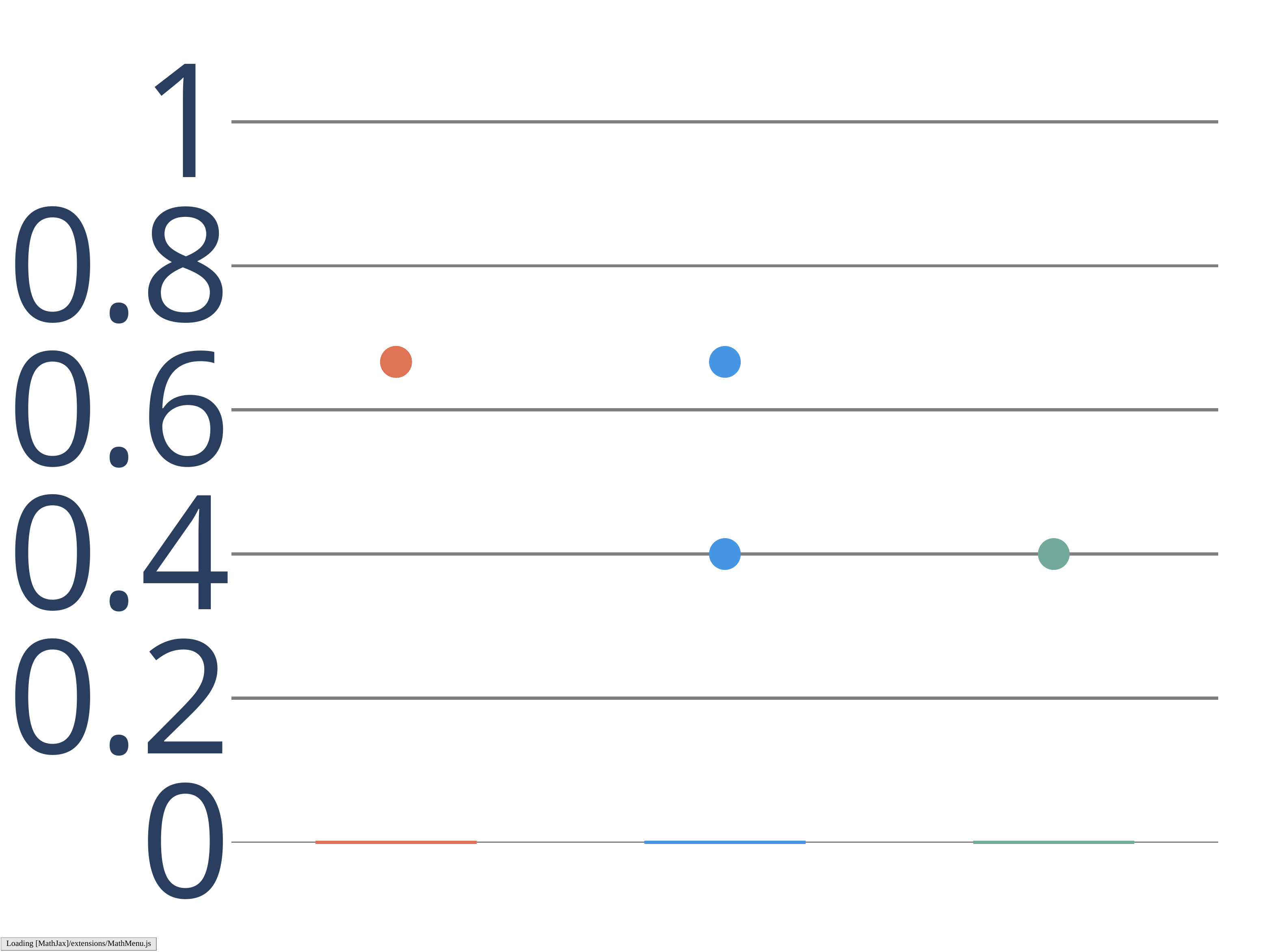}
                                        &
                                        &
                                        &
                                        & \includegraphics[scale=0.025]{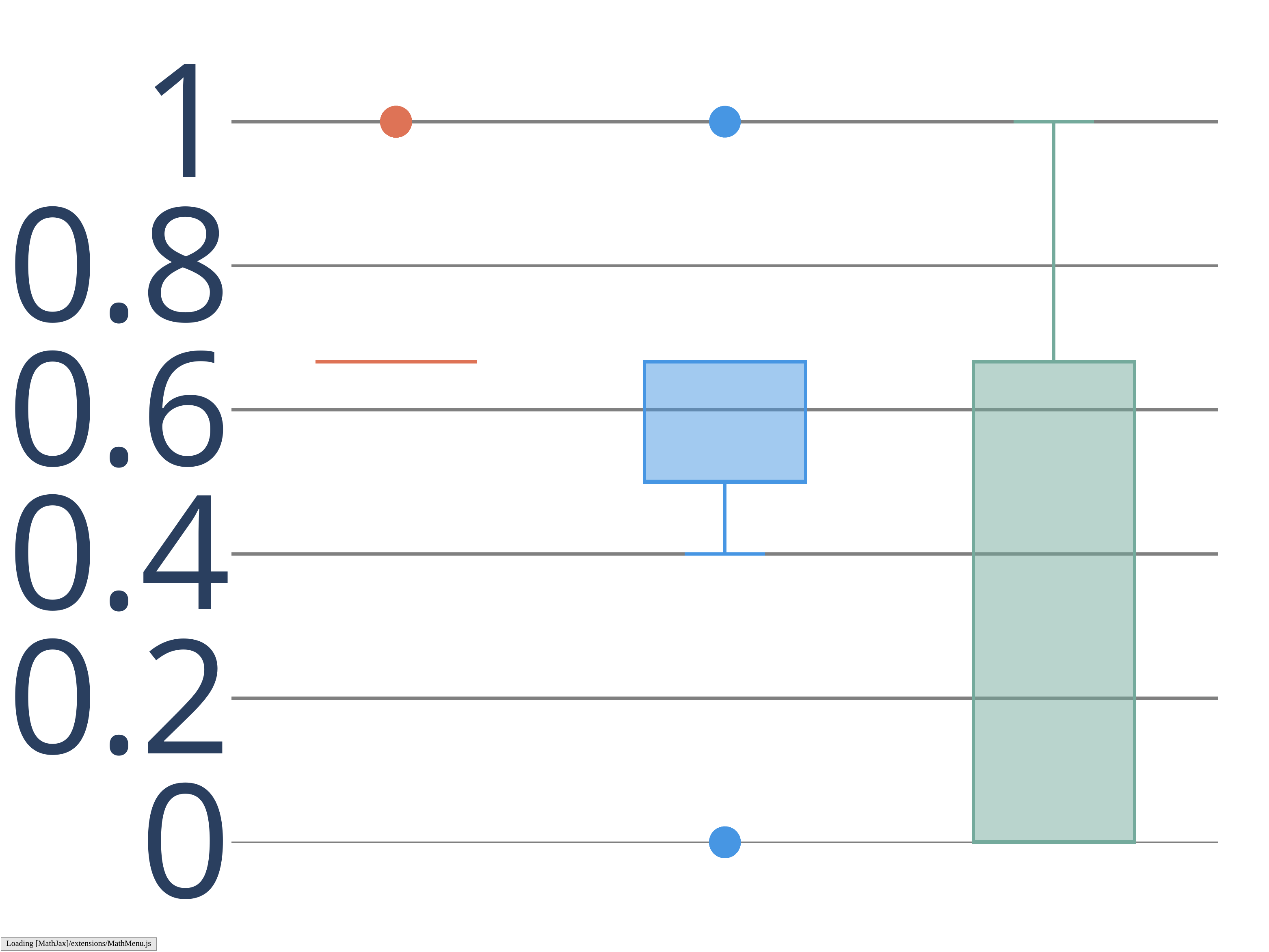}\\
                                    & \textbf{Overplotting}
                                        &
                                        &
                                        &
                                        &
                                        &
                                        &
                                        &
                                        &
                                        &
                                        & \includegraphics[scale=0.025]{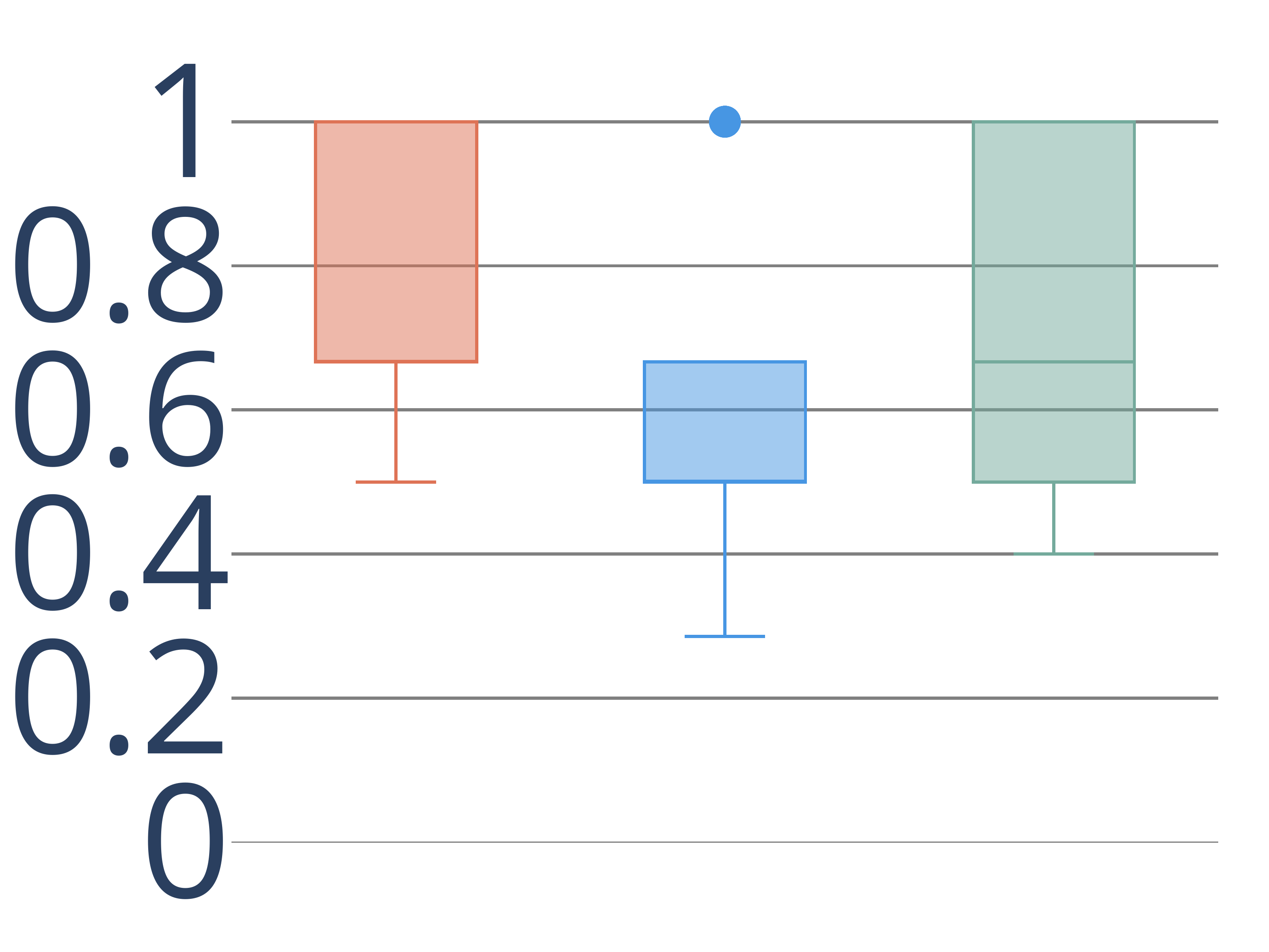}\\
                                    & \textbf{Inappropriate Order} 
                                        & \includegraphics[scale=0.025]{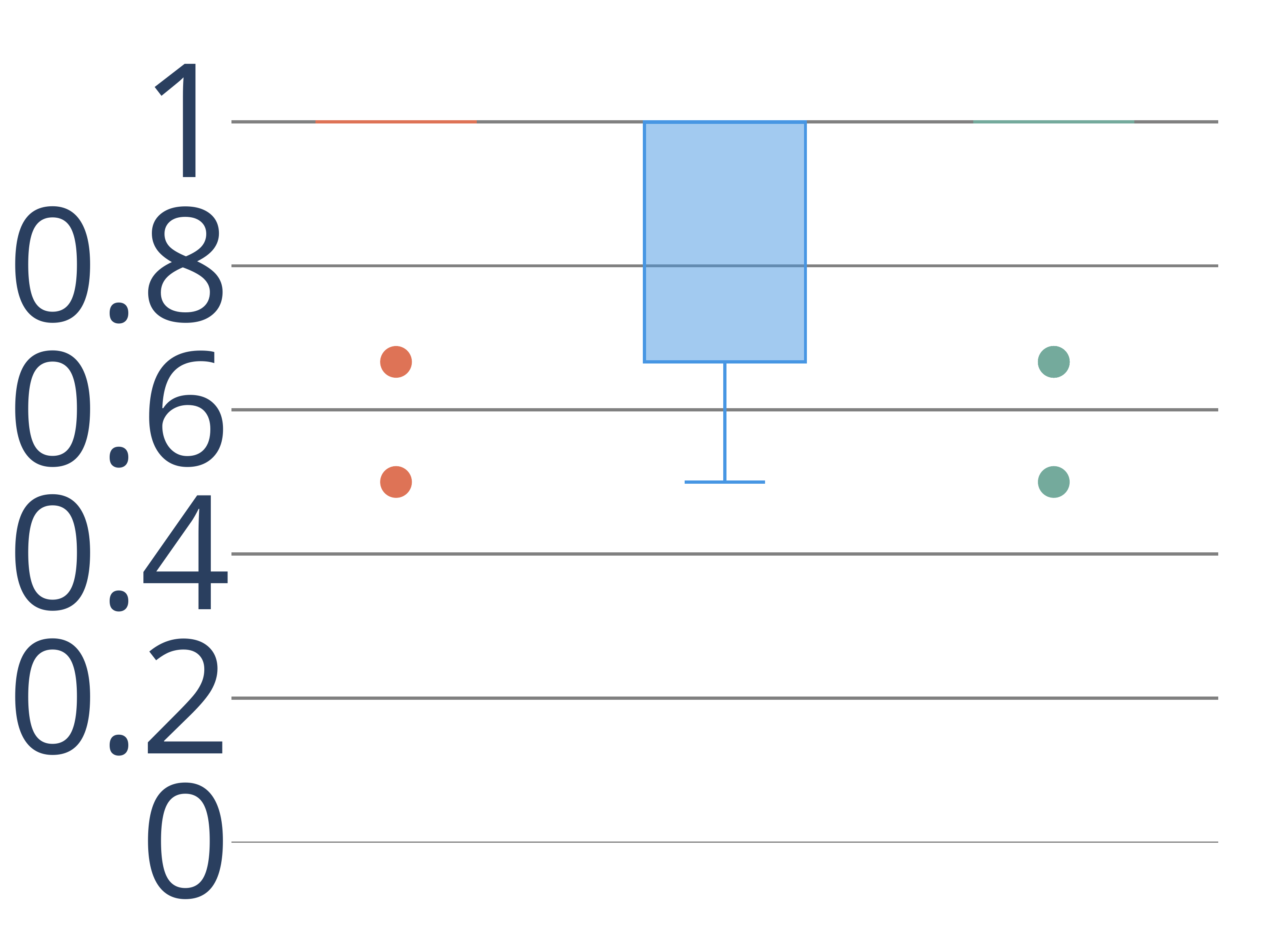} 
                                        & \includegraphics[scale=0.025]{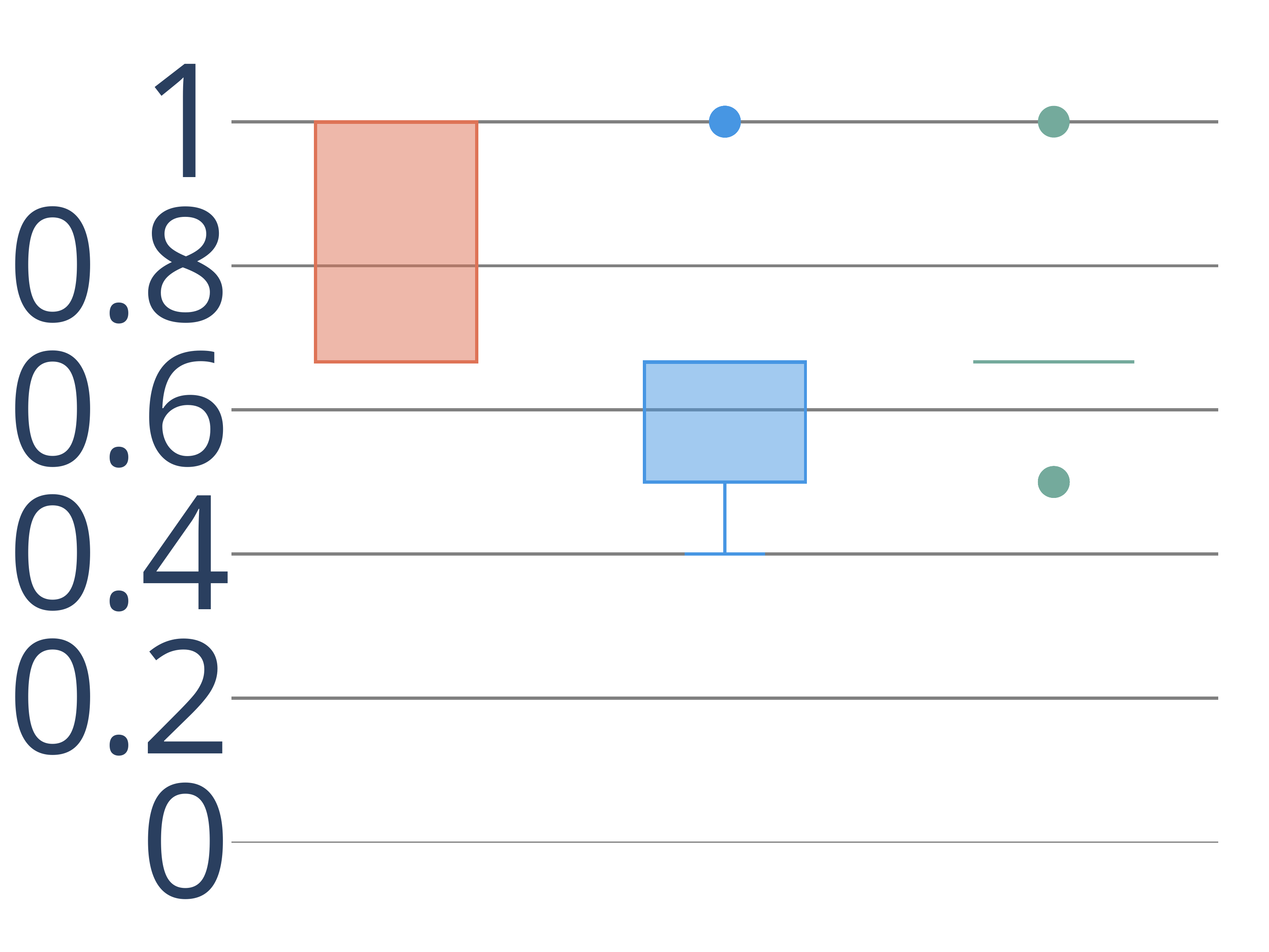} 
                                        & \includegraphics[scale=0.025]{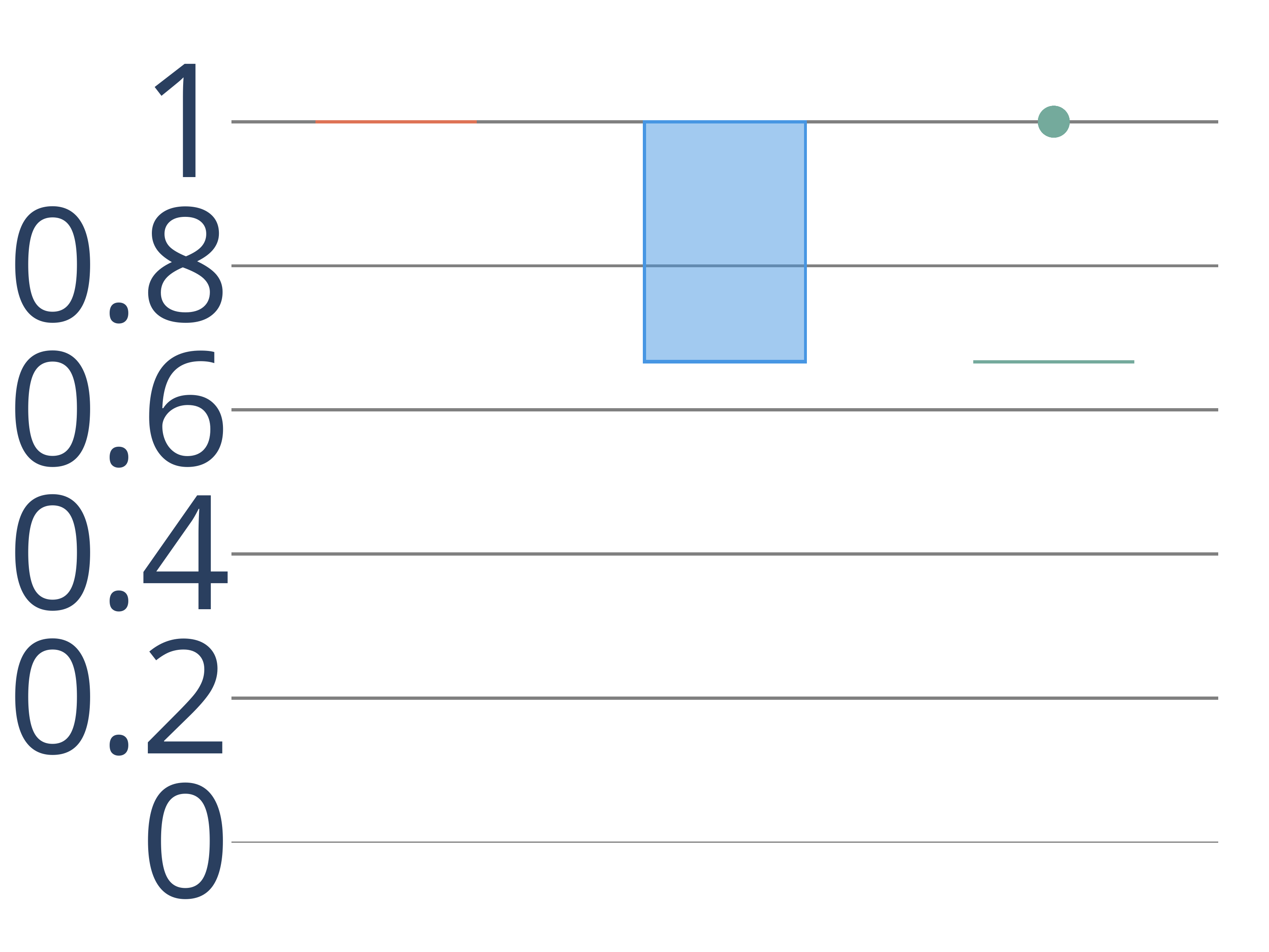} 
                                        & \includegraphics[scale=0.025]{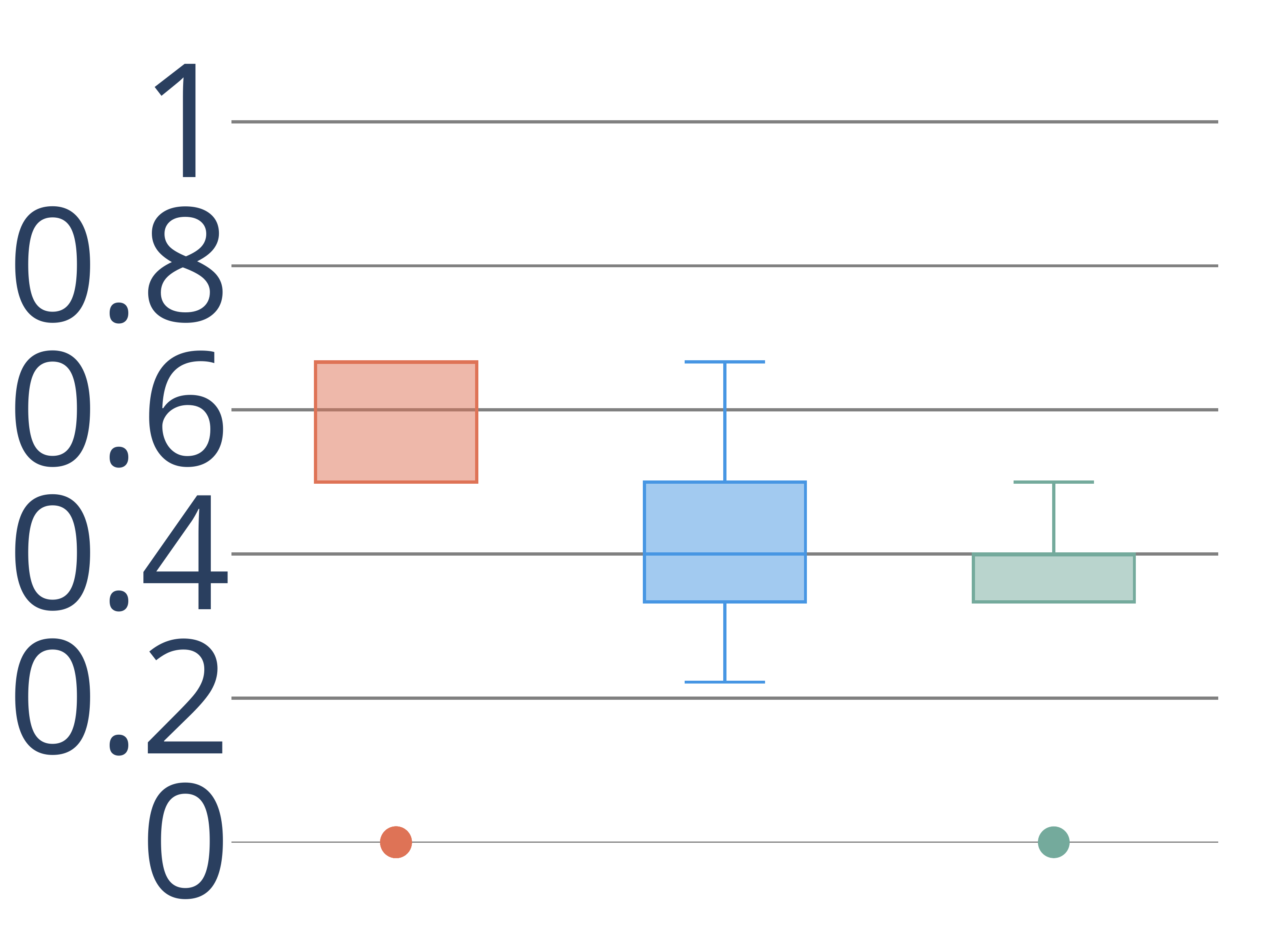} 
                                        & 
                                        & \includegraphics[scale=0.025]{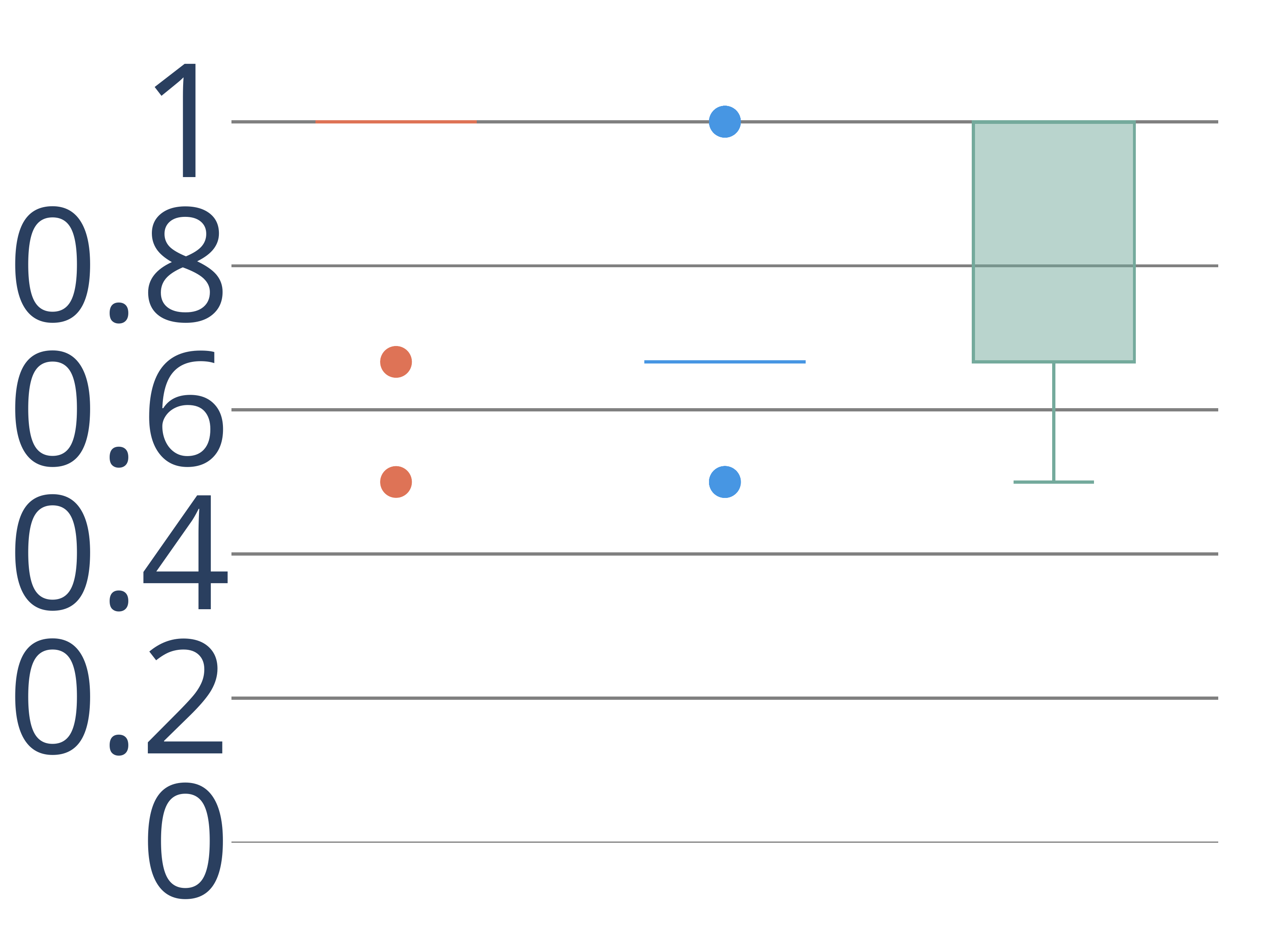} 
                                        & 
                                        & 
                                        & 
                                        & \includegraphics[scale=0.025]{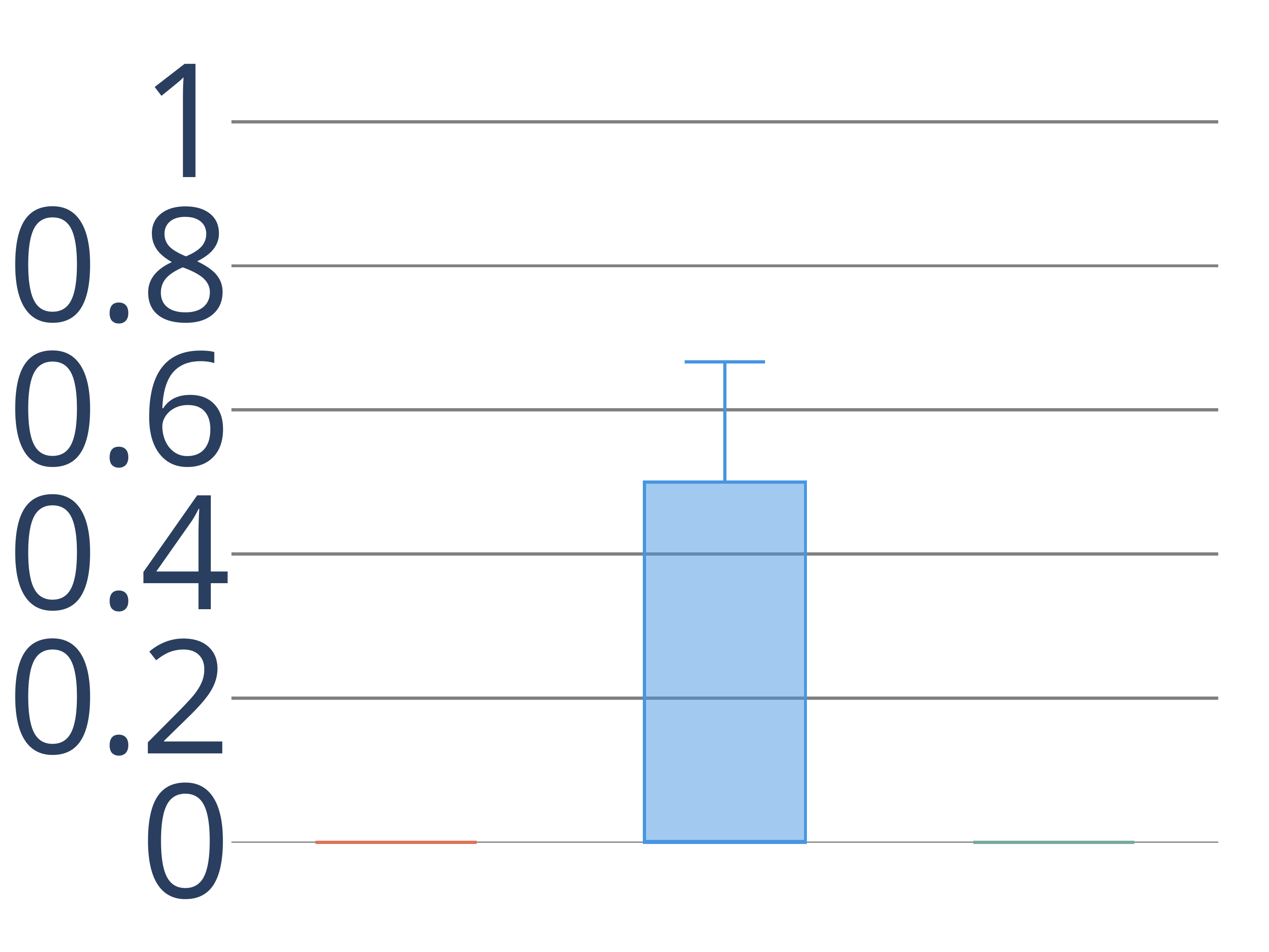}\\
        \hline
        \SetCell[r=4]{c,manuAnnotate} \rotatebox[origin=c]{90}{\textbf{Manipulated Annotation}}
                                    & \textbf{Deceptive Labeling} 
                                        & 
                                        & 
                                        & \includegraphics[scale=0.025]{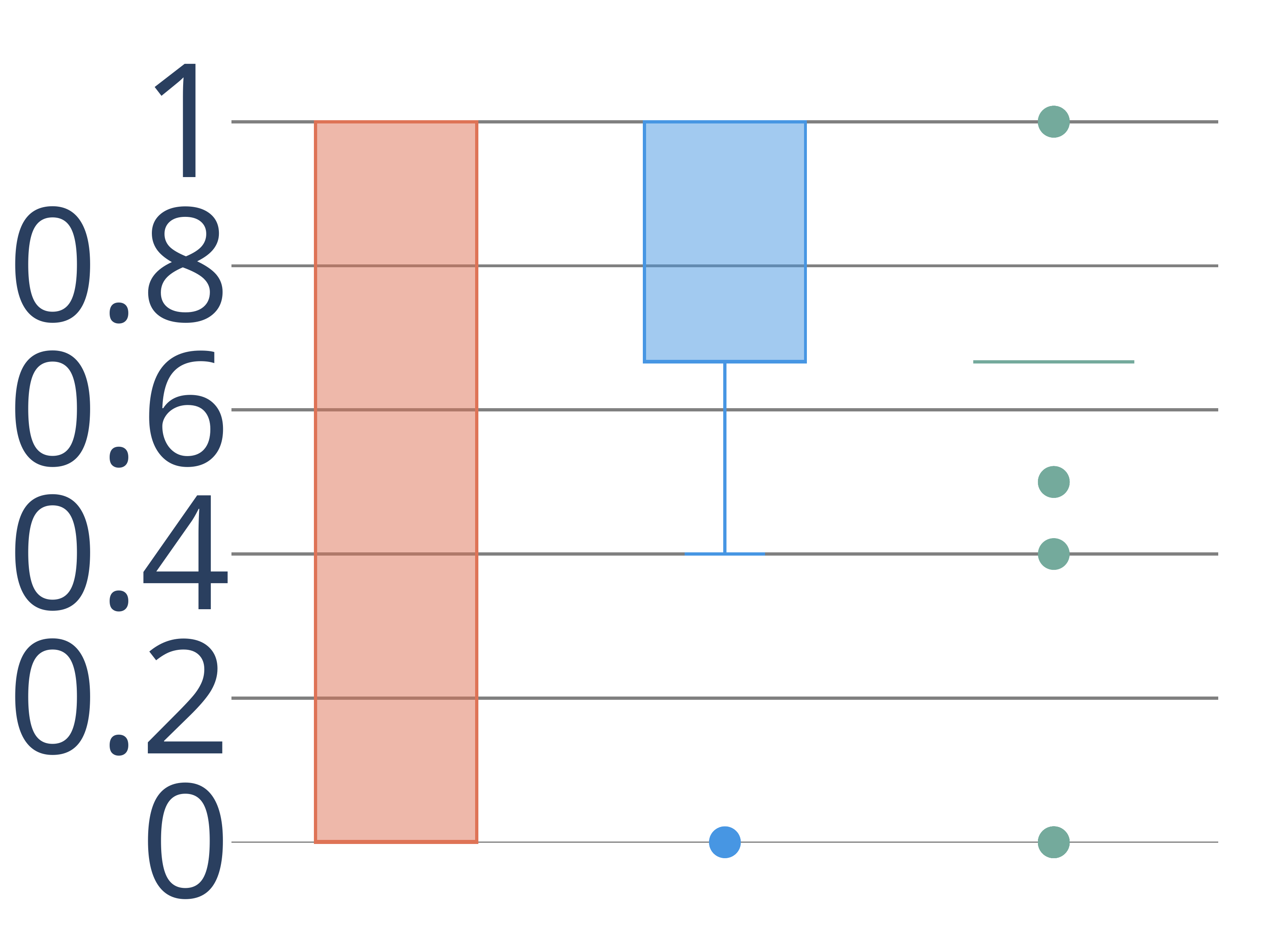} 
                                        & 
                                        & 
                                        & \includegraphics[scale=0.025]{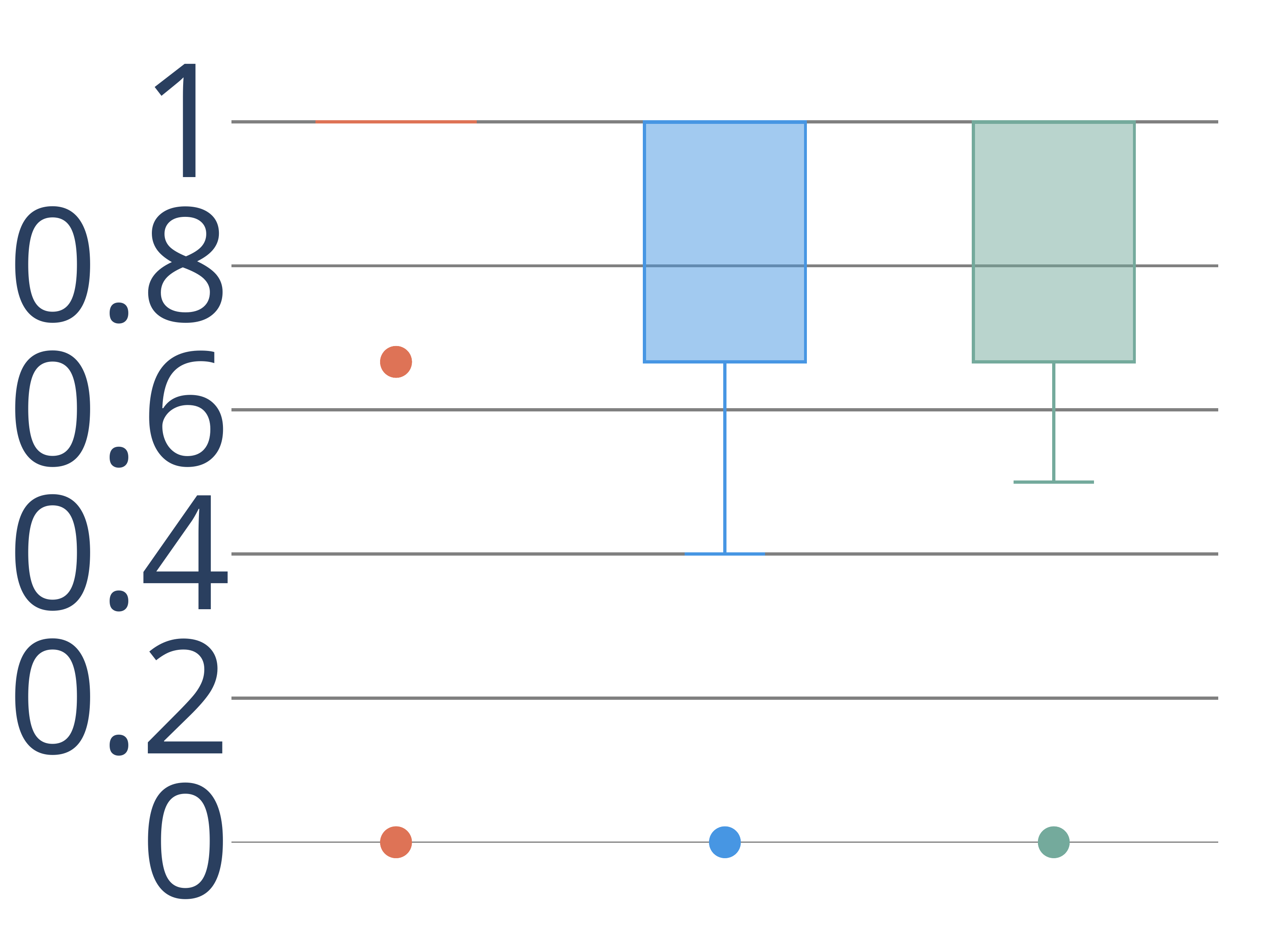} 
                                        & \includegraphics[scale=0.025]{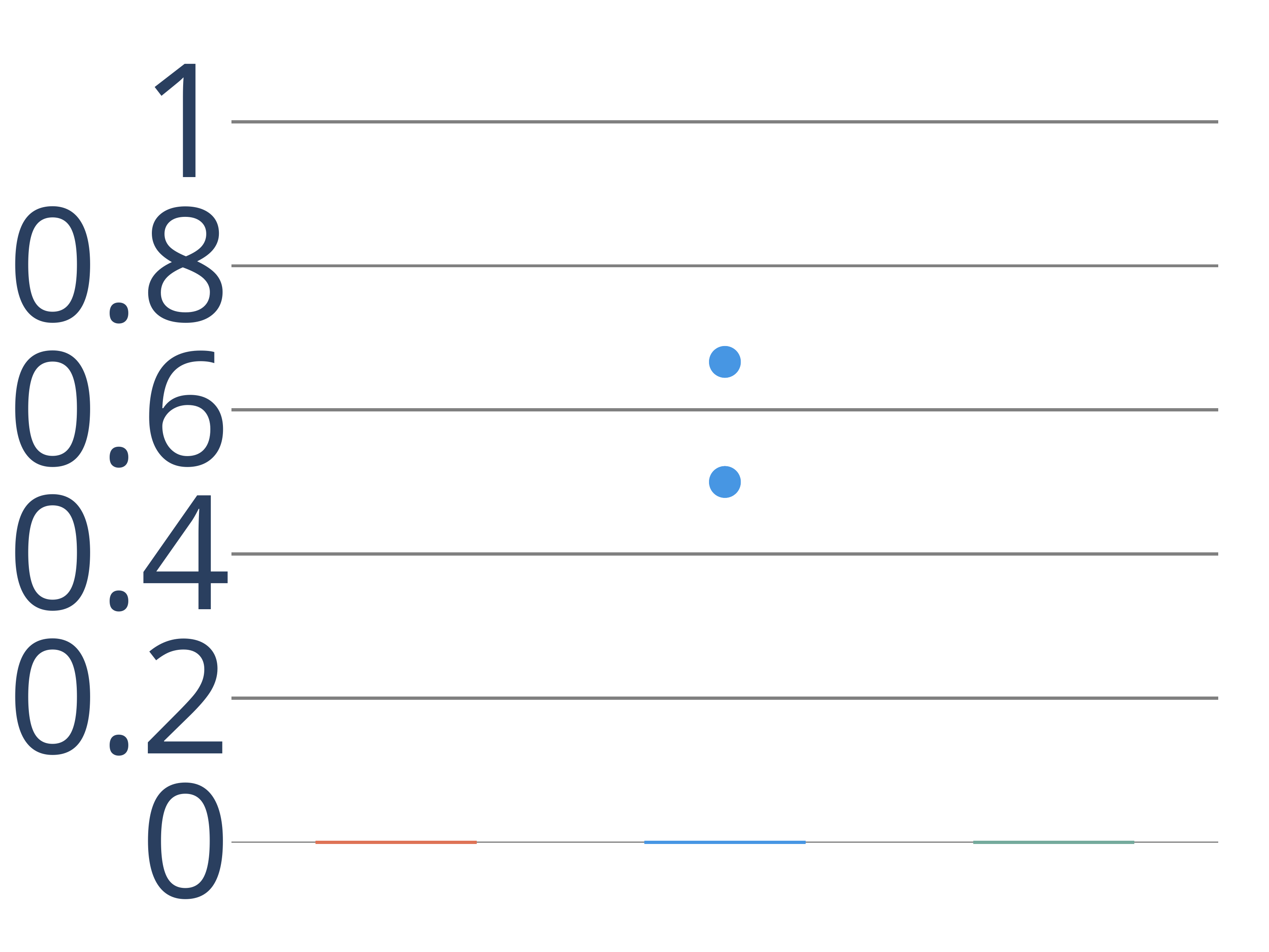} 
                                        & 
                                        & 
                                        &\\
                                    & \makecell[l]{\textbf{Lack of Labeling}\\ \textit{Lack of legend}} 
                                        & 
                                        & 
                                        & 
                                        & 
                                        & 
                                        & 
                                        & 
                                        & \includegraphics[scale=0.025]{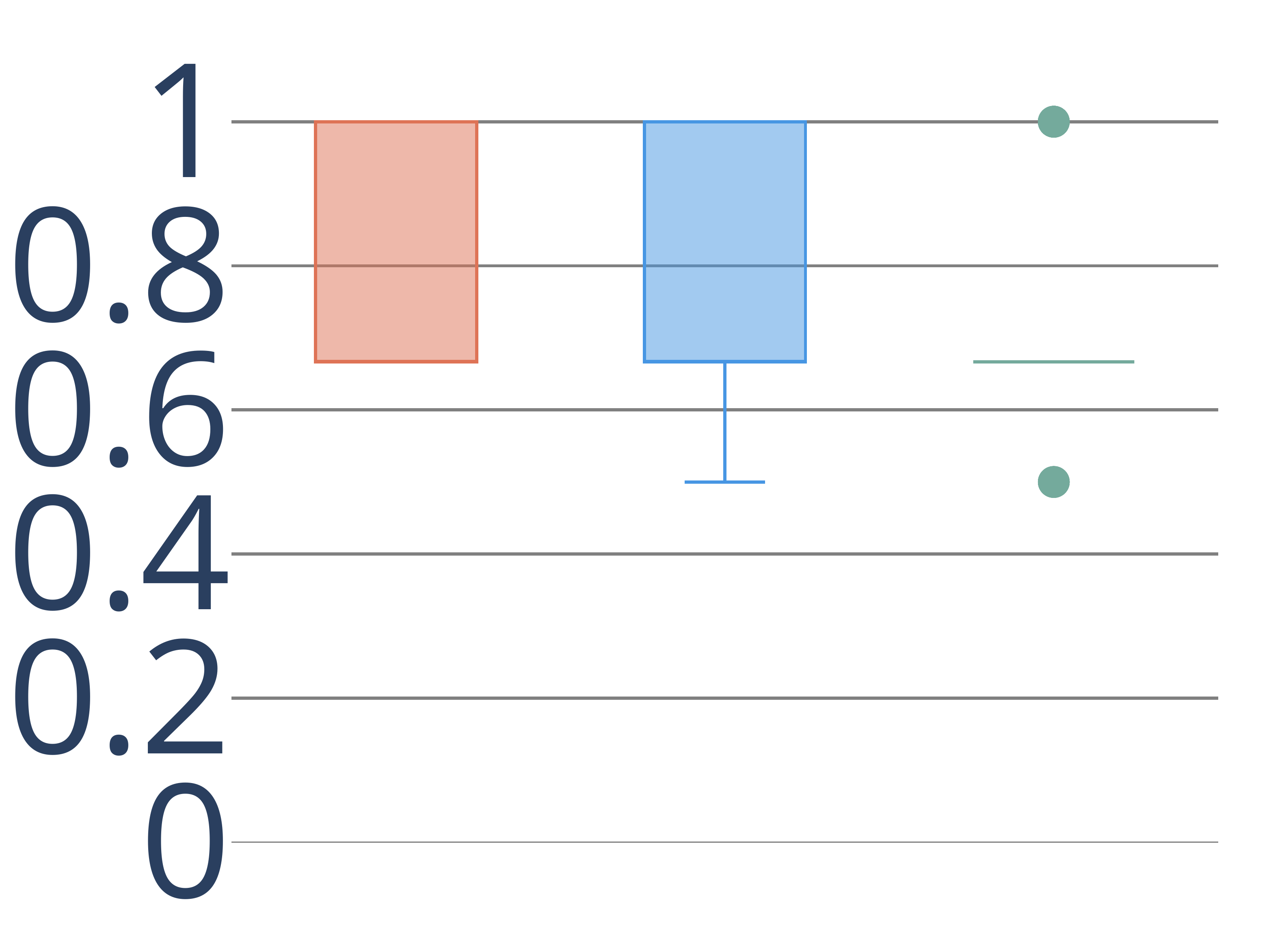}
                                        & \includegraphics[scale=0.025]{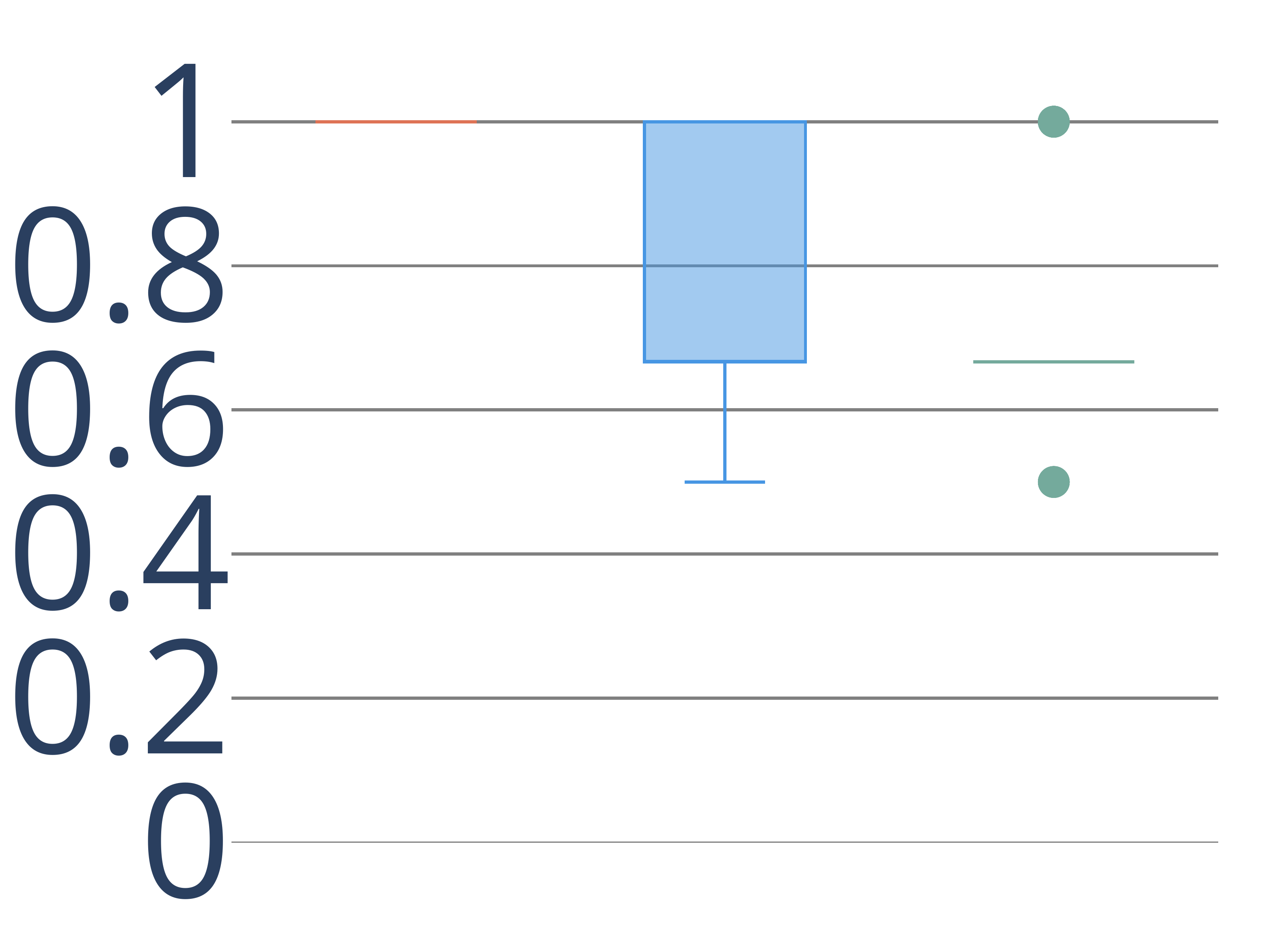} 
                                        &\\
                                    & \makecell[l]{\textbf{Lack of Labeling}\\ \textit{Lack of scales}} 
                                        & 
                                        & \includegraphics[scale=0.025]{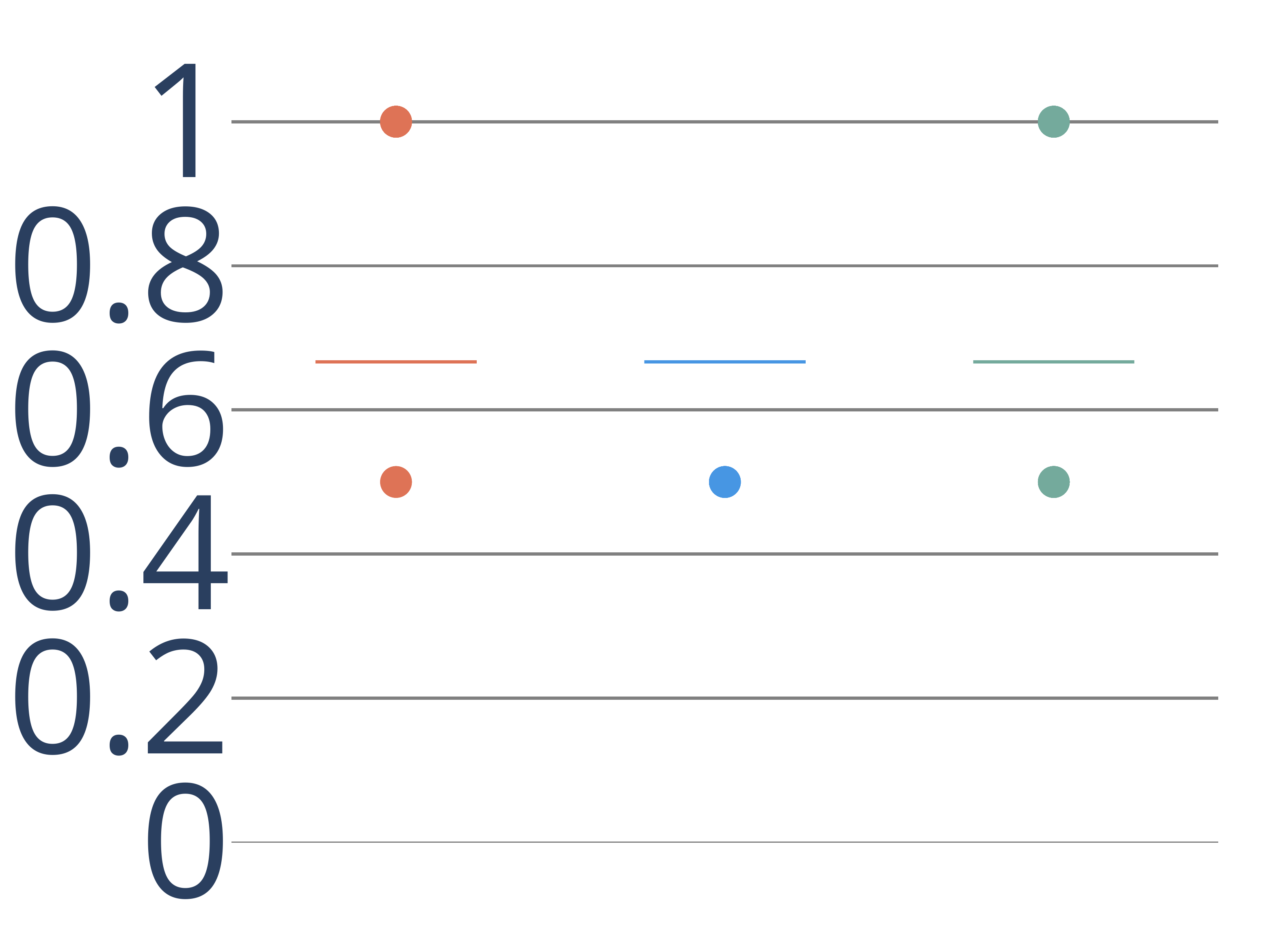} 
                                        & \includegraphics[scale=0.025]{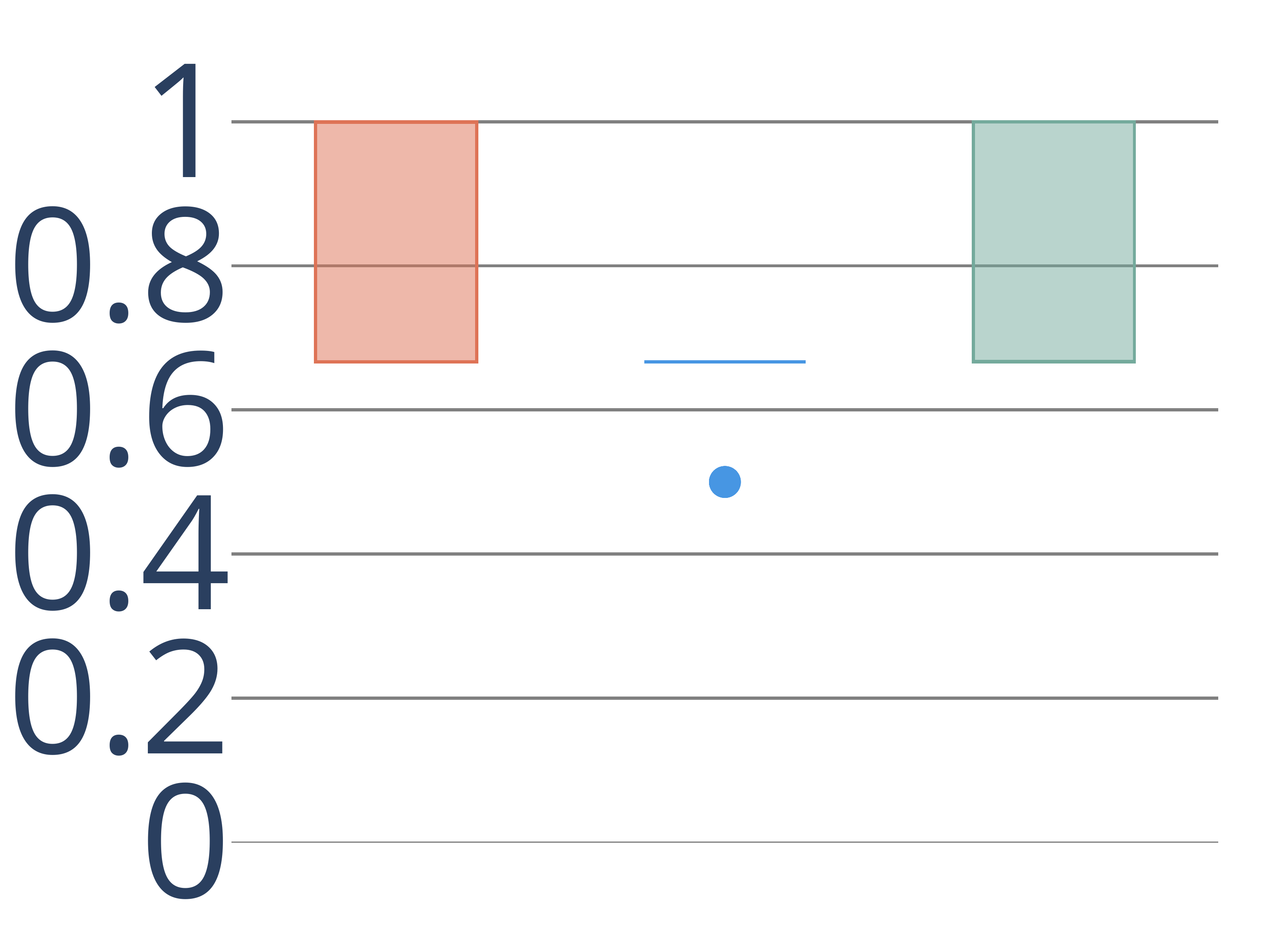} 
                                        & 
                                        & 
                                        & \includegraphics[scale=0.025]{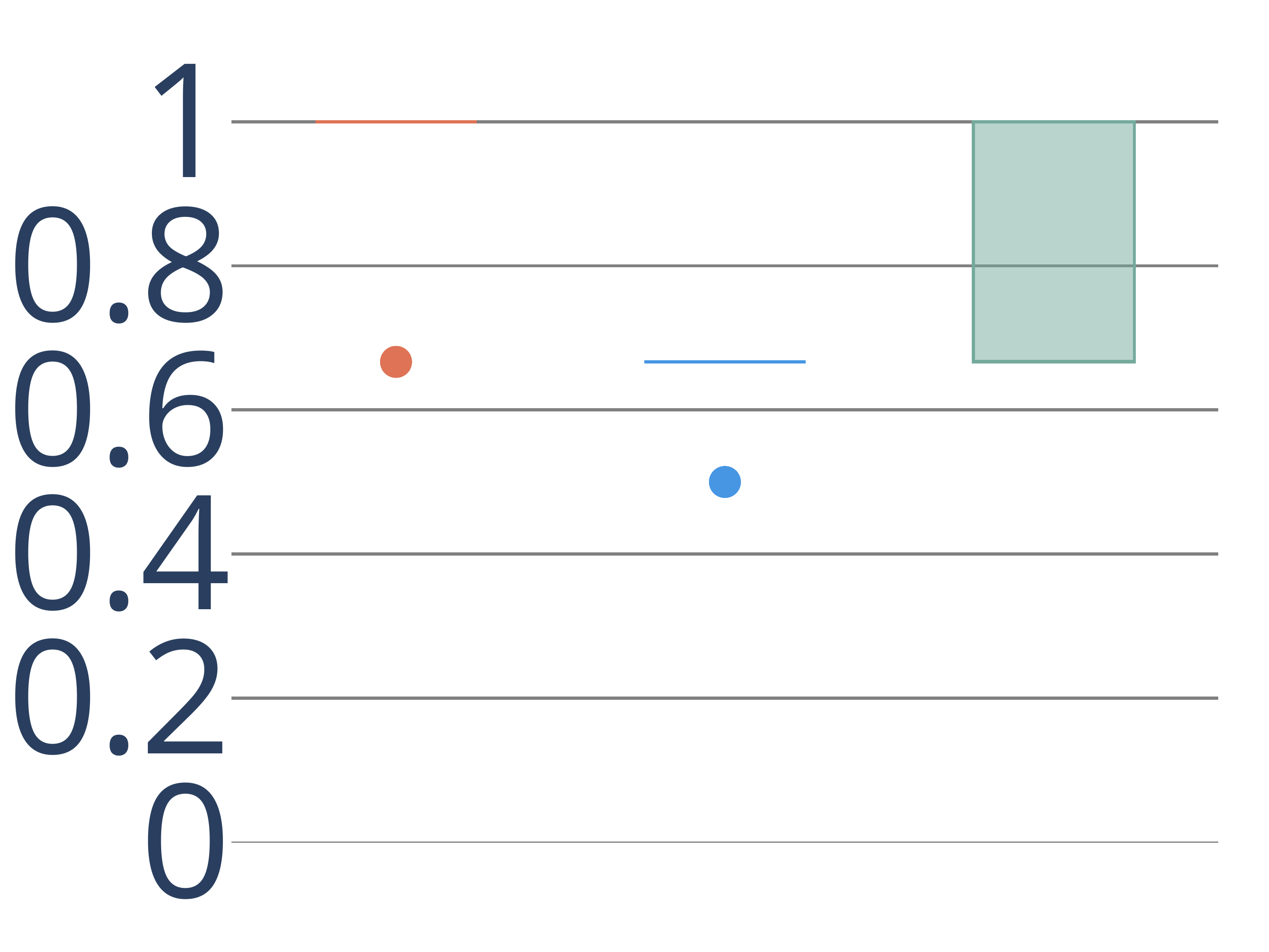} 
                                        & 
                                        & 
                                        & 
                                        &\\
                                    & \makecell[l]{\textbf{Inappropriate}\\ \textbf{Aggregation}} 
                                        & 
                                        & \includegraphics[scale=0.025]{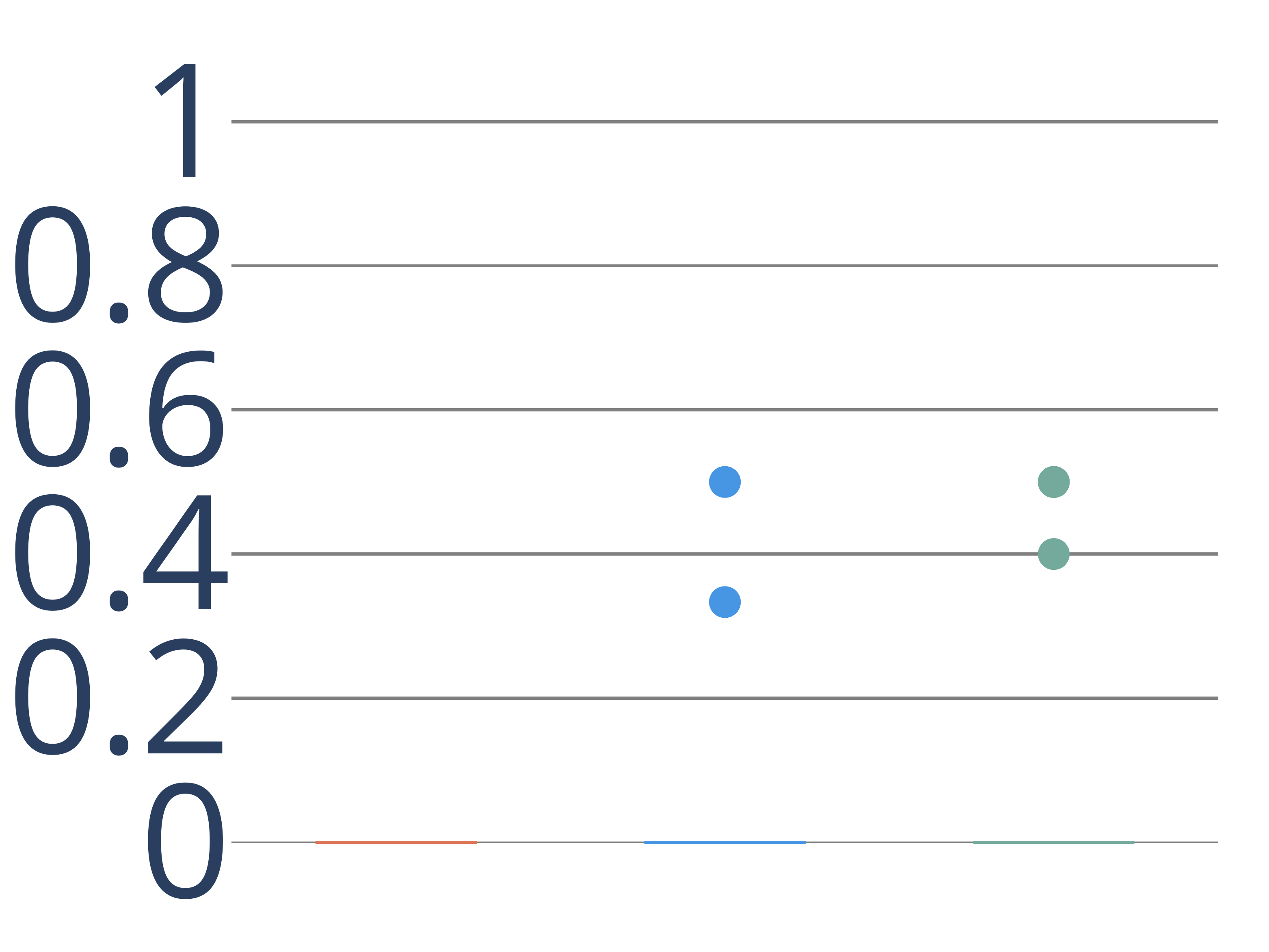} 
                                        & \includegraphics[scale=0.025]{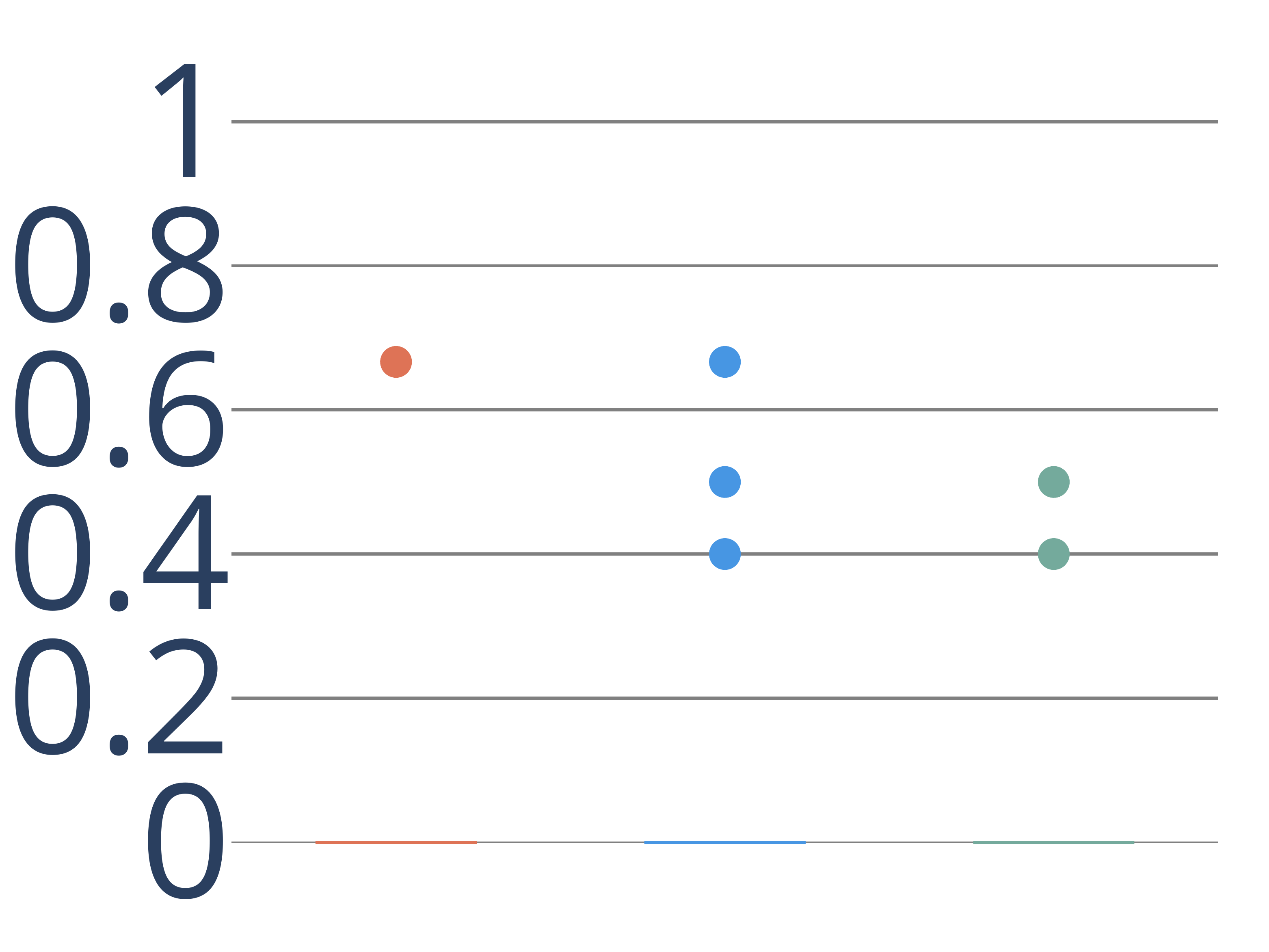} 
                                        & 
                                        & 
                                        & \includegraphics[scale=0.025]{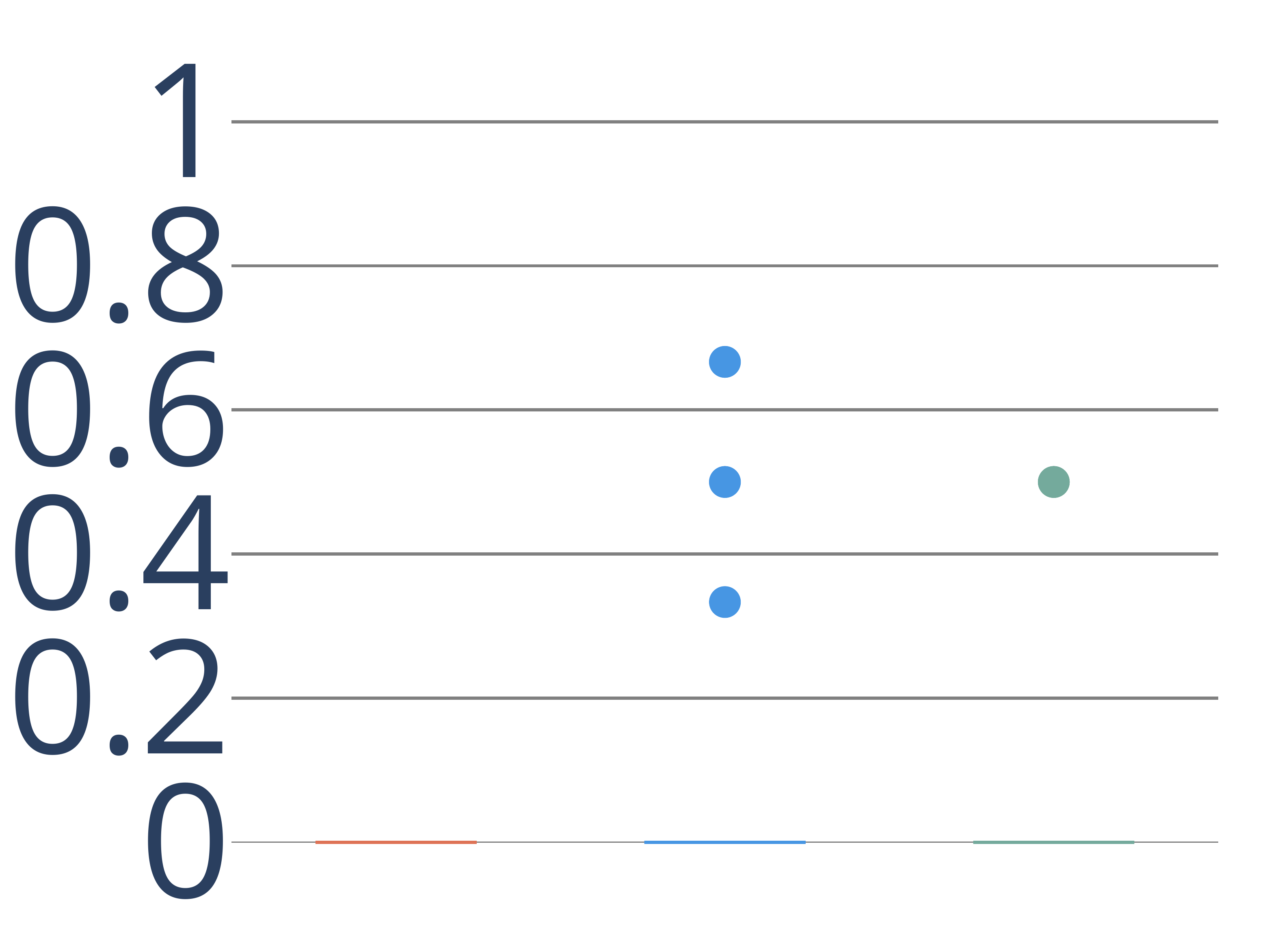} 
                                        & 
                                        & 
                                        & \includegraphics[scale=0.025]{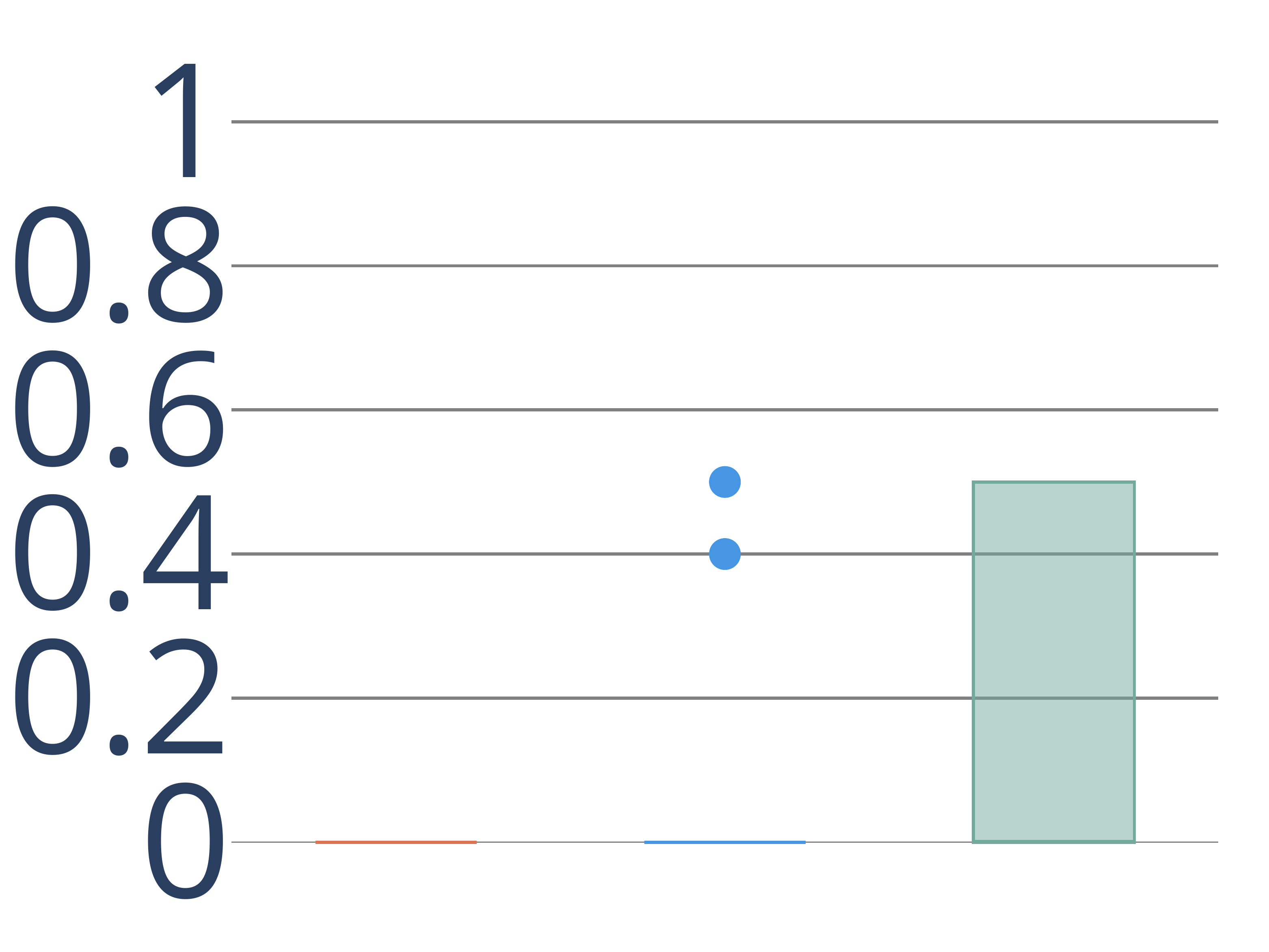} 
                                        & \includegraphics[scale=0.025]{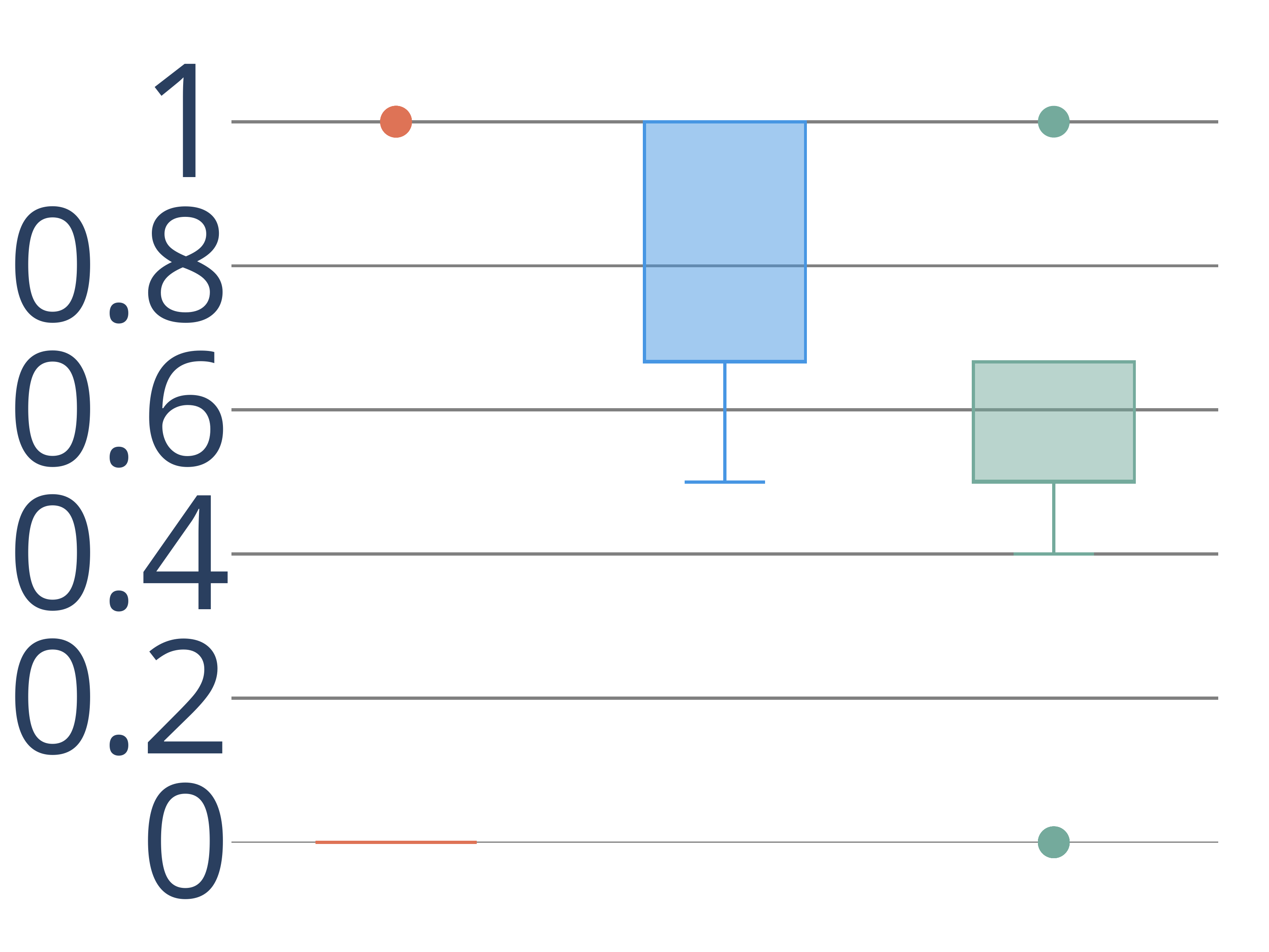}\\
        \hline
        \SetCell[r=4]{c,manuVisEncode} \rotatebox[origin=c]{90}{\textbf{Manipulated Visual Encoding}}
                                    & \makecell[l]{\textbf{Data-visual}\\ \textbf{Disproportion}} 
                                        & 
                                        & 
                                        & \includegraphics[scale=0.025]{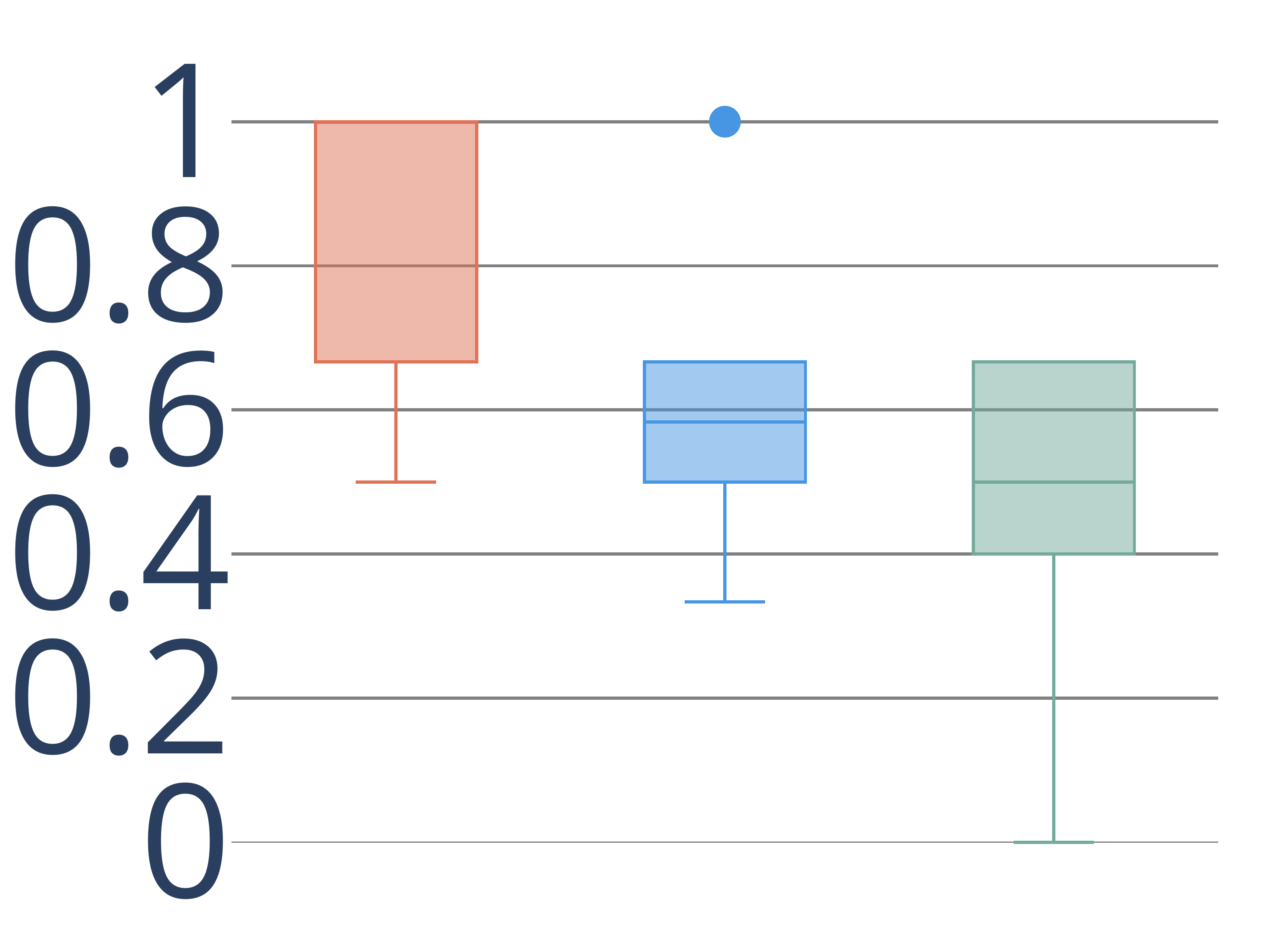} 
                                        & 
                                        & 
                                        & \includegraphics[scale=0.025]{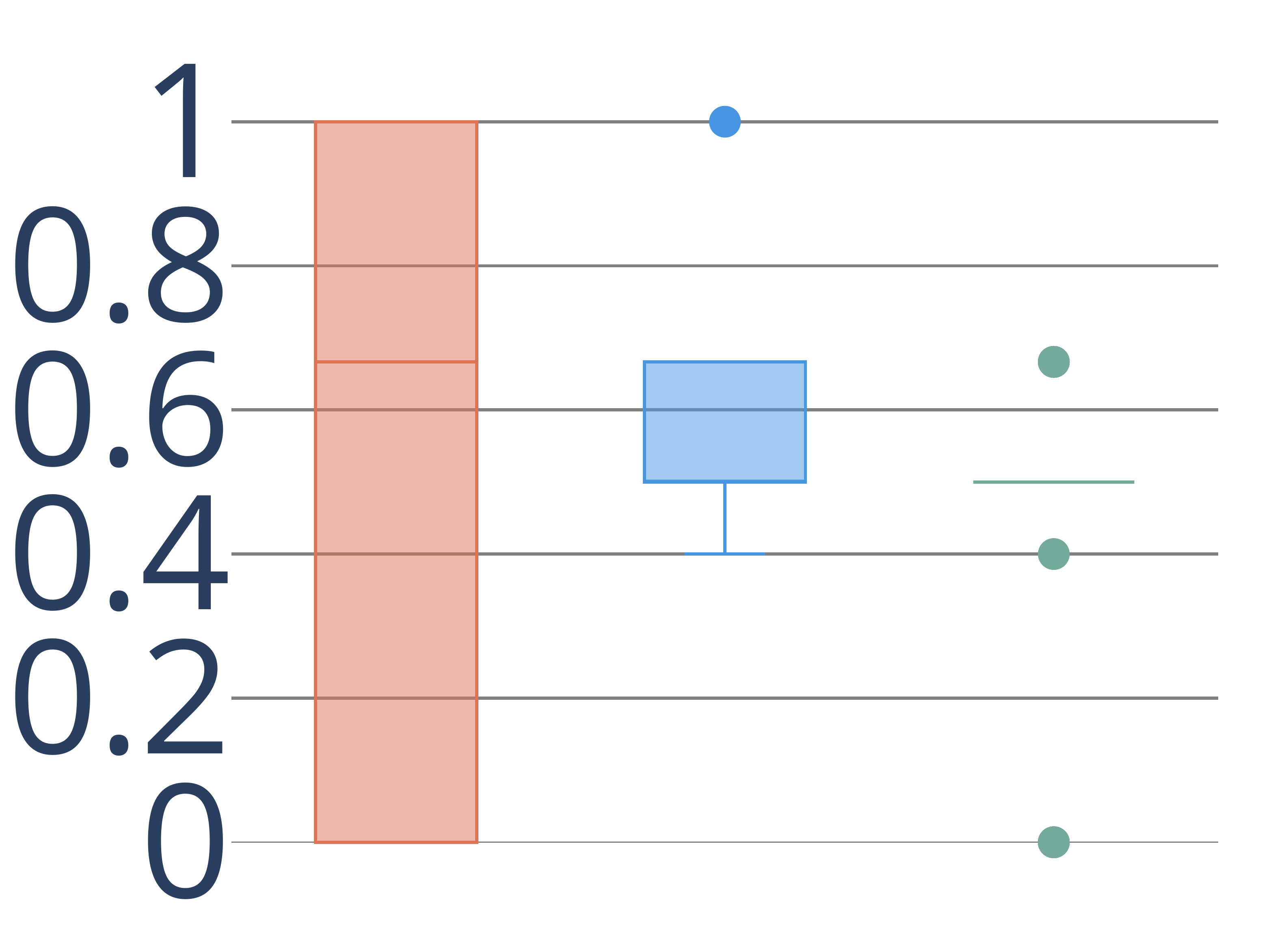} 
                                        & \includegraphics[scale=0.025]{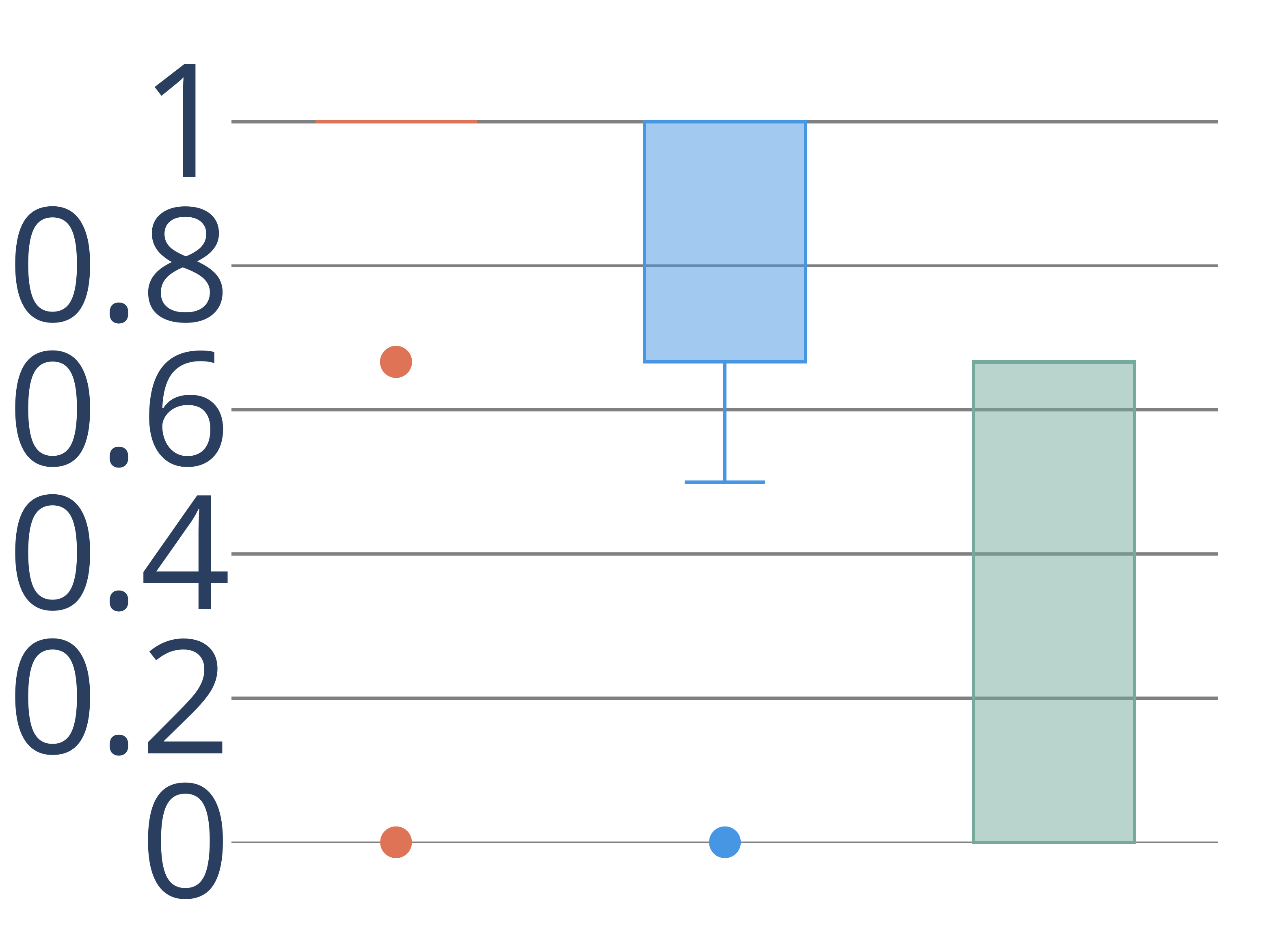} 
                                        & 
                                        & 
                                        & \includegraphics[scale=0.025]{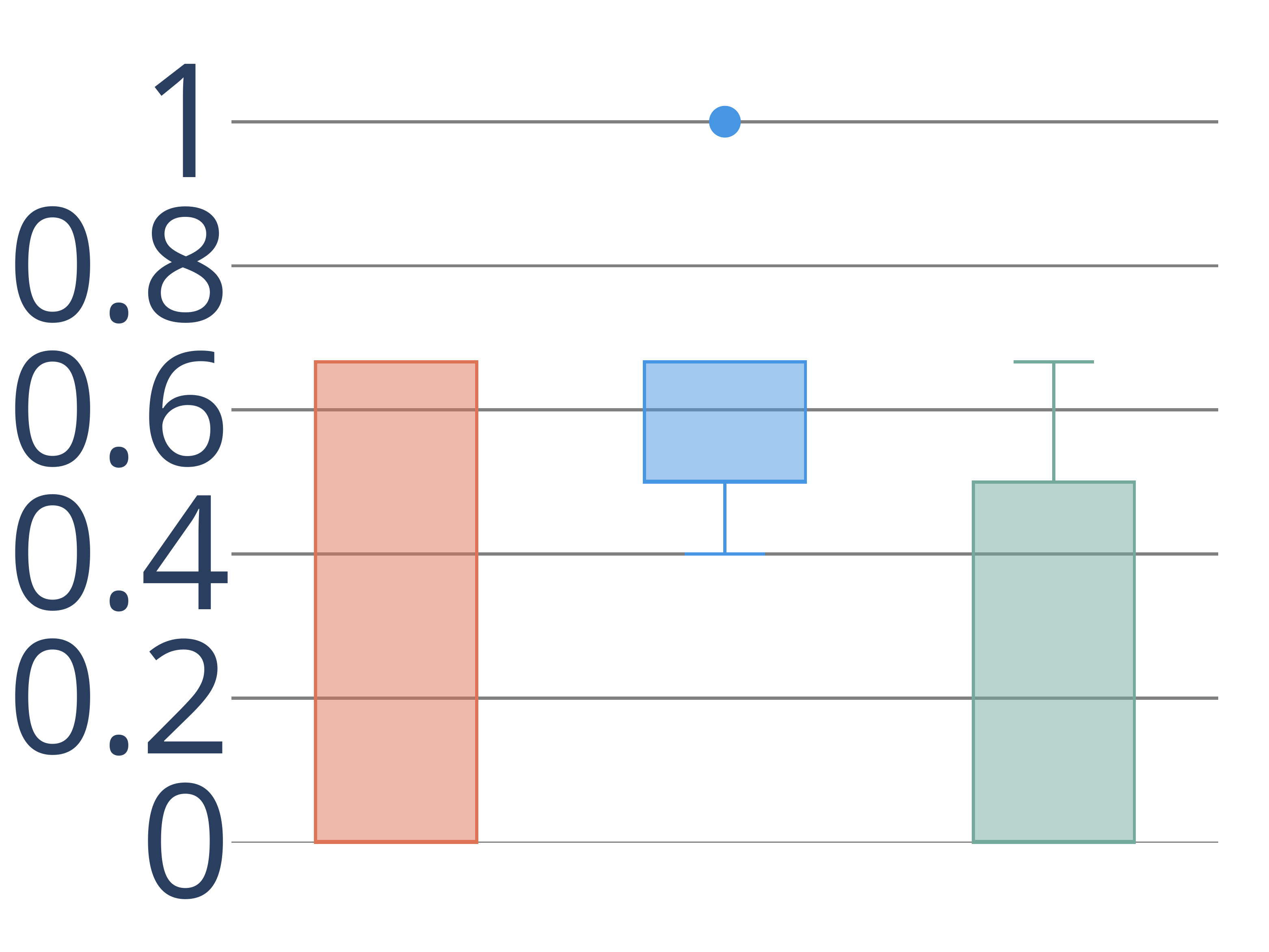}\\
                                    & \textbf{Dual Encoding} 
                                        & 
                                        &
                                        & \includegraphics[scale=0.025]{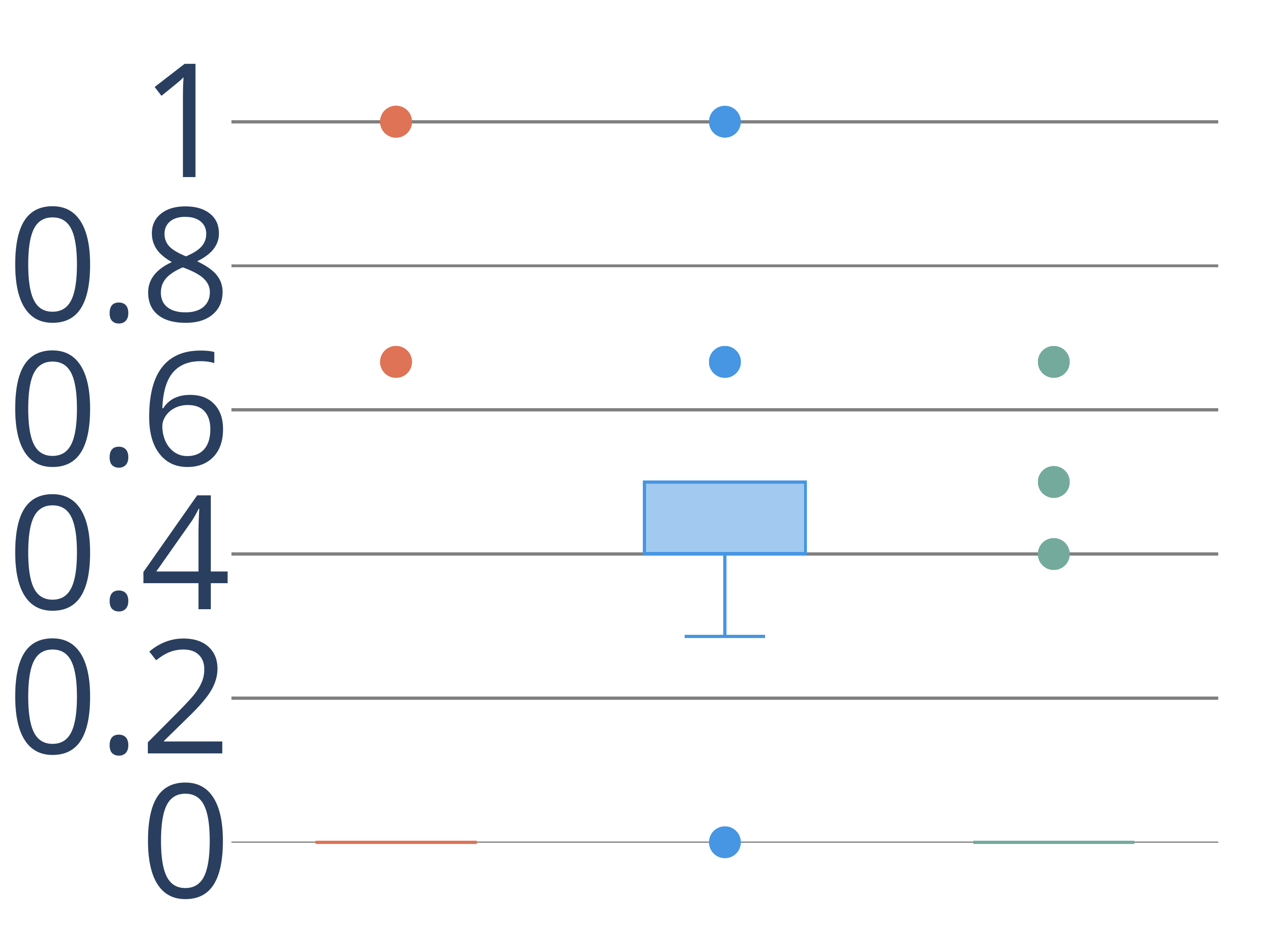}  
                                        & 
                                        & 
                                        & 
                                        & \includegraphics[scale=0.025]{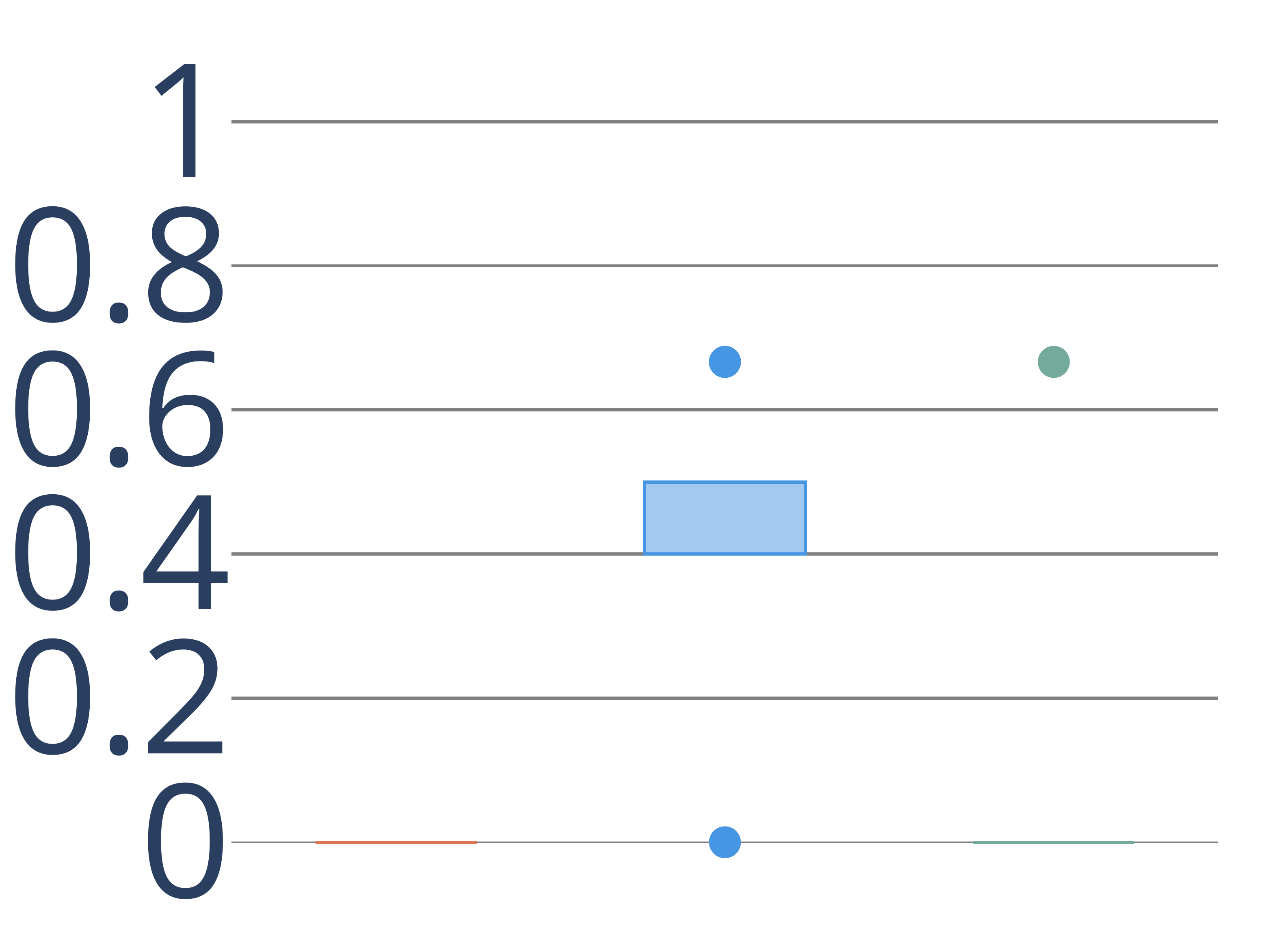} 
                                        & 
                                        & 
                                        & \includegraphics[scale=0.025]{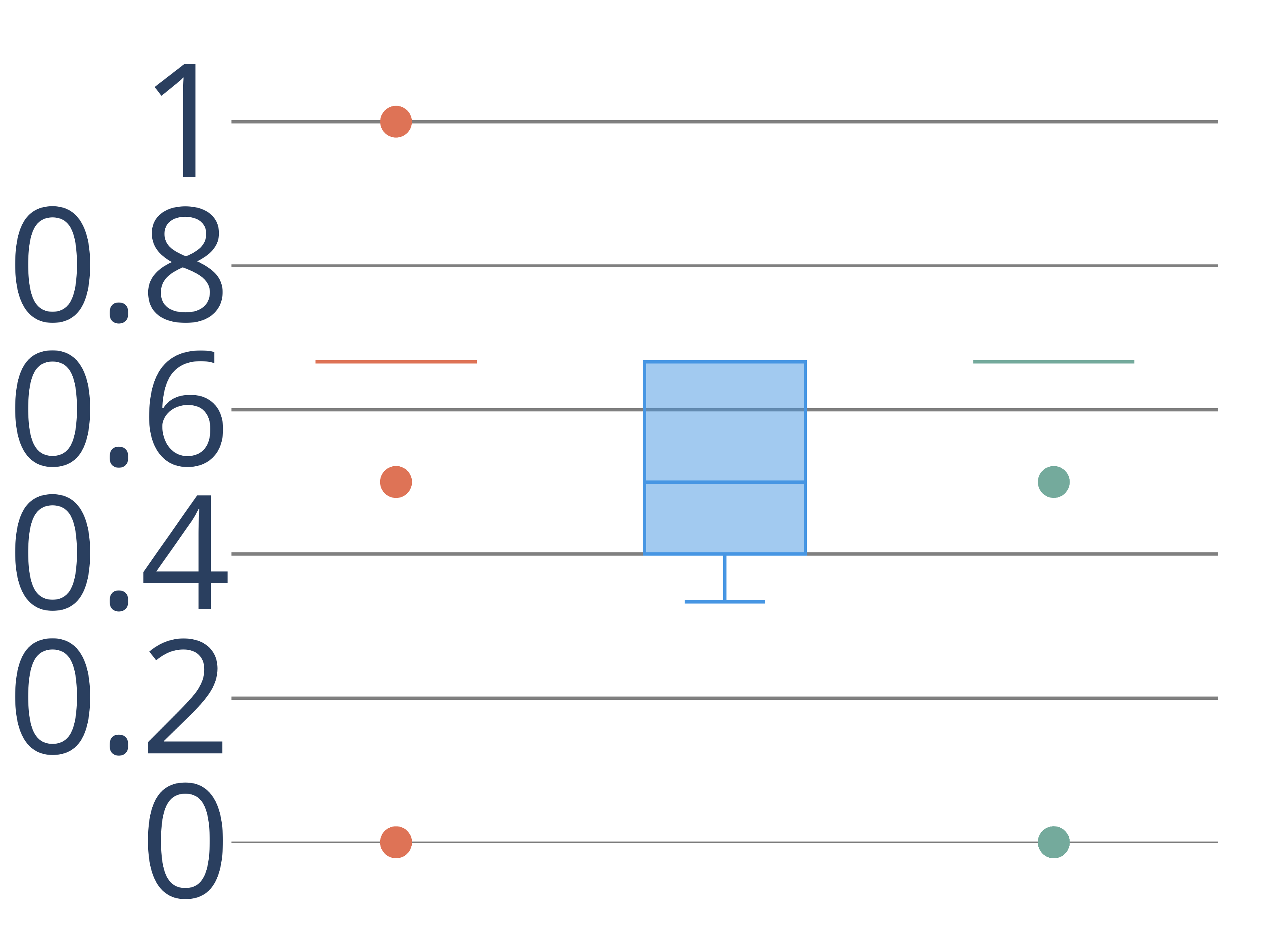}\\
                                    & \makecell[l]{\textbf{Mismatched Encoding:}\\ \textit{Continuous encoding of}\\ \textit{categorical data}} 
                                        & 
                                        & \includegraphics[scale=0.025]{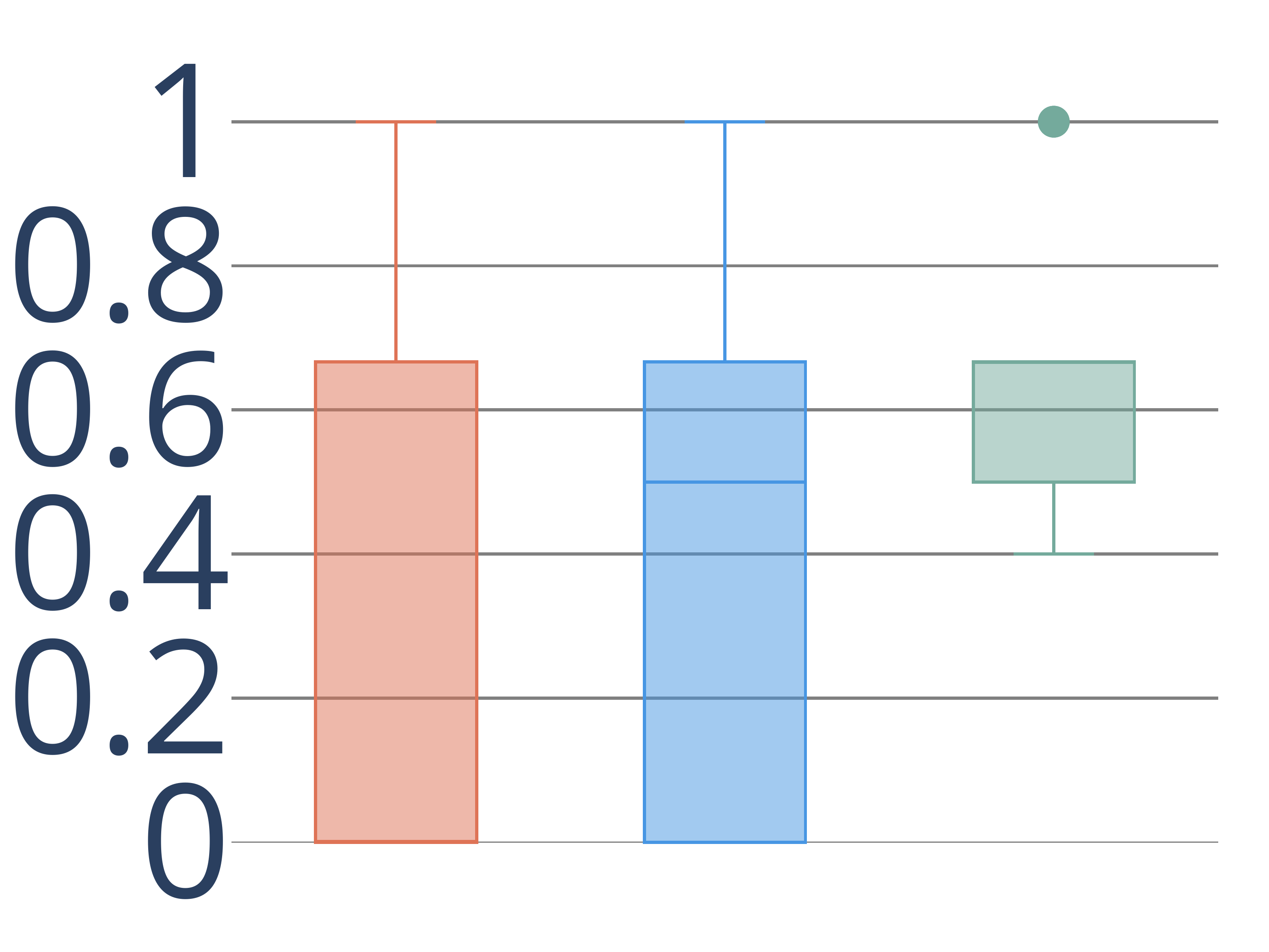} 
                                        & 
                                        & 
                                        & 
                                        & \includegraphics[scale=0.025]{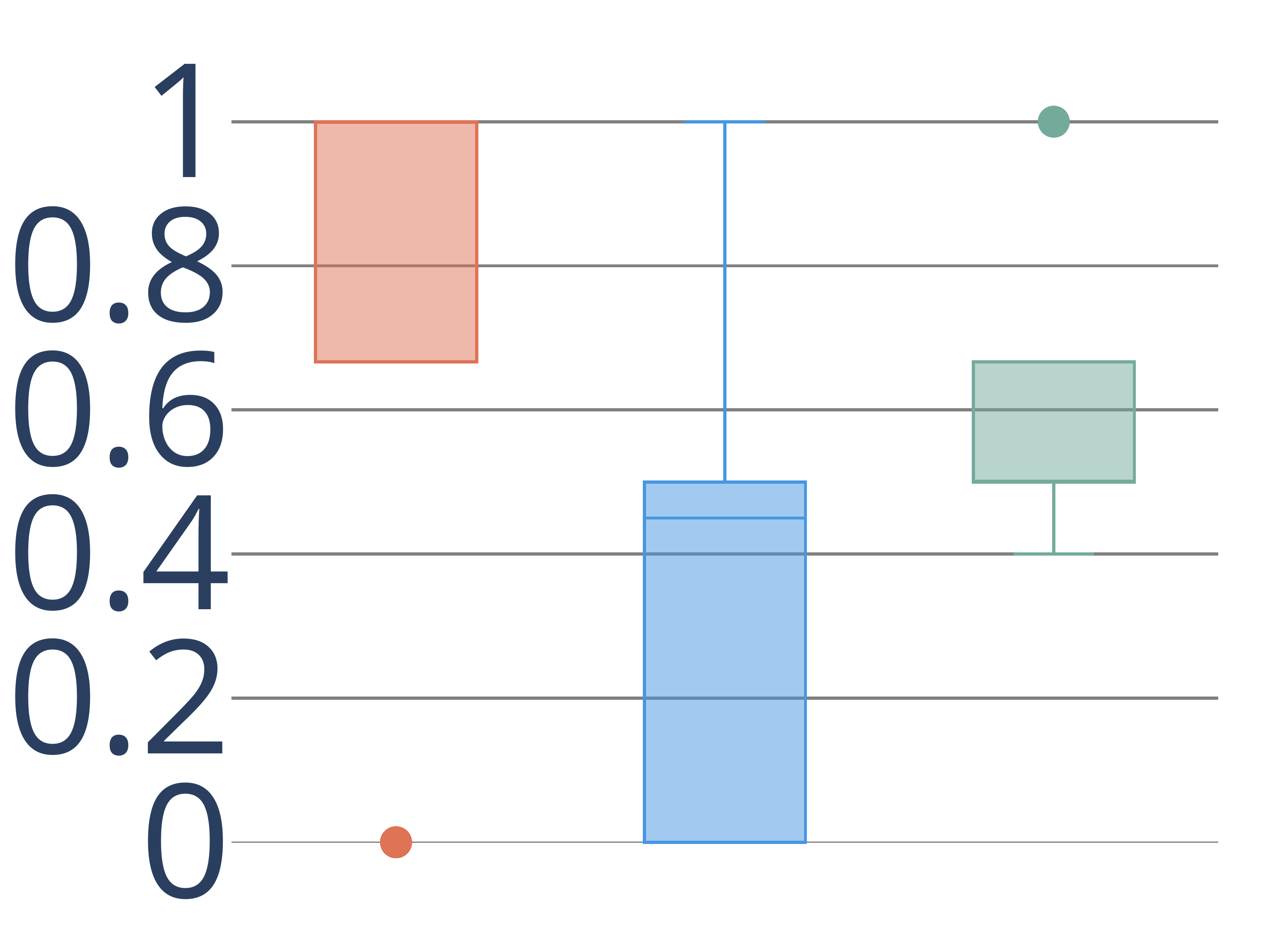} 
                                        & \includegraphics[scale=0.025]{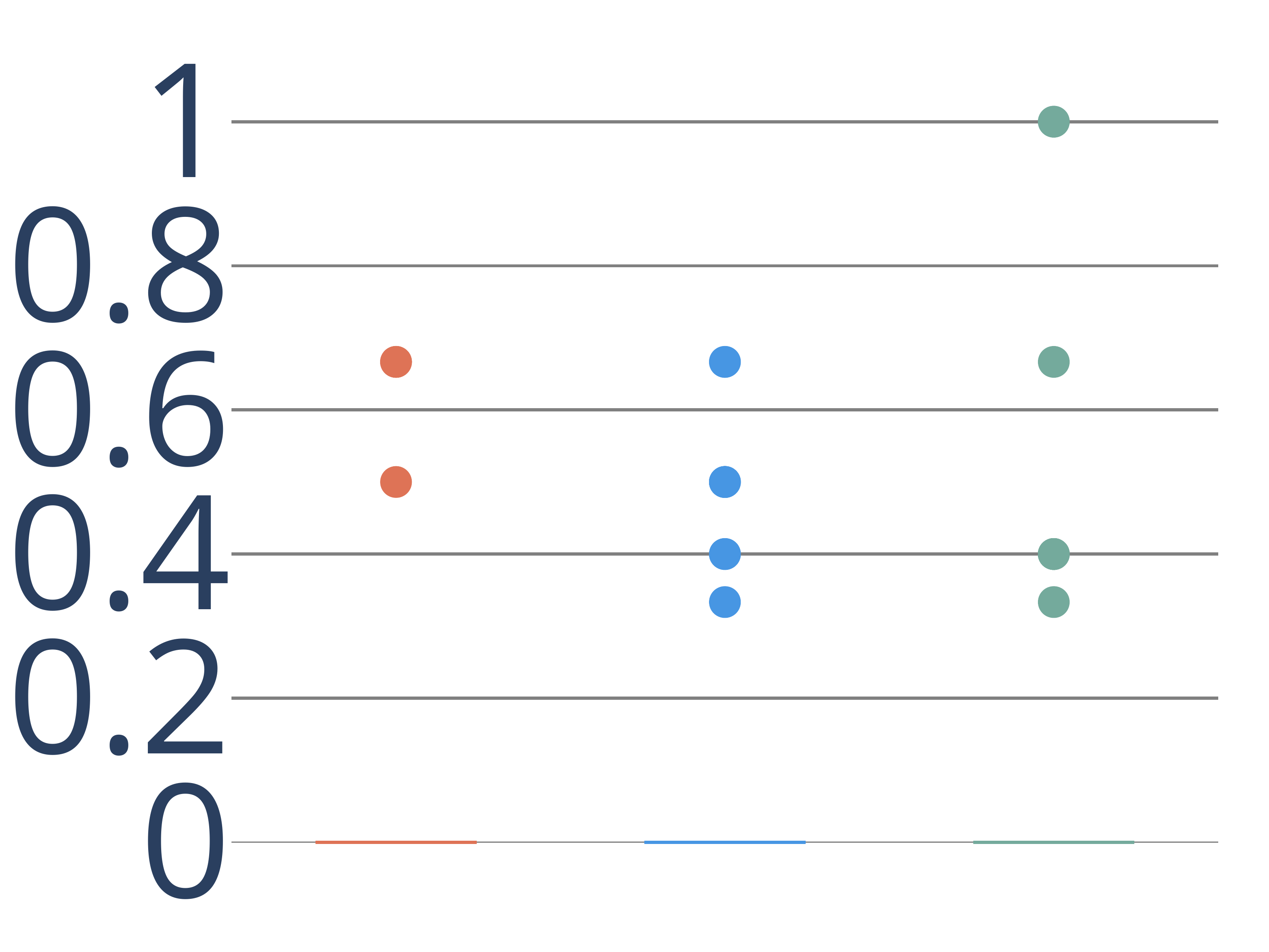} 
                                        & 
                                        & 
                                        &\\
                                    & \makecell[l]{\textbf{Mismatched Encoding:}\\ \textit{Categorical encoding of}\\ \textit{continuous data}} 
                                        & 
                                        & 
                                        & 
                                        & \includegraphics[scale=0.025]{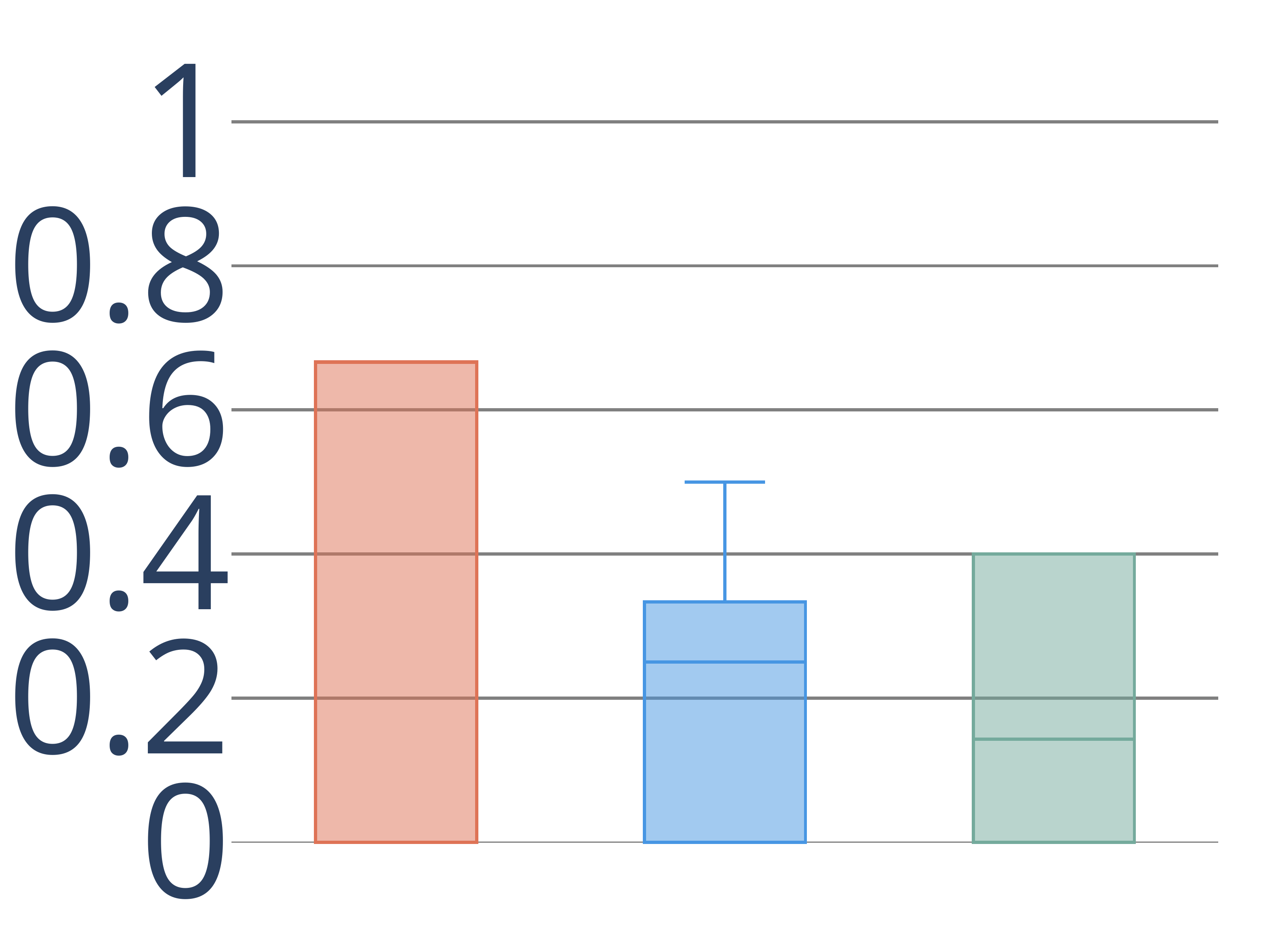} 
                                        & \includegraphics[scale=0.025]{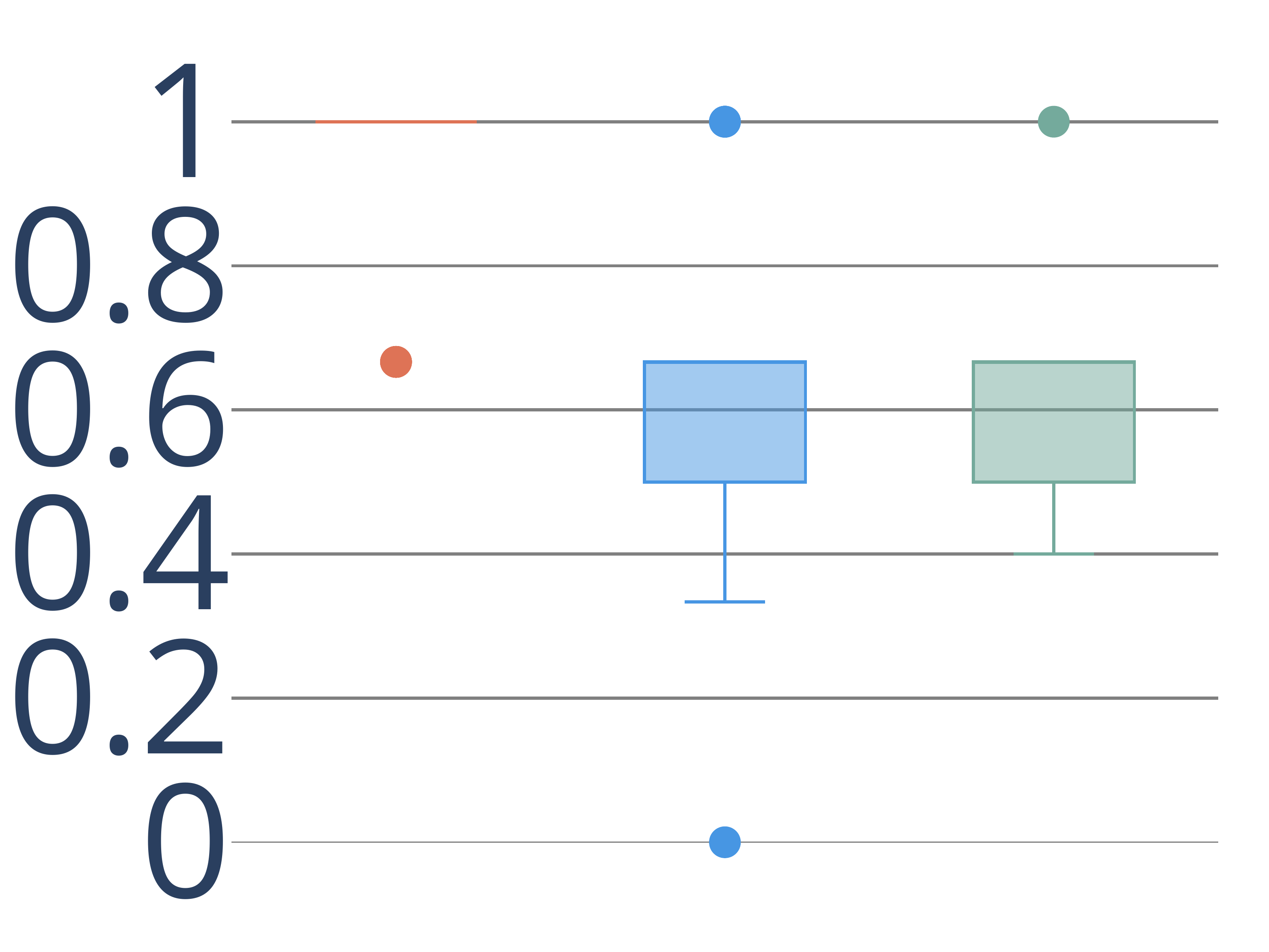} 
                                        & 
                                        & 
                                        & 
                                        & 
                                        &\\
        \hline
        \SetCell[r=6]{c,manuScale} \rotatebox[origin=c]{90}{\textbf{Manipulated Scale}}
                                    & \makecell[l]{\textbf{Inappropriate}\\ \textbf{Scale Range}} 
                                        & 
                                        & 
                                        & \includegraphics[scale=0.025]{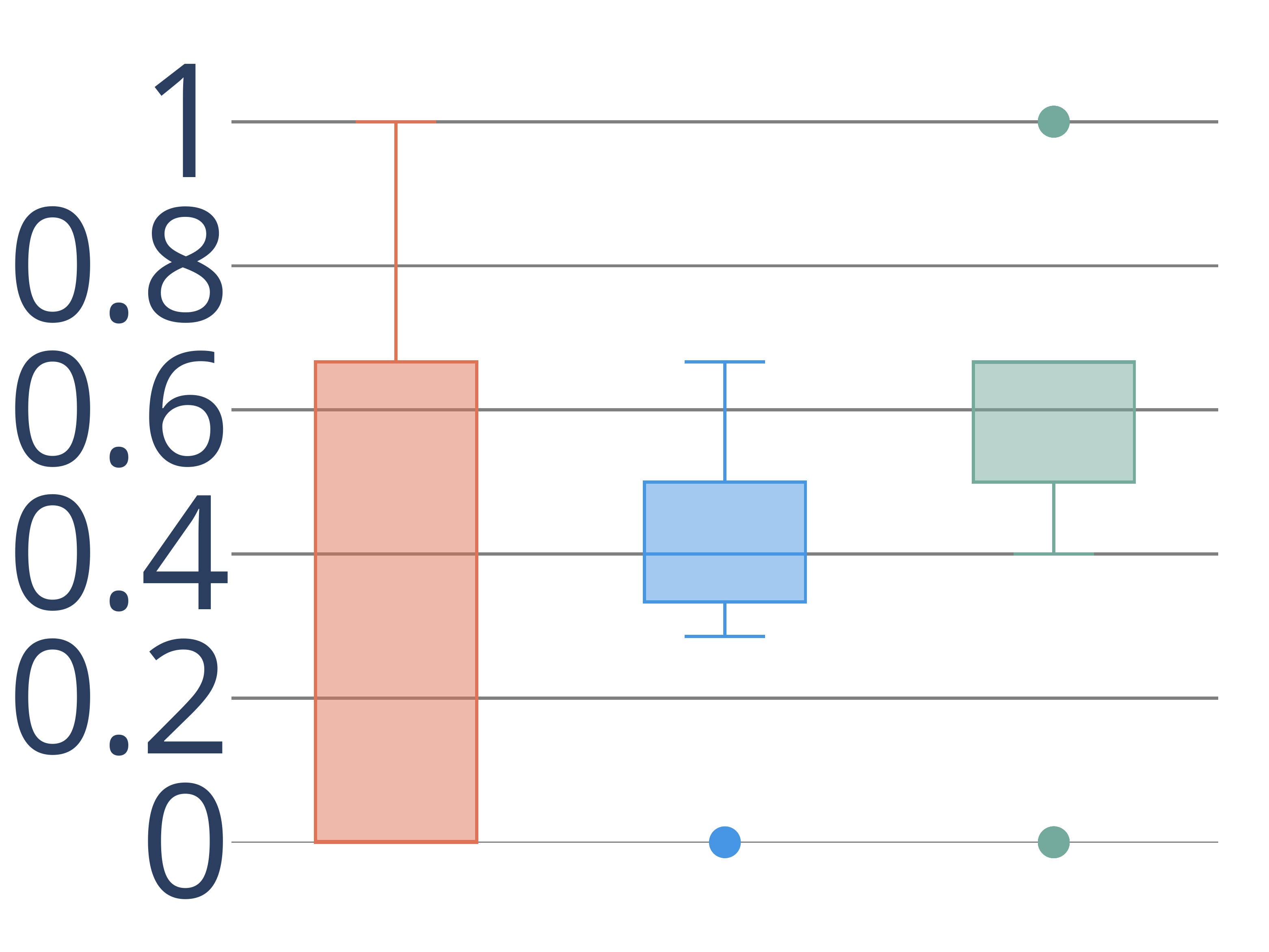} 
                                        & \includegraphics[scale=0.025]{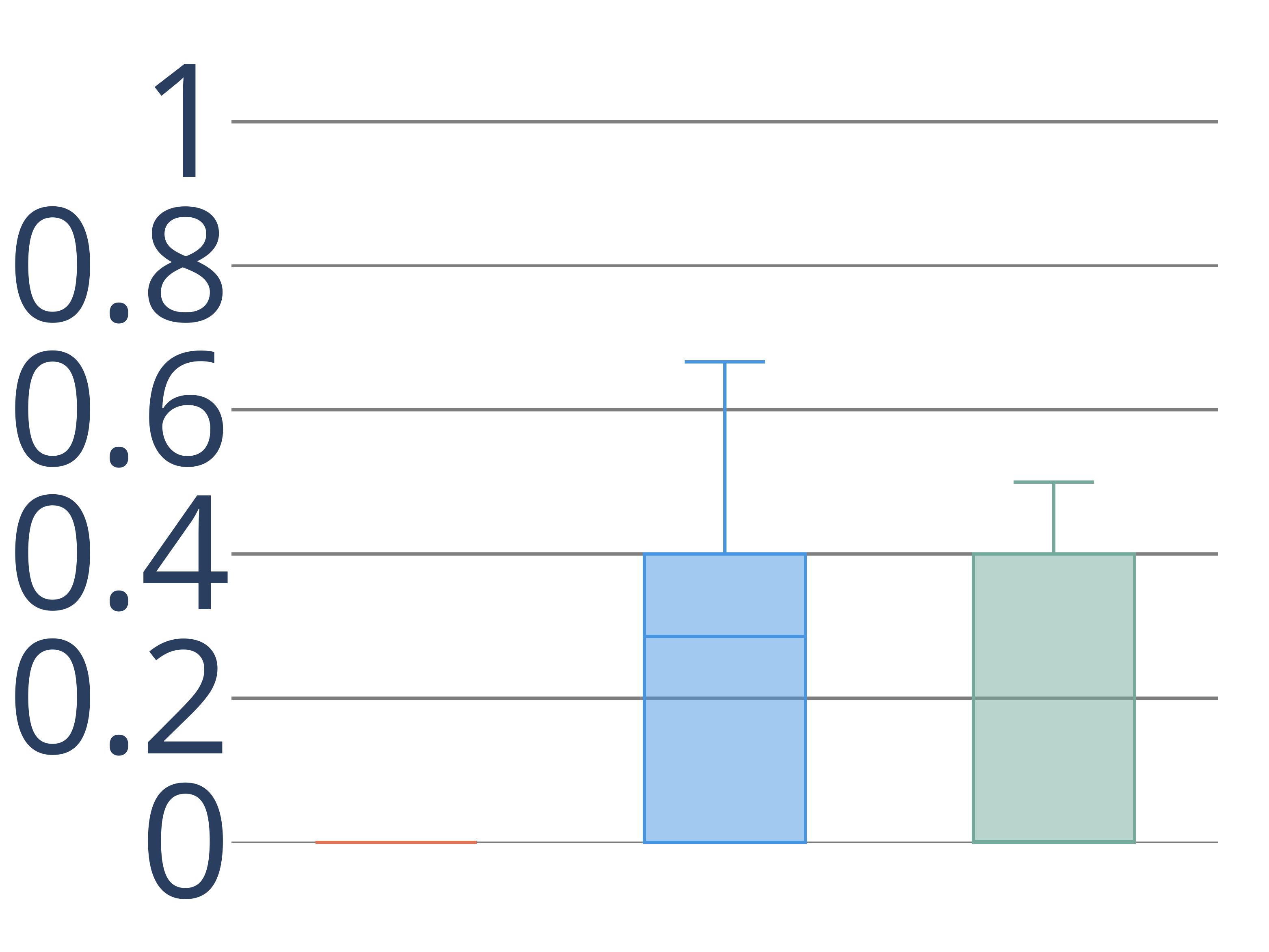} 
                                        & 
                                        & \includegraphics[scale=0.025]{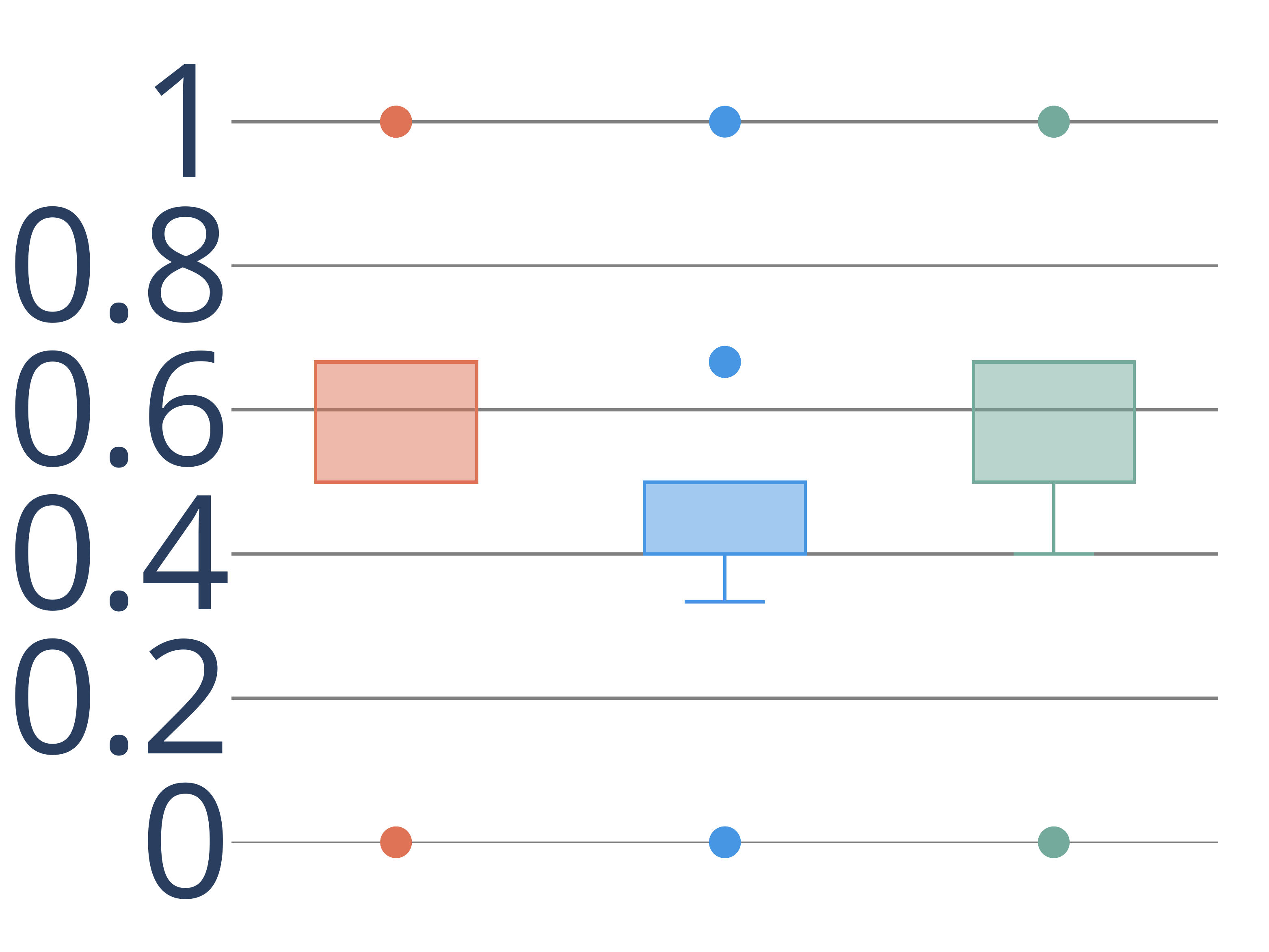} 
                                        & 
                                        & \includegraphics[scale=0.025]{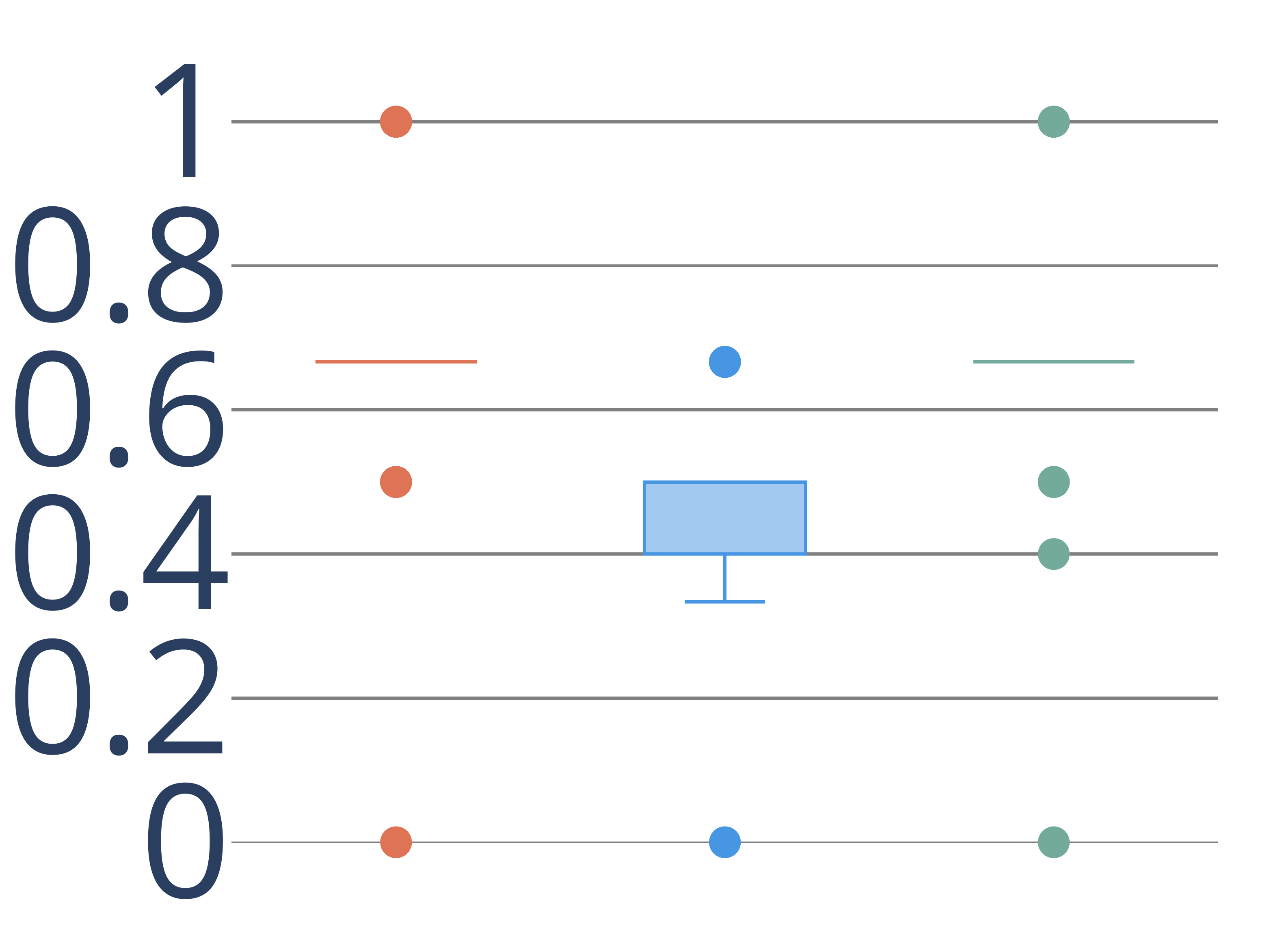} 
                                        & \includegraphics[scale=0.025]{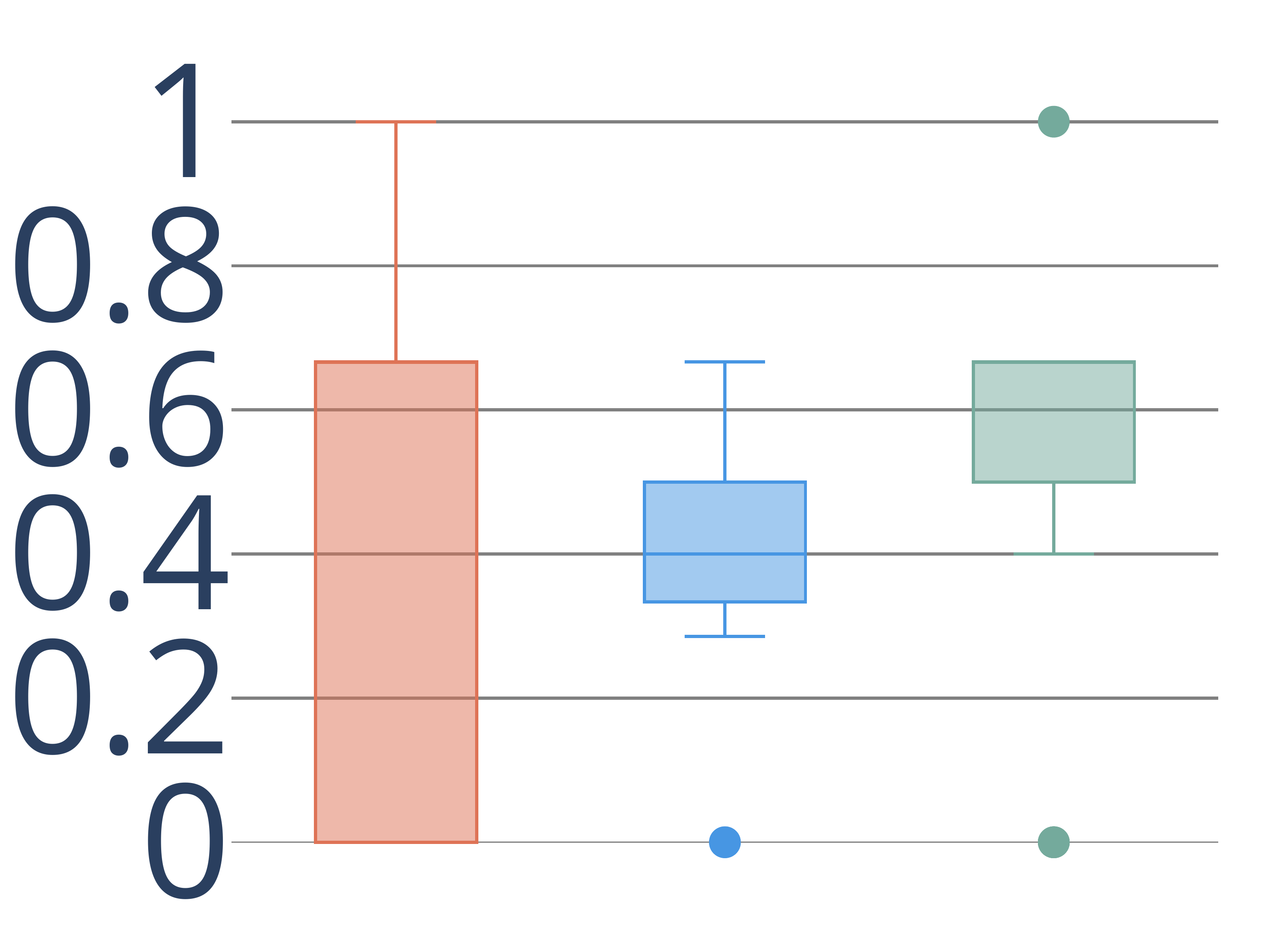} 
                                        &\\
                                    & \makecell[l]{\textbf{Inappropriate Scale}\\ \textbf{Functions}} 
                                        & 
                                        & 
                                        & \includegraphics[scale=0.025]{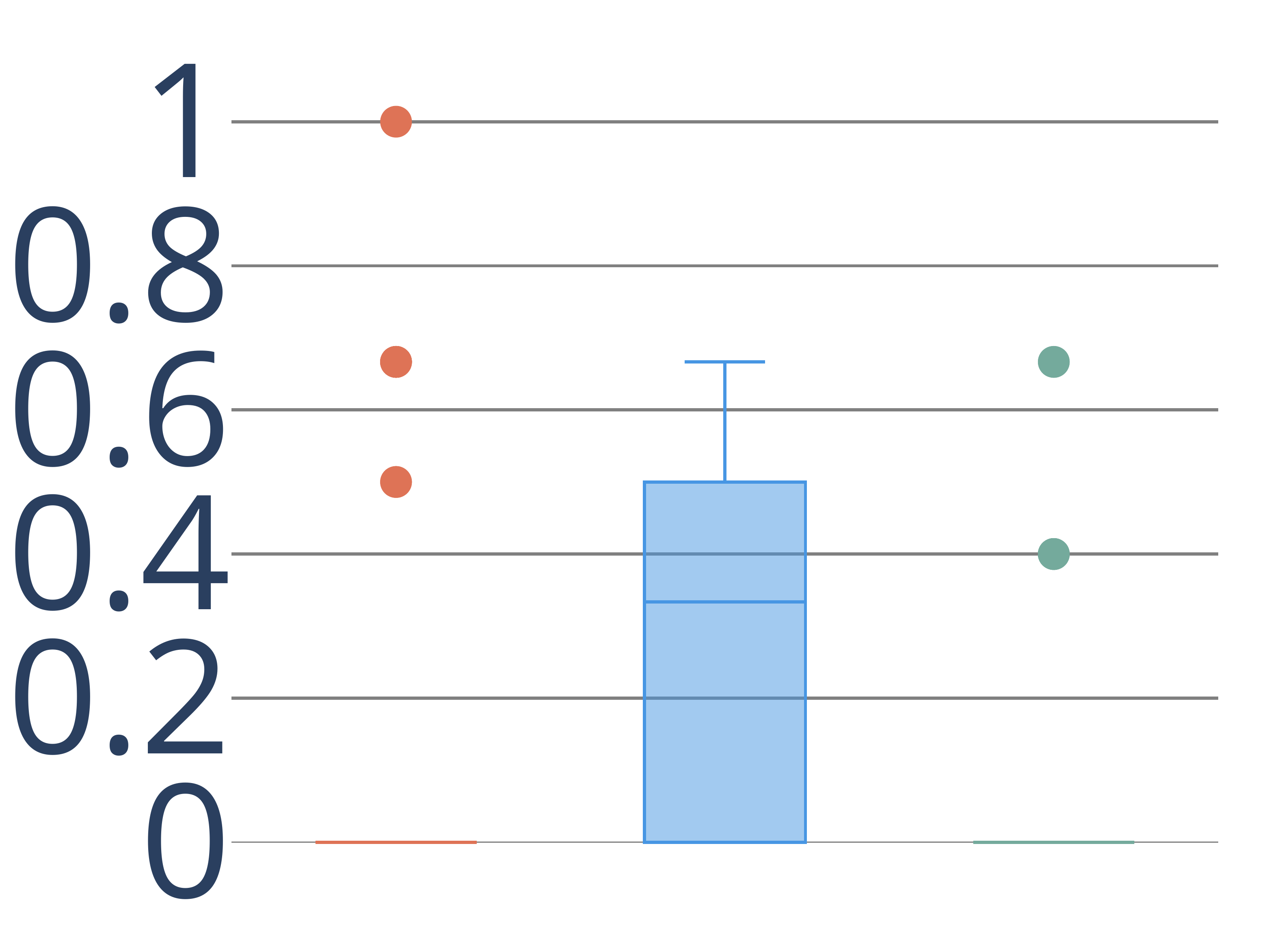} 
                                        & 
                                        & 
                                        & \includegraphics[scale=0.025]{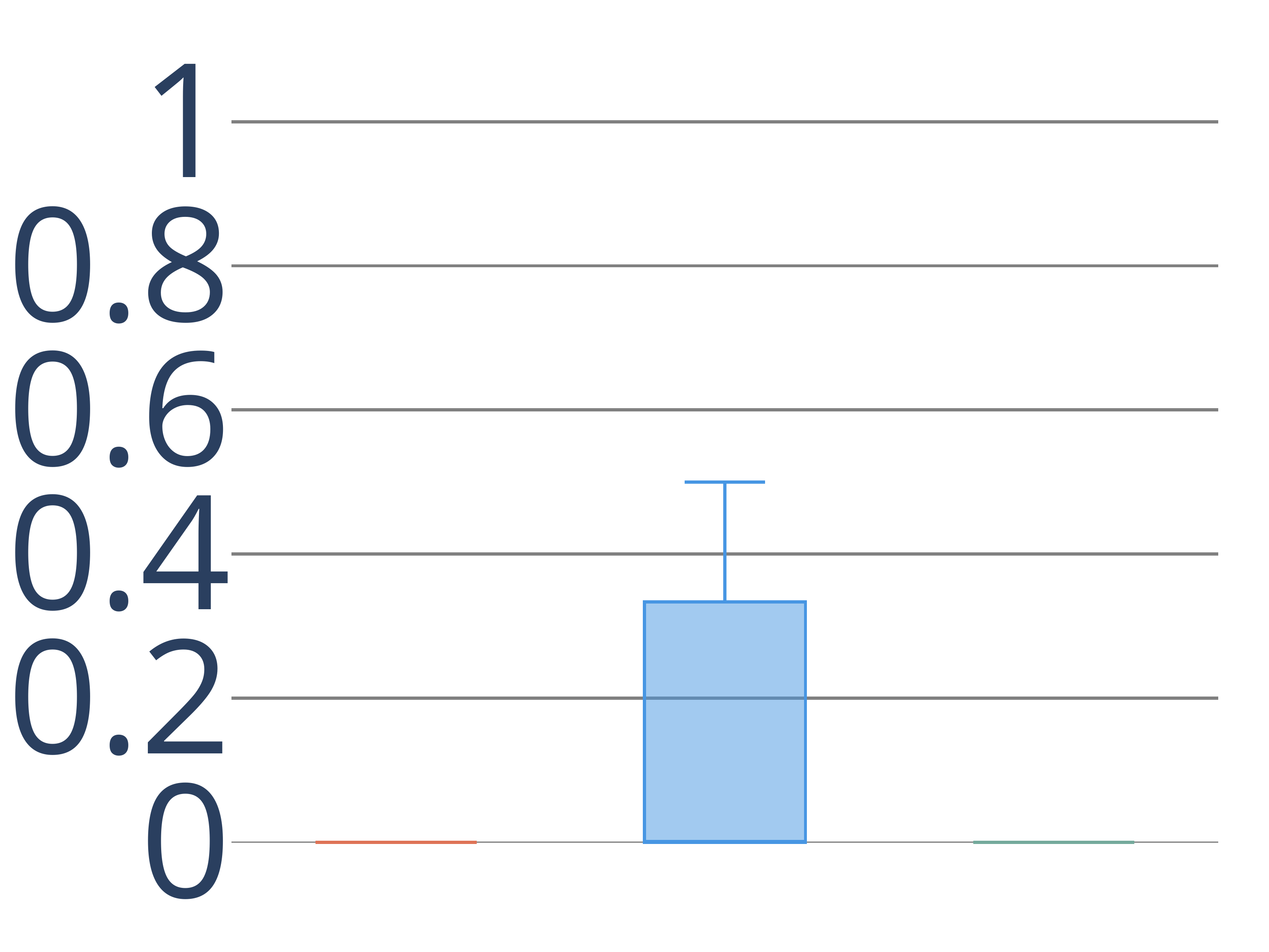} 
                                        & \includegraphics[scale=0.025]{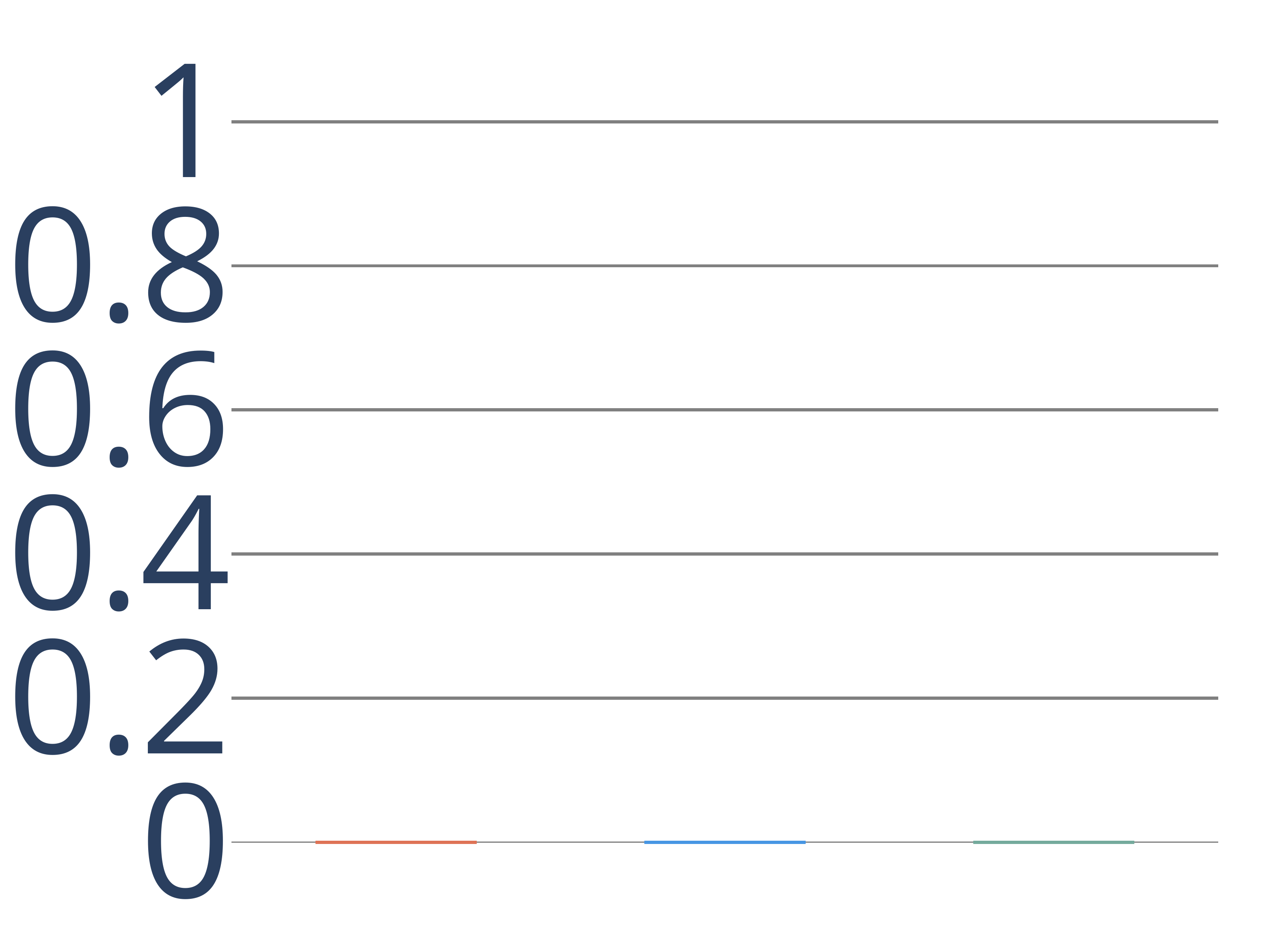} 
                                        & 
                                        & 
                                        &\\
                                    & \makecell[l]{\textbf{Unconventional Scale}\\ \textbf{Directions}} 
                                        & 
                                        & \includegraphics[scale=0.025]{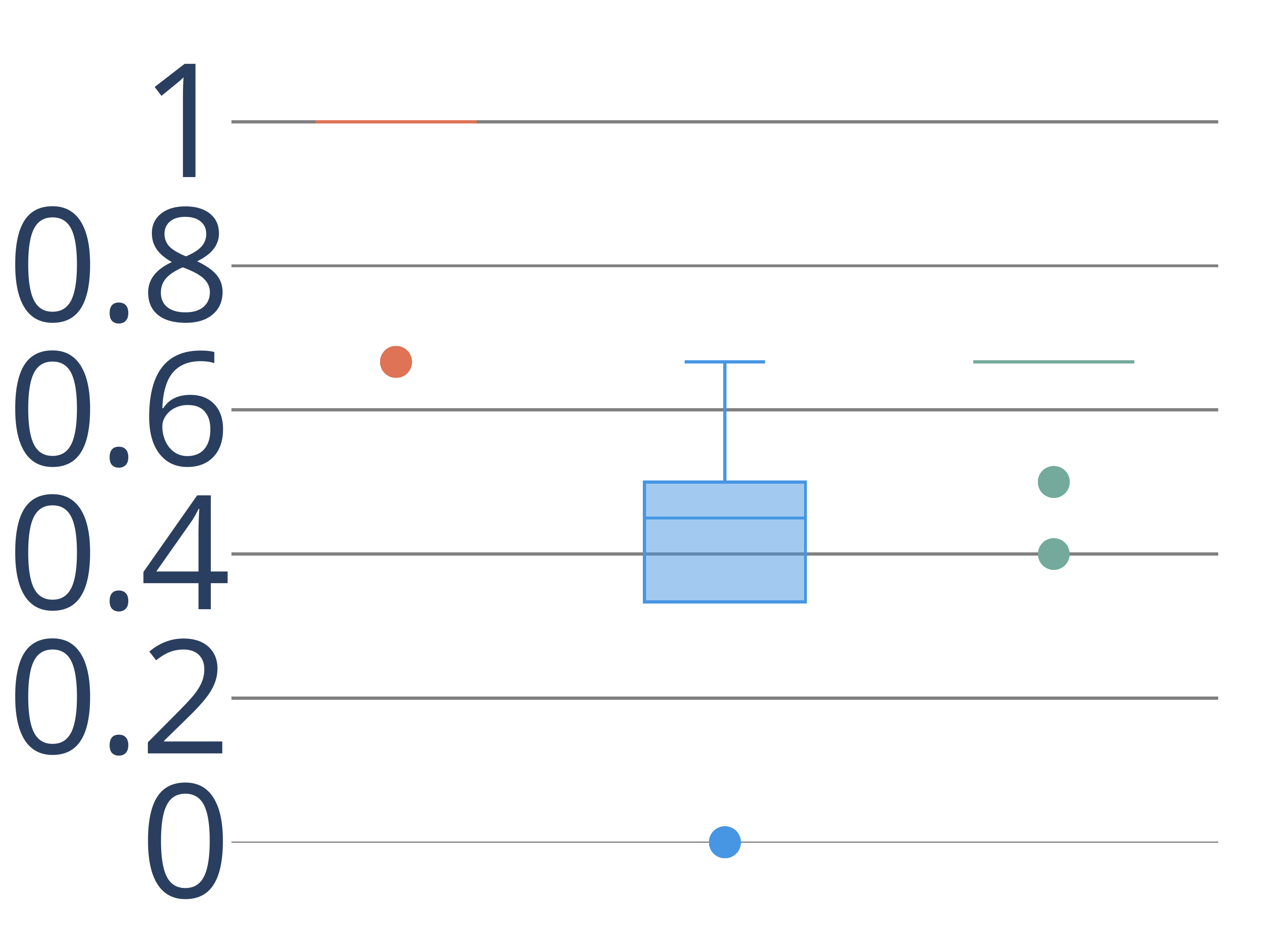} 
                                        & \includegraphics[scale=0.025]{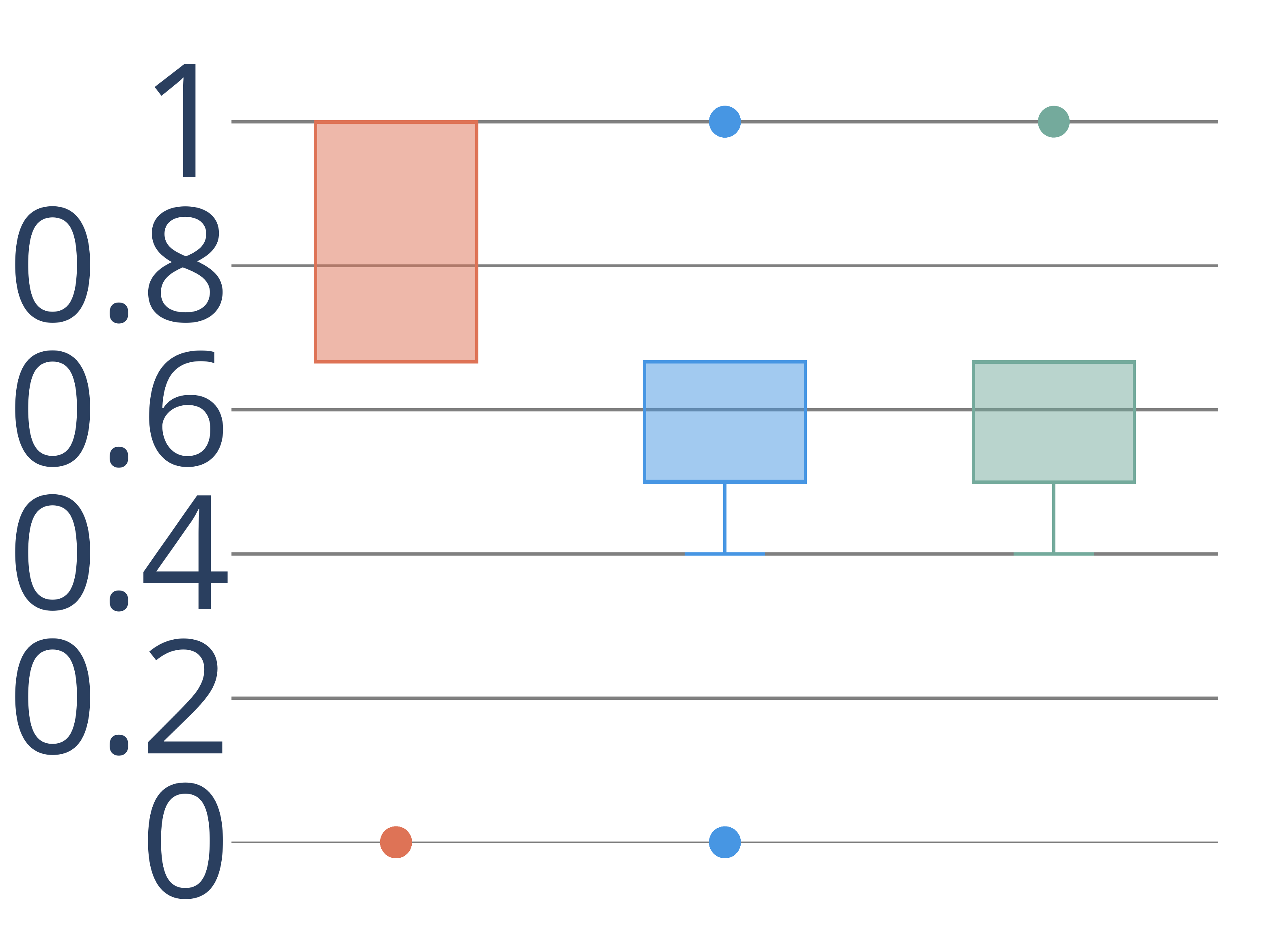} 
                                        & \includegraphics[scale=0.025]{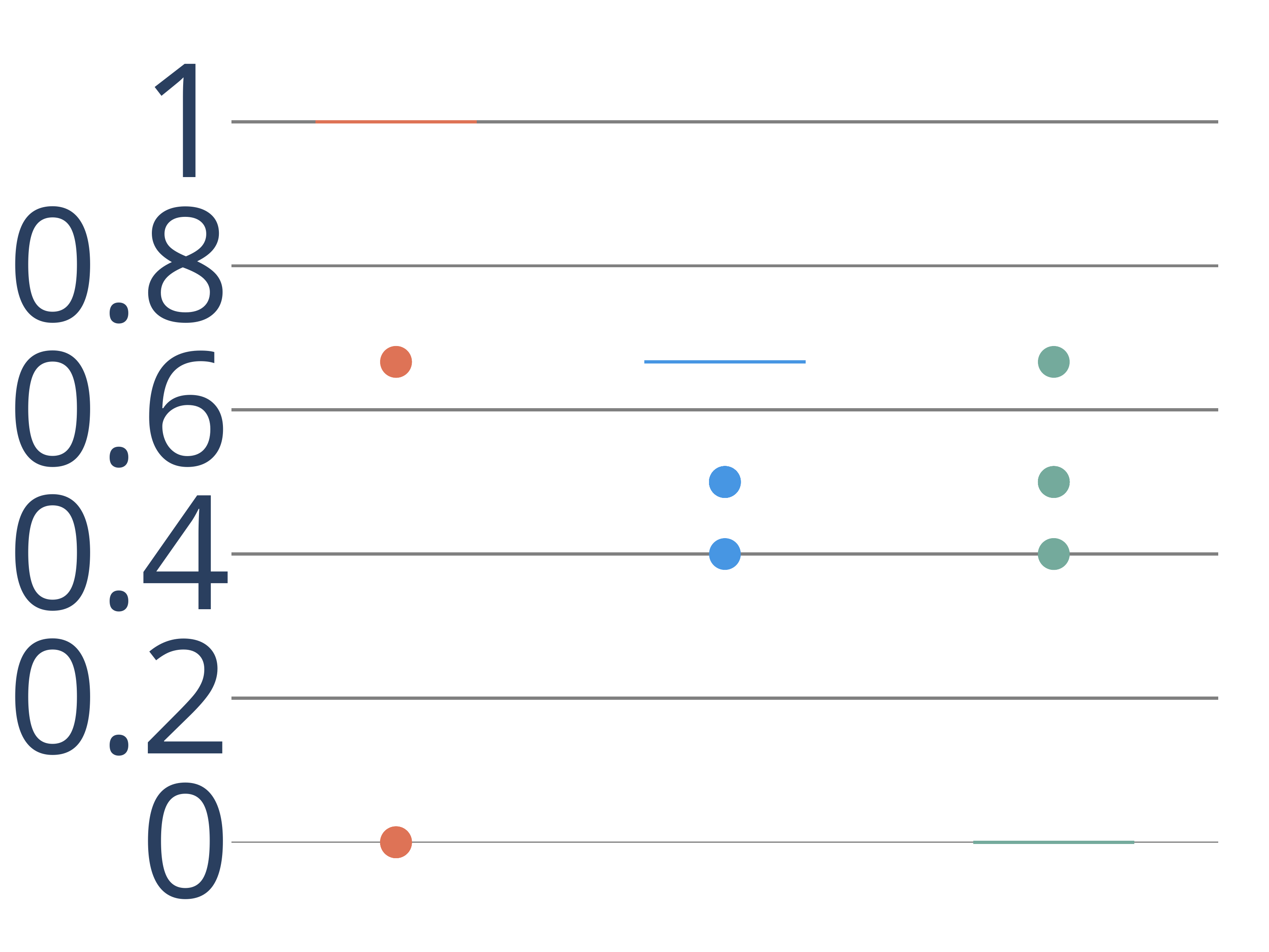} 
                                        & 
                                        & \includegraphics[scale=0.025]{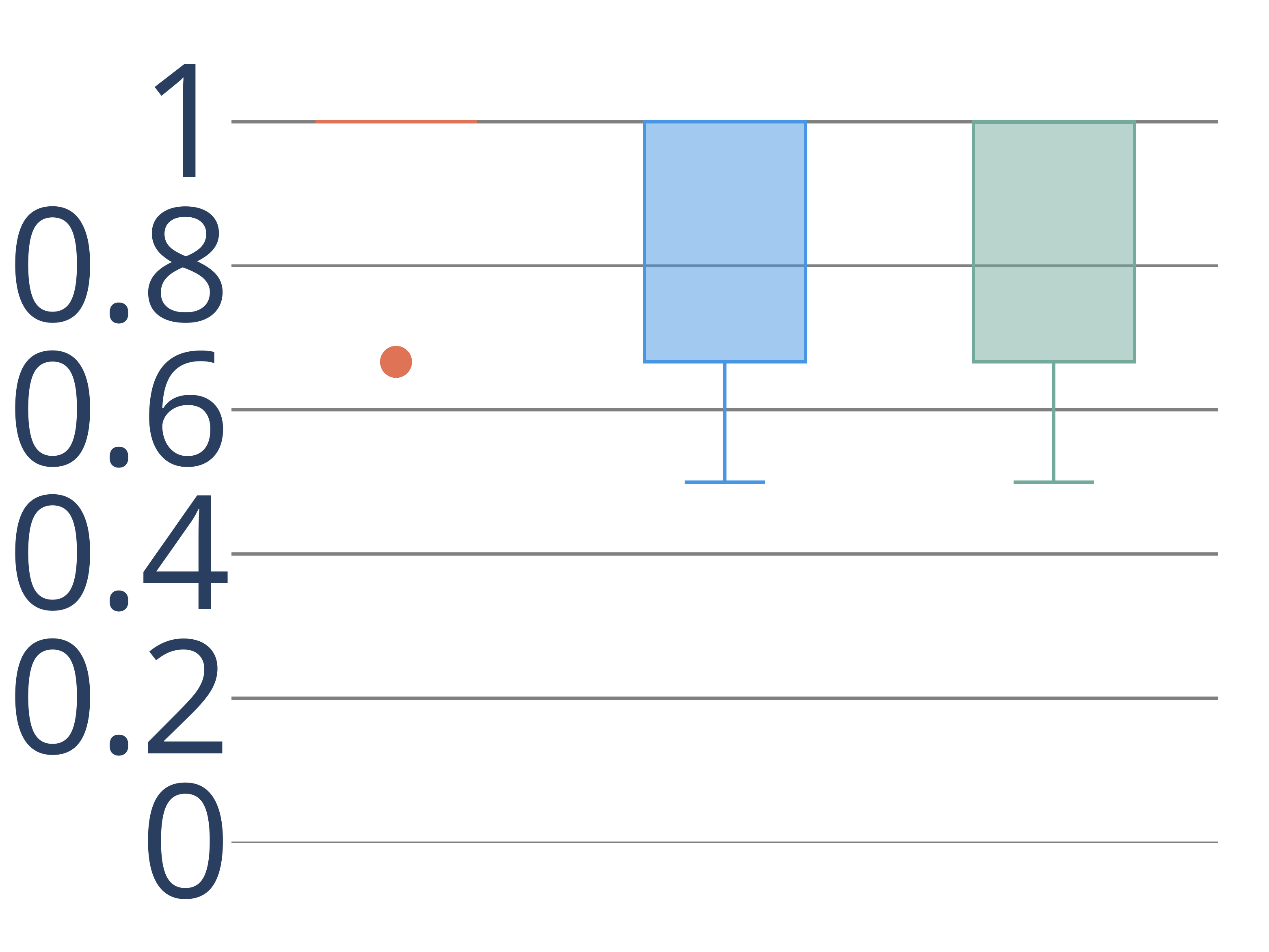} 
                                        & 
                                        & 
                                        & 
                                        & \includegraphics[scale=0.025]{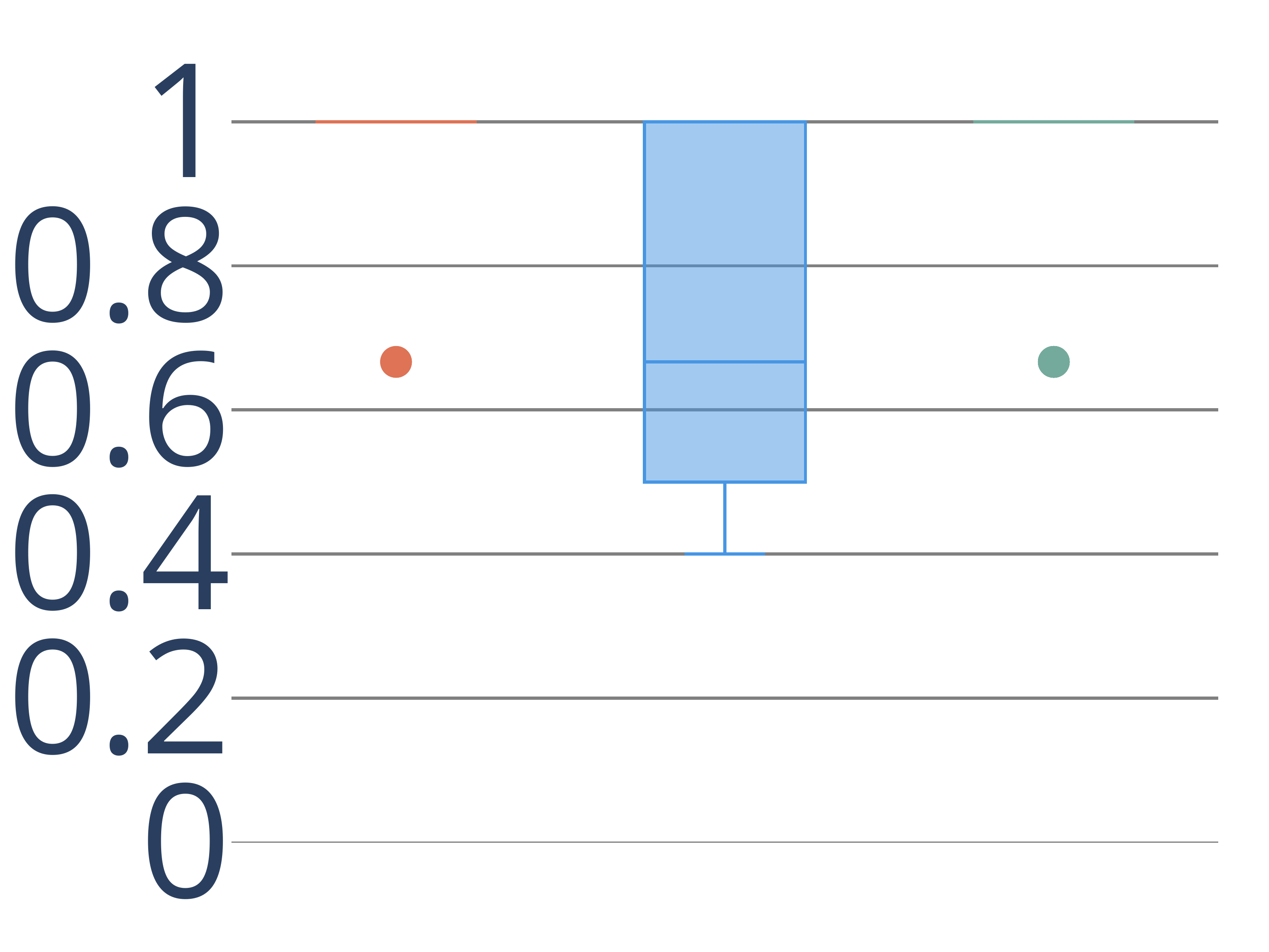}\\
                                    & \makecell[l]{\textbf{Misuse of Cumulative}\\ \textbf{Relationship}} 
                                        & \includegraphics[scale=0.025]{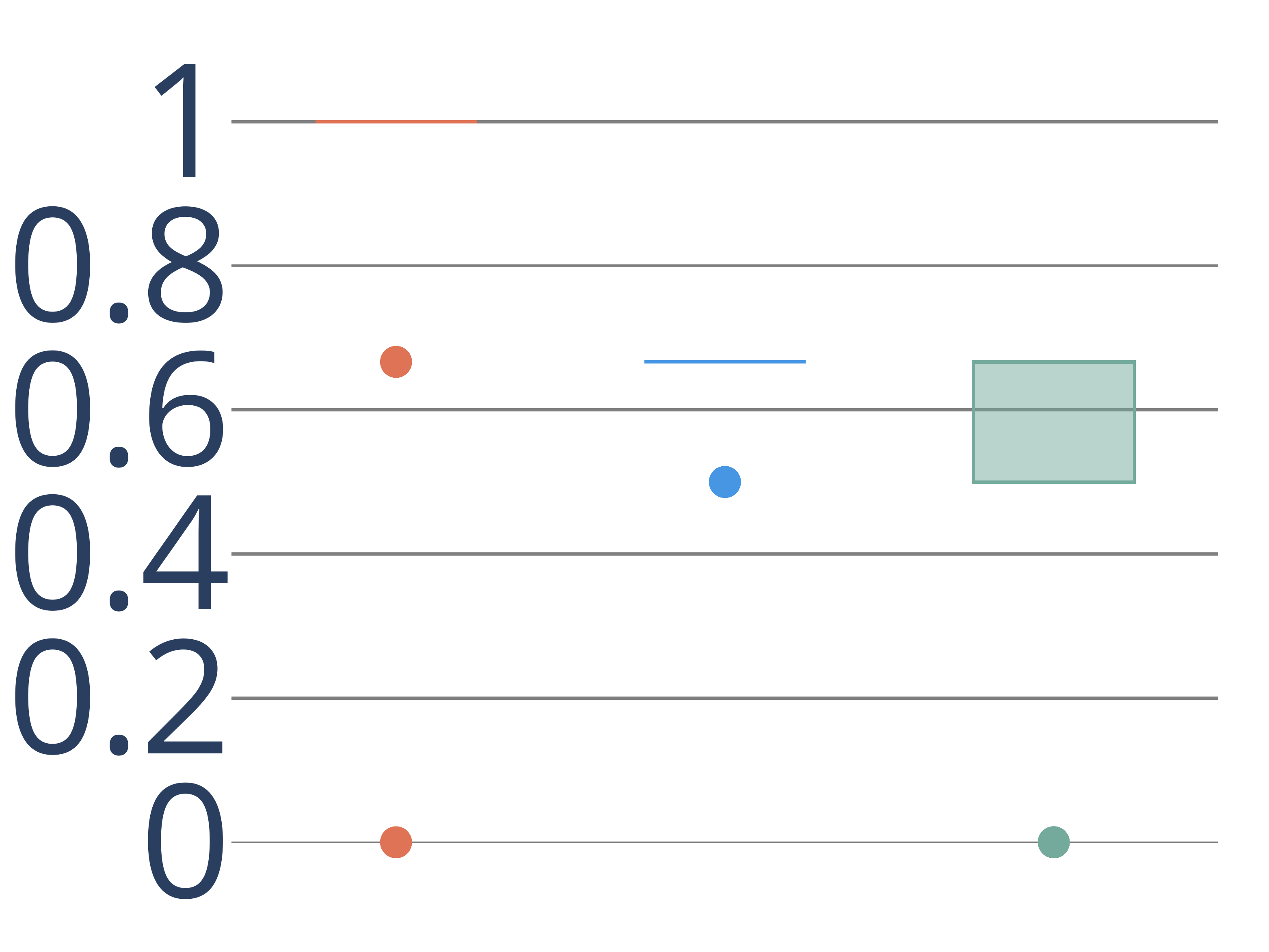} 
                                        & 
                                        & 
                                        & 
                                        & 
                                        & 
                                        & 
                                        & \includegraphics[scale=0.025]{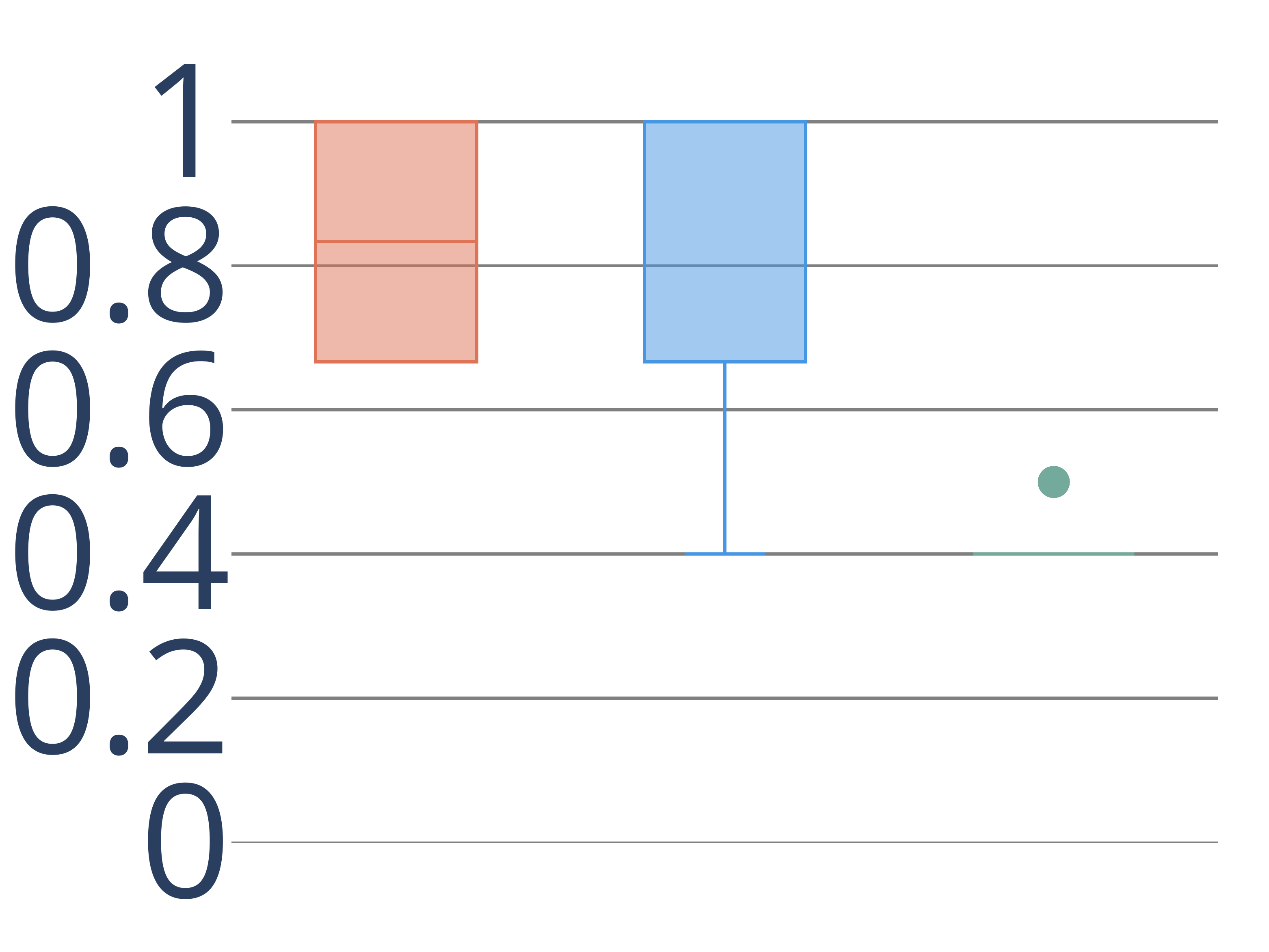} 
                                        & \includegraphics[scale=0.025]{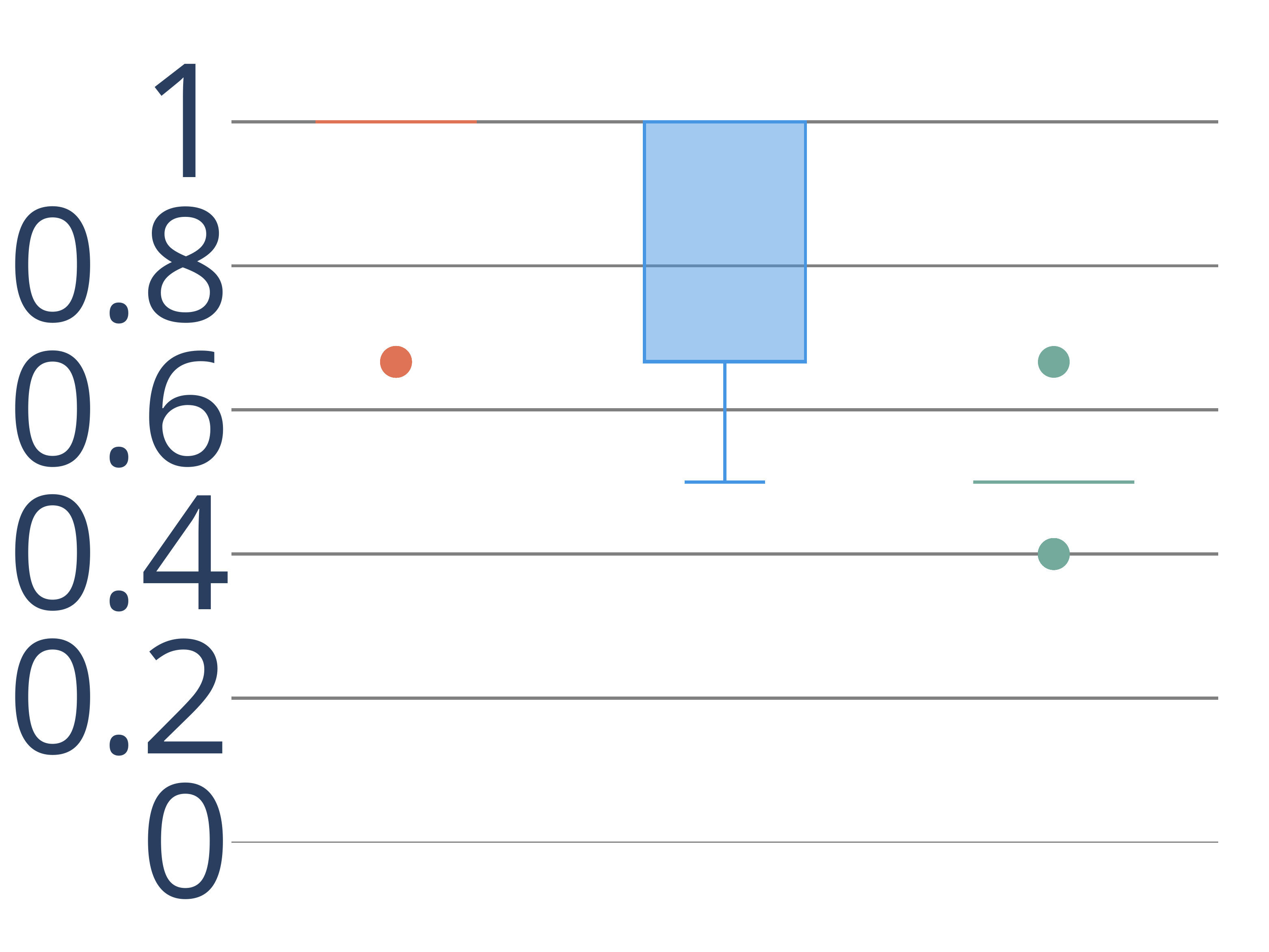} 
                                        &\\
                                    & \textbf{Exceeding the Canvas} 
                                        & 
                                        & \includegraphics[scale=0.025]{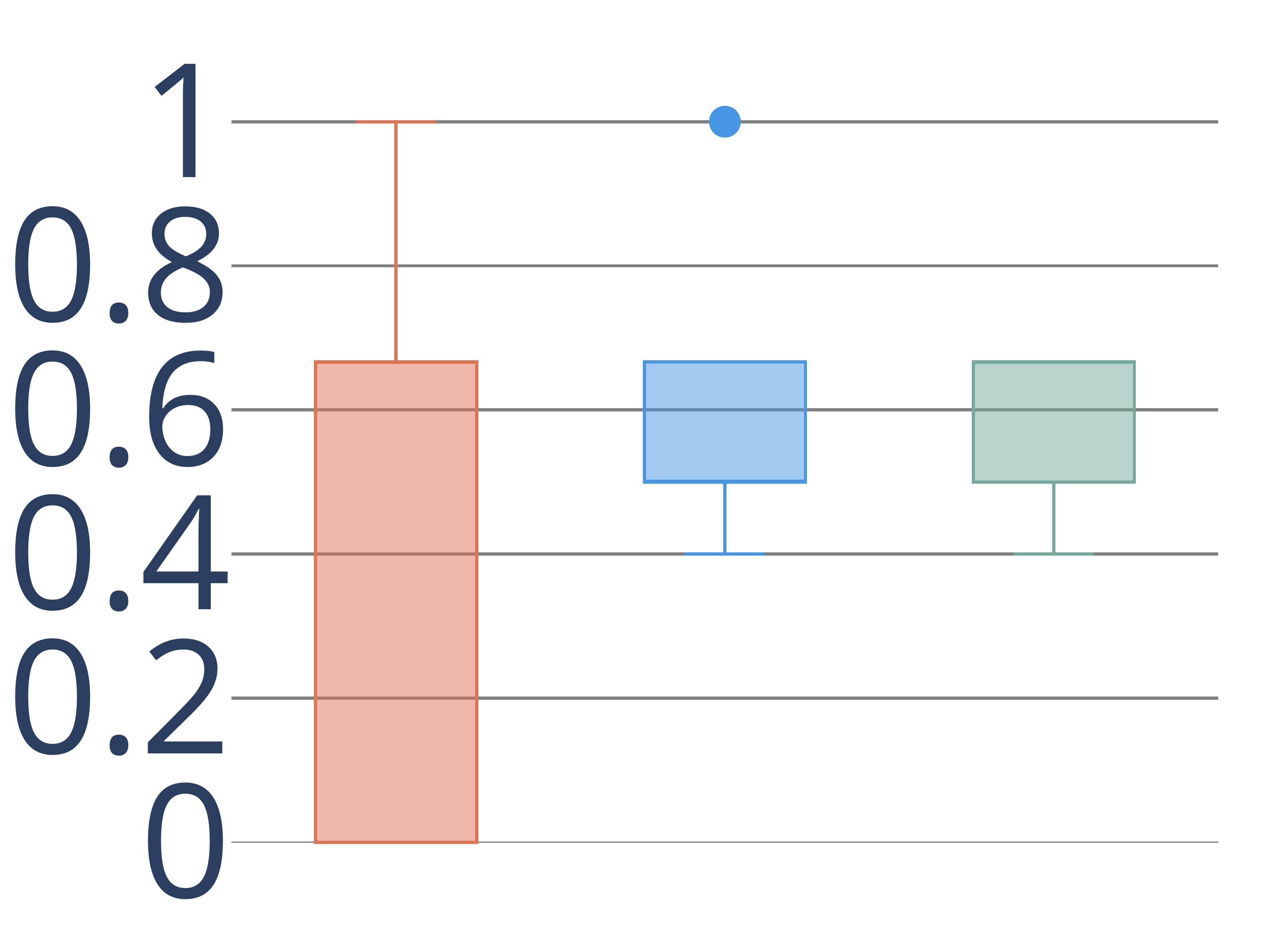} 
                                        & \includegraphics[scale=0.025]{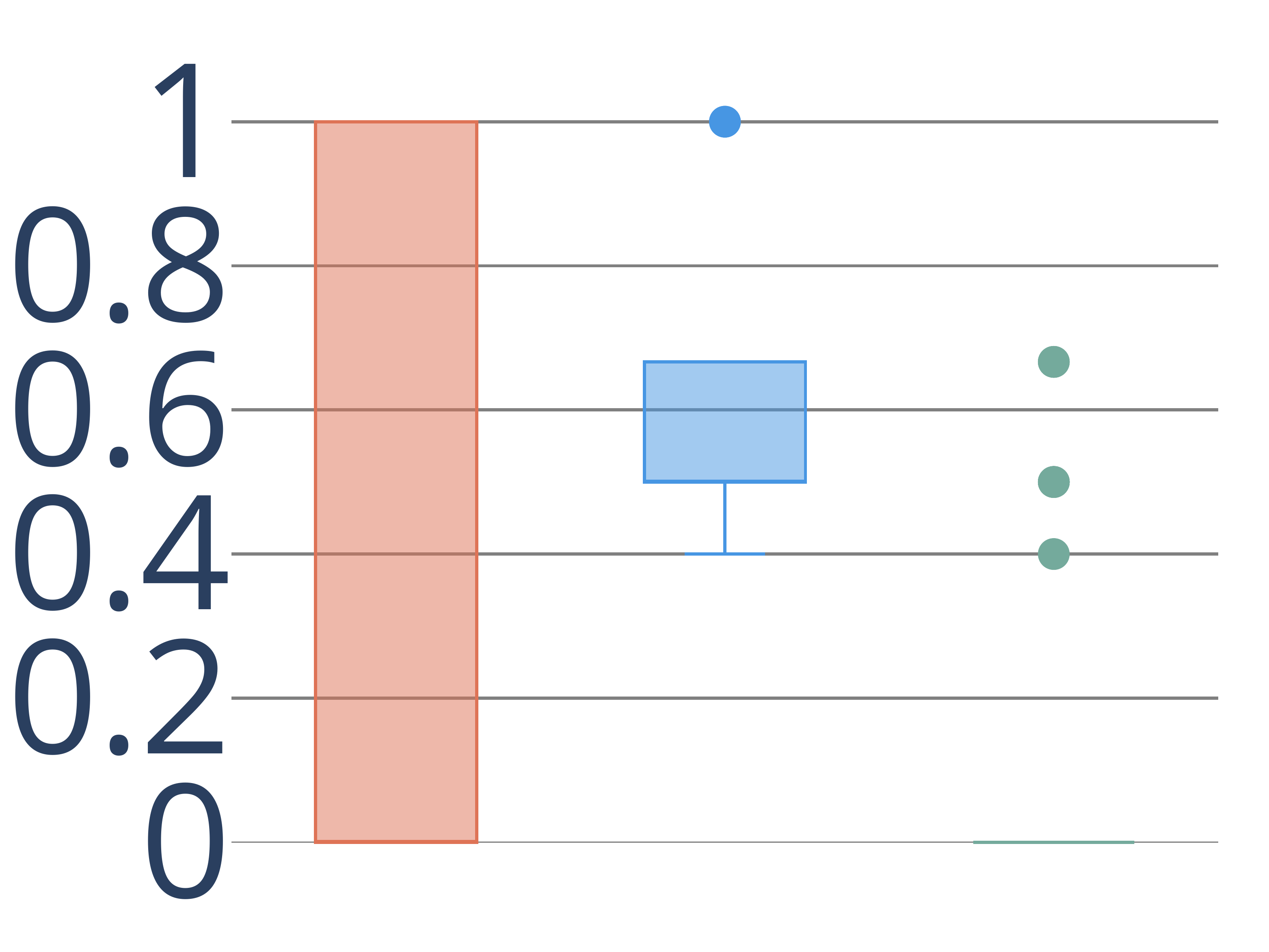} 
                                        & 
                                        & 
                                        & \includegraphics[scale=0.025]{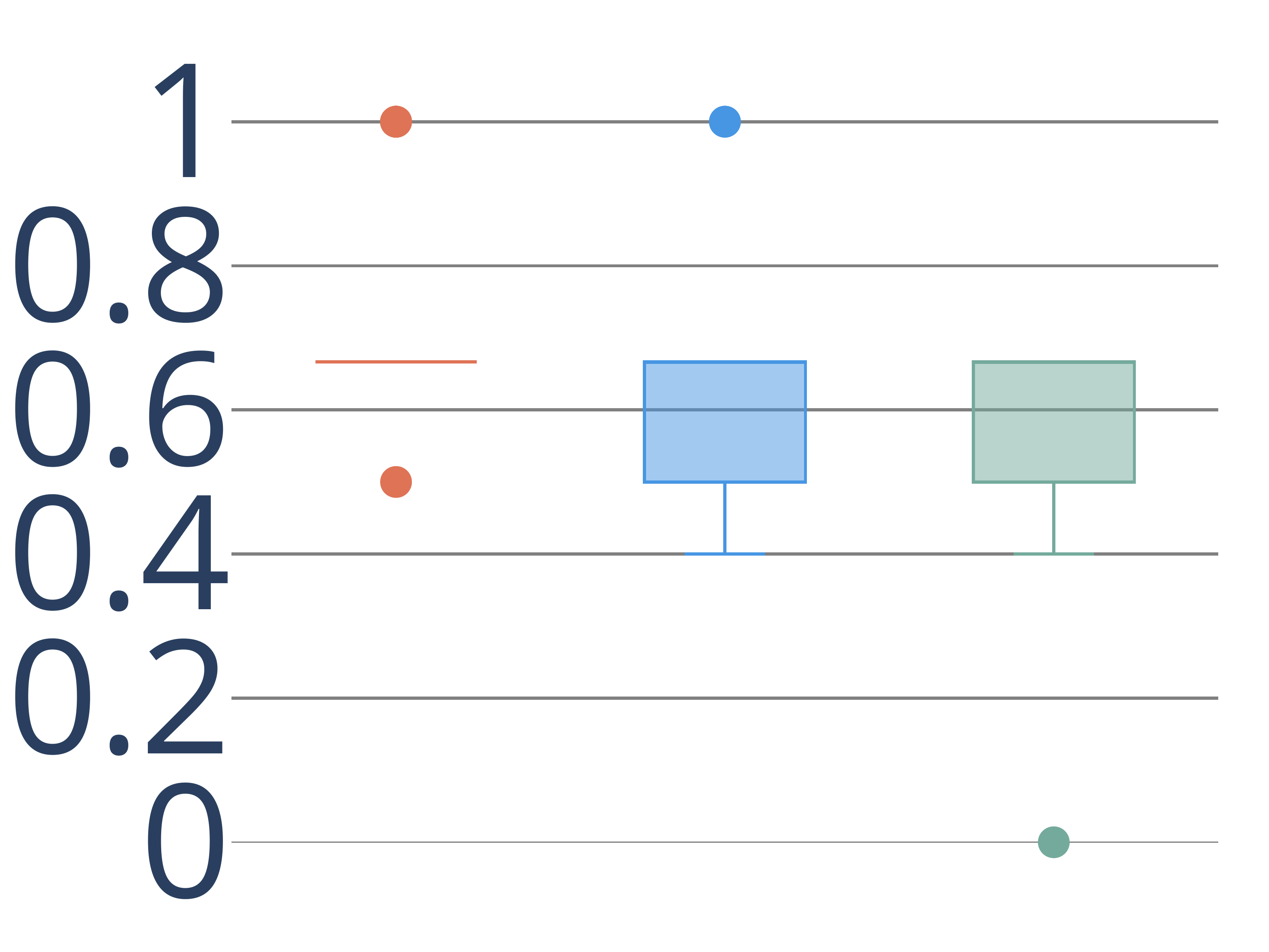} 
                                        & 
                                        & 
                                        & 
                                        & 
                                        & \\
                                    & \textbf{Small Size} 
                                        & 
                                        & 
                                        & 
                                        & \includegraphics[scale=0.025]{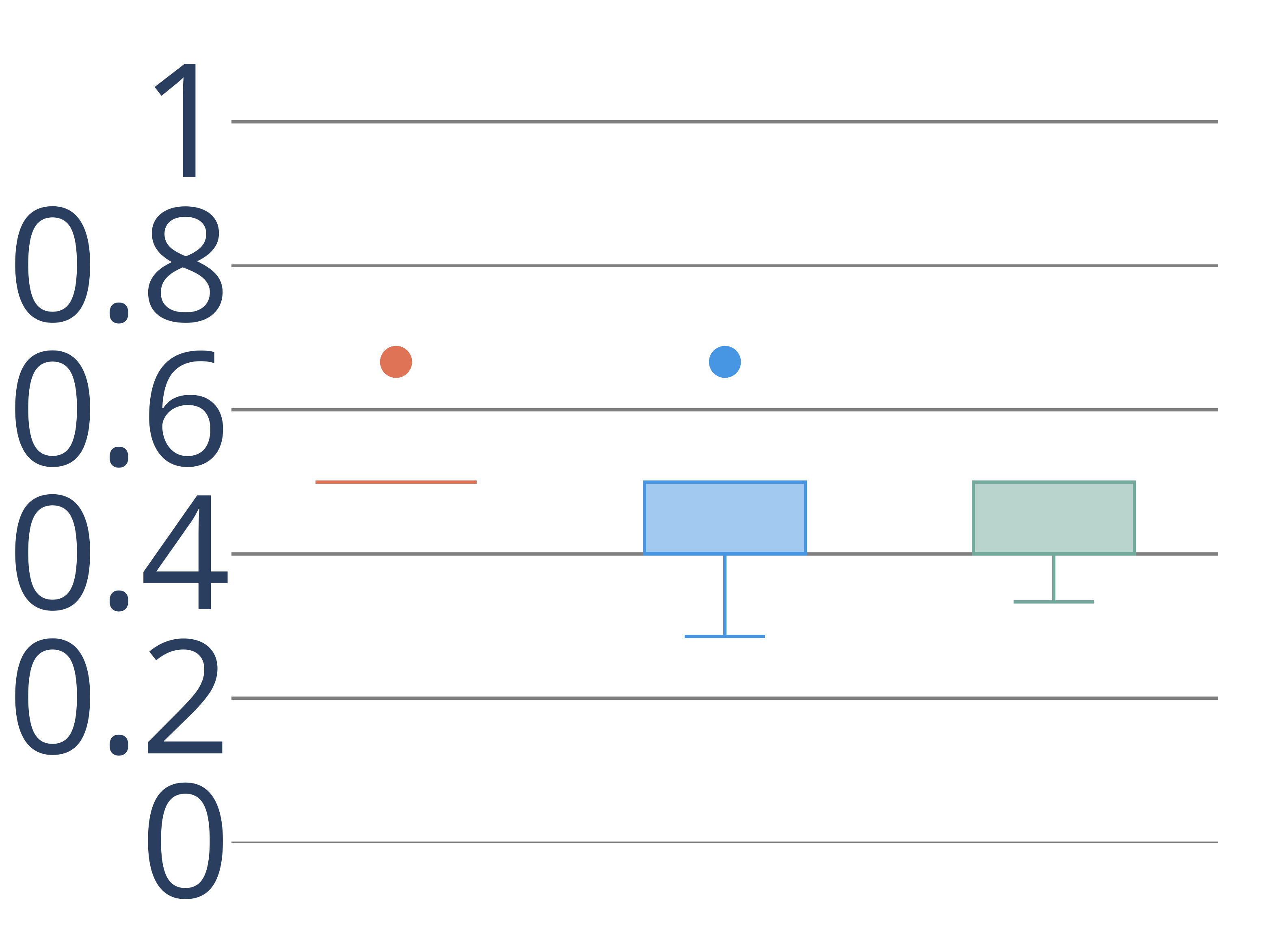} 
                                        & 94
                                        & \includegraphics[scale=0.025]{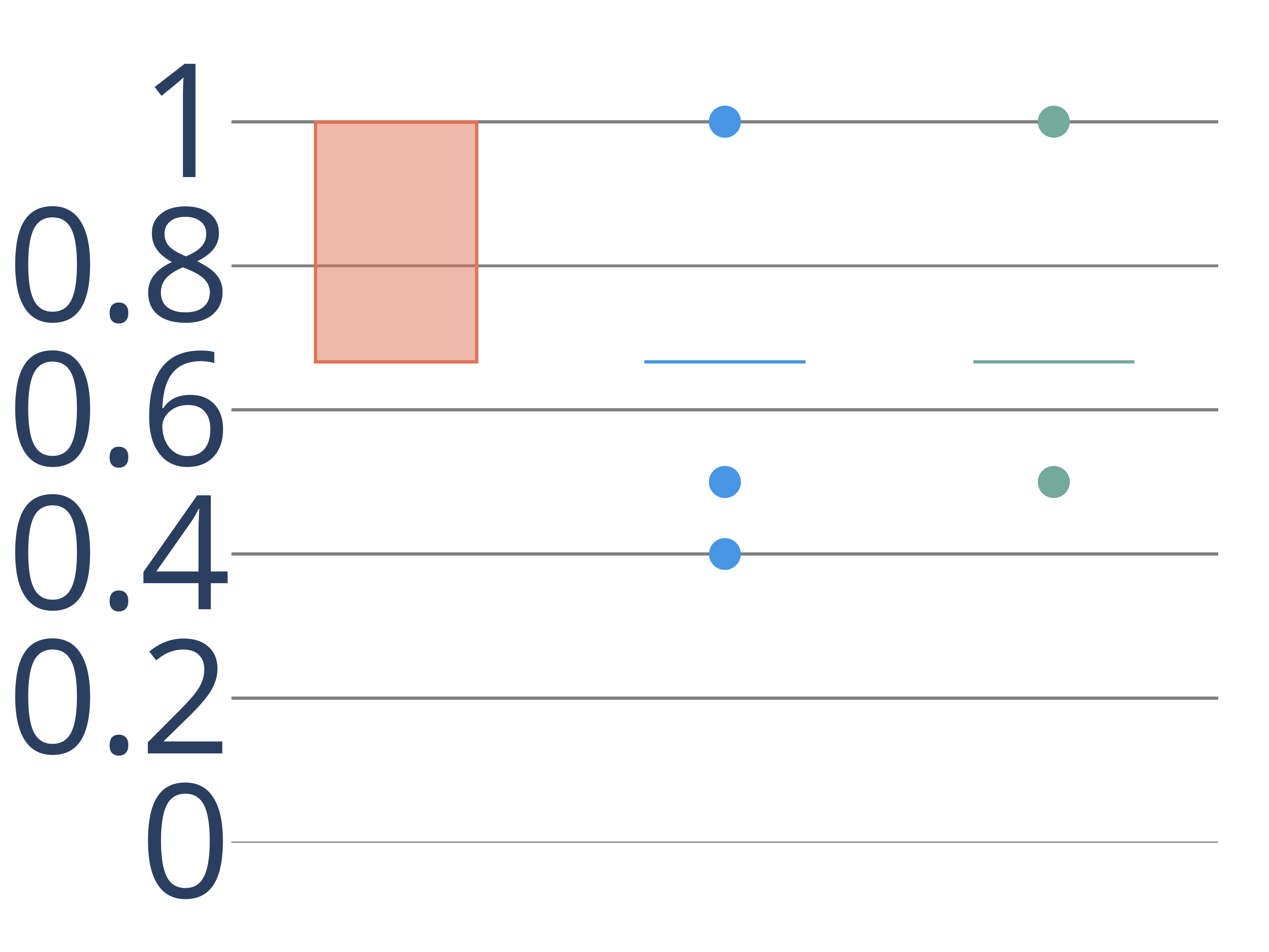} 
                                        & 
                                        & 
                                        & 
                                        & \includegraphics[scale=0.025]{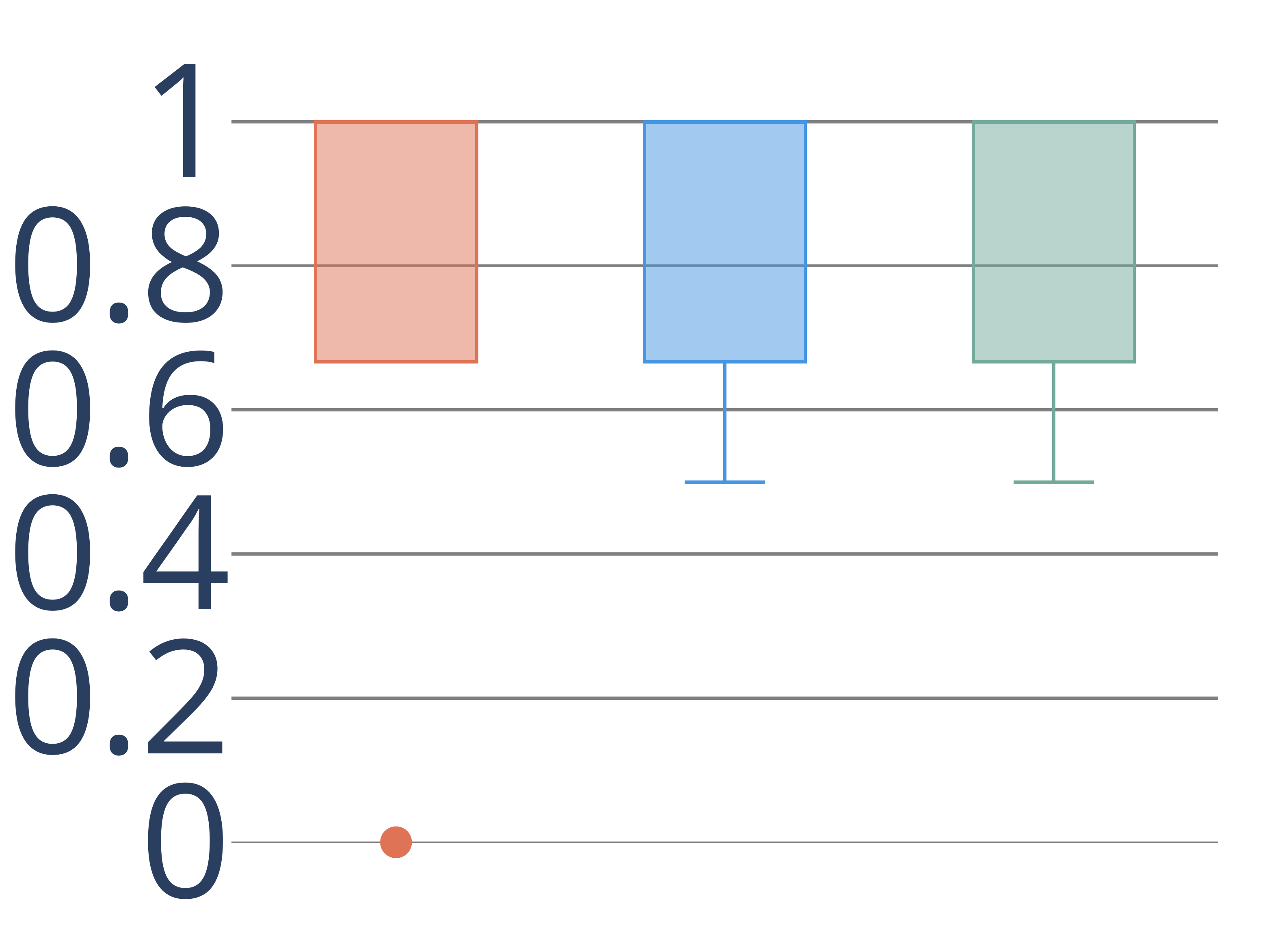}\\
        \hline
    \end{tblr}
    \label{tab:results_dsCoef_ex3}
\end{table*}

\begin{figure*}[]
    \centering
    \begin{subfigure}{.494\textwidth}
        \includegraphics[width=\textwidth]{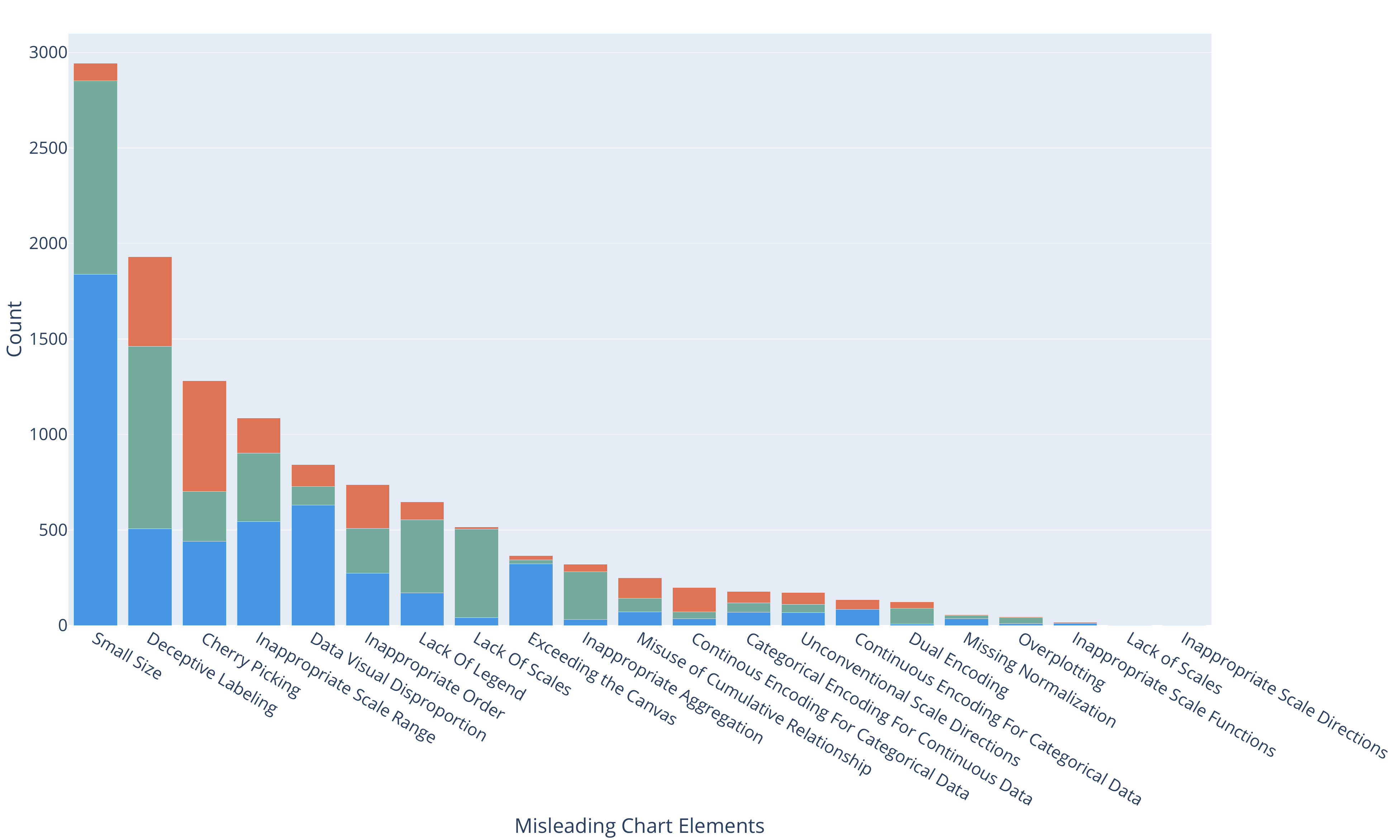}\vspace{-1ex}
        \caption{Charts with misleading elements}
        \label{fig:paretoMislead}
    \end{subfigure}
    \begin{subfigure}{.494\textwidth}
        \includegraphics[width=\textwidth]{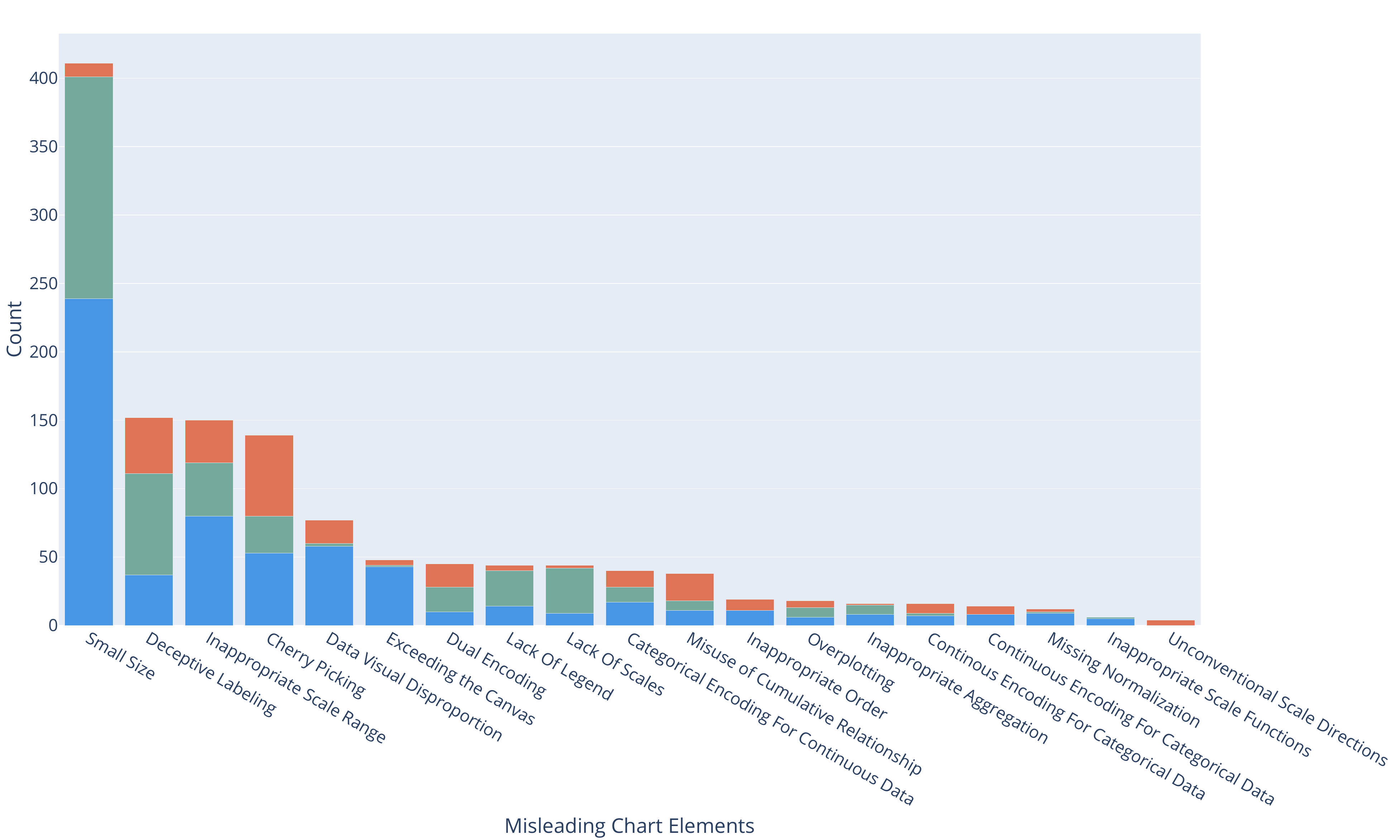}\vspace{-1ex}
        \caption{Charts without misleading elements}
        \label{fig:paretoCorrected}
    \end{subfigure}
    \caption{Pareto chart of hallucinated misleading chart elements for charts containing (a) misleading elements and (b) no misleading for \claudeFig, \gptFig, and \geminiFig.}
    \label{fig:paretosGen}
\end{figure*}

\begin{figure*}[]
    \centering
    \begin{subfigure}{.494\textwidth}
        \includegraphics[width=\textwidth]{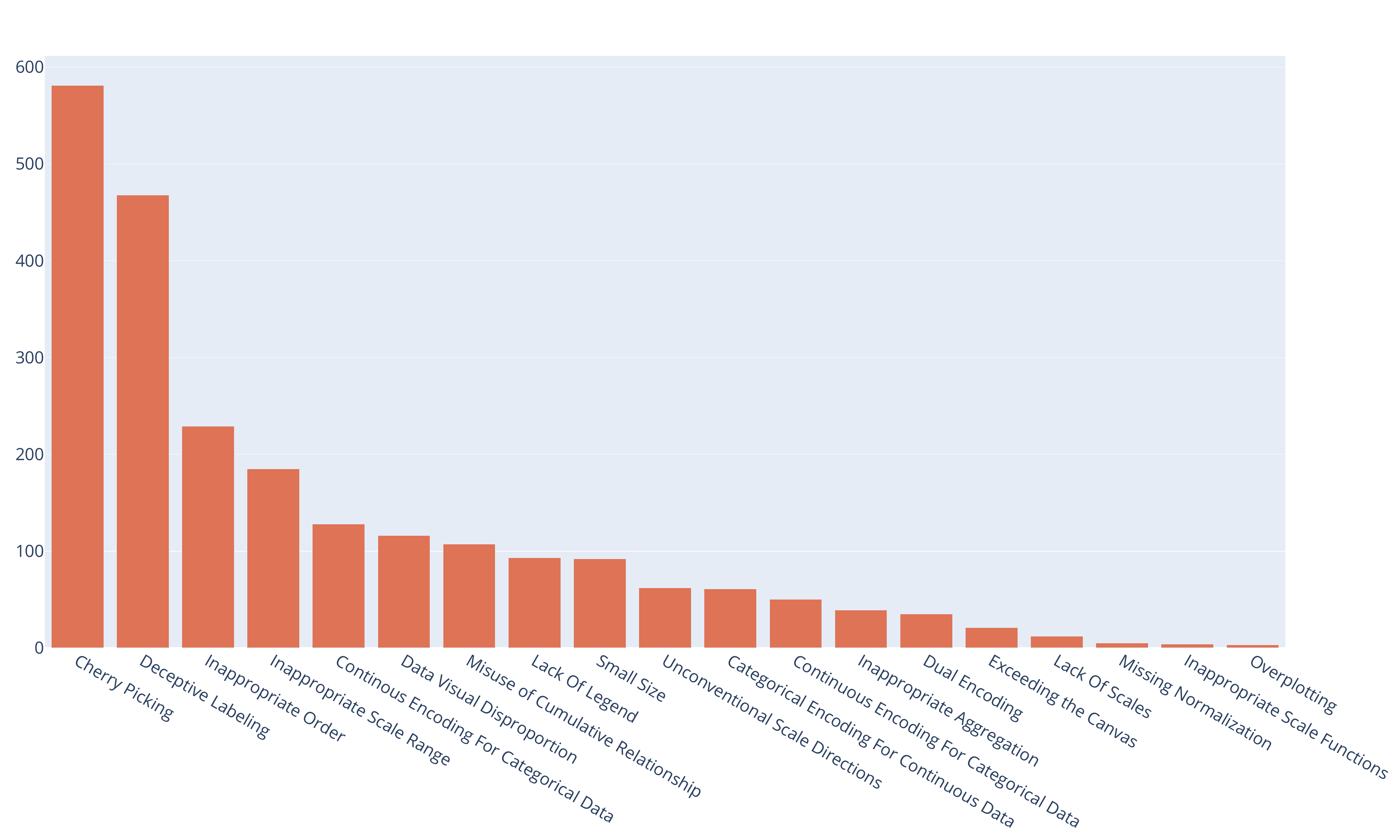}\vspace{-1ex}
        \caption{Charts with misleading elements}
        \label{fig:paretoMisleadClaude}
    \end{subfigure}
    \begin{subfigure}{.494\textwidth}
        \includegraphics[width=\textwidth]{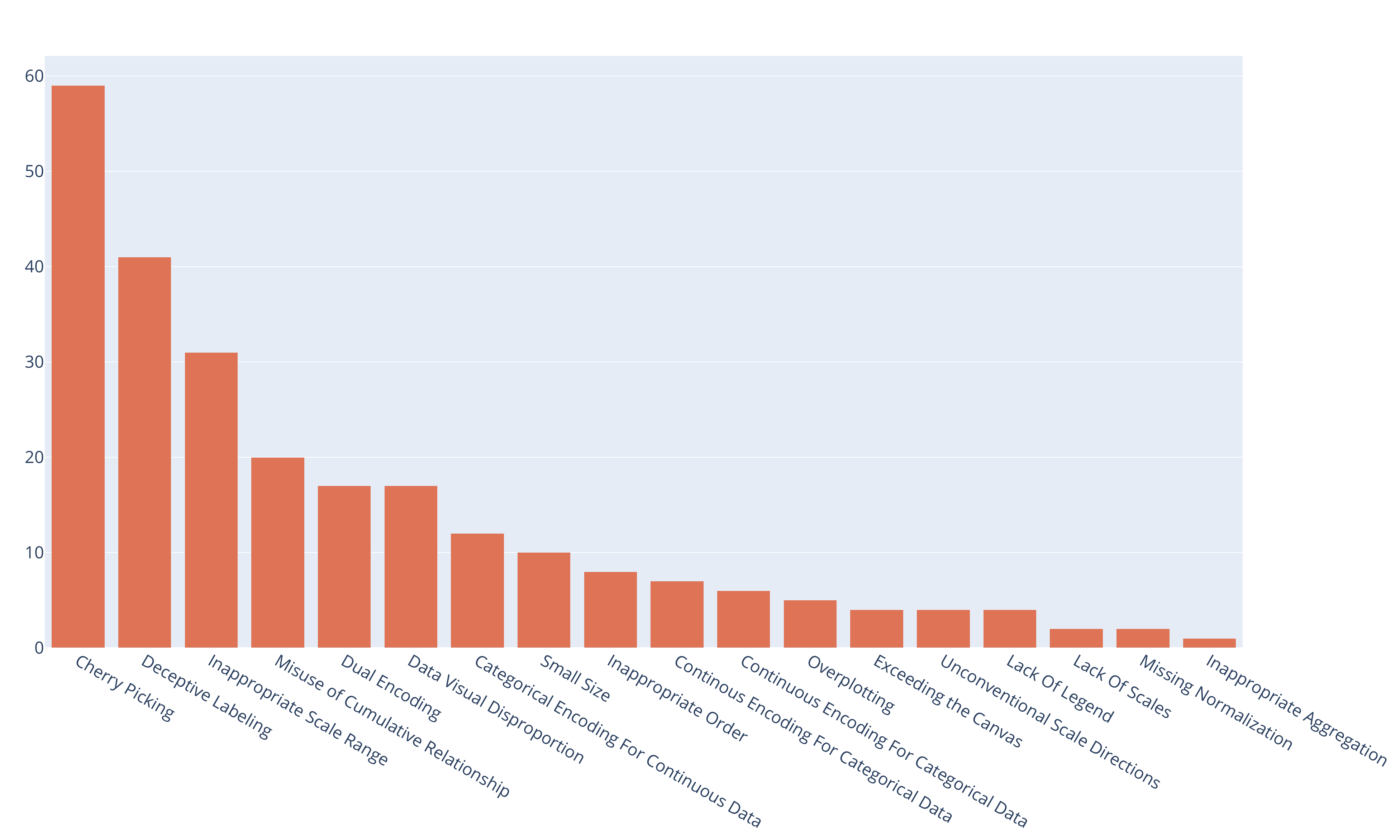}\vspace{-1ex}
        \caption{Charts without misleading elements}
        \label{fig:paretoCorrectedClaude}
    \end{subfigure}
    \caption{Pareto chart of hallucinated misleading chart elements for charts containing (a) misleading elements and (b) no misleading for \claudeFig.}
    \label{fig:paretosClaude}
\end{figure*}

\begin{figure*}[]
    \centering
    \begin{subfigure}{.494\textwidth}
        \includegraphics[width=\textwidth]{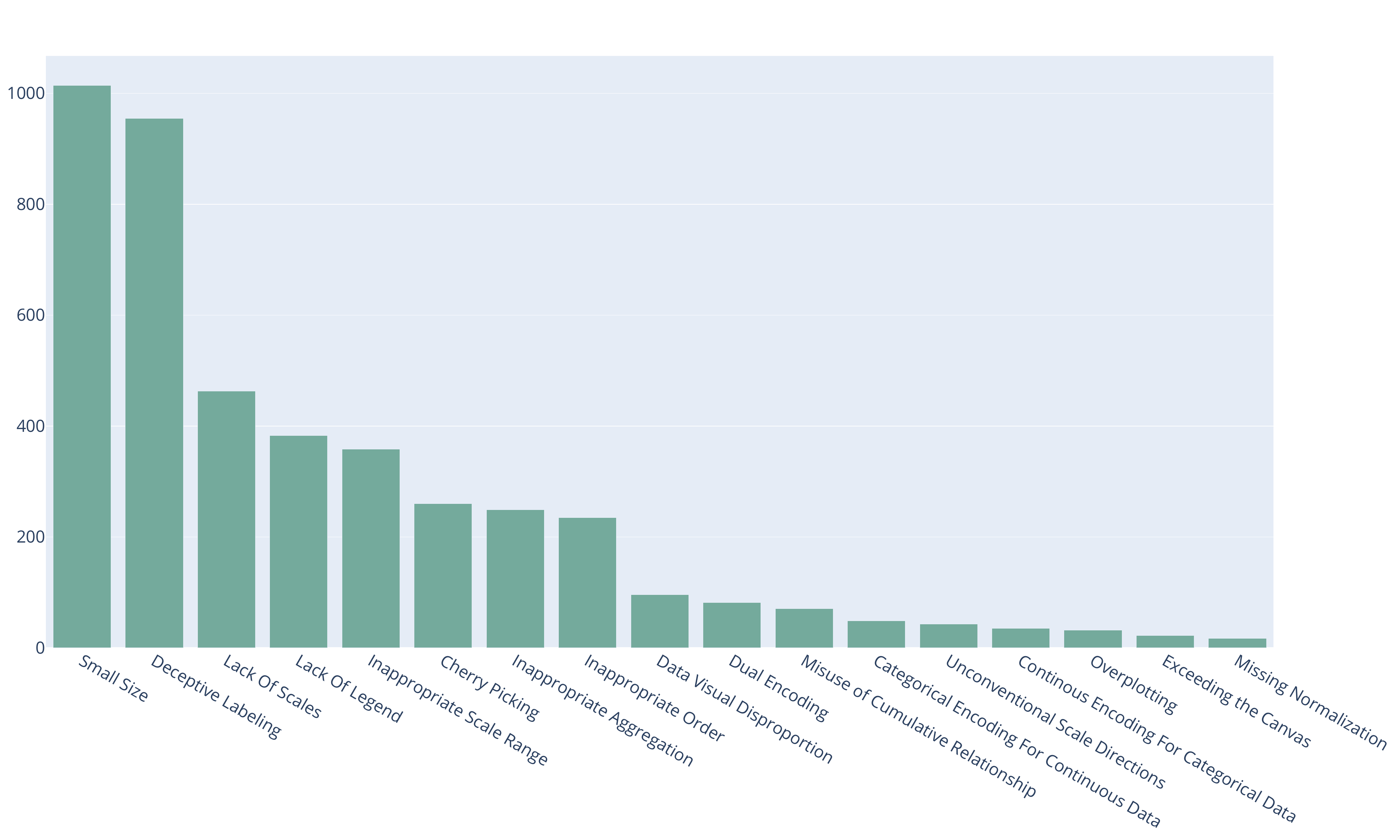}\vspace{-1ex}
        \caption{Charts with misleading elements}
        \label{fig:paretoMisleadGPT}
    \end{subfigure}
    \begin{subfigure}{.494\textwidth}
        \includegraphics[width=\textwidth]{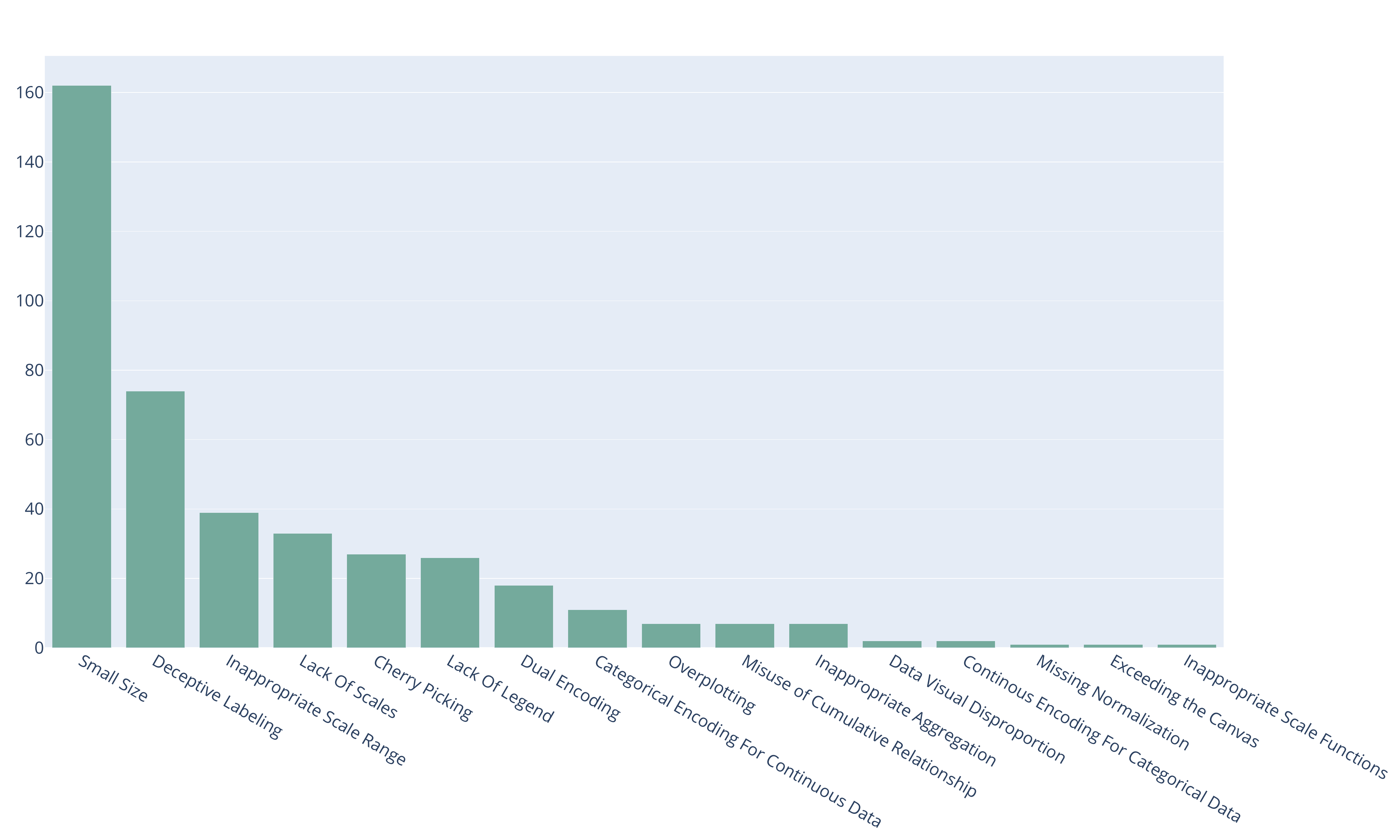}\vspace{-1ex}
        \caption{Charts without misleading elements}
        \label{fig:paretoCorrectedGPT}
    \end{subfigure}
    \caption{Pareto chart of hallucinated misleading chart elements for charts containing (a) misleading elements and (b) no misleading for \gptFig.}
    \label{fig:paretosGPT}
\end{figure*}

\begin{figure*}[]
    \centering
    \begin{subfigure}{.494\textwidth}
        \includegraphics[width=\textwidth]{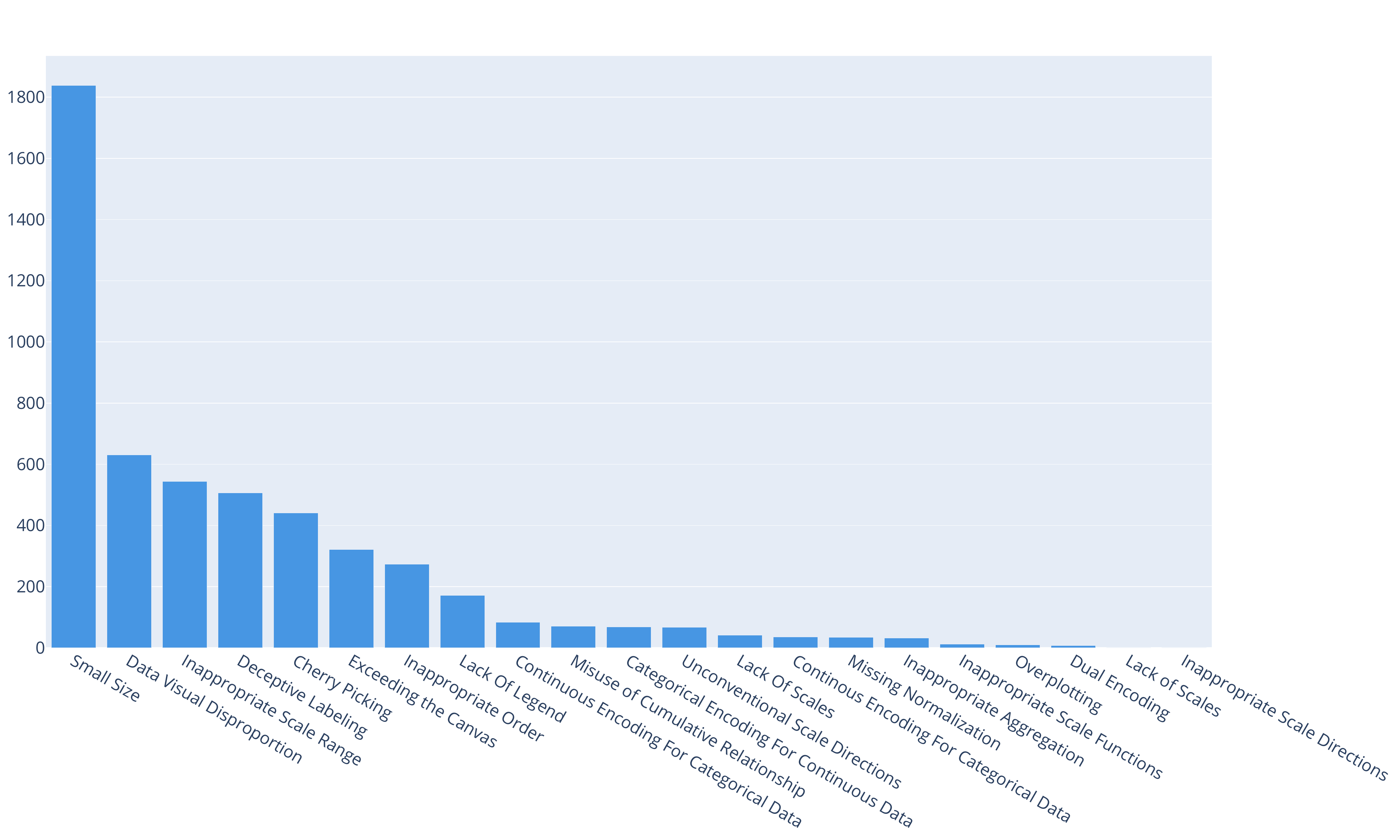}\vspace{-1ex}
        \caption{Charts with misleading elements}
        \label{fig:paretoMisleadGemini}
    \end{subfigure}
    \begin{subfigure}{.494\textwidth}
        \includegraphics[width=\textwidth]{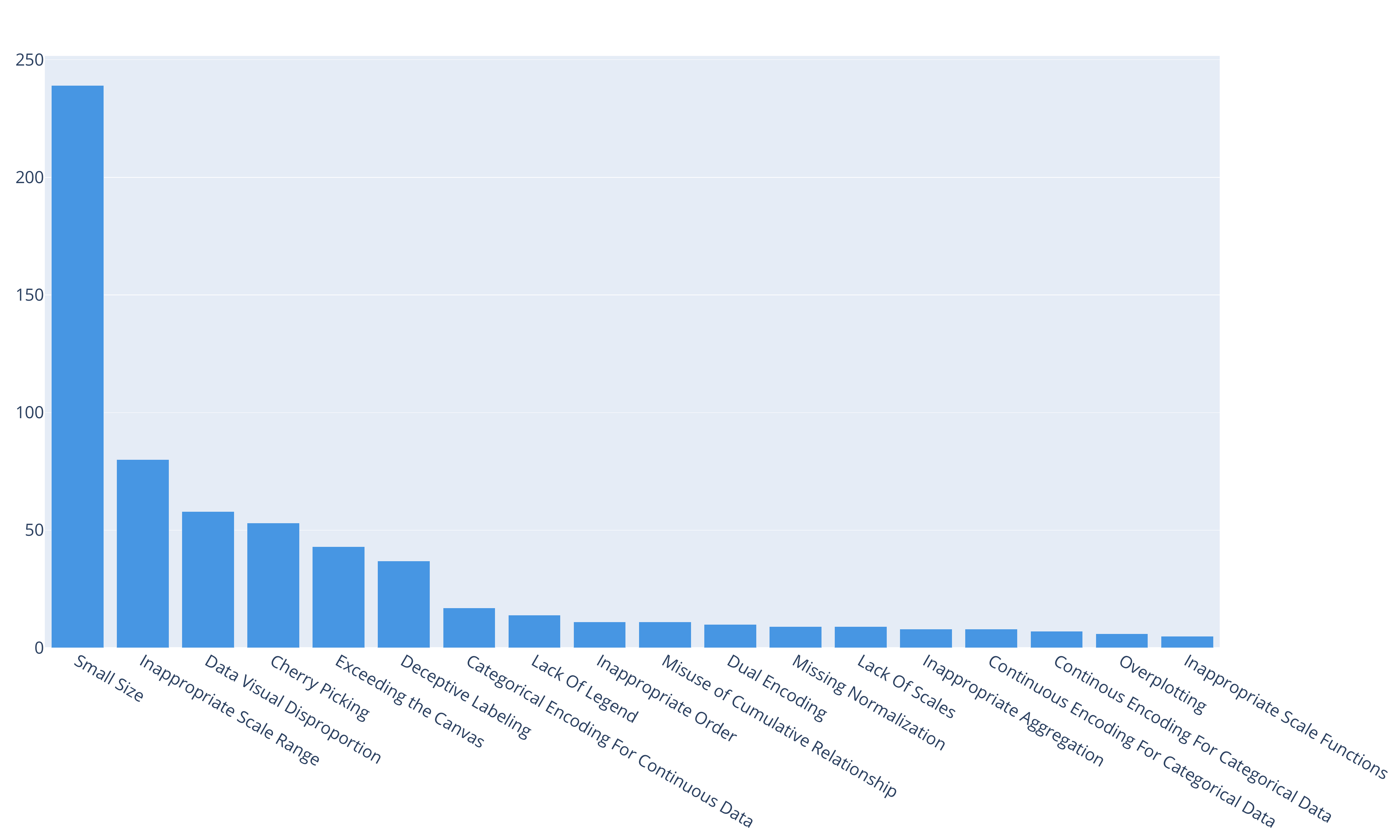}\vspace{-1ex}
        \caption{Charts without misleading elements}
        \label{fig:paretoCorrectedGemini}
    \end{subfigure}
    \caption{Pareto chart of hallucinated misleading chart elements for charts containing (a) misleading elements and (b) no misleading for \geminiFig.}
    \label{fig:paretosGemini}
\end{figure*}

\end{document}